\documentstyle[12pt,a4]{report}
\makeatletter
\def\@normalsize{\@setsize\normalsize{14.5pt}\xiipt\@xiipt
\abovedisplayskip 12pt plus3pt minus7pt%
\belowdisplayskip \abovedisplayskip
\abovedisplayshortskip  \z@ plus3pt%   
\belowdisplayshortskip  6.5pt plus3.5pt minus3pt}
\def\small{\@setsize\small{13.6pt}\xipt\@xipt
\abovedisplayskip 11pt plus3pt minus6pt%
\belowdisplayskip \abovedisplayskip
\abovedisplayshortskip  \z@ plus3pt%   
\belowdisplayshortskip  6.5pt plus3.5pt minus3pt
\def\@listi{\parsep 4.5pt plus 2pt minus 1pt
            \itemsep \parsep
            \topsep 9pt plus 3pt minus 5pt}}
\def\footnotesize{\@setsize\footnotesize{12pt}\xpt\@xpt
\abovedisplayskip 10pt plus2pt minus5pt%
\belowdisplayskip \abovedisplayskip
\abovedisplayshortskip  \z@ plus3pt%   
\belowdisplayshortskip  6pt plus3pt minus3pt
\def\@listi{\topsep 6pt plus 2pt minus 2pt\parsep 3pt plus 2pt minus 1pt
\itemsep \parsep}}
\def\scriptsize{\@setsize\scriptsize{9.5pt}\viiipt\@viiipt}
\def\tiny{\@setsize\tiny{7pt}\vipt\@vipt}
\def\large{\@setsize\large{18pt}\xivpt\@xivpt}
\def\Large{\@setsize\Large{22pt}\xviipt\@xviipt}
\def\LARGE{\@setsize\LARGE{25pt}\xxpt\@xxpt}
\def\huge{\@setsize\huge{30pt}\xxvpt\@xxvpt}

\@normalsize  % Choose the normalsize font.
\makeatother
\makeatletter
\newbox\sstrutbox
\setbox\sstrutbox=\hbox{\vrule height4.25pt depth1.75pt width\z@}
\newcommand{\sstrut}{\relax\ifmmode\copy\sstrutbox\else\unhcopy\sstrutbox\fi}
\makeatother
\newcommand{\comment}[1]{}
\newcommand{\arrow}{\rightarrow}
\newcommand{\iso}{\cong}

\newcommand{\indexed}[1]{$\scriptstyle #1$-indexed}

\newcommand{\myfigure}[3]{\begin{figure}
\begin{center}
\fbox{\parbox{.95\textwidth}{#3\caption{#2}\label{#1}}}
\end{center}
\end{figure}}
\newcommand{\anno}[1]{[#1]}

\newcommand{\eqvelm}[1]{\langle #1\rangle}

\newcommand{\pair}[1]{\langle #1\rangle}
\newif\ifinlogicrule\inlogicrulefalse
\newlength{\logiclength}
\newcommand{\logicrule}[2]{%
\ifinlogicrule
  \settowidth{\logiclength}{$\displaystyle\frac{#1}{#2}$}%
  \makebox[\logiclength]{$\displaystyle\frac{\quad #1\quad}{\quad #2\quad}$}%
\else
  \inlogicruletrue
  \displaystyle\frac{\quad #1\quad}{\quad #2\quad}
  \inlogicrulefalse
\fi}
\newcommand{\defeq}{\stackrel{\rm def}{=}}

\let\implies=\Rightarrow

\newcommand{\dimplies}{\mathrel{\rlap{$\implies$}\,\implies}}
\newsavebox{\commutebox}
\savebox{\commutebox}{\rule[-2.5pt]{0pt}{8pt}\begin{picture}(10,10)
\put(5,5){\oval(8,8)[tr]}
\put(5,5){\oval(8,8)[l]}
\put(8.5,5){\vector(0,-1){2}}
\end{picture}}
\newcommand\commute{\usebox{\commutebox}}
\newcommand{\sqdiagram}[9]{
\setlength{\unitlength}{1mm}
\begin{picture}(60,30)
\put(10,25){\makebox(0,0){$#1$}}
\put(25,27){\makebox(0,0)[b]{$#2$}}
\put(40,25){\makebox(0,0){$#3$}}
\put(15,25){\vector(1,0){20}}
\put(8,15){\makebox(0,0)[r]{$#4$}}
\put(25,15){\makebox(0,0){\commute}}
\put(42,15){\makebox(0,0)[l]{$#5$}}
\put(10,22.5){\vector(0,-1){15}}
\put(40,22.5){\vector(0,-1){15}}
\put(10,5){\makebox(0,0){$#6$}}
\put(25,3){\makebox(0,0)[t]{$#7$}}
\put(15,5){\vector(1,0){20}}
\put(40,5){\makebox(0,0){$#8$}}
#9
\end{picture}}
\newcommand{\ardiagram}[3]{
\setlength{\unitlength}{1mm}
\begin{picture}(50,10)(0,2.5)
\put(10,5){\makebox(0,0)[r]{$#1$}}
\put(12,5){\vector(1,0){26}}
\put(25,7){\makebox(0,0)[b]{$#2$}}
\put(40,5){\makebox(0,0)[l]{$#3$}}
\end{picture}}
\let\lleq=\sqsubseteq

\let\lub=\sqcup
\let\Lub=\bigsqcup
\newcommand{\natrightarrow}{\stackrel{\cdot}{\rightarrow}}

\newcommand{\subrightarrow}[1]{\smash{\mathop{\longrightarrow}
\limits^{\scriptstyle #1}}}
\newlength{\rubberlength}
\newcommand{\rubberrightarrow}[1]{%
\settowidth{\rubberlength}{$\scriptstyle\;\;\; #1\;\;\;$}%
\stackrel{#1}{\makebox[\rubberlength]{\rightarrowfill}}}
\mathchardef\headofarrow"3A29

\newcommand{\coloneq}{\mathrel{::=}}
\newcommand{\minfixop}{\underline \mu}
\newcommand{\maxfixop}{\overline \mu}
\newcommand{\minfix}[2]{\minfixop #1.(#2)}
\newcommand{\maxfix}[2]{\maxfixop #1.(#2)}
\newcommand{\reduce}{\triangleright}
\newcommand{\reducen}{\stackrel{\star}{\triangleright}}
\makeatletter
\def\@listi{\leftmargin\leftmargini}
\def\@listii{\leftmargin\leftmarginii
   \labelwidth\leftmarginii\advance\labelwidth-\labelsep}
\def\@listiii{\leftmargin\leftmarginiii
    \labelwidth\leftmarginiii\advance\labelwidth-\labelsep}
\mathcode`\:="603A

\newif\if@qed\@qedfalse
\def\@begintheorem#1#2{\global\@qedfalse \rm \trivlist \item[\hskip
\labelsep{\bf #1\ #2:}]}
\newsavebox{\qedbox}
\savebox{\qedbox}{\rm[\hspace{-0.1em}]}
\def\qed{\ifvmode \nobreak \usebox{\qedbox}\else \unskip
~\usebox{\qedbox}\fi \global\@qedtrue}
\def\@endtheorem{\if@qed \else \qed \fi \endtrivlist \global\@qedfalse}

\let\endseparateproof=\@endtheorem

\let\endcontdefinition=\@endtheorem
\def\leaderfill{\leaders\hbox to 1em{\hss.\hss}\hfill}
\def\thenotationlabel#1{#1\leaderfill}
\def\thenotation{\list{}{\leftmargin\thenotationmargin
       \labelwidth\leftmargin
       \let\makelabel\thenotationlabel}}

\newdimen\thenotationmargin
\def\Hom#1#2#3{{\rm Hom}_{#1}(#2,#3)}
\def\cC{{\cal C}}
\def\cD{{\cal D}}
\def\cE{{\cal E}}
\def\adjoint{\mapstochar\relbar}
\def\fnum@figure{{\bf Figure \thefigure}}
\def\eqalign#1{\,\vcenter{\openup\jot\m@th
  \ialign{\strut\hfil$\displaystyle{##}$&$\displaystyle{{}##}$\hfil
      \crcr#1\crcr}}\,}
\def\eqalignno#1{\displ@y \tabskip\@centering
  \halign to\displaywidth{\hfil$\displaystyle{##}$\tabskip\z@skip
    &$\displaystyle{{}##}$\hfil\tabskip\@centering
    &\llap{$##$}\tabskip\z@skip\crcr
    #1\crcr}}
\def\leqalignno#1{\displ@y \tabskip\@centering
  \halign to\displaywidth{\hfil$\displaystyle{##}$\tabskip\z@skip
    &$\displaystyle{{}##}$\hfil\tabskip\@centering
    &\kern-\displaywidth\rlap{$##$}\tabskip\displaywidth\crcr
    #1\crcr}}
\def\interdisplayline#1{%
	\noalign{\hbox{\kern\displayindent 
		\hbox to\displaywidth{#1\hfil}}\penalty
		\predisplaypenalty}}
\def\cite{\@ifnextchar [{\@tempswatrue\@citex}{\@tempswafalse\@citex[]}}
\def\@citex[#1]#2{\if@filesw\immediate\write\@auxout{\string\citation{#2}}\fi
  \def\@citea{}\@cite{\@for\@citeb:=#2\do
    {\@citea\def\@citea{,\ }\@ifundefined
       {b@\@citeb}{{\bf ?}\@warning
       {Citation `\@citeb' on page \thepage \space undefined}}%
       {\csname b@\@citeb\endcsname}%
    }}{#1}}
\long\def\@makefntext#1{\parindent 0em\noindent\llap{$^{\@thefnmark}$}#1}
\def\]{\relax\ifmmode $$%%$$ BRACE MATCH HACK
        \else \@badmath \fi\ignorespaces}
\newif\ifbbold
\bboldtrue
\bboldfalse
\ifbbold
\font\fivbb=msym5
\font\sixbb=msym6
\font\sevbb=msym7
\font\egtbb=msym8
\font\ninbb=msym9
\font\tenbb=msym10
\font\elvbb=msym10 \@halfmag
\font\twlbb=msym10 \@magscale1
\font\frtnbb=msym10 \@magscale2
\font\svtnbb=msym10 \@magscale3
\font\twtybb=msym10 \@magscale4
\newfam\bbfam
\def\@vpt{\def\pbb{\fam\bbfam\fivbb}\textfont\bbfam\fivbb
    \scriptfont\bbfam\fivbb \scriptscriptfont\bbfam\fivbb}
\def\@vipt{\def\pbb{\fam\bbfam\sixbb}\textfont\bbfam\sixbb
    \scriptfont\bbfam\sixbb \scriptscriptfont\bbfam\sixbb}
\def\@viipt{\def\pbb{\fam\bbfam\sevbb}\textfont\bbfam\sevbb
    \scriptfont\bbfam\fivbb \scriptscriptfont\bbfam\fivbb}
\def\@viiipt{\def\pbb{\fam\bbfam\egtbb}\textfont\bbfam\egtbb
    \scriptfont\bbfam\sixbb \scriptscriptfont\bbfam\fivbb}
\def\@ixpt{\def\pbb{\fam\bbfam\ninbb}\textfont\bbfam\ninbb
   \scriptfont\bbfam\sixbb \scriptscriptfont\bbfam\fivbb}
\def\@xpt{\def\pbb{\fam\bbfam\tenbb}\textfont\bbfam\tenbb
   \scriptfont\bbfam\sevbb \scriptscriptfont\bbfam\fivbb}
\def\@xipt{\def\pbb{\fam\bbfam\elvbb}\textfont\bbfam\elvbb
   \scriptfont\bbfam\egtbb \scriptscriptfont\bbfam\sixbb}
\def\@xiipt{\def\pbb{\fam\bbfam\twlbb}\textfont\bbfam\twlbb
   \scriptfont\bbfam\egtbb \scriptscriptfont\bbfam\sixbb}
\def\@xivpt{\def\pbb{\fam\bbfam\frtnbb}\textfont\bbfam\frtnbb
   \scriptfont\bbfam\tenbb \scriptscriptfont\bbfam\sevbb}
\def\@xviipt{\def\pbb{\fam\bbfam\svtnbb}\textfont\bbfam\svtnbb
   \scriptfont\bbfam\twlbb   \scriptscriptfont\bbfam\tenbb}
\def\@xxpt{\def\pbb{\fam\bbfam\twtybb}\textfont\bbfam\twtybb
   \scriptfont\bbfam\frtnbb \scriptscriptfont\bbfam\twlbb}
\def\@xxvpt{\def\pbb{\fam\bbfam\twtybb}\textfont\bbfam\twtybb
   \scriptfont\bbfam\twtybb \scriptscriptfont\bbfam\svtnbb}

\let\cal\bb
\fi
\begingroup \catcode `|=0 \catcode `[= 1
\catcode`]=2 \catcode `\{=12 \catcode `\}=12
\catcode`\\=12
|gdef|@xcomputer#1\end{computer}[#1|end[computer]]
|gdef|@xbcomputer#1\end{bcomputer}[#1|end[bcomputer]]
|endgroup
\def\activeul{\catcode`/=\active \catcode`<=\active \catcode`>=\active}
\newif\iful
\def\ulescape{\iful\tt\ulfalse\else\it\ultrue\fi}
\def\ltescape{\iful$<$\else{\tt <}\fi}
\def\gtescape{\iful$>$\else{\tt >}\fi}
{\activeul\global\let/=\ulescape \global\let<=\ltescape \global\let>=\gtescape}
\def\computer{\quote \ulfalse \footnotesize \@verbatim \activeul
\frenchspacing\@vobeyspaces \@xcomputer}

\def\bcomputer{\quote \ulfalse \@verbatim \activeul
\frenchspacing\@vobeyspaces \@xbcomputer}

\makeatother
\newtheorem{theorem}{Theorem}[section]
\newtheorem{lemma}[theorem]{Lemma}
\newtheorem{corollary}[theorem]{Corollary}
\newtheorem{definition}[theorem]{Definition}
\newtheorem{example}[theorem]{Example}
\newtheorem{proposition}[theorem]{Proposition}
\makeatletter
\def\verbatim{\def\baselinestretch{1.0} \@normalsize
\@verbatim \frenchspacing\@vobeyspaces \@xverbatim}

\def\computer{\quote \ulfalse \def\baselinestretch{1.0} \footnotesize
\@verbatim \activeul \frenchspacing\@vobeyspaces \@xcomputer}

\def\bcomputer{\quote \ulfalse \def\baselinestretch{1.0} \@normalsize
\@verbatim \activeul \frenchspacing\@vobeyspaces \@xbcomputer}
\makeatother
\title{A Categorical Programming Lanuage}
\author{Tatsuya Hagino\\ Department of Computer Science\\
		University of Edinburgh}
\begin{document}
\pagenumbering{roman}
\begin{titlepage}
\null
\vfil
\vskip 60pt
\begin{center}
\begin{LARGE}
A Categorical Programming Language \\
\end{LARGE}
\vskip 4em
\begin{Large}
Tatsuya Hagino \\
\end{Large}
\end{center}
\vfil
\begin{center}
Doctor of Philosophy \\		% name of degree
University of Edinburgh \\
1987 \\      			% year of presentation
\end{center}
\vskip 2em
\null
\end{titlepage}

\thispagestyle{empty}
\null
\vfill
\begin{tabular}{l}
{\bf Author's current address:} \\
\qquad Tatsuya Hagino \\
%\qquad Data Processing Center \\
%\qquad Kyoto University \\
%\qquad Kyoto 606 \\
\qquad Faculty of Environment and Information Studies \\
\qquad Keio University \\
\qquad Endoh 5322, Fujisawa city, Kanagawa \\
\qquad Japan 252-0882 \\
\qquad E-mail: {\tt hagino@sfc.keio.ac.jp} \\
\end{tabular}
\vskip 2em
\null
\clearpage

\chapter*{Abstract}

\write16{Start Abstract}
\begingroup
\def\baselinestretch{1}
\small
\normalsize
\vskip -2em
A theory of data types and a programming language based on category
theory are presented.

Data types play a crucial role in programming.  They enable us to write
programs easily and elegantly.  Various programming languages have been
developed, each of which may use different kinds of data types.
Therefore, it becomes important to organize data types systematically so
that we can understand the relationship between one data type and
another and investigate future directions which lead us to discover
exciting new data types.

There have been several approaches to systematically organize data
types: algebraic specification methods using algebras, domain theory
using complete partially ordered sets and type theory using the connection
between logics and data types.  Here, we use category theory.  Category
theory has proved to be remarkably good at revealing the nature of
mathematical objects, and we use it to understand the true nature
of data types in programming.

We organize data types under a new categorical notion of {\it
$F,G$-dialgebras} which is an extension of the notion of adjunctions as
well as that of $T$-algebras.  $T$-algebras are also used in domain
theory, but while domain theory needs some primitive data types, like
products, to start with, we do not need any.  Products, coproducts and
exponentiations (i.e.\ function spaces) are defined exactly like in
category theory using adjunctions.  $F,G$-dialgebras also enable us to
define the natural number object, the object for finite lists and other
familiar data types in programming.  Furthermore, their symmetry allows
us to have the dual of the natural number object and the object for
infinite lists (or lazy lists).

We also introduce a functional programming language in a categorical
style.  It has no primitive data types nor primitive control structures.
Data types are declared using $F,G$-dialgebras and each data type is
associated with its own control structure.  For example, natural numbers
are associated with primitive recursion.  We define the meaning of the
language operationally by giving a set of reduction rules.  We also
prove that any computation in this programming language terminates using
Tait's computability method.

A specification language to describe categories is also included.  It is
used to give a formal semantics to $F,G$-dialgebras as well as to give a
basis to the categorical programming language we introduce.
\par
\endgroup
\normalsize
\write16{End Abstract}

\chapter*{Acknowledgements}

The greatest thanks go to Professor Rod Burstall who, first of all,
accepted me as a Ph.\ D. student, then, supervised me during my Ph.\ D.
study and, further, gave me an opportunity to continue working in
Edinburgh.  He gave me not only useful advices but also much-needed
mental support.  I would also thank to his wife, Sissi Burstall, who
invited me to their house often and made my stay in Edinburgh very
pleasant.

I am also grateful to Professor Reiji Nakajima for having encouraged me
to come to Edinburgh and to Professor Heisuke Hironaka for having helped
me to solve the financial problem for studying in Edinburgh.

Many lectures I attended in the first year enlightened me a lot: domain
theory, operational semantics, denotational semantics, algebraic
specification, category theory, and so on.  I did not know what category
theory really is before I came to Edinburgh.  Thanks to Andrzej
Tarlecki, Edmund Robinson and John Cartmell for having helped me to
overcome the initial difficulty of category theory.  John Cartmell also
helped me a lot through discussions in the early stage of the thesis.  I
am also in debt to Furio Honsell who introduced me to Tait's method for
proving normalization theorems of lambda calculi when I stuck in the
normalization proof of CPL.  Many other people in Edinburgh helped me
with their comments and through discussions.  I would especially like to
thank Bob McKay and Paul Taylor for reading the early drafts and
discovering some disastrous mistakes.

This thesis is written using \LaTeX\ on a Sun workstation and printed by
an Apple LaserWriter.  I would like to thank George Cleland and Hugh
Stabler for providing wonderful computing facilities and software to
Laboratory of Foundation of Computer Science.

My Ph.\ D. study at the University of Edinburgh has been funded by the
Educational Project for Japanese Mathematical Scientists, Harvard
University, by the Overseas Research Students Award and by the Science
and Engineering Research Council Research Fellowship.

\tableofcontents

\pagenumbering{arabic}
\pagestyle{headings}
\chapter{Introduction}
\label{ch-introduction}

This is an exploration of data types through category theory.  It is an
attempt to achieve better understanding of data types, their uniform
classification, and discovery of a new world of data types.  Data types
have been with us since the very first programming languages.  Even some
machine languages now have some concept of data types, but early
programming languages had only a fixed number of data types, like
integers, reals and strings, and/or a fixed number of data type
constructors, like array constructors and record constructors.  When we
gradually realized how important data types were, programming languages
started having richer and richer data types.  A number of programming
languages now allow us to define our own data types.  Some might even
say that the richer they are, the better the programming languages are.
Programming languages can be classified by the way how they handle
data types.

There is no question about the importance of data types.  Much research
in this area has produced various kinds of data types, so varied that
one cannot capture them all.  We now need to systematically organize
data types.  We want to know the connection between one data type and
another.  We want to know the reason why those data types are with us
and while some other data types are not.  After getting a clear view of
data types, we might find the future direction to discover other
important data types.

There have been already some attempts to organize data types.  We can
name some of the important ones: {\it Domain theory} is one, {\it
algebraic specification} is another and {\it type theory} is one where a
lot of research is going on at the moment.  In this thesis, we will
present yet another attempt to organize data types.  We do so by using
category theory.  We call our data types {\it Categorical Data Types}
(or {\it CDT} for short).

One might ask ``Why category theory?''  Category theory is known as
highly abstract mathematics.  Some call it abstract nonsense.  It chases
abstract arrows and diagrams, proves nothing but about those arrows and
diagrams, rarely talks about what arrows are for and often concepts go
beyond one's imagination.  However, when this `abstract nonsense' works,
it is like magic.  One may discover a simple theorem actually means very
deep things and some concepts beautifully unify and connect things which
are unrelated before.

In ordinary mathematics, whether we are aware or not, we are in the
world of set theory.  Mathematics has been so well developed with set
theory that we can hardly do anything without it.  Therefore, it is very
natural that semantics of programming languages is generally based on set
theory.  Note that it is often said that most of programming languages do
not have set theoretic semantics and, therefore, domain theory has been
developed, but this does not contradict with ``semantics based on set
theory'', because domain theory itself is based on set theory.  A domain
is a set with certain properties.

Set theory is a powerful tool, but sometimes this power disfigures
beautiful objects so that we cannot directly see their natural
properties.  For example, in set theory it is not easy to see either
the duality between injective and surjective functions or the duality
between cartesian products and disjoint sums.  It is in category theory
that these dualities come out clearly.  Category theory concentrates
on the outer behaviour of objects.  It does not care what is in an
object, whereas set theory is all about what is in an object.  It is
interesting to know that seeing from the outside reveals the nature of
an object more naturally than seeing its inside.  For example, one of
the most important concepts discovered by category theory is {\it
adjunction} (or {\it adjoint situation}), which is strikingly simple but
very beautiful and unifies various concepts under the same name.

Our slogan is: ``{\it category theory can provide a better and more
natural understanding of mathematical objects than set theory\/}'', so
we use it to guide our tour around the world of data types.  Note that
we do not mean to abandon set theory by this.  We will still heavily
rely on it, but our intuition should not be obstructed by it.

\section{Backgrounds}
\label{sec-backgrounds}

\subsection{Algebraic Specification Methods}
\label{ssec-alg-spec}

Algebraic specification methods were first developed to describe what
programs do.  They are not like operational semantics or denotational
semantics.  These semantics also describe what programs do but in a
different way.  They describe it by giving meaning to each part of
programs.  They need to know how programs are written, that is, they
need actual codes.  Whereas, algebraic specification methods never talk
about how programs are implemented.  They describe their behaviour
abstractly viewing from outside.

This abstract view point, seeing from outside, led to the discovery of
{\it abstract data types} (see e.g.\ \cite{goguen-thatcher-wagner-78}).
It is interesting to know that algebraic specification methods were
started to describe programs but it also developed a theory of data
types.  Since algebraic specification methods try to describe things
from an outside point of view, they cannot talk about the concrete
nature of data types which programs handle.  Therefore, the data types
also needed to be abstracted and, thus, abstract data types have been
developed.  We may divide algebraic specification methods into two:
specification of data types and specification of programs.
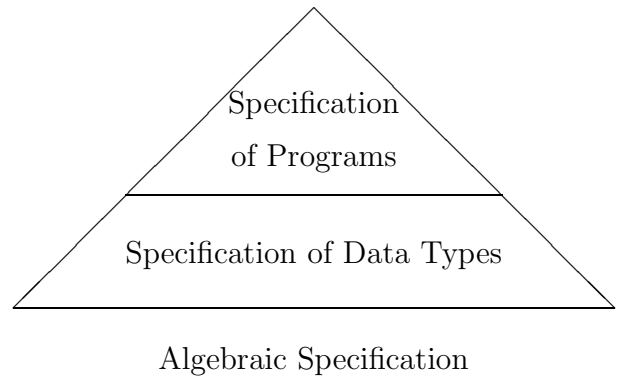
\begin{displaymath}
\setlength{\unitlength}{1mm}
\begin{picture}(95,50)
\put(45,37){\makebox(0,0){Specification}}
\put(45,30){\makebox(0,0){of Programs}}
\put(45,17){\makebox(0,0){Specification of Data Types}}
\put(45,3){\makebox(0,0){Algebraic Specification}}
\put(45,50){\line(-1,-1){40}}
\put(45,50){\line(1,-1){40}}
\put(20,25){\line(1,0){50}}
\put(5,10){\line(1,0){80}}
\end{picture}
\end{displaymath}
It is the former, specification of data types, that concerns us in this
thesis.

Algebraic specification methods got their method of describing data
types from abstract algebras in mathematics.  Mathematicians have been
using abstract algebras for about a century.  Abstract algebras are only
concerned with concrete real algebras insofar as they satisfy some laws.
For example, a set with a binary operation is a group when the operation
is associative, there is an identity and every element is invertible.  A
real set and a real operation can be anything, integers and $+$,
general linear matrixes and their multiplication, and so on.  Any theorem
established for general groups can be applied to any real groups.  There
are various kinds of abstract algebras: groups, rings, fields, and so
on.  Those abstract algebras can be presented uniformly by {\it universal
algebras}.  Algebraic specification methods use a many-sorted version of
universal algebras.

Na\"{\i}vely speaking, an algebraic specification is a triple $(S,\Sigma,E)$,
where $S$ is a set of sorts, $\Sigma$ is a $S^\ast\times S$-indexed set
of operations and $E$ is a set of equations over $\Sigma$.  For example,
in an algebraic specification language CLEAR
\cite{burstall-goguen-80,burstall-goguen-81} a specification of lists
may be as follows.
\begin{bcomputer}
constant List =
    theory
        sorts element, list
        opns nil : list
             cons : element, list -> list
             head : list -> element
             tail : list -> list
        eqns all e : element, l : list, head(cons(e,l)) = e
             all e : element, l : list, tail(cons(e,l)) = l
     endth
\end{bcomputer}
$S$ is $\{\; {\tt element}, {\tt list} \;\}$, $\Sigma_{\tt list}$ is
$\{\; {\tt nil} \;\}$, $\Sigma_{{\tt list}\;{\tt element}}$ is $\{\;
{\tt head} \;\}$, and so on.  $E$ consists of the two equations above.

There are several problems about this specification as we will see
immediately after we say what a specification means.  An algebraic
specification $(S,\Sigma,E)$ defines a class of many sorted algebras
each of which, say $A$, consists of an $S$-sorted set $|A|$ and functions
$f_A: |A|_{s_1} \times\ldots\times |A|_{s_n} \longrightarrow |A|_s$ for each
$f \in \Sigma_{s_1\ldots s_n s}$ which satisfy the equations in $E$.

The first problem of the above specification is that not only lists
satisfy it but also many of other data types as well.  There is actually
no way to make it describe only lists so long as we stick to first order
methods.  We need something of second order.  The way algebraic
specification methods usually obtain this is to put {\it data
constraints}.  We rely on the categorical fact that {\it the initial
algebra is unique up to isomorphism}.  In this case, we put a data
constraint onto the sort `{\tt list}', but not to `{\tt element}'
because if we put a data constraint onto `{\tt element}' then the `{\tt
element}' sort would be empty.

The second problem is that `{\tt head}' and `{\tt tail}' are partial
functions.  The specification does not say what is `{\tt head(nil)}' or
what is `{\tt tail(nil)}'.  In order to fix this problem, we have to
introduce, for example,  error algebras or go into partial algebras.

The third problem is that although we put a data constraint on `{\tt
list}' it is not immediately obvious that `{\tt nil}' and `{\tt cons}'
can construct all the lists.  Some algebraic specification
languages do distinguish these constructors from the others.

The fourth problem is about the sort `{\tt element}'.  We actually need
it as a parameter.  When we use this specification, `{\tt element}'
denotes a particular data type defined by another specification and we
need a way to plug in any specification of `{\tt element}' into this
specification.  Actually, CLEAR has this facility.  `{\tt List}' can be
defined as `{\tt procedure}' taking parametrized type `{\tt element}'.
However, this new specification no longer corresponds to a class of
algebras but to something one level higher.

Many other problems there might be, but most of them have been solved in
one way or another.  The important point we would like to make is that
the na\"{\i}ve idea of
\begin{displaymath}
\mbox{algebraic specification} = \mbox{universal algebra}
\end{displaymath}
does not work well and we have to put a lot of other ideas into
algebraic specification methods.  One might wonder why so many
complications are needed to define everyday objects like lists.

In CDT, we stick to the very simple relation
\begin{displaymath}
\mbox{categorical data type} = \mbox{$F,G$-dialgebra}
\end{displaymath}
$F,G$-dialgebras can be seen as an extension of universal algebras (see
section~\ref{sec-what-cdt}).  We do not need to introduce meta arguments
or any other complicated ideas into CDT in order to define lists or
other basic data types.

\subsection{Domain Theory}
\label{ssec-domain-theory}

Domain theory was started with denotational semantics
\cite{stoy-77,scott-76}.  In order to give denotational semantics to
programs, we need several domains to which the denotations are mapped.  Those
domains are often interwoven and recursively defined.  The
most famous example of this is the following $D$.
\begin{displaymath}
D \iso D \rightarrow D
\end{displaymath}
This domain $D$ was necessary to give denotational semantics to the
untyped lambda calculus.  In general, we would like to solve the
following domain equation:
\begin{displaymath}
D \iso F(D)
\end{displaymath}
where $F(D)$ is a domain expression involving $D$.

Though domains are mathematical objects and not necessarily
representable in computers, the idea of recursively defined data types
has been adopted into several programming languages.  For example, we
can have a domain $L$ for lists of $A$ elements by solving\footnote{We
have to say what kind of domains we are dealing with.  Let us say in
this thesis that a domain is a complete partially ordered set with the
least element and a function between domains needs to be continuous and
strict.}
\begin{displaymath}
L \iso 1 + A\times L,
\end{displaymath}
and in the original version of ML \cite{gordon-milner-wordsworth-79}, we
could define the data type for lists just like the same.
\begin{bcomputer}
abstype 'a list = unit + 'a # 'a list
    with ...
\end{bcomputer}
On the other hand, some domains cannot be represented in the same way.
For example, we can have a domain $I$ for infinite lists of $A$ elements
by solving
\begin{displaymath}
I \iso A \times I_\bot
\end{displaymath}
where $I_\bot$ is the lifting of $I$ by adding the new least element,
but we cannot define infinite lists in ML in a similar way.

Comparing with algebraic specification methods, in domain theory we
can define data types easily and there is no complication of
parametrized data types, but we have some difficulty of defining
operations over data types.  In algebraic specification methods we
define operations together with data types, but in domain theory we have
to define them using the isomorphisms of domain equations.

If an algebraic specification $(S,\Sigma,E)$ has no equational
constraints (i.e.\ $E = \emptyset$), the initial algebra can be given by
solving the following domain equations.
\begin{displaymath}
|A|_s \iso \sum_{f \in \Sigma_{s_1\ldots s_n s}} |A|_{s_1}\times \ldots
\times |A|_{s_n}
\end{displaymath}
By this connection, we can see the possibility of combining algebraic
specification methods and domain theory together.  Actually, data types
in the current Standard ML \cite{milner-84,harper-macqueen-milner-86}
are defined in this mixed fashion (see also section~\ref{sec-ml-cpl}).

Categorically, we can go the other way round.  If $F(D)$ is a covariant
functor, the initial fixed point of $F(D)$ can be characterized as the
initial $F$-algebra.  A $F$-algebra is a categorical generalization of
an ordinary algebra.  The main idea we borrow from domain theory is this
connection between initial fixed points and initial algebras.

After becoming familiar with category theory, one can notice the dual
connection between final fixed points and final co-algebras.  People
rarely talked about them until recently \cite{arbib-manes-80}.  One of
the reasons is that co-algebras are not so popular and another reason is
that final fixed points are often the same as initial fixed points in
domain theory.  However, in CDT we will use this dual connection as
well.  Final co-algebras give us some very intersting data types like
infinite lists.  We defined infinite lists by the initial fixed point of
\begin{displaymath}
I \iso A \times I_\bot
\end{displaymath}
Actually, what we were doing using the lifting $I_\bot$ is to get the final
fixed point of
\begin{displaymath}
I \iso A \times I.
\end{displaymath}

\section{Basic Category Theory}
\label{sec-category-theory}

This section is to roughly introduce some categorical concepts we
will use in the rest of this thesis.  The author refers to category
theory text books like \cite{maclane-71}, \cite{arbib-manes-75} and
\cite{lambek-scott-86} for a more detailed account of category theory.

A {\it category} $\cC$ is given by
\begin{itemize}
\item a collection of {\it objects} $|\cC|$,
\item for any pair of objects $A$ and $B$, a collection
$\Hom{\cC}{A}{B}$ of {\it morphisms} from domain $A$ to codomain $B$,
(we write $f: A \rightarrow B$ for $f \in \Hom{\cC}{A}{B}$)
\item for any objects $A$, $B$ and $C$, an operation called {\it
composition} denoted by `$\circ$' from $\Hom{\cC}{B}{C}\times
\Hom{\cC}{A}{B}$ to $\Hom{\cC}{A}{C}$ which is associative,
\begin{displaymath}
(f\circ g)\circ h = f\circ (g\circ h)
\end{displaymath}
\item for any object $A$, an {\it identity} morphism ${\bf I}_A: A \rightarrow
A$ such that for any $f: B \rightarrow A$ and any $g: A \rightarrow C$
\begin{displaymath}
{\bf I}_A\circ f = f \qquad \mbox{and} \qquad g\circ {\bf I}_A = g
\end{displaymath}
\end{itemize}

Two objects $A$ and $B$ are called {\it isomorphic} if there are two
morphisms $f: A \rightarrow B$ and $g: B \rightarrow A$ such that
\begin{displaymath}
f\circ g = {\bf I}_B \qquad \mbox{and} \qquad g\circ f = {\bf I}_A.
\end{displaymath}
$f$ and $g$ are called {\it isomorphisms}.

The {\it opposite category} $\cC^{\rm op}$ of a category $\cC$ is defined
by reversing the direction of all the morphisms in $\cC$.
\begin{displaymath}
\Hom{\cC^{\rm op}}{A}{B} = \Hom{\cC}{B}{A}
\end{displaymath}
We may write $\cC^{\rm op}$ morphism $f^{\rm op}: A \rightarrow B$ for
$\cC$ morphism $f: B \rightarrow A$.

The {\it product category} $\cC\times \cD$ of a category $\cC$ and a
category $\cD$ is given as
\begin{itemize}
\item a $\cC\times \cD$ object is $\pair{A,B}$ for a $\cC$ object $A$
and a $\cD$ object $B$
\item a $\cC\times \cD$ morphism from $\pair{A,B}$ to $\pair{A',B'}$ is
$\pair{f,g}$ for a $\cC$ morphism $f: A \rightarrow B$ and a $\cD$
morphism $g: A \rightarrow B$.
\end{itemize}

A {\it covariant functor} $F$ from a category $\cC$ to a category $\cD$
(we write $F: \cC \rightarrow \cD$) is given by
\begin{itemize}
\item associating a $\cD$ object $F(A)$ for every $\cC$ object $A$
\item associating a $\cD$ morphism $F(f): F(A) \rightarrow F(B)$ for
every $\cC$ morphism $f: A \rightarrow B$ such that
\begin{displaymath}
F({\bf I}_A) = {\bf I}_{F(A)} \qquad \mbox{and} \qquad F(f\circ g) =
F(f)\circ F(g)
\end{displaymath}
\end{itemize}
A {\it contravariant functor} is defined in a similar way except that
$F(f): F(B) \rightarrow F(A)$.

A {\it natural transformation} $\alpha$ from a covariant functor $F: \cC
\rightarrow \cD$ to a covariant functor $G: \cC \rightarrow \cD$ (we write
$\alpha: F \natrightarrow G$) is given by
\begin{itemize}
\item associating a $\cD$ morphism $\alpha_A: F(A) \rightarrow G(A)$ for
every $\cC$ object $A$ such that for any $\cC$ morphism $f: A
\rightarrow B$ the following diagram commutes.
\begin{displaymath}
\sqdiagram{F(A)}{\alpha_A}{G(A)}{F(f)}{G(f)}{F(B)}{\alpha_B}{G(B)}{}
\end{displaymath}
\end{itemize}
When every $\alpha_A$ is an isomorphism, we call $\alpha$ {\it natural
isomorphism}.

Two functors $F: \cC \rightarrow \cD$ and $G: \cD \rightarrow \cC$ are
called {\it adjoints} if there exists a natural isomorphism
\begin{displaymath}
\psi_{A,B}: \Hom{\cD}{F(A)}{B} \stackrel{\iso}{\longrightarrow}
\Hom{\cC}{A}{G(B)}.
\end{displaymath}
$F$ is called the left adjoint functor of $G$ and $G$ is called the
right adjoint functor of $F$.  We also call $\psi_{A,B}$ (or its inverse
$\psi_{A,B}^{-1}$) {\it factorizer} or {\it mediating morphism}.

\section{Development of Categorical Data Types}
\label{sec-develop-CDT}

The motivation of CDT was to adopt the categorical way of defining data
types into specification languages.  Anybody educated using set theory
has quite a shock when he first sees the way category theory
works.  It gives a totally different point of view to things which are
familiar.  Things which were vaguely connected suddenly are fitted into
systematic places.  It seems that the nature of things is finally
revealed.

There are many beautiful concepts discovered through category theory,
but here we concentrate only one of them, namely {\it adjunction} (or
{\it adjoint situation}).  In~\cite{maclane-71},
one will find many equivalent forms of the definition of adjunction (we
gave one of them in section~\ref{sec-category-theory}).  One may be first
at a loss for chosing the definition.  Adjunction is so versatile
that it can be seen in a number of different forms and it is sometimes
difficult to understand it if one sticks to a particular form of the
definition.  The form is not important if the spirit is understood.
Adjunction can be regarded as a property of two functors or because of
the unique correspondence between two functors it can be seen as
defining one of them from the other.  It is the latter which is
important to us because it is a typical way of defining things in category
theory.  Let us see an example.

Using set theory, we can define what the product of two sets is, what
the product of two groups is, what the product of two topological
spaces is, and so on.  Each definition is obviously deferent from the
others but all of them are called by the same name, {\it product}.  Why
is that so?  Is there any common property which all the different kinds
of products should satisfy?  Can we give the general definition of {\it
product}?  Category theory can give an affirmative answer to these
questions.  The categorical definition of products is
\begin{quote}
For object $A$ and $B$, the product $A\times B$ is an object such that
there are two morphisms
\begin{displaymath}
\pi_1: A\times B \longrightarrow A \qquad \mbox{and} \qquad \pi_2: A\times B
\longrightarrow B
\end{displaymath}
and for any given two morphisms
\begin{displaymath}
f: C \longrightarrow A \qquad \mbox{and} \qquad g: C \longrightarrow B
\end{displaymath}
there is a unique morphism $h: C \longrightarrow A\times B$ such that the
following diagram commutes.
\begin{displaymath}
\setlength{\unitlength}{1mm}
\begin{picture}(70,40)(0,2.5)
\put(5,35){\makebox(0,0){$A$}}
\put(35,35){\makebox(0,0){$A\times B$}}
\put(65,35){\makebox(0,0){$B$}}
\put(17,37){\makebox(0,0)[b]{$\pi_1$}}
\put(53,37){\makebox(0,0)[b]{$\pi_2$}}
\put(27.5,35){\vector(-1,0){20}}
\put(42.5,35){\vector(1,0){20}}
\put(35,5){\makebox(0,0){$C$}}
\put(17.5,17.5){\makebox(0,0){$f$}}
\put(52.5,17.5){\makebox(0,0){$g$}}
\put(35,20){\makebox(0,0){$h$}}
\put(22.5,27.5){\makebox(0,0){\commute}}
\put(47.5,27.5){\makebox(0,0){\commute}}
\put(32.5,7.5){\vector(-1,1){25}}
\put(37.5,7.5){\vector(1,1){25}}
\multiput(35,7.5)(0,5){2}{\line(0,1){3}}
\put(35,22.5){\line(0,1){3}}
\put(35,27.5){\vector(0,1){5}}
\end{picture}
\end{displaymath}
It is easily shown that any two objects satisfy this definition are
isomorphic.  We may write $\pair{f,g}$ for $h$.
\end{quote}
This definition is general enough to cover the definition of products for
sets, groups, topological spaces, and so on.  We no longer need to define
products for each individual case.

The generality should not be bound only in mathematics.  Why should it not
equally be appropriate to the definition of products in programming
languages?  The product data type of type $A$ and type $B$ is usually
defined as a type of records whose first component is of type $A$ and
the second one is of type $B$, but this definition is like one in set
theory.  It assumes too much about how elements of data types are
represented.  It is not acceptable as an abstract description of the
product data type.  If the product data type is defined as an abstract
data type, how can we present it?

We can directly adopt the categorical definition of products.  There are five
ingredients in the definition.
\begin{displaymath}
\begin{tabular}{rl}
1. & two given objects $A$ and $B$, \\
2. & the object $A\times B$ we are defining, \\
3. & two morphisms $\pi_1: A\times B \longrightarrow A$ and $\pi_2: A
\times B \longrightarrow B$, \\
4. & $\pair{f,g}: C \longrightarrow A\times B$ for $f: C \longrightarrow A$ and
$g: C \longrightarrow B$, and \\
5. & the commutative diagram. \\
\end{tabular}
\end{displaymath}
We may write these down as follows
\begin{displaymath}
\begin{tabular}{l}
object $A\times B$ is \\
$\qquad \pi_1: A \times B \longrightarrow A$ \\
$\qquad \pi_2: A \times B \longrightarrow B$ \\
$\qquad \pair{f,g}: C \longrightarrow A\times B$ for $f: C
\longrightarrow A$ and $g: C \longrightarrow B$ \\
\quad where \\
$\qquad \pi_1\circ \pair{f,g} = f$ \\
$\qquad \pi_2\circ \pair{f,g} = g$ \\
$\qquad \pi_1\circ h = f \land \pi_2\circ h = g \implies h = \pair{f,g}$
\\
end object. \\
\end{tabular}
\eqno(*)
\end{displaymath}
Can we call this a categorical definition of the product data type
constructor?  Although this is an exact copy of the categorical
definition, it has somehow lost the spirit of category theory; its beauty;
its simplicity.  The categorical definition of products we gave is in a
disguised form of adjunction.  The definition could have been sufficient
to just say that the product functor is the right adjoint of the
diagonal functor.  The previous definition expands this into plain words
so that there are a lot of duplications.  One of them is that the type
of $\pair{~,~}$ can be deduced from the type of $\pi_1$ and $\pi_2$.  If
$f$ and $g$ has the same type as $\pi_1$ and $\pi_2$ except replacing
$A\times B$ by $C$, $\pair{f,g}$ is a morphism from $C$ to $A\times B$.
Another duplication is that the commutative diagram can also be deduced
from the rest.  There are no other trivial ways to make diagrams
involving $\pi_1$, $\pi_2$, $f$, $g$ and $\pair{f,g}$.  Therefore, the
definition of product data types can be written simply as
\begin{displaymath}
\begin{tabular}{l}
object $A\times B$ is \\
$\qquad \pi_1: A\times B \longrightarrow A$ \\
$\qquad \pi_2: A\times B \longrightarrow B$ \\
end object. \\
\end{tabular}
\end{displaymath}
This supplies the minimal information to get back to $(*)$.  Now, we
have to use the fact that $A\times B$ is defined by adjunction (it was
not necessary in $(*)$).  Let us indicate this by saying it is a {\it
right object} as well as declaring $\pair{~,~}$.
\begin{displaymath}
\begin{tabular}{l}
right object $A\times B$ with $\pair{~,~}$ is \\
$\qquad \pi_1: A\times B \longrightarrow A$ \\
$\qquad \pi_2: A\times B \longrightarrow B$ \\
end object \\
\end{tabular}
\end{displaymath}
This is the declaration of the product data type constructor in CDT
(except for minor changes).

Let us examine the generality and simplicity of this declaration
mechanism through examples.  Let us try exponentials $B^A$.  The functor
$\bullet^A$ is defined as the right adjoint functor of $\bullet \times A$.  The
definition in CDT is
\begin{displaymath}
\begin{tabular}{l}
right object $B^A$ with ${\rm curry}(~)$ is \\
$\qquad {\rm eval}: B^A \times A \longrightarrow B$ \\
end object \\
\end{tabular}
\end{displaymath}
We can derive the usual definition of exponentials from this definition.
First, the type of ${\rm curry}(~)$ should be
\begin{displaymath}
\logicrule{f: C \times A \longrightarrow B}{{\rm curry}(f): C \longrightarrow B^A}.
\end{displaymath}
The type of $f$ is obtained from the type of `eval' just replacing $B^A$
by $C$.  The commutative diagram which ${\rm curry}(f)$ gives can be
obtained by connecting
\begin{displaymath}
\ardiagram{B^A \times A}{\rm eval}{B} \qquad \mbox{and} \qquad
\ardiagram{C \times A}{f}{B}
\end{displaymath}
by ${\rm curry}(f): C \longrightarrow B^A$.  The only way to connect them
together results
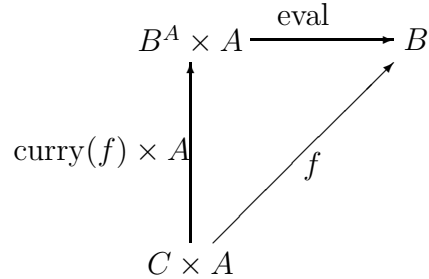
\begin{displaymath}
\setlength{\unitlength}{1mm}
\begin{picture}(55,45)
\put(20,5){\makebox(0,0){$C\times A$}}
\put(20,20){\makebox(0,0)[r]{${\rm curry}(f)\times A$}}
\put(35,20){\makebox(0,0)[tl]{$f$}}
\put(20,35){\makebox(0,0){$B^A\times A$}}
\put(50,35){\makebox(0,0){$B$}}
\put(35,37){\makebox(0,0)[b]{eval}}
\put(28,35){\vector(1,0){19}}
\put(20,8){\vector(0,1){24}}
\put(23,8){\vector(1,1){24}}
\end{picture}
\end{displaymath}
The morphism denoted by ${\rm curry}(f)$ is the unique one which
makes this diagram commute.  Thus, we recovered the ordinary definition
of exponentials.

We said `right object' for products and exponentials.  It is natural to
think that we also have `{\it left object\/}' as dual.  The dual of
products are coproducts.  Let us define them in CDT.
\begin{displaymath}
\begin{tabular}{l}
left object $A+B$ with $[~,~]$ is \\
$\qquad \nu_1: A \longrightarrow A+B$ \\
$\qquad \nu_2: B \longrightarrow A+B$ \\
end object \\
\end{tabular}
\end{displaymath}
The type of $[~,~]$ can be obtained from the type of $\nu_1$ and
$\nu_2$.
\begin{displaymath}
\logicrule{f: A \longrightarrow C \qquad g: B \longrightarrow C}{[f,g]:
A+B \longrightarrow C}
\end{displaymath}
Note that $[f,g]$ goes from $A+B$ to $C$ not the other way round as it
would be if it were a right object.  The name `left object' came from
the fact that $A+B$ is in the left hand side of $\longrightarrow$.
Remember that $A\times B$ was in the right hand side of
$\longrightarrow$ for $\pair{~,~}$.  A natural way of connecting $f$
and $g$ with $\nu_1$ and $\nu_2$ by $[f,g]$ gives us the ordinary
commutative diagram which $[f,g]$ should satisfy.
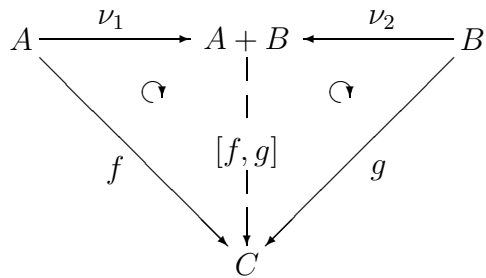
\begin{displaymath}
\setlength{\unitlength}{1mm}
\begin{picture}(70,40)(0,2.5)
\put(5,35){\makebox(0,0){$A$}}
\put(35,35){\makebox(0,0){$A+B$}}
\put(65,35){\makebox(0,0){$B$}}
\put(17,37){\makebox(0,0)[b]{$\nu_1$}}
\put(53,37){\makebox(0,0)[b]{$\nu_2$}}
\put(7.5,35){\vector(1,0){20}}
\put(62.5,35){\vector(-1,0){20}}
\put(35,5){\makebox(0,0){$C$}}
\put(17.5,17.5){\makebox(0,0){$f$}}
\put(52.5,17.5){\makebox(0,0){$g$}}
\put(35,20){\makebox(0,0){$[f,g]$}}
\put(22.5,27.5){\makebox(0,0){\commute}}
\put(47.5,27.5){\makebox(0,0){\commute}}
\put(7.5,32.5){\vector(1,-1){25}}
\put(62.5,32.5){\vector(-1,-1){25}}
\multiput(35,32.5)(0,-5){2}{\line(0,-1){3}}
\put(35,17.5){\line(0,-1){3}}
\put(35,12.5){\vector(0,-1){5}}
\end{picture}
\end{displaymath}

We demonstrated that we can express basic categorical constructs in CDT.
Those constructs, or data types, are primitives in ordinary programming
languages.  Can we declare more familiar data types?  In fact, the `left
object' declaration gives all those which can be defined by algebraic
methods with no equations.  `Without equations' seems that we cannot
define much, but actually it gives us all the important data types of
ordinary programming languages.  For example, natural numbers can be
defined as
\begin{displaymath}
\begin{tabular}{l}
left object nat with ${\rm pr}(~,~)$ is \\
$\qquad {\rm zero}: 1 \longrightarrow {\rm nat}$ \\
$\qquad {\rm succ}: {\rm nat} \longrightarrow {\rm nat}$ \\
end object \\
\end{tabular}
\end{displaymath}
This is very much like a specification of natural numbers in algebraic
specification methods except that we do not have the predecessor
function or plus or times and that we have something called ${\rm
pr}(~,~)$.  From analogy of the types of $[~,~]$ and $\pair{~,~}$, the
type of ${\rm pr}(~,~)$ should be
\begin{displaymath}
\logicrule{f: 1 \longrightarrow C \qquad g: C \longrightarrow C}{{\rm
pr}(f,g): {\rm nat} \longrightarrow C}.
\end{displaymath}
We also obtain the diagram characterizing `nat' as we did for products
and others.
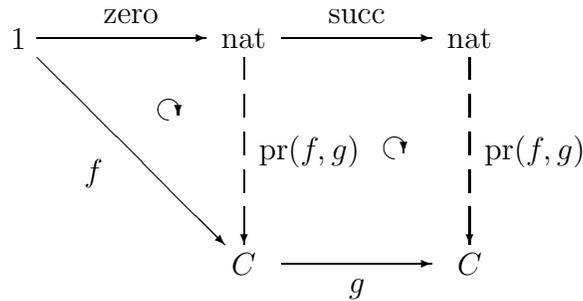
\begin{displaymath}
\setlength{\unitlength}{1mm}
\begin{picture}(75,45)(0,2.5)
\put(5,40){\makebox(0,0){1}}
\put(20,42){\makebox(0,0)[b]{zero}}
\put(35,40){\makebox(0,0){nat}}
\put(50,42){\makebox(0,0)[b]{succ}}
\put(65,40){\makebox(0,0){nat}}
\put(15,22){\makebox(0,0){$f$}}
\put(25,30){\makebox(0,0){\commute}}
\put(37,25){\makebox(0,0)[l]{${\rm pr}(f,g)$}}
\put(55,25){\makebox(0,0){\commute}}
\put(67,25){\makebox(0,0)[l]{${\rm pr}(f,g)$}}
\put(35,10){\makebox(0,0){$C$}}
\put(65,10){\makebox(0,0){$C$}}
\put(50,8){\makebox(0,0)[t]{$g$}}
\put(7.5,40){\vector(1,0){22.5}}
\put(40,40){\vector(1,0){20}}
\put(7.5,37.5){\vector(1,-1){25}}
\put(40,10){\vector(1,0){20}}
\multiput(35,37.5)(0,-5){4}{\line(0,-1){3}}
\put(35,17.5){\vector(0,-1){5}}
\multiput(65,37.5)(0,-5){4}{\line(0,-1){3}}
\put(65,17.5){\vector(0,-1){5}}
\end{picture}
\end{displaymath}
This is exactly the definition of `nat' being a natural number object in
category theory and it is well-known that we can define all the primitive
recursive functions using ${\rm pr}(~,~)$.  For example, the addition
function can be defined by
\begin{displaymath}
{\rm add} \defeq {\rm eval}\circ \pair{{\rm pr}({\rm curry}({\rm
\pi_2}),{\rm curry}({\rm succ}\circ{\rm eval}))\circ \pi_1,\pi_2}.
\end{displaymath}

As another example, we give the definition of lists in CDT.  It is
\begin{displaymath}
\begin{tabular}{l}
left object ${\rm list}(A)$ with ${\rm prl}(~,~)$ is \\
$\qquad {\rm nil}: 1 \longrightarrow {\rm list}(A)$ \\
$\qquad {\rm cons}: A \times {\rm list}(A) \longrightarrow {\rm list}(A)$ \\
end object \\
\end{tabular}
\end{displaymath}
The type of ${\rm prl}(~,~)$ is
\begin{displaymath}
\logicrule{f: 1 \longrightarrow C \qquad g: A \times C \longrightarrow
C}{{\rm prl}(f,g): {\rm list}(A) \longrightarrow C}.
\end{displaymath}
The diagram is
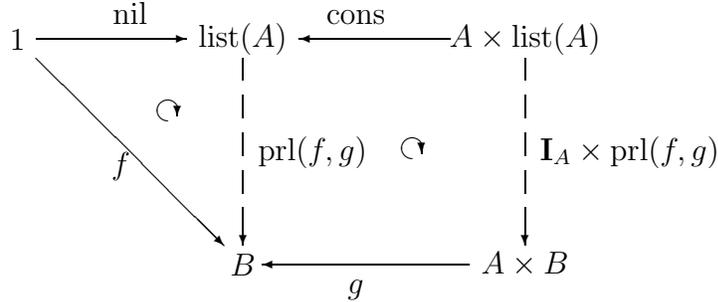
\begin{displaymath}
\setlength{\unitlength}{1mm}
\begin{picture}(100,35)(0,2.5)
\put(5,35){\makebox(0,0){1}}
\put(35,35){\makebox(0,0){${\rm list}(A)$}}
\put(72.5,35){\makebox(0,0){$A \times {\rm list}(A)$}}
\put(35,5){\makebox(0,0){$B$}}
\put(72.5,5){\makebox(0,0){$A\times B$}}
\put(20,37){\makebox(0,0)[b]{nil}}
\put(50,37){\makebox(0,0)[b]{cons}}
\put(20,20){\makebox(0,0)[tr]{$f$}}
\put(25,25){\makebox(0,0){\commute}}
\put(37,20){\makebox(0,0)[l]{${\rm prl}(f,g)$}}
\put(57.5,20){\makebox(0,0){\commute}}
\put(74.5,20){\makebox(0,0)[l]{${\bf I}_A\times{\rm prl}(f,g)$}}
\put(50,3){\makebox(0,0)[t]{$g$}}
\put(7.5,35){\vector(1,0){20}}
\put(7.5,32.5){\vector(1,-1){25}}
\put(62.5,35){\vector(-1,0){20}}
\put(65,5){\vector(-1,0){27.5}}
\multiput(35,32.5)(0,-5){4}{\line(0,-1){3}}
\put(35,12.5){\vector(0,-1){5}}
\multiput(72.5,32.5)(0,-5){4}{\line(0,-1){3}}
\put(72.5,12.5){\vector(0,-1){5}}
\end{picture}
\end{displaymath}
Remember that our definition of lists in CLEAR had `head' and `tail',
but we do not declare them here.  We can define them by ${\rm prl}(~,~)$.
\begin{displaymath}
\begin{array}{rll}
{\rm head} & {} \defeq {\rm prl}(\nu_2,\nu_1\circ \pi_1) & \qquad :\;
{\rm list}(A) \longrightarrow A + 1 \\[1ex]
{\rm tail} & {} \defeq [\nu_1\circ \pi_2,\nu_2]\circ {\rm
prl}(\nu_2,\nu_1\circ \pair{\pi_1,[{\rm cons},{\rm nil}]}) & \qquad : \;
{\rm list}(A) \longrightarrow {\rm list}(A) + 1 \\
\end{array}
\end{displaymath}
`Without equations' is not a disadvantage to define everyday data types.

By the connection between initial fixed points and $F$-algebras, we can
define the initial fixed point $D$ of a covariant functor $F(X)$ as
follows.
\begin{displaymath}
\begin{tabular}{l}
left object $D$ with $\psi(~)$ is \\
$\qquad \alpha: F(D) \longrightarrow D$ \\
end object \\
\end{tabular}
\end{displaymath}
$\alpha$ gives one direction of the isomorphism between $D$ and $F(D)$,
and	 $\psi(~)$ gives unique arrows.
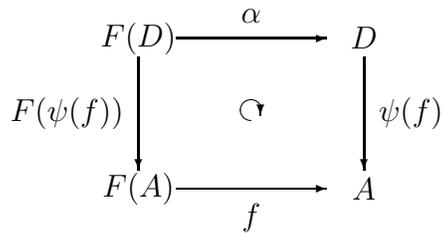
\begin{displaymath}
\sqdiagram{F(D)}{\alpha}{D}{F(\psi(f))}{\psi(f)}{F(A)}{f}{A}{}
\end{displaymath}

We have been using `left object' more than `right object', but they are
dual and there are equally as many right objects as left objects.  Just
they are not familiar in ordinary programming languages.  For example,
the following definition gives the data type for infinite lists.
\begin{displaymath}
\begin{tabular}{l}
right object ${\rm inflist}(A)$ with ${\rm fold}(~,~)$ is \\
$\qquad {\rm hd}: {\rm inflist}(A) \longrightarrow A$ \\
$\qquad {\rm tl}: {\rm inflist}(A) \longrightarrow {\rm inflist}(A)$ \\
end object \\
\end{tabular}
\end{displaymath}
This gives the final fixed point of $I \iso A\times I$.

We have devised, based on category theory, a simple way of defining data
types.  The next question is whether we can adopt this method into
ordinary programming languages.  The answer is negative.  Although what
we can define in this way is far less than what we can define using
algebraic specification languages with equations, we still have some
strange things that can be defined in this way.  Let us see an example.
We defined ${\rm list}(A)$ as a parametrized data type but in fact it is
a functor.
\begin{displaymath}
{\rm list}(f):  {\rm list}(A) \longrightarrow {\rm list}(B)
\end{displaymath}
for a morphism $f: A \longrightarrow B$ is often called {\it map
function}.  In LISP it is `MAPCAR' and in ML it is `map'.  The general
declaration mechanism of CDT allows us to define the left and right
adjoint functors of ${\rm list}(A)$ which could not exist in the world
of programming.  We need to put a restriction to prevent these objects.
The restriction will come out of a notion of computability in our
setting.  Interestingly, it turns out the category should be
\begin{displaymath}
\mbox{cartesian closed} + \mbox{initial fixed points} + \mbox{final fixed
points}
\end{displaymath}
We might see the similarity between this and the connection of lambda
calculus and cartesian closed categories.

By putting this computability restriction, we can regard CDT as not only
a device of defining data types but also a programming language.  We
program in a categorical fashion; there are no concrete data but
morphisms; programs are also morphisms; there are no variables in
programs.  The computation in this language is reduction from morphisms
to canonical ones.  For example, we can reduce
\begin{displaymath}
{\rm add}\circ\pair{{\rm succ}\circ{\rm zero},{\rm succ}\circ{\rm
zero}} \implies {\rm succ}\circ {\rm succ}\circ {\rm zero}
\end{displaymath}
which corresponds to the calculation of $1+1 = 2$.

We summerize the characteristics of CDT as follows.
\begin{enumerate}
\item CDT uses categorical characterization of data types.  We do not
need to say things explicitly.  All the equations are automatically
generated for definitions.
\item CDT needs no primitive data types.  Ordinary programming languages
(e.g.\ PASCAL, LISP, ML) have primitive data types: natural numbers,
lists, records, and so on, but CDT does not.  They can be defined.  Thus,
CDT is analogous to algebraic specification methods where we can specify
them as well.  However, algebraic specification methods cannot specify
higher order data types (i.e.\ function spaces or exponentials in
a categorical term) nor can they specify products without using
equations.
\item CDT can not only define products without explicitly mentioning
equations but also can define exponentials.
\item CDT is symmetric in the sense that we can define initial algebras
(or initial fixed points) as well as final co-algebras (or final fixed
points).
\item Algebraic specification methods use initiality implicitly and
do not use the unique homomorphisms between the initial algebras and the
others, whereas CDT has explicit access to the unique morphisms.  This
gives the power of programming without going though equational
characterization as it is necessary in algebraic specification methods.
\item Domain theory does not use the initiality explicitly either.  The
reason for this is that recursion in ordinary programming languages
provide all the power of programming.
\item CDT defines functors.  Functors are thought to be parametrized
data types.  Algebraic specification methods usually introduce
parametrization later, but in CDT functorial behaviour of parametrized
data types is treated at the base level.
\end{enumerate}

\section{In This Thesis}
\label{sec-in-thesis}

The theory of categorical data types is divided into three: {\it Categorical
Specification Language}, {\it Categorical Data Types} and {\it
Categorical Programming Language}.
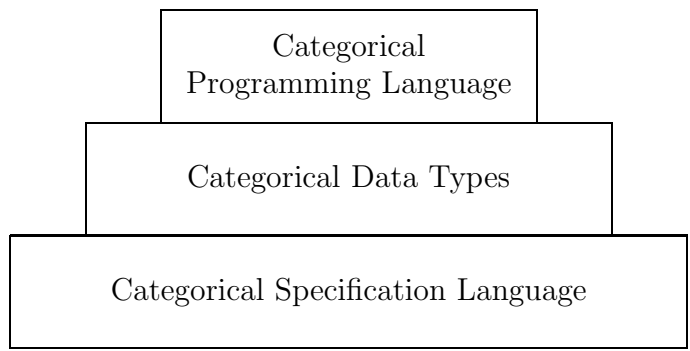
\begin{displaymath}
\setlength{\unitlength}{1mm}
\begin{picture}(100,55)
\put(25,50){\line(1,0){50}}
\put(50,45){\makebox(0,0){Categorical}}
\put(50,40){\makebox(0,0){Programming Language}}
\put(25,50){\line(0,-1){15}}
\put(75,50){\line(0,-1){15}}
\put(15,35){\line(1,0){70}}
\put(50,27.5){\makebox(0,0){Categorical Data Types}}
\put(15,35){\line(0,-1){15}}
\put(85,35){\line(0,-1){15}}
\put(5,20){\line(1,0){90}}
\put(50,12.5){\makebox(0,0){Categorical Specification Language}}
\put(5,20){\line(0,-1){15}}
\put(95,20){\line(0,-1){15}}
\put(5,5){\line(1,0){90}}
\end{picture}
\end{displaymath}
Chapter~\ref{ch-csl} is about Categorical Specification Language (CSL
for short).  CSL is a specification language.  It is an extension of
ordinary algebraic specification languages.  Whereas algebraic
specifications specify algebras, it specifies categories.  In order to
specify categories, CSL has to handle functors, natural transformations
and factorizers (or mediating morphisms).  A CSL signature declares some
functor names and their types (i.e.\ variances), some natural transformation
names and their types and some factorizer names and their types.  A CSL
sentence is a conditional equation of functors, natural transformations
and factorizers.  A CSL model is a category equipped with functors,
natural transformations and factorizers which have right types as are
specified in the signature and which satisfy the sentences.

Chapter~\ref{ch-cdt} gives the main idea of Categorical Data Types.  The
difference between CSL and CDT is that whereas CSL declares functors,
natural transformations and factorizers separately and connects them by
sentences, CDT declare them together in a style of adjoint declarations.
The semantics of CDT will be given informally in terms of
$F,G$-algebras and formally in terms of CSL.  There are some examples
of data types we can define in CDT.

Chapter~\ref{ch-cpl} is about Categorical Programming Language (CPL for
short).  CPL is a functional programming language which adopts the
categorical declaration mechanism of data types from CDT.  In order to
define the notion of computation in CPL, we have to put some
restrictions to CDT.  We will introduce the notion of {\it elements} and
{\it canonical elements} and present reduction rules to reduce
elements to their equivalent canonical elements.  We will also prove
that any reduction in CPL terminates using Tait's computability method.

Each of those three languages, CSL, CDT and CPL, characterizes a
category of data types in different ways.
\begin{displaymath}
\begin{tabular}{|r|l|l|}
\cline{2-3}
\multicolumn{1}{c|}{} & \multicolumn{1}{c|}{\bf Syntax} &
\multicolumn{1}{c|}{\bf Semantics} \\
\hline
CSL & Signature and Sentences & Models \\
\hline
CDT & Adjoint Declarations & Freeness and Co-Freeness \\
\hline
CPL & Restricted Adjoint Declarations & Operational \\
\hline
\end{tabular}
\end{displaymath}

In chapter~\ref{ch-application}, we will investigate the real
consequences of our study in CDT.  Section~\ref{sec-imp-cpl} is about an
implementation of CPL as a real programming language.
Section~\ref{sec-lambda-calculus} is about the connection between CDT
and typed lambda calculi, and finally in section~\ref{sec-ml-cpl} we
will attempt to extend ML incorporating the CDT data type declaration
mechanism.

\section{Comparison with Other Works}
\label{sec-comparison}

Systematic studies of data types have already been carried out by
various people in various contexts: ADJ in the context of initial
algebras \cite{goguen-thatcher-wagner-78}, Plotkin, Smyth and Lehmann in
the context of domains \cite{lehmann-smyth-81,smyth-plotkin-82} and
Martin-L\"of in the context of type theory \cite{martin-lof-79}.  This
thesis is about a study of the same subject in the context of category
theory.  We do not try to extend the traditional approaches as
Parasaya-Ghomi did for algebraic specification methods to include higher
order types \cite{ghomi-82}, nor do we try to unify two approaches
together like \cite{dybjer-83}, but we just directly use categorical
methods of defining things.

Categorical Programming Language in chapter~\ref{ch-cpl} might resemble
Categorical Abstract Machine (CAM) by Curien \cite{curien-86}, but he is
only interested in cartesian closed categories whereas CPL deals with a
class of different categories.  Moreover, the reduction rules in CPL is
systematically generated for products, coproducts, exponentials, natural
numbers, and so on.  We do not give any special reduction rules for any
data types we define.  CPL can be seen as
\begin{displaymath}
{\rm CPL} = {\rm CAM} + \mbox{initial data types} + \mbox{final data types}.
\end{displaymath}
Actually, CAM can be absorbed into `initial data types' and `final data
types', so in CPL we do not need to start with a particular set of
reduction rules for cartesian closed categories.  CPL has an ability to
define cartesian closed categories and the introduction of data types
also gives the control structure over those data types.  Here is another
slogan: ``{\it control structures in programming languages come out of
the structure of data types\/}''.
\begin{displaymath}
\begin{tabular}{rcl}
\multicolumn{1}{c}{Data Types} & & \multicolumn{1}{c}{Control Structures} \\
boolean & $\leftrightarrow$ & {\tt if} statement \\
disjoint union & $\leftrightarrow$ & {\tt case} statement \\
natural number & $\leftrightarrow$ & primitive recursion, {\tt for}
statement \\
product & $\leftrightarrow$ & pairing \\
function space & $\leftrightarrow$ & function call \\
\end{tabular}
\end{displaymath}

Barr and Wells uses {\it sketches} in \cite{barr-wells-85} to describe
algebras categorically.  It is more powerful than ordinary algebraic
specification methods because sketches can use any kind of limits
whereas algebraic specification methods uses only products.  It is
interesting to investigate CSL by sketches.

\chapter{Categorical Specification Language}
\label{ch-csl}

CDT can be seen from various points of view and can be presented in many
ways.  In this chapter, we present it as a specification language for
categories (we call the specification language {\it Categorical
Specification Language} or {\it CSL} for short).  This is not the way
originated, and it is difficult to recognize natural properties of data
types in this way.  We will give an alternative and more intuitive
definition of CDT in chapter~\ref{ch-cdt}.  However, the aim of CSL is
to give mathematically rigorous background for the more intuitive
presentation of CDT.

In applicative functional programming languages like ML, it is natural
to see that their data types and functions form a category; each data
type is an object; each function is a morphism; we have an identity
function; and two functions can be composed in a usual way.  We can also
treat other programming languages including procedural ones semantically
as defining domains and functions, and we can see that they form a
category.  These categories associated with programming languages
reflect the characteristics of the programming languages.  Thus, the
study of data types can be carried out by examining these categories.
CSL is a specification language for these categories.

Usually, a category is given by defining what an object consists of
(e.g.\ a set for {\bf Set}, the category of sets) and what a morphism
between objects is (e.g.\ a set function for {\bf Set}), but this is not
the way CSL works.  We are trying to understand a category in an
abstract manner; we do not say what an object is; instead, we specify
how it is constructed through its relationship among other objects.  We
saw in chapter~\ref{ch-introduction} an ML data type for lists and it
was a parameterized data type.  We can now see it as a data type
constructor; given a data type it constructs a new data type for the
lists of the given data type.  Categorically, constructors of objects
are functors and they provide structures for categories.  Remember that
a cartesian closed category is a category with three functors, the
terminal (constant) functor, the product functor and the exponential
functor.  Thus, CSL specifies a category equipped with some functors.
Because the properties of functors are often described in terms of their
interaction with natural transformations and factorizers (e.g.\ the
binary product functor is explained with two natural transformations,
$\pi_1$ and $\pi_2$, and factorizer $\pair{~,~}$), CSL also specifies
natural transformations and factorizers.

Let us make a comparison with algebraic specification languages like
CLEAR \cite{burstall-goguen-80,burstall-goguen-81}.  An algebraic
specification consists of declarations of some sorts and some operations
on these sorts.  Sorts are their data types.  A model of an algebraic
specification is a many-sorted algebra.  On the other hand, a CSL
specification consists of declarations of some functors, some natural
transformations between them and some factorizers.  Functors correspond
to sorts, and natural transformations and factorizers correspond to
operations.  A model of a CSL specification is a category equipped with
some functors, some natural transformations between them and some
factorizers.  It is abstract in the sense that the specification does
not distinguish between equivalent categories (an algebraic
specification does not distinguish isomorphic algebras).  Note that, in
general, the various models of a specification are not equivalent (e.g.\
not all the cartesian categories are equivalent).

As CSL specifies functors, the treatment of parametrized data types is
different from algebraic specification languages.  It specifies one
level higher objects.  The concept of parameterized data types and how
to combine them play very essential roles in algebraic specification
languages, but parameterized data types are treated in their meta-level
(one level higher than the level treating algebras), that is
specifications themselves are parameterized rather than dealing with
parameterized sorts in specifications.  In CSL, on the other hand,
parameterized data types are the basic objects in specifications.  In
one specification, several parameterized data types can be declared and
their relationship is directly specified.  Therefore, combining
specifications does not play as important a role as it does in algebraic
specification languages.

Our goal in this chapter is to define the specification language CSL.  In
section~\ref{sec-fun-calc}, we will introduce several notations for
dealing with functors which will be necessary later.  In
section~\ref{sec-csl-sig}, we will define the CSL signatures and in
section~\ref{sec-cls-mod} the CSL structures.  The definition of CSL
sentences and the CSL satisfaction relation will be in
section~\ref{sec-csl-sen} which follows section~\ref{sec-fun-calc-2} in
which we will introduce expressions involving natural transformations
and factorizers.  Finally, in section~\ref{sec-free-cat} we will show
that there is an initial CSL structure for each CSL theory.

\section{A Functorial Calculus}
\label{sec-fun-calc}

Before giving the definition of CSL signatures, we will look at some
aspects of functors.  This will be a kind of a functorial calculus though
not as abstract as \cite{kelly-72} is.  \cite{kelly-72} develops a
calculus of combining functors as we will do in this section, but for
the purpose of solving the coherence problems, and he treats many
variable functors quite extensively.  However, he does not say much about
mixed variant functors which we are interested in.  We will also
generalize variances to include free-variance and fixed-variance for
uniform treatment.

Functors are very much like ordinary functions except that functors have
variances.  Let $F$ be a unary functor $\cC \rightarrow \cC$ and $G$ be
a binary functor $\cC\times \cC \rightarrow \cC$.  Then, we can combine
them to get more complex functors:
\begin{displaymath}
G(F(X),Y) \qquad F(G(X,F(Y))) \qquad G(F(X),G(X,Y)) \qquad \ldots
\end{displaymath}
We call them {\it functorial expressions}.  Of course, not every such
expression denotes a functor.  For example, $G(X,X)$ is not a proper
functor if $G$ is covariant in one argument and contravariant in the
other.  It is a functor if $G$ is covariant in both arguments or
contravariant in both arguments, or if $G$ does not depend on one of the
arguments.

In order to cope with these situations uniformly, we introduce two new
variances: fixed-variance and free-variance.  We say that a functor
$F(X)$ is {\it fixed-variant} in $X$ if $F$ is not functorial in $X$, that is,
$F$ maps objects to objects, but not morphisms.  We also say that a
functor $F(X)$ is {\it free-variant} in $X$ if $F$ does not depend on $X$.
Therefore, when $G:\cC \times \cC \rightarrow \cC$ is covariant in the
first argument and contravariant in the second, $G(X,X)$ is a
fixed-variant functor.

Let us introduce the symbols for variances.
\begin{definition}
Let {\bf Var} be the set of variances $\{ +,\; -,\; \bot,\; \top \}$:
$+$ for covariance, $-$ for contravariance, $\bot$ for free-variance
and $\top$ for fixed-variance.
\end{definition}

The next definition is extending the notion of opposite categories.
\begin{definition}
Let $\cC$ be a category.
\begin{enumerate}
\item $\cC^+$ is $\cC$ itself.
\item $\cC^-$ is the opposite category of $\cC$.
\item $\cC^\bot$ is the category which has only one object and only one
morphism (i.e.\ the identity of the one object).  We may call the
category {\it one point category}.
\item $\cC^\top$ is the category which has the same objects as $\cC$ but
no morphisms except identities. \qed
\end{enumerate}
\end{definition}
In this way, we can regard the variances as functions mapping categories
to categories (i.e.\ ${\bf Cat} \rightarrow {\bf Cat}$, where ${\bf
Cat}$ is the category of (small) categories) or we can even regard them
as a monoid acting on {\bf Cat}.

\begin{definition}
\label{def-var-monoid}
$\pair{{\bf Var},\bullet}$ is a commutative monoid with unit $+$, where
the monoid operation $\bullet: {\bf Var}\times {\bf Var} \rightarrow
{\bf Var}$ is defined by the following table.
\begin{displaymath}
\begin{array}{c|c|c|c|c}
\;\bullet\; & \;\bot\; & \;+\; & \;-\; & \;\top\; \\
\hline
\bot & \bot & \bot & \bot & \bot \\
\hline
+ & \bot & + & - & \top \\
\hline
- & \bot & - & + & \top \\
\hline
\top & \bot & \top & \top & \top
\end{array}
\end{displaymath}
\end{definition}

\begin{proposition}
\label{prop-var-monoid}
$\pair{{\bf Var},\bullet}$ is a monoid acting on {\bf Cat}, that is,
$\cC^{u\bullet v} = (\cC^u)^v$ for any $u,v \in {\bf Var}$. \\
{\bf Proof:} We have to show $(\cC^-)^- = \cC$ (i.e.\ the opposite of
opposite is itself) and so on, but they are trivial.
\end{proposition}

Since we will deal with many variable functors and they are functors
from products of categories $\cC \times \cdots \times \cC$ to a category
$\cC$, {\bf Var} action for product categories should be investigated.
\begin{proposition}
\label{prop-var-prod}
{\bf Var} action on {\bf Cat} distributes over products, that is, for
any categories $\cC$ and $\cD$
\begin{displaymath}
(\cC \times \cD)^u \iso \cC^u \times \cD^u.
\end{displaymath}
{\bf Proof:} In case $u$ is $-$, it says that the opposite of the
product category is isomorphic to the product of the opposite categories
and it is the case because $\pair{f,g}^{\rm op}: \pair{A,B} \rightarrow
\pair{C,D}$ $\iff$ $\pair{f,g}: \pair{C,D} \rightarrow \pair{A,B}$
$\iff$ $f: C \rightarrow A$ and $g: D \rightarrow B$ $\iff$ $f^{\rm op}:
A \rightarrow C$ and $g^{\rm op}: B \rightarrow D$ $\iff$ $\pair{f^{\rm
op},g^{\rm op}}: \pair{A,B} \rightarrow \pair{C,D}$.  The other cases
are trivial.
\end{proposition}

We need one more preparation before talking about mixed variant
functors.  The category $\cC^\top$ can be embedded into $\cC^+$ as well
as into $\cC^-$ and they themselves are embedded into $\cC^\bot$, that
is we have the following embedding functors.
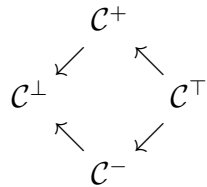
\begin{displaymath}
\begin{array}{ccccc}
& & \cC^+ & & \\
& \swarrow & & \nwarrow & \\
\cC^\bot & & & & \cC^\top \\
& \nwarrow & & \swarrow & \\
& & \cC^- & &
\end{array}
\end{displaymath}
We introduce a partial order on ${\bf Var}$ to respect these
embeddings.
\begin{definition}
$\lleq$ is a partial order on {\bf Var} such that $\bot \lleq {+} \lleq
\top$ and $\bot \lleq {-} \lleq \top$.
\end{definition}
From the definition, it is clear that
\begin{proposition}
If $u \lleq v$, there is an embedding functor $e_{u,v}: \cC^v
\rightarrow \cC^u$.
\end{proposition}

Now we can start talking about mixed variant functors and their
calculus.  As a function is associated with an arity (simply a natural
number), a mixed variant functor over a particular category $\cC$ is
associated with a {\it varity} which is a sequence of variances.  For
example, binary functor $G: \cC \times \cC \rightarrow \cC$ which is
contravariant in the first argument and covariant in the second is a
functor with a varity $-+$.
\begin{definition}
\label{def-pseudo-functor}
A mixed variant functor $F$ of varity $v_1\ldots v_n$ is a (covariant)
functor from $\cC^{v_1}\times \cdots \times \cC^{v_n}$ to $\cC$.
\end{definition}
When we are given a \indexed{{\bf Var}^\ast} set $\Gamma$ of primitive
mixed variant functors, where $F \in \Gamma_{v_1\ldots v_n}$ is a
functor of varity $v_1\ldots v_n$, we would like to establish how
to combine these primitive functors and get more complex
functors like $H(G(X,Y),F(X))$.

Firstly, we extend the action of variances on categories to that on
functors.  For example, from a contravariant functor $F: \cC^-
\rightarrow \cC$ we get a covariant functor $F^-: \cC \rightarrow \cC^-$
as $F^-(f: A \rightarrow B) \defeq F(f^{\rm op}: B \rightarrow A)$.
\begin{definition}
\label{def-fun-var-act}
For a functor $F: \cC \rightarrow \cD$,
\begin{enumerate}
\item a functor $F^+ :\cC^+ \rightarrow \cD^+$ is $F$ itself,
\item a functor  $F^- :\cC^- \rightarrow \cD^-$ is given by
$F^-(A) = A$ for an object $A$ in $\cC$ and $F^-(f^{\rm op}) =
F(f)^{\rm op}$ for a morphism $f$ in $\cC$,
\item a functor $F^\bot:\cC^\bot \rightarrow \cD^\bot$ is the
identity functor since both $\cC^\bot$ and $\cD^\bot$ are
the one point category, and
\item a functor $F^\top:\cC^\top \rightarrow \cD^\top$ has the same
object mapping as $F$ but no morphism mapping.
\end{enumerate}
\begin{displaymath}
\begin{picture}(65,50)
\put(10,25){\makebox(0,0){$F:\cC \rightarrow \cD$}}
\put(50,45){\makebox(0,0){$F^\top: \cC^\top \rightarrow \cD^\top$}}
\put(50,25){\makebox(0,0){$F^-: \cC^- \rightarrow \cD^-$}}
\put(50,5){\makebox(0,0){$F^\bot: \cC^\bot \rightarrow \cD^\bot$}}
\put(15,30){\vector(4,3){20}}
\put(20,25){\vector(1,0){15}}
\put(15,20){\vector(4,-3){20}}
\put(80,0){\makebox(0,0)[b]{\qed}}
\end{picture}
\end{displaymath}
\end{definition}
This definition is forced from the definition of $\cC^u$, so ${\bf Var}$
action has more structure.

\begin{proposition}
For a functor $F: \cC \rightarrow \cD$, $(F^u)^v = F^{u\bullet v}$. \\
{\bf Proof:} We have to check this for all the combinations of $u$ and
$v$.  For example, $(F^-)^\top = F^\top$ is true because $F^-$ only
changes the mapping of morphisms but $(F^-)^\top$ forgets it.
\end{proposition}

We have the two propositions which give us the basis for combining mixed
variant functors.
\begin{proposition}
\label{prop-fun-comp}
For functors $F: \cC^{u_1}\times\cdots\times\cC^{u_n} \rightarrow
\cC^{v_1}\times\cdots\times\cC^{v_m}$ and $G:
\cC^{v_1}\times\cdots\times\cC^{v_m} \rightarrow \cC$ (i.e.\ varity
$v_1\ldots v_m$), $G\circ F$ is a functor of
$\cC^{u_1}\times\cdots\times\cC^{u_n} \rightarrow \cC$ (i.e. varity
$u_1\ldots u_n$).
\begin{displaymath}
\begin{picture}(110,10)
\put(15,5){\makebox(0,0){$\cC^{u_1}\times\cdots\times\cC^{u_n}$}}
\put(40,7){\makebox(0,0)[b]{$F$}}
\put(65,5){\makebox(0,0){$\cC^{v_1}\times\cdots\times\cC^{v_m}$}}
\put(90,7){\makebox(0,0)[b]{$G$}}
\put(105,5){\makebox(0,0){$\cC$}}
\put(30,5){\vector(1,0){20}}
\put(80,5){\vector(1,0){20}}
\end{picture}
\end{displaymath}
{\bf Proof:} Trivial from the definition of composition of functors.
\end{proposition}

\begin{proposition}
\label{prop-fun-pair}
For functors $F_1: \cC^{u_1}\times\cdots\times\cC^{u_n} \rightarrow
\cC^{v_1}, \ldots, F_m:\cC^{u_1}\times\cdots\times\cC^{u_n} \rightarrow
\cC^{v_m}$, $\pair{F_1,\ldots,F_m}$ is a functor of
$\cC^{u_1}\times\cdots\times\cC^{u_n} \rightarrow
\cC^{v_1}\times\cdots\times\cC^{v_m}$.
\begin{displaymath}
\begin{picture}(145,50)
\put(15,25){\makebox(0,0){$\cC^{u_1}\times\cdots\times\cC^{u_n}$}}
\put(35,40){\makebox(0,0)[b]{$F_1$}}
\put(35,35){\makebox(0,0){$\vdots$}}
\put(35,27){\makebox(0,0)[b]{$F_i$}}
\put(35,20){\makebox(0,0){$\vdots$}}
\put(35,10){\makebox(0,0)[t]{$F_m$}}
\put(50,45){\makebox(0,0){$\cC^{v_1}$}}
\put(50,35){\makebox(0,0){$\vdots$}}
\put(50,25){\makebox(0,0){$\cC^{v_i}$}}
\put(50,15){\makebox(0,0){$\vdots$}}
\put(50,5){\makebox(0,0){$\cC^{v_m}$}}
\put(20,30){\vector(2,1){25}}
\put(30,25){\vector(1,0){15}}
\put(20,20){\vector(2,-1){25}}
\put(60,25){\makebox(0,0){$\implies$}}
\put(80,25){\makebox(0,0){$\cC^{u_1}\times\cdots\times\cC^{u_n}$}}
\put(105,27){\makebox(0,0)[b]{$\pair{F_1,\ldots,F_m}$}}
\put(130,25){\makebox(0,0){$\cC^{v_1}\times\cdots\times\cC^{v_m}$}}
\put(95,25){\vector(1,0){20}}
\end{picture}
\end{displaymath}
{\bf Proof:} It is trivial from the definition of products in {\bf Cat}.
\end{proposition}
The two propositions allow us to combine functors only if the source and
target categories match exactly.  For example, $F: \cC \rightarrow \cC$
and $G: \cC^- \rightarrow \cC$ cannot be composed into $G\circ F$.
Therefore, we have to first convert functors of
$\cC^{v_1}\times\cdots\times\cC^{v_n} \rightarrow \cC$ into those of
$\cC^{v'_1}\times\cdots\times\cC^{v'_n} \rightarrow \cC^u$.  There are
two ways to do so.

Firstly, from definition~\ref{def-fun-var-act}, functor $F$ of
varity $v_1\ldots v_n$, that is $F$ is a functor of
$\cC^{v_1}\times\cdots\times\cC^{v_n} \rightarrow \cC$, into
\begin{displaymath}
F^u:(\cC^{v_1}\times\cdots\times\cC^{v_n})^u \rightarrow \cC^u \iso
\cC^{v_1\bullet u}\times\cdots\times\cC^{v_n\bullet u} \rightarrow \cC^u
\end{displaymath}
The isomorphism is from proposition~\ref{prop-var-prod} and
proposition~\ref{prop-var-monoid}.

The other way of conversion is using embedding functors and coercing
functors into greater variances (e.g.\ covariant functor can be a fixed
variant functor).
\begin{definition}
\label{def-fun-coerce}
If $u_1 \lleq v_1,\ldots, u_n \lleq v_n$ and $F$ is a functor of
varity $u_1\ldots u_n$, we can coerce it to a functor of varity
$v_1\ldots v_n$ by
\begin{displaymath}
F|_{u_1\ldots u_n}^{v_1\ldots v_n} \defeq
F\circ(e_{u_1,v_1}\times\cdots\times e_{u_n,v_n}).
\end{displaymath}
\begin{displaymath}
\begin{picture}(70,30)
\put(25,25){\makebox(0,0){$\cC^{v_1}\times\cdots\times\cC^{v_n}$}}
\put(23,15){\makebox(0,0)[r]{$e_{u_1,v_1}\times\cdots\times
e_{u_n,v_n}$}}
\put(25,5){\makebox(0,0){$\cC^{u_1}\times\cdots\times\cC^{u_n}$}}
\put(50,7){\makebox(0,0)[b]{$F$}}
\put(65,5){\makebox(0,0){$\cC$}}
\put(25,22.5){\vector(0,-1){15}}
\put(40,5){\vector(1,0){22.5}}
\end{picture}
\end{displaymath}
We may write $F|^{v_1\ldots v_n}$ when $u_1\ldots u_n$ is obvious.
\end{definition}

Let us now define the composition of mixed variant functors.
\begin{definition}
\label{def-mixed-fun-comp}
Let $F$ be a functor of varity $u_1\ldots u_n$ and
$G_1,\ldots,G_n$ be functors of varity $v_{11}\ldots
v_{1m},\ldots,v_{n1}\ldots v_{nm}$, respectively.  Then we have a
functor $F[G_1,\ldots,G_n]$ of varity $w_1\ldots w_m$ where $w_i
\defeq u_1\bullet v_{1i} \lub \cdots \lub u_n\bullet
v_{ni}$\footnote{{\bf Var} is a commutative semiring with unit: $\lub$ as
its addition and $\bullet$ as its multiplication.  If we express
varities as vectors, then varity of $F[G_1,\ldots,G_n]$ can be computed
by the following matrix multiplication.
\begin{displaymath}
(w_1,\ldots,w_m) = (u_1,\ldots,u_n) \left(
\begin{array}{ccc}
v_{11} & \ldots & v_{1m} \\
\vdots & \ddots & \vdots \\
v_{n1} & \ldots & v_{nm} \\
\end{array}
\right)
\end{displaymath}
}.  The definition of the functor is
\begin{displaymath}
F[G_1,\ldots,G_n] \defeq F \circ \pair{G_1^{u_1}|^{w_1\ldots
w_m},\ldots,G_n^{u_n}|^{w_1\ldots w_m}}
\end{displaymath}
{\bf Proof of well-definedness:} $G_i$ is a functor of
$\cC^{v_{i1}}\times\cdots\times\cC^{v_{im}} \rightarrow \cC$.
$G_i^{u_i}$ is a functor of $\cC^{u_i\bullet
v_{i1}}\times\cdots\times\cC^{u_i\bullet v_{im}} \rightarrow \cC^{u_i}$.
Since $u_i\bullet v_{i1} \lleq w_1, \ldots, u_i\bullet v_{im} \lleq
w_m$, from definition~\ref{def-fun-coerce}
$G_i^{u_i}|^{w_1\ldots w_m}$ is a functor of
$\cC^{w_1}\times\cdots\times\cC^{w_m} \rightarrow \cC^{u_i}$.  From
proposition~\ref{prop-fun-pair} 
$\pair{G^{u_1}|^{w_1\ldots w_m},\ldots,G^{u_n}|^{w_1\ldots w_m}}$ is a
functor of $\cC^{w_1}\times\cdots\times\cC^{w_m} \rightarrow
\cC^{u_1}\times\cdots\times\cC^{u_n}$.  Therefore, from
proposition~\ref{prop-fun-comp},
\begin{displaymath}
F\circ \pair{G^{u_1}|^{w_1\ldots w_m},\ldots,G^{u_n}|^{w_1\ldots w_m}}:
\cC^{w_1}\times\cdots\times\cC^{w_m} \rightarrow \cC
\end{displaymath}
is a functor of varity $w_1\ldots w_m$.
\end{definition}
The variances of $G_1,\ldots,G_n$ are appropriately modified according
to the varity of $F$ and then the least upper bound is taken so that we
can pair them together.

We need some lemmas to show the associativity of the composition.
\begin{lemma}
Let $u \lleq v$ and $F$ be a functor of varity $w_1\ldots w_n$.
\begin{enumerate}
\item For any $w$, $u\bullet w \lleq v\bullet w$ and $(e_{u,v})^w =
e_{u\bullet w,v\bullet w}$.
\item The following diagram commutes.
\begin{displaymath}
\begin{picture}(85,30)
\put(30,25){\makebox(0,0){$\cC^{v\bullet
w_1}\times\cdots\times\cC^{v\bullet w_n}$}}
\put(28,15){\makebox(0,0)[r]{$e_{u,v}^{w_1}\times\cdots\times e_{u,v}^{w_n}$}}
\put(30,5){\makebox(0,0){$\cC^{u\bullet
w_1}\times\cdots\times\cC^{u\bullet w_n}$}}
\put(60,27){\makebox(0,0)[b]{$F^v$}}
\put(60,15){\makebox(0,0){\commute}}
\put(60,7){\makebox(0,0)[b]{$F^u$}}
\put(75,25){\makebox(0,0){$\cC^v$}}
\put(77,15){\makebox(0,0)[l]{$e_{u,v}$}}
\put(75,5){\makebox(0,0){$\cC^v$}}
\put(47.5,25){\vector(1,0){22.5}}
\put(47.5,5){\vector(1,0){22.5}}
\put(30,22.5){\vector(0,-1){15}}
\put(75,22.5){\vector(0,-1){15}}
\end{picture}
\end{displaymath}
In other words, the action of {\bf Var} on functors is natural with
respect to the partial order $\lleq$.
\item $e_{u,v}\circ F^v = F^u|^{w_1\bullet v\ldots w_n\bullet v}$
\item $F|^{u_1\ldots u_n}|^{v_1\ldots v_n} = F|^{u_1\lub v_1\ldots
u_n\lub v_n}$
\item $(F\circ G)^u = F^u\circ G^u$
\item $\pair{F_1,\ldots,F_n}^u = \pair{F_1^u,\ldots,F_n^u}$
\item $(F|^{v_1\ldots v_n})^u = F^u|^{v_1\bullet u\ldots v_n\bullet u}$
\end{enumerate}
{\bf Proof:} We have to check any pairs of $u$ and $v$ 1 and 2 hold from
the definitions.  3 follows 2. 4, 5 and 6 are
easy to show and 7 follows them.
\begin{displaymath}
\displaylines{
\qquad (F|^{v_1\ldots v_n})^u = (F\circ (e_{w_1,v_1} \times \cdots \times
e_{w_n,v_n}))^u \hfill \cr
\qquad \llap{${}={}$} F^u \circ (e_{w_1,v_1} \times \cdots \times
e_{w_n,v_n})^u = F^u \circ ((e_{w_1,v_1})^u \times \cdots \times
(e_{w_n,v_n})^u) \hfill \cr
\qquad \llap{${}={}$} F^u \circ (e_{w_1\bullet u,v_1\bullet u} \times
\cdots \times e_{w_n\bullet u,v_n\bullet u}) = F^u|^{v_1\bullet u\ldots
v_n\bullet u} \hfill {\qed} \cr}
\end{displaymath}
\end{lemma}

\begin{proposition}
\label{prop-fun-comp-assoc}
Let $F$, $G_1,\ldots,G_n$ and $H_1,\ldots,H_m$ be functors of the
following varities:
\begin{displaymath}
\begin{array}{c}
F: u_1\ldots u_n \\
G_1: v_{11}\ldots v_{1m},\quad \ldots \quad, G_n: v_{n1}\ldots v_{nm} \\
H_1: w_{11}\ldots w_{1l},\quad \ldots \quad, H_n: w_{m1}\ldots w_{ml} \\
\end{array}
\end{displaymath}
Then, the following equality between functors holds:
\begin{displaymath}
(F[G_1,\ldots,G_n])[H_1,\ldots,H_m] =
F[G_1[H_1,\ldots,H_m],\ldots,G_n[H_1,\ldots,H_m]].
\end{displaymath}
{\bf Proof:} Let varity $a_1\ldots a_m$, $b_{11}\ldots
b_{1l},\ldots,b_{n1}\ldots b_{nl}$ and $c_1\ldots c_l$ be
\begin{displaymath}
\displaylines{
\qquad (a_1,\ldots,a_m) \defeq (u_1,\ldots,u_n) \left(
\begin{array}{ccc}
v_{11} & \ldots & v_{1m} \\
\vdots & \ddots & \vdots \\
v_{n1} & \ldots & v_{nm} \\
\end{array}
\right), \hfill \cr
\qquad \left(
\begin{array}{ccc}
b_{11} & \ldots & b_{1l} \\
\vdots & \ddots & \vdots \\
b_{n1} & \ldots & b_{nl} \\
\end{array}
\right) \defeq  \left(
\begin{array}{ccc}
v_{11} & \ldots & v_{1m} \\
\vdots & \ddots & \vdots \\
v_{n1} & \ldots & v_{nm} \\
\end{array}
\right) \left(
\begin{array}{ccc}
w_{11} & \ldots & w_{1l} \\
\vdots & \ddots & \vdots \\
w_{m1} & \ldots & w_{ml} \\
\end{array}
\right), \quad \mbox{and} \hfill \cr
\qquad (c_1,\ldots,c_l) \defeq (u_1,\ldots,u_n) \left(
\begin{array}{ccc}
v_{11} & \ldots & v_{1m} \\
\vdots & \ddots & \vdots \\
v_{n1} & \ldots & v_{nm} \\
\end{array}
\right) \left(
\begin{array}{ccc}
w_{11} & \ldots & w_{1l} \\
\vdots & \ddots & \vdots \\
w_{m1} & \ldots & w_{ml} \\
\end{array}
\right). \hfill \cr}
\end{displaymath}
Then,
\begin{displaymath}
\displaylines{
\qquad (F[G_1,\ldots,G_n])[H_1,\ldots,H_m] \hfill \cr
\qquad \llap{${}={}$} F\circ \pair{G_1^{u_1}|^{a_1\ldots
a_m},\ldots,G_n^{u_n}|^{a_1\ldots a_m}} \circ
\pair{H_1^{a_1}|^{c_1\ldots c_l},\ldots,H_m^{a_m}|^{c_1\ldots c_l}}
\hfill \cr
\qquad \llap{${}={}$} F\circ \pair{G_1^{u_1}|^{a_1\ldots
a_m} \circ \pair{H_1^{a_1}|^{c_1\ldots c_l},\ldots},\ldots} \hfill \cr
\qquad \llap{${}={}$} F\circ \pair{G_1^{u_1} \circ (e_{v_{11}\bullet
u_1,a_1}\times\cdots\times e_{v_{1m}\bullet u_1,a_m}) \circ
\pair{H_1^{a_1}|^{c_1\ldots c_l},\ldots},\ldots} \hfill \cr
\qquad \llap{${}={}$} F\circ \pair{G_1^{u_1} \circ
\pair{e_{v_{11}\bullet u_1,a_1} \circ H_1^{a_1}|^{c_1\ldots
c_l},\ldots},\ldots} \hfill \cr
\qquad \llap{${}={}$} F\circ \pair{G_1^{u_1} \circ
\pair{H^{v_11\bullet u_1}|^{w_{11}\bullet v_1\bullet u_1\ldots
w_{1l}\bullet v_1\bullet u_1}|^{c_1\ldots c_l},\ldots},\ldots} \hfill \cr
\qquad \llap{${}={}$} F\circ \pair{G_1^{u_1} \circ
\pair{H^{v_11\bullet u_1}|^{c_1\ldots c_l},\ldots},\ldots} \hfill \cr}
\end{displaymath}
whereas
\begin{displaymath}
\displaylines{
\qquad F[G_1[H_1,\ldots,H_m],\ldots,G_n[H_1,\ldots,H_m]] \hfill \cr
\qquad \llap{${}={}$} F \circ \pair{(G_1\circ
\pair{H_1^{v_{11}}|^{b_{11}\ldots
b_{1l}},\ldots,H_m^{v_{1m}}|^{b_{11}\ldots b_{1l}}})^{u_1}|^{c_1\ldots
c_l},\ldots} \hfill \cr
\qquad \llap{${}={}$} F \circ \pair{G_1^{u_1} \circ
\pair{H_1^{v_{11}\bullet u_1}|^{b_{11}\bullet u_1\ldots
b_{1l}\bullet u_1}|^{c_1\ldots c_l},\ldots},\ldots} \hfill \cr
\qquad \llap{${}={}$} F \circ \pair{G_1^{u_1} \circ
\pair{H_1^{v_{11}\bullet u_1}|^{c_1\ldots c_l},\ldots},\ldots} \hfill \cr}
\end{displaymath}
Therefore, the proposition holds.
\end{proposition}
Mixed variant functors form an algebraic theory but with an extra
structure.  Mixed variant functors are functions between four sorts,
$+$, $-$, $\bot$ and $\top$.

Finally, we can show what functorial expressions like $H(G(X,Y),F(X))$
mean.  Let $\Gamma$ be a \indexed{{\bf Var}^\ast} set of primitive
functor names\footnote{From now on, we distinguish names (or
symbols) from what they denote.}.  Then, we define
\begin{displaymath}
\displaylines{
\qquad {\bf FE}(\Gamma) \defeq \parbox[t]{4.5in}{the set of terms
constructed from $\Gamma$ and variables like term algebras.} \hfill \cr
\qquad {\bf CFE}(\Gamma) \defeq \{\; \lambda(X_1,\ldots,X_n).E \;\mid\;
\parbox[t]{3in}{$E\in {\bf FE}(\Gamma)$ and $X_1,\ldots,X_n$
includes all the variables in $E$ $\;\}$} \hfill \cr}
\end{displaymath}
We call elements of ${\bf FE}(\Gamma)$ {\it functorial expressions} over
$\Gamma$ and call elements of ${\bf CFE}(\Gamma)$ {\it closed functorial
expressions} over $\Gamma$.  As we can convert any lambda closed term in
a term algebra to a morphism in the corresponding algebraic theory, we
can convert any closed functorial expression into functors.  Let $\xi$
be an assignment of functor symbols in $\Gamma_{v_1\ldots v_n}$ to real
functors of varity $v_1\ldots v_n$ in $\cC$, that is, $\xi$ is a
\indexed{{\bf Var}^\ast} function such that
\begin{displaymath}
\xi_{v_1\ldots v_n}: \Gamma_{v_1\ldots v_n} \rightarrow {\bf
Funct}(\cC^{v_1}\times\cdots\times\cC^{v_n},\cC)
\end{displaymath}
where ${\bf Funct}(\cD,\cE)$ is the category of functors from $\cD$ to
$\cE$.  We may write $\xi F$ for $\xi_{v_1\ldots v_n}(F)$.  We can
extend assignment $\xi$ to that over ${\bf CFE}(\Gamma)$ as follows
\begin{displaymath}
\xi(\lambda(X_1,\ldots,X_n).E) \defeq \xi_{X_1,\ldots,X_n}(E)
\end{displaymath}
where $\xi_{X_1,\ldots,X_n}$ assigns expressions in ${\bf
FE}(\Gamma)$ which have $X_1,\ldots,X_n$ as variables to functors
of $n$ variables and is defined by
\begin{displaymath}
\displaylines{
\qquad \xi_{X_1,\ldots,X_n}(X_i) \defeq \Pi^n_i:
\cC^\bot\times\cdots\cC^+\times\cdots\cC^\bot \rightarrow \cC \hfill \cr
\qquad \xi_{X_1,\ldots,X_n}(F(E_1,\ldots,E_n)) \defeq \xi
F[\xi_{X_1,\ldots,X_n}(E_1),\ldots,\xi_{X_1,\ldots,X_n}(E_n)] \hfill \cr}
\end{displaymath}
where $\Pi^n_i$ is the $i$th projection of the $n$ fold product of $\cC$.

\begin{example}
Let $\Gamma$ consist of a unary functor symbol $F$ which is covariant, a
functor symbol $G$ of varity $++$ and a functor symbol $H$ of varity
$-+$.  Let $\xi$ be an assignment of $\Gamma$.  Then,
$\lambda(X,Y).H(G(X,Y),F(X))$ denotes the following functor by $\xi$.
\begin{displaymath}
\xi(\lambda(X,Y).H(G(X,Y),F(X))) = \xi H[\xi
G[\Pi^2_1,\Pi^2_2],\xi F[\Pi^2_1]]
\end{displaymath}
The varity of $\Pi^2_1$ is $+\bot$, that of $\xi F$ is $+$ and,
therefore, from definition~\ref{def-mixed-fun-comp}, $\xi F[\Pi^2_1]$
has varity $+\bot$.  The varity of $\Pi^2_1$ is $+\bot$, that of
$\Pi^2_2$ is $\bot+$, that of $\xi G$ is $++$ and from
definition~\ref{def-mixed-fun-comp} the varity of $\xi
G[\Pi^2_1,\Pi^2_2]$ is
\begin{displaymath}
({+}, {+}) \left(
\begin{array}{ccc}
{+} &\; & {\bot} \\
{\bot} &\; & {+} \\
\end{array}
\right) = ({+}\bullet{+}\lub{+}\bullet{\bot},
{+}\bullet{\bot}\lub{+}\bullet{+}) = ({+}\lub{\bot}, {\bot}\lub{+}) =
({+}, {+})
\end{displaymath}
Finally, the varity of the whole functor is
\begin{displaymath}
({-}, {+}) \left(
\begin{array}{ccc}
{+} &\; & {+} \\
{+} &\; & {\bot} \\
\end{array}
\right) = ({-}\bullet{+}\lub{+}\bullet{+},
{-}\bullet{+}\lub{+}\bullet{\bot}) = ({-}\lub{+}, {-}\lub{\bot}) =
({\top}, {-}),
\end{displaymath}
so it is fixed-variant in $X$ and contravariant in $Y$.
\end{example}

The last proposition in this section is to establish the relationship
between the syntactic substitution of functorial expressions and the
composition of functors defined in
definition~\ref{def-mixed-fun-comp}.
\begin{proposition}
\label{prop-fun-comp-subst}
Let
\begin{displaymath}
\lambda(X_1,\ldots,X_n).E \qquad \mbox{and} \qquad
\lambda(Y_1,\ldots,Y_m).E_1,\; \ldots,\; \lambda(Y_1,\ldots,Y_m).E_n
\end{displaymath}
be closed functorial expressions.  Then,
\begin{displaymath}
\displaylines{
\qquad \xi(\lambda(X_1,\ldots,X_n).E)[\xi(\lambda(Y_1,\ldots,Y_m).E_1),\ldots,
\xi(\lambda(Y_1,\ldots,Y_m).E_n)]\hfill \cr
\hfill {} = \xi(\lambda(Y_1,\ldots,Y_m).E[E_1/X_1,\ldots,E_n/X_n])
\qquad \cr}
\end{displaymath}
where $E[E_1/X_1,\ldots,E_n/X_n]$ is $E$ in which $X_1,\ldots,X_n$ are
replaced by $E_1,\ldots,E_n$, respectively. \\
{\bf Proof:} We can easily prove it by structural induction on $E$ using proposition~\ref{prop-fun-comp-assoc}
\end{proposition}
From this correspondence, for closed functorial expression $K$ of $n$
variables and $L_1,\ldots,L_n$ of $m$ variables, we write
\begin{displaymath}
K[L_1,\ldots,L_n]
\end{displaymath}
for a closed functorial expression of $m$ variables which is obtained by
replacing $n$ variables in $K$ by $L_1,\ldots,L_n$.  Then, the
proposition is
\begin{displaymath}
\xi(K[L_1,\ldots,L_n]) = \xi K[\xi L_1,\ldots,\xi L_n].
\end{displaymath}

\section[Signatures of Categorical Specification Language]{Signatures of
Categorical Specification\\ Language}
\label{sec-csl-sig}

A specification normally consists of a {\it signature}, which says what
kind of sorts there are and what kinds of operations there are, and a
set of {\it sentences} (or {\it equations}), which give properties of
the operations.  A specification defines a class of {\it models} which
have what the signature says and satisfies the sentences.  Therefore, in
order to define a specification language, we have to define what its
signatures are, what its sentences are and what its models are.
However, it is often convenient to define, first, models without being
constrained by sentences and, then, define a {\it satisfaction relation}
between a statement and a model determining whether the statement is
true in the model or not.  A model which satisfies all the sentences is
called a {\it theory model}.  See {\it institutions}
\cite{goguen-burstall-83} for a categorical abstract definition of what
specification languages are.

A CSL signature will be divided into three parts; in the first part,
we will declare some names for functors, which will serve as parameterized
data types (or data type constructors); in the second part, we will
declare some names for natural transformations, which will serve as
polymorphic functions over the parameterized data types; and in the
third part, we will declare some names for factorizers (or mediating
morphisms), which will be necessary to put initial or final constrains
on the data types.

The first part can be presented as a \indexed{{\bf Var}^\ast} set
$\Gamma$.  $F$ in $\Gamma_{v_1\ldots v_n}$ is said to have varity
$v_1\ldots v_n$.  We may write $F(v_1,\ldots,v_n)$ to indicate this.

$\Gamma$ looks almost like an equational signature for algebraic
specification languages.  It is as if {\bf Var} were the set of sorts
and $\Gamma$ were a set of operations over the sorts.  The only
difference is that $\Gamma$ is not a \indexed{{\bf Var}^\ast\times{\bf
Var}} set but simply a \indexed{{\bf Var}^\ast} set.  This is because we
can apply {\bf Var} to functors to get other functors as we explained in
definition~\ref{def-fun-var-act}, so it is sufficient to give $\Gamma$ as
a \indexed{{\bf Var}^\ast} set.

Note that $\Gamma$ is one sorted.  Each signature describes only one
category.  However, because we are dealing with one level higher
objects, it has the power to describe more than one data type (or sort)
inside one signature.  We will illustrate this later in this section.

In the second part, a CSL signature introduces some symbols for natural
transformations.  Normally, a natural transformation is defined as
follows: given two functors $F,G:\cC \rightarrow \cD$, a natural
transformation $\alpha: F \natrightarrow G$ is a function which assigns
to each $A \in |\cC|$ a morphism $\alpha_A: F(A) \rightarrow G(A)$ such
that for any morphism $f: A \rightarrow B$ in $\cC$ the following
diagram commutes.
\begin{displaymath}
\sqdiagram{F(A)}{\alpha_A}{G(A)}{F(f)}{G(f)}{F(B)}{\alpha_B}{G(B)}{}
\end{displaymath}

As we have seen in section~\ref{sec-fun-calc}, closed functorial
expressions provide more complicated functors constructed from primitive
functors in $\Gamma$.  Let $\Lambda$ be ${\bf CFE}(\Gamma)$.  We index
the set of natural transformations by two closed functorial expressions,
that is, the second part of a CSL signature is given by a
\indexed{\Lambda\times \Lambda} set $\Delta$; $\alpha \in \Delta_{K,L}$
will denote a natural transformation $\xi K \natrightarrow \xi L$, where
$\xi K$ and $\xi L$ will be the denotations of $K$ and $L$,
respectively.  We may write $\alpha: K \natrightarrow L$ to indicate
this.  Since a natural transformation should go between functors of the
same number of variables, $\Delta_{K,L}$ should be empty when $K$ and
$L$ have different number of variables.  Even if they have the same
number of variables, their variances may be different.  In that case, we
take the least upper bound of two variances.

The third and the final part of a CSL signature introduces symbols for
factorizers (or mediating morphisms).  In general, a factorizer is an
isomorphism associated with an adjunction.  If $F \adjoint G$ where $F:
\cC \rightarrow \cD$ and $G: \cD \rightarrow \cC$, the factorizer $\psi$
gives the following natural isomorphism between hom sets.
\begin{displaymath}
\psi: \Hom{\cD}{F(A)}{B} \stackrel{\iso}{\longrightarrow} \Hom{\cC}{A}{G(B)}
\end{displaymath}
For example, the factorizer associated with the binary product functor
`prod' ($: \cC \times \cC \rightarrow \cC$) is `pair' which gives for
any two morphisms $f: C \rightarrow A$ and $g: C \rightarrow B$ a
morphism ${\rm pair}(f,g): C \rightarrow {\rm prod}(A,B)$.  We can write
this situation as the following rule.
\begin{displaymath}
\logicrule{C \rubberrightarrow{f} A \qquad C \rubberrightarrow{g}
B}{C \rubberrightarrow{{\rm pair}(f,g)} {\rm prod}(A,B)}
\end{displaymath}

In CSL, factorizers are given by a \indexed{(\Lambda\times
\Lambda)^\ast\times(\Lambda\times \Lambda)} set $\Psi$ ($\Lambda =
{\bf CFE}(\Gamma)$). The index $\scriptstyle (\Lambda\times \Lambda)^\ast$
specifies the type of morphisms to which a factorizer can be applied and
the index $\scriptstyle \Lambda\times \Lambda$ specifies the type of morphisms
obtained by applying the factorizer.  For
\begin{displaymath}
\psi \in \Psi_{\pair{K_1,L_1}\ldots\pair{K_m,L_m},\pair{K,L}},
\end{displaymath}
we may write it as the following rule.
\begin{displaymath}
\logicrule{f_1: K_1 \rightarrow L_1 \quad\ldots\quad f_m: K_m
\rightarrow L_m}{\psi(f_1,\ldots,f_m): K \rightarrow L}
\end{displaymath}
where $f_1,\ldots,f_n$ are auxiliary names of morphisms introduced for
this rule.  As we have restricted the indexed set $\Delta$, $\Psi$ should be
indexed by functors of the same number of variables, and also we take
the least upper bound of the variances and regard it as the overall
variance.

Hence, we come to the definition of CSL signatures.
\begin{definition}
A {\it CSL signature} is a triple $\pair{\Gamma,\Delta,\Psi}$, where $\Gamma$ is a \indexed{{\bf Var}^\ast} set, $\Delta$ is a
\indexed{\Lambda\times \Lambda} set (where $\Lambda = {\bf
CFE}(\Gamma)$) for natural transformations and $\Psi$ is a
\indexed{(\Lambda\times \Lambda)^\ast\times(\Lambda\times \Lambda)} set for
factorizers.
%\footnote{If we had introduced the terminology of dependent
%data types \cite{burstall-lampson-84}, we could have presented CSL
%signatures as the following {\it dependent product}:
%\begin{displaymath}
%\displaylines{
%\qquad \Gamma\in{\bf Set}({\bf Var}^\ast) \depprod ({\bf Set}({\bf
%CFE}(\Gamma)\times{\bf CFE}(\Gamma)) \times {\bf Set}(({\bf
%CFE}(\Gamma)\times{\bf CFE}(\Gamma))^\ast \times {} \hfill \cr
%\hfill ({\bf CFE}(\Gamma)\times{\bf CFE}(\Gamma)))) \qquad \cr}
%\end{displaymath}
%where ${\bf Set}(I)$ denotes the class of all $I$-indexed sets.}
As a restriction to the triple, $\Delta_{K,L}$ should be empty if $K$ and $L$
has the different number of variables and
$\Psi_{\pair{\pair{K_1,L_1}\ldots\pair{K_m,L_m},\pair{K,L}}}$ should
also be empty if $K_i$, $L_i$, $K$ and $L$ do not have the same number
of variables.
\end{definition}

As an example, we will give a CSL signature for cartesian closed
categories.
\begin{example}
\label{ex-ccc-sig}
A cartesian closed category can be characterized as a category having
three special functors: terminal object `1' (which is a constant
functor), binary product `prod' and exponential `exp'.  Therefore,
\begin{displaymath}
\Gamma_{()} = \{ {\rm 1} \}, \qquad \Gamma_{{+}{+}} = \{ {\rm prod} \}, \qquad
\Gamma_{{-}{+}} = \{ {\rm exp} \},
\end{displaymath}
where `$()$' denotes the empty string in ${\bf Var}^\ast$.  The rest of
$\Gamma_s$ are empty.  We sometimes write the index set $\Gamma$ as
\begin{displaymath}
\{ {\rm 1},\; {\rm prod}(+,+),\; {\rm exp}(-,+) \}.
\end{displaymath}

The product functor `prod' is associated with two natural
transformations which give projection morphisms.
\begin{displaymath}
\displaylines{
\pi_1: \lambda(X,Y).{\rm prod}(X,Y) \natrightarrow \lambda(X,Y).X\cr
\pi_2: \lambda(X,Y).{\rm prod}(X,Y) \natrightarrow \lambda(X,Y).Y\cr}
\end{displaymath}
If there is no ambiguity, we may write them by listing their components
as follows:
\begin{displaymath}
\displaylines{
{\pi_1}_{A,B}: {\rm prod}(A,B) \rightarrow A\cr
{\pi_2}_{A,B}: {\rm prod}(A,B) \rightarrow B\cr}
\end{displaymath}
We might even omit subscripts from ${\pi_1}_{A,B}$ and ${\pi_2}_{A,B}$.
The exponential functor `exp' is associated with one natural
transformation which gives evaluation morphisms.
\begin{displaymath}
{\rm ev}: \lambda(X,Y).{\rm prod}({\rm exp}(X,Y),X) \natrightarrow
\lambda(X,Y).Y
\end{displaymath}
Note that the variance of these two functors is $\top+$.  Therefore,
$\Delta$ is
\begin{displaymath}
\begin{array}{l@{\;=\;}l}
\Delta_{\lambda(X,Y).{\rm prod}(X,Y),\lambda(X,Y).X} & \{ \pi_1 \} \\
\Delta_{\lambda(X,Y).{\rm prod}(X,Y),\lambda(X,Y).Y} & \{ \pi_2 \} \\
\Delta_{\lambda(X,Y).{\rm prod}({\rm exp}(X,Y),X),\lambda(X,Y).Y} & \{
{\rm ev} \}
\end{array}
\end{displaymath}
For any other combinations of closed functorial expression $K$ and $L$,
$\Delta_{K,L}$ is empty.

Finally, there are three factorizers for the three functors: `!' for
`1', `pair' for `prod' and `curry' for `exp'.
\begin{displaymath}
\logicrule{}{{\rm !}: \lambda(X).X \rightarrow \lambda(X).{\rm 1}}
\end{displaymath}
\begin{displaymath}
\logicrule{f: \lambda(X,Y,Z).Z \rightarrow \lambda(X,Y,Z).X \qquad
g: \lambda(X,Y,Z).Z \rightarrow \lambda(X,Y,Z).Y}{{\rm
pair}(f,g): \lambda(X,Y,Z).Z \rightarrow \lambda(X,Y,Z).{\rm
prod}(X,Y)}
\end{displaymath}
\begin{displaymath}
\logicrule{h: \lambda(X,Y,Z).{\rm prod}(Z,X) \rightarrow
\lambda(X,Y,Z).Y}{{\rm curry}(h): \lambda(X,Y,Z).Z
\rightarrow \lambda(X,Y,Z).{\rm exp}(X,Y)}
\end{displaymath}
If there is no ambiguity, we might write these rules down as in
figure~\ref{fig-ccc-sig}, where we summarize the definition of the CSL
signature for cartesian closed categories.
\myfigure{fig-ccc-sig}{CSL Signature for Cartesian Closed Categories}{
\begin{center}
\begin{tabular}{rl}
{\small Functors ($\Gamma$)} & ${\rm 1}\quad{\rm prod}({+},{+})\quad{\rm
exp}({-},{+})$ \\[2ex]
{\small Natural Transformations ($\Delta$)} & $\pi_1:{\rm prod}(A,B)
\rightarrow A$ \\
& $\pi_2:{\rm prod}(A,B) \rightarrow B$ \\
& ${\rm ev}:{\rm prod}({\rm exp}(A,B),A) \rightarrow B$ \\[2ex]
{\small Factorizers ($\Psi$)} & $\logicrule{}{{\rm !}:A
\rightarrow {\rm 1}}$ \\[2ex]
& $\logicrule{f:C \rightarrow A \qquad g: C \rightarrow
B}{{\rm pair}(f,g):C \rightarrow {\rm prod}(A,B)}$ \\[2ex]
& $\logicrule{h:{\rm prod}(C,A) \rightarrow B}{{\rm
curry}(h):C \rightarrow {\rm exp}(A,B)}$ \\
\end{tabular}
\end{center}
}
We will omit the tedious formal definition of $\Psi$ as an indexed set.
\end{example}

As we mentioned earlier, a CSL signature is one sorted, but because it
handles higher objects, it is no less powerful than a many sorted
equational signature.  Let us demonstrate this.  An equational signature
is given by a pair $\pair{S,\Sigma}$ where $S$ is a set (of sort names)
and $\Sigma$ is an \indexed{S^\ast\times S} set (of operator names).  We
will translate it to a corresponding CSL signature
$\pair{\Gamma,\Delta,\Psi}$.  Since functors play a role of sorts, we
declare a constant functor for each sort in $S$.
\begin{displaymath}
\Gamma_{()} = S,
\end{displaymath}
and we have a binary product functor to deal with sequences of
sorts,\footnote{We could have {\it n}-ary product functors for all natural
numbers as well, but since we can represent them using a binary
one, we only declare the binary one.} and the terminal object for constants.
\begin{displaymath}
\Gamma_{{+}{+}} = \{\; {\rm prod} \;\} \qquad \mbox{and} \qquad \Gamma_{()}
= \{\; 1 \;\}
\end{displaymath}
Operations will be translated to natural transformations: for each
operation $o \in \Sigma_{s_1\ldots s_n,s}$, we have a natural
transformation of the same name.
\begin{displaymath}
o: {\rm prod}(s_1,{\rm prod}(\ldots,{\rm prod}(s_{n-1},s_n)))
\rightarrow s
\end{displaymath}
that is,
\begin{displaymath}
\Delta_{\lambda().{\rm prod}(s_1,{\rm prod}(\ldots,{\rm
prod}(s_{n-1},s_n))),\lambda().s} = \Sigma_{s_1\ldots s_n,s}
\end{displaymath}
We also have two projections for `prod' in $\Delta$ and $\Psi$ has only
one factorizer `pair' for pairing.  It is easy to see that the CSL
signature $\pair{\Gamma,\Delta,\Psi}$ corresponds to the equational
signature $\pair{S,\Sigma}$.
\begin{proposition}
A many-sorted equational signature can be represented by a CSL
signature.
\end{proposition}
It is interesting to know that we represented a many-sorted equational
signature by a category with products because an {\it algebraic
theory} can exactly be given as a category with products \cite{lawvere-63}.

Let us make CSL signatures form a category by extending a pre-CSL
signature morphism in a natural way.  Intuitively, a signature morphism
(not only in CSL but in general) does some renamings of symbols and/or
some mergings of symbols.
\begin{definition}
A {\it CSL signature morphism} $\sigma$ from a CSL signature
$\pair{\Gamma,\Delta,\Psi}$ to a CSL signature
$\pair{\Gamma',\Delta',\Psi'}$ is a triple
$\pair{\sigma',\sigma'',\sigma'''}$ consisting of
\begin{enumerate}
\item a \indexed{{\bf Var}^\ast} function $\sigma'_{v_1\ldots v_n}:
\Gamma_{v_1\ldots v_n} \rightarrow \Gamma'_{v_1\ldots v_n}$ for mapping
functor names,
\item a \indexed{\Lambda\times \Lambda} function 
(where $\Lambda = {\bf CFE}(\Gamma)$) $\sigma''_{K,L}:
\Delta_{K,L} \rightarrow \Delta'_{\sigma' K,\sigma' L}$ for mapping
natural transformation names, and
\item a \indexed{(\Lambda\times \Lambda)^\ast\times(\Lambda\times
\Lambda)} function
\begin{displaymath}
\displaylines{
\quad\sigma'''_{\pair{\pair{K_1,L_1}\ldots\pair{K_m,L_m},\pair{K,L}}}:\hfill\cr
\hfill \Psi_{\pair{\pair{K_1,L_1}\ldots\pair{K_m,L_m},\pair{K,L}}} \rightarrow
\Psi'_{\pair{\pair{\sigma' K_1,\sigma' L_1}\ldots\pair{\sigma' K_m,\sigma'
L_m},\pair{\sigma' K,\sigma' L}}}\quad\cr}
\end{displaymath}
for mapping factorizer names.
\end{enumerate}
We often write $\sigma F$ for $\sigma'_{v_1\ldots v_n}(F)$, $\sigma
\alpha$ for $\sigma''_{K,L}(\alpha)$ and $\sigma \psi$ for
$\sigma'''_{\pair{\pair{K_1,L_1}\ldots,\pair{K,L}}}(\psi)$.
\end{definition}

\begin{definition}
\label{def-csig}
The {\it category of CSL signatures} {\bf CSig} has CSL signatures as its
objects and CSL signature morphisms as its morphisms.  The identity
morphism on a CSL signature $\pair{\Gamma,\Delta,\Psi}$ consists of the
corresponding identity functions for the components and the composition
of CSL morphisms is given by combining the component-wise compositions
as indexed functions.  This clearly forms a category.
\end{definition}

\section{Structures of Categorical Specification Language}
\label{sec-cls-mod}

In this section, we will define {\it CSL structures}.  A CSL signature
specifies symbols for functors, natural transformations and factorizers,
so intuitively, a CSL structure is a category associated with these
functors, natural transformations and factorizers.
\begin{definition}
\label{def-csl-mod}
Given a CSL signature $\pair{\Gamma,\Delta,\Psi}$, a {\it CSL structure}
$\pair{\cC,\xi}$ is a small category $\cC$ together with an assignment $\xi$
\begin{enumerate}
\item $\xi$ assigns each functor name of varity $v_1\ldots v_n$ to a
functor $\cC^{v_1}\times\cdots\times\cC^{v_n} \rightarrow
\cC$.\footnote{We could say
$\xi$ is a \indexed{{\bf Var}^\ast} function
\begin{displaymath}
\xi_{v_1\ldots v_n}: \Gamma_{v_1\ldots v_n} \rightarrow {\bf
Funct}(\cC^{v_1}\times\cdots\times\cC^{v_n},\cC).
\end{displaymath}}
As we have seen in section~\ref{sec-fun-calc}, $\xi$ can be extended to
the assignment of closed functorial expressions to functors.
\item $\xi$ assigns each natural transformation name $\alpha \in
\Delta_{K,L}$, where $K$ and $L$ are closed functorial expressions of $n$
variables, to a set of $\cC$ morphisms
\begin{displaymath}
\xi\alpha_{A_1,\ldots,A_n}: \xi K(A_1,\ldots,A_n) \rightarrow \xi
L(A_1,\ldots,A_n)
\end{displaymath}
for arbitrary $\cC$ objects $A_1,\ldots,A_n$.\footnote{We could
express it as a \indexed{\Lambda\times \Lambda} function (where $\Lambda
= {\bf CFE}(\Gamma)$)
\begin{displaymath}
\xi_{K,L}: \Delta_{K,L} \rightarrow {\bf Nat}(\xi
K|^{\top\ldots\top},\xi L|^{\top\ldots\top}).
\end{displaymath}}
\item Each factorizer symbol $\psi \in
\Psi_{\pair{\pair{K_1,L_1}\ldots,\pair{K,L}}}$, where
$K_1,L_1$,\ldots, $K_m,L_m$, $K,L$ are closed functorial expressions of
$n$ variables, is assigned to a set of {\bf Set} functions
\begin{displaymath}
\displaylines{
\qquad \xi\psi_{A_1,\ldots,A_n}: \prod_{i=1}^m \Hom{\cC}{\xi
K_i(A_1,\ldots,A_n)}{\xi L_i(A_1,\ldots,A_n)} \hfill \cr
\hfill {} \rightarrow \Hom{\cC}{\xi K(A_1,\ldots,A_n)}{\xi
L(A_1,\ldots,A_n)} \qquad \cr}
\end{displaymath}
for arbitrary $\cC$ objects $A_1,\ldots,A_n$. \qed
\end{enumerate}
\end{definition}
Note that we do not assign a natural transformation symbol to a natural
transformation, but it is mapped to a set of morphisms and whether they
form a natural transformation or not is left to be stated by equations.

\begin{example}
Let $\pair{\Gamma,\Delta,\Psi}$ be the CSL signature for cartesian
closed categories defined in example~\ref{ex-ccc-sig}.  Then, any
cartesian closed category is a CSL structure of this signature by obvious
assignment of the symbols to the functors, natural transformations and
factorizers.  However, the converse is not true.  We can have a category
which has three functors, three natural-transformation-look-alikes and
three factorizer-look-alikes.  CSL structures do not require mapping natural
transformation symbols to natural transformations but only to a set of
morphisms, nor do they require mapping factorizer symbols to
factorizers.  The factorizer-look-alikes may not give unique morphisms
or may not commute some diagrams.
\end{example}

CSL structure morphisms are defined simply as a kind of homomorphisms.  They
keep the structure nicely.
\begin{definition}
Given a CSL signature $\pair{\Gamma,\Delta,\Psi}$, a {\it CSL structure
morphism} from a CSL structure $\pair{\cC,\xi}$ to another
$\pair{\cD,\zeta}$ is a covariant functor $T:\cC \rightarrow \cD$ such
that
\begin{enumerate}
\item for any $F$ in $\Gamma_{v_1\ldots v_n}$
\begin{displaymath}
T\circ \xi F = \zeta F \circ (T^{v_1}\times\cdots\times T^{v_n})
\end{displaymath}
holds,
\begin{displaymath}
\begin{picture}(70,30)
\put(25,25){\makebox(0,0){$\cC^{v_1}\times\cdots\times\cC^{v_n}$}}
\put(23,15){\makebox(0,0)[r]{$T^{v_1}\times\cdots\times T^{v_n}$}}
\put(25,5){\makebox(0,0){$\cD^{v_1}\times\cdots\times\cD^{v_n}$}}
\put(50,27){\makebox(0,0)[b]{$\xi F$}}
\put(50,15){\makebox(0,0){\commute}}
\put(50,3){\makebox(0,0)[t]{$\zeta F$}}
\put(65,25){\makebox(0,0){$\cC$}}
\put(65,5){\makebox(0,0){$\cD$}}
\put(25,22.5){\vector(0,-1){15}}
\put(40,25){\vector(1,0){22.5}}
\put(40,5){\vector(1,0){22.5}}
\put(65,22.5){\vector(0,-1){15}}
\end{picture}
\end{displaymath}
\item for any $\alpha \in \Delta_{K,L}$ and for any $\cC$ objects
$A_1,\ldots,A_n$
\begin{displaymath}
T(\xi \alpha_{A_1,\ldots,A_n}) = \zeta \alpha_{T(A_1),\ldots,T(A_n)},
\end{displaymath}
and
\item for any $\psi \in
\Psi_{\pair{K_1,L_1}\ldots\pair{K_m,L_m},\pair{K,L}}$, for any $\cC$
objects $A_1,\ldots,A_n$ and for any $\cC$ morphisms $f_i:\xi
K_i(A_1,\ldots,A_n) \rightarrow \xi L_i(A_1,\ldots,A_n)$ ($i=1,\ldots,m$),
\begin{displaymath}
T(\xi \psi_{A_1,\ldots,A_n}(f_1,\ldots,f_m)) = \zeta
\psi_{T(A_1),\ldots,T(A_n)}(T(f_1),\ldots,T(f_m)). \qed
\end{displaymath}
\end{enumerate}
\end{definition}

Hence, the category of CSL structures is:
\begin{definition}
\label{def-CMod}
For a CSL signature $\pair{\Gamma,\Delta,\Psi}$, the category of CSL
structures, ${\bf CMod}(\pair{\Gamma,\Delta,\Psi})$, has CSL structures
as objects and CSL structure morphisms as morphisms; the identity morphism
on $\pair{\cC,\xi}$ is the identity functor ${\bf I}_\cC$ and the
composition of morphisms is the composition of their underlying
functors.
\end{definition}

\section{Functorial Calculus (revisit)}
\label{sec-fun-calc-2}

In section~\ref{sec-fun-calc} we saw functorial expressions denote
functors.  In this section we will see an expression involving natural
transformation symbols and factorizer symbols denote a set of
morphisms.\footnote{Some may want it to denote a natural transformation,
but the treatment of factorizers seems very complicated to do so.}
For example, under the signature of cartesian closed categories given in
example~\ref{ex-ccc-sig}, what should the following expression denote?
\begin{displaymath}
\pi_1\circ {\rm prod}({\rm curry}(\pi_2),{\bf I})
\end{displaymath}
We even have a problem for expressions like $\pi_1\circ\pi_2$ because
$\pi_1$ and $\pi_2$ are not projections of the same product: $\pi_2$ is
of ${\rm prod}(A,{\rm prod}(B,C))$ and $\pi_1$ is of ${\rm prod}(B,C)$.
Actually, natural transformations are polymorphic like ML functions
are.\footnote{Conversely, ML polymorphic functions are natural
transformations.  The author has not yet seen the definite statement of
this, but it is a folklore among computer scientists dealing with
category theory.  This fact is rarely used in practice, but it sometimes
helps to understand the behaviour of polymorphic functions.  For
example, any ML function $f$ of type $\alpha {\rm list} \rightarrow {\rm
int}$ should never depend on elements in the list but only to the length
of list.  As another example, if $f$ is of type $\alpha {\rm list}
\rightarrow \alpha {\rm list}$ and if we apply it to an integer list
$[1,4,3,5]$ and get $[5,1,3]$, we know the result of applying $f$ to
$[2,8,6,10]$ (each element is doubled) without actually applying it.
The result should be $[10,2,6]$, i.e.\ each element of the result is
doubled as well.} As we defined in definition~\ref{def-csl-mod}, $\pi_1$
denotes a set of morphisms.
\begin{displaymath}
{\xi \pi_1}_{A,B}: \xi {\rm prod}(A,B) \rightarrow A
\end{displaymath}
We have to figure out for each occurrence of $\pi_1$ what $A$ and $B$
are.

\begin{definition}
\label{def-exp}
For a CSL signature $\pair{\Gamma,\Delta,\Psi}$, we have a set ${\bf
Exp}(\Gamma,\Delta,\Psi)$ of {\it CSL expressions} defined by the
following BNF.
\begin{displaymath}
e \coloneq {\bf I} \;\mid\; e_1\circ e_2 \;\mid\; \alpha \;\mid\;
\psi(e_1,\ldots,e_m) \;\mid\; F(e_1,\ldots,e_n) \;\mid\; f
\end{displaymath}
where $\alpha \in \Delta$, $\psi \in \Psi$ and $f$ is a variable for
morphisms.  We also have a set ${\bf AExp}(\Gamma,\Delta,\Psi)$ of {\it
CSL annotated expressions} defined by the following BNF.
\begin{displaymath}
e \coloneq {}
\begin{array}[t]{l}
{\bf I}\anno{K} \;\mid\; e_1\circ e_2 \;\mid\;
\alpha\anno{K_1,\ldots,K_n} \;\mid\;
\psi\anno{K_1,\ldots,K_n}(e_1,\ldots,e_m) \;\mid\; \\
F(e_1,\ldots,e_n) \;\mid\; f \\
\end{array}
\end{displaymath}
where $K_1,\ldots,K_n$ are closed functorial expressions over $\Gamma$.
\end{definition}
It is trivial to see that for each annotated expression there is a
corresponding expression (i.e.\ forgetting all the annotations,
$\anno{\ldots}$), which we call {\it skeleton} of the annotated
expression.  We are going to type-check an expression first and, then,
we determine its denotation.  Annotated expressions are used to remember
the type-checking information inside expressions.  We will give the typing
rules for annotated expressions and show that every expression has the
most general annotated expression and we take the type of this annotated
expression as the type of the expression.

First, we define the notion of unification.
\begin{definition}
A closed functorial expression $K$ of $n$ variables is said to be {\it
more general} than a closed functorial expression $K'$ of $m$ variables
if there are closed functorial expressions $K_1,\ldots,K_n$ of $m$
variables such that
\begin{displaymath}
K[K_1,\ldots,K_n] \equiv K'
\end{displaymath}
where $\equiv$ is the equivalence relation ignoring variable renaming.
(Trivially, $K \equiv K'$ implies $\xi K = \xi K'$.)
\end{definition}

\begin{definition}
Closed functorial expressions $K$ and $L$ are said to be {\it unifiable}
when there is a closed functorial expression $K'$ such that $K'$ is less
general than $K$ as well as $L$.
\end{definition}

\begin{proposition}
If closed functorial expressions $K$ and $L$ are unifiable, then there
is a most general unification, that is, there exist $K_1,\ldots,K_n$ and
$L_1,\ldots,L_m$ such that $K[K_1,\ldots,K_n] \equiv L[L_1,\ldots,L_m]$
and for any $K'$ which is less general than $K$ and $K'$ there are
$K'_1,\ldots,K'_l$ such that $K' \equiv
K[K_1,\ldots,K_n][K'_1,\ldots,K'_l]$. \\
{\bf Proof:} Same as ordinary unification of terms.
\end{proposition}

Let us now define the type of annotated expressions.
\begin{definition}
\label{def-anno-exp-type}
Let $\pair{\Gamma,\Delta,\Psi}$ be a CSL signature.  An annotated
expression $e$ has a type $\rho \vdash e: K \rightarrow L$ when it can
be derived from the following rules, where $\rho$ is a given assignment
of each morphism variable to its type and $K$ and $L$ are closed
functorial expressions.
\begin{enumerate}
\item For the identity, $\rho \vdash {\bf I}\anno{K}: K \rightarrow K$.
\item For the composition,
\begin{displaymath}
\logicrule{\rho \vdash e_1: K' \rightarrow L \qquad \rho \vdash e_2: K
\rightarrow K'}{\rho \vdash e_1\circ e_2: K \rightarrow L}.
\end{displaymath}
\item For a natural transformation $\alpha \in \Delta_{K,L}$, where $K$
and $L$ are of $n$ variables,
\begin{displaymath}
\rho \vdash \alpha\anno{K_1,\ldots,K_n}: K[K_1,\ldots,K_n] \rightarrow
L[K_1,\ldots,K_n]
\end{displaymath}
\item For a factorizer $\psi \in
\Psi_{\pair{\pair{K_1,L_1}\ldots\pair{K_m,L_m},\pair{K,L}}}$, where
$K_i,L_i,K,L$ are of $n$ variables,
\begin{displaymath}
\logicrule{\rho \vdash e_i: K_i[K'_1,\ldots,K'_n] \rightarrow
L_i[K'_1,\ldots,K'_n]}{\rho \vdash
\psi\anno{K'_1,\ldots,K'_n}(e_1,\ldots,e_m): K[K'_1,\ldots,K'_n] \rightarrow
L[K'_1,\ldots,K'_n]}
\end{displaymath}
\item For a functor $F \in \Gamma_{v_1\ldots v_n}$
\begin{displaymath}
\logicrule{\rho \vdash e_i: K_i \subrightarrow{v_i} L_i}{\rho \vdash
F(e_1,\ldots,e_n): F[K_1,\ldots,K_n] \rightarrow  F[L_1,\ldots,L_n]}
\end{displaymath}
where $e_i: K_i \subrightarrow{v_i} L_i$ is $e_i: K_i \rightarrow L_i$
if $v_i$ is $+$ or $\bot$, $e_i: L_i \rightarrow K_i$ if $v_i$ is $-$
and ${\bf I}: K_i \rightarrow K_i$ if $v_i$ is $\top$.
\item For a morphism variable $f$, $\rho \vdash f: \rho(f)$. \qed
\end{enumerate}
\end{definition}

\begin{definition}
We say an annotated expression $e$ is {\it more general} than $e'$ if
there exist closed functorial expressions $K_1,\ldots,K_n$ such that $e'
\equiv e[K_1,\ldots,K_n]$, where $e[K_1,\ldots,K_n]$ is $e$ with
all its annotations being composed with $K_1,\ldots,K_n$, e.g.\
$\alpha\anno{L_1,\ldots,L_m}[K_1,\ldots,K_n]$ is
$\alpha\anno{L_1[K_1,\ldots,K_n],\ldots,L_m[K_1,\ldots,K_n]}$.
\end{definition}

\begin{proposition}
If an annotated expression $e$ has a type $\rho \vdash e: K \rightarrow L$,
$e[K_1,\ldots,K_n]$ has the following type.
\begin{displaymath}
\rho[K_1,\ldots,K_n] \vdash e[K_1,\ldots,K_n]: K[K_1,\ldots,K_n] \rightarrow
L[K_1,\ldots,K_n]
\end{displaymath}
where $\rho[K_1,\ldots,K_n](f)$ is $K'[K_1,\ldots,K_n] \rightarrow
L'[K_1,\ldots,K_n]$ when $\rho(f)$ is $K' \rightarrow L'$. \\
{\bf Proof:} It can easily proved from definition~\ref{def-anno-exp-type}
by structural induction on $e$.
\end{proposition}

\begin{proposition}
\label{prop-exp-type}
Let $e \in {\bf Exp}(\Gamma,\Delta,\Psi)$ be an expression of a CSL
signature $\pair{\Gamma,\Delta,\Psi}$.  If there exists an annotated
expression $e'$ whose skeleton is $e$ and if it has a type, then there
exists a most general annotated expression which has a type and whose
skeleton is $e$. \\
{\bf Proof:} It can be proved by structural induction on $e$.  Here, we
present an algorithm of calculating a most general annotated expression.
\begin{enumerate}
\item If $e$ is ${\bf I}$, the most general annotated expression is
${\bf I}\anno{I}$, where $I$ is the identity closed functorial expression
$\lambda(X).X$.
\item If $e$ is $e_1\circ e_2$, from induction hypothesis, we have the
most general annotated expressions for $e_1$ and $e_2$.
\begin{displaymath}
\rho_1 \vdash e'_1: K \rightarrow K' \qquad \rho_2 \vdash e'_2: L
\rightarrow L'
\end{displaymath}
We unify $K$ with $L'$ achieving $K[K_1,\ldots,K_n] \equiv
L'[L_1,\ldots,L_m]$.  The most general annotated expression for $e$
and its type is
\begin{displaymath}
\rho \vdash e'_1[K_1,\ldots,K_n]\circ e'_2[L_1,\ldots,L_m]:
L[L_1,\ldots,L_m] \rightarrow K'[K_1,\ldots,K_n]
\end{displaymath}
where $\rho$ is the result of combining $\rho_1[K_1,\ldots,K_n]$ and
$\rho_2[L_1,\ldots,L_m]$.
\item If $e$ is a natural transformation $\alpha \in \Delta_{K,L}$, the
most general annotated expression and its type is
\begin{displaymath}
\rho \vdash \alpha\anno{\Pi^n_1,\ldots,\Pi^n_n}: K \rightarrow L,
\end{displaymath}
where $\Pi^n_i$ is the closed functorial expression
$\lambda(X_1,\ldots,X_n).X_i$.
\item If $e$ is $\psi(e_1,\ldots,e_m)$ for a factorizer $\psi \in
\Psi_{\pair{\pair{K_1,L_1}\ldots,\pair{K,L}}}$, from
induction hypothesis, we have annotated expressions $e'_i$ such that
\begin{displaymath}
\rho_i \vdash e'_i: K'_i \rightarrow L'_i.
\end{displaymath}
We unify $K'_i$ with $K_i$ and $L'_i$ with $L_i$.  If the unification is
successful, we have the most general annotated expression for $e$.
\begin{displaymath}
\displaylines{
\qquad \rho \vdash
\psi\anno{J_1,\ldots,J_n}(e'_1[J'_{11},\ldots],\ldots,e'_m[J'_{1m},\ldots]):
\hfill \cr
\hfill K[J_1,\ldots,J_n] \rightarrow L[J_1,\ldots,J_n] \qquad \cr}
\end{displaymath}
where $J_1,\ldots,J_n,J'_{11},\ldots$ are the substitution for $K_i$ and
$L_i$ obtained from the unification and $\rho$ is the result of
combining $\rho_i[J'_{1i},\ldots]$.
\item If $e$ is $F(e_1,\ldots,e_n)$, let the most general annotated
expressions for $e_i$ be
\begin{displaymath}
\rho_i \vdash e'_i: K_i \subrightarrow{v_i} L_i.
\end{displaymath}
Then, the most general annotated expression for $e$ and its type is
\begin{displaymath}
\rho \vdash F(e'_1,\ldots,e'_n): F[K_1,\ldots,K_n] \rightarrow
F[L_1,\ldots,L_n].
\end{displaymath}
\item If $e$ is a morphism variable $f$, its most general annotated
expression is itself and has the following type.
\begin{displaymath}
\rho \vdash f: \lambda(X,Y).X \rightarrow \lambda(X,Y).Y \rlap{\qquad
\qquad \qed}
\end{displaymath}
\end{enumerate}
\end{proposition}
From this proposition,
\begin{definition}
We define the type of an expression $e\in {\bf Exp}(\Gamma,\Delta,\Psi)$
to be the type of the most general annotated expression $e'$ whose
skeleton is $e$.
\end{definition}

Let us finally define the denotation of expressions in ${\bf
Exp}(\Gamma,\Delta,\Psi)$ when a CSL structure $\pair{\cC,\xi}$ is given.
Since each expression $e$ is associated uniquely to the most general
annotated expression $e'$ by proposition~\ref{prop-exp-type}, we can
define the denotation of $e$ to be that of $e'$ and define the
denotation of annotated expressions.
\begin{definition}
\label{def-anno-deno}
Let $\pair{\Gamma,\Delta,\Psi}$ be a CSL signature and $\pair{\cC,\xi}$
be a CSL structure.  For an annotated expression $e \in {\bf
AExp}(\Gamma,\Delta,\Psi)$ of type $\rho \vdash e: K \rightarrow L$
(where $K$ and $L$ are of $l$ variables), we define its denotation, $\xi
e$, to be a set of morphisms
\begin{displaymath}
(\xi e)_{A_1,\ldots,A_l}: \xi K(A_1,\ldots,A_l) \rightarrow \xi
L(A_1,\ldots,A_n)
\end{displaymath}
for any $\cC$ objects $A_1,\ldots,A_l$ and for any morphism variable
assignment $\omega$ (where $\omega(f,A_1,\ldots,A_l)$ gives a morphism
of type $\xi K'(A_1,\ldots,A_l) \rightarrow \xi L'(A_1,\ldots,A_l)$ when
$\rho(f)$ is $K' \rightarrow L'$).
\begin{enumerate}
\item For the identity,
\begin{displaymath}
(\xi {\bf I}\anno{K})_{A_1,\ldots,A_l} \defeq {\bf I}_{{\xi K}[A_1,\ldots,A_l]}
\end{displaymath}
\item For compositions,
\begin{displaymath}
(\xi e_1\circ e_2)_{A_1,\ldots,A_l} \defeq (\xi e_1)_{A_1,\ldots,A_l}
\circ (\xi e_2)_{A_1,\ldots,A_l}
\end{displaymath}
\item For natural transformations,
\begin{displaymath}
(\xi \alpha\anno{K_1,\ldots,K_n})_{A_1,\ldots,A_l} \defeq
\xi\alpha_{\xi K_1(A_1,\ldots,A_l),\ldots,\xi K_n(A_1,\ldots,A_l)}
\end{displaymath}
\item For factorizers,
\begin{displaymath}
\displaylines{
\qquad (\xi \psi\anno{K_1,\ldots,K_n}(e_1,\ldots,e_m))_{A_1,\ldots,A_l}
\defeq \hfill \cr
\hfill \psi_{\xi K_1(A_1,\ldots,A_l),\ldots,\xi K_n(A_1,\ldots,A_l)}((\xi
e_1)_{A_1,\ldots,A_l},\ldots,(\xi e_m)_{A_1,\ldots,A_l}) \qquad \cr}
\end{displaymath}
\item For functors,
\begin{displaymath}
(\xi F(e_1,\ldots,e_n))_{A_1,\ldots,A_l} \defeq \xi F((\xi
e_1)_{A_1,\ldots,A_l},\ldots,(\xi e_n)_{A_1,\ldots,A_l})
\end{displaymath}
\item For morphism variables,
\begin{displaymath}
(\xi f)_{A_1,\ldots,A_n} \defeq \omega(f,A_1,\ldots,A_l)
\end{displaymath}
\end{enumerate}
{\bf Proof of well-definedness:} We have to show, for example, for a
natural transformation $\alpha \in \Delta_{K,L}$, $(\xi
\alpha\anno{K_1,\ldots,K_n})_{A_1,\ldots,A_l}$ is a morphism from
\begin{displaymath}
\xi K[K_1,\ldots,K_n](A_1,\ldots,A_l) \qquad \mbox{to} \qquad \xi
L[K_1,\ldots,K_n](A_1,\ldots,A_l).
\end{displaymath}
This holds because from definition~\ref{def-csl-mod} $\xi\alpha_{\xi
K_1(A_1,\ldots,A_l),\ldots,\xi K_n(A_1,\ldots,A_l)}$
is a morphism from $\xi K(\xi K_1(A_1,\ldots,A_l),\ldots))$ to $\xi
L(\xi K_1(A_1,\ldots,A_l),\ldots)$ and from
proposition~\ref{prop-fun-comp-subst} it is from $\xi
K[K_1,\ldots,K_n](A_1,\ldots,A_l)$ to $\xi
L[K_1,\ldots,K_n](A_1,\ldots,A_l)$.
\end{definition}

\begin{example}
Let $\pair{\Gamma,\Delta,\Psi}$ be a CSL signature for cartesian closed
categories presented in example~\ref{ex-ccc-sig} and $\pair{\cC,\xi}$ be
a CSL structure of this signature where $\cC$ is a cartesian closed category
and $\xi$ is the standard assignment (i.e.\ the product symbol to the
product functor and so on).  Let us find out the denotation of ${\rm
ev}\circ{\rm pair}(f,\pi_2)$.  First, we have to find out the
corresponding most general annotated expression and its type by the
algorithm used to prove proposition~\ref{prop-exp-type}.
\begin{enumerate}
\item ${\rm ev}\circ{\rm pair}(f,\pi_2)$ is given by composing ${\rm ev}$
and ${\rm pair}(f,\pi_2)$, so we need to calculate the most general
annotated expressions for these two sub-expressions first.
\begin{enumerate}
\item ${\rm ev}$ is a natural transformation, and its most general
annotated expression is
\begin{displaymath}
\rho_1 \vdash {\rm ev}\anno{\Pi^2_1,\Pi^2_2}: {\rm prod}[{\rm exp},\Pi^2_1]
\rightarrow \Pi^2_2.
\end{displaymath}
\item ${\rm pair}(f,\pi_2)$ is given by applying the factorizer ${\rm pair}$
to $f$ and $\pi_2$.
\begin{enumerate}
\item `$f$' is a morphism variable, its most general annotated expression is
\begin{displaymath}
\rho_2 \vdash f: \Pi^2_1 \rightarrow \Pi^2_2.
\end{displaymath}
\item $\pi_2$ is a natural transformation, and its most general
annotated expression is
\begin{displaymath}
\rho_3 \vdash \pi_2\anno{\Pi^2_1,\Pi^2_1}: {\rm prod} \rightarrow \Pi^2_2.
\end{displaymath}
\end{enumerate}
${\rm pair}$ has the type
\begin{displaymath}
\pair{\pair{\Pi^3_3,\Pi^3_1}\pair{\Pi^3_3,\Pi^3_2},\pair{\Pi^3_3,{\rm
prod}[\Pi^3_1,\Pi^3_2]}}
\end{displaymath}
By unification, we get the most general annotated expression for ${\rm
pair}(f,\pi_2)$.
\begin{displaymath}
\displaylines{
\qquad \rho_4 \vdash {\rm pair}\anno{\Pi^3_3,\Pi^3_2,{\rm
prod}[\Pi^3_1,\Pi^3_2]}(f,\pi_2\anno{\Pi^3_1,\Pi^3_2}): \hfill \cr
\hfill {\rm prod}[\Pi^3_1,\Pi^3_2] \rightarrow {\rm
prod}[\Pi^3_3,\Pi^3_2] \qquad \cr}
\end{displaymath}
where $\rho_4$ maps $f$ to ${\rm prod}[\Pi^3_1,\Pi^3_2] \rightarrow
\Pi^3_3$.
\end{enumerate}
Unifying ${\rm prod}[\Pi^3_3,\Pi^3_2]$ and ${\rm prod}[{\rm
exp},\Pi^2_1]$, we get the most general annotated expression
for ${\rm ev} \circ {\rm pair}(f,\pi_2)$.
\begin{displaymath}
\displaylines{
\qquad \rho_5 \vdash {\rm ev}\anno{\Pi^3_2,\Pi^3_3} \circ {\rm pair}\anno{{\rm
exp}[\Pi^3_2,\Pi^3_3],\Pi^3_2,{\rm
prod}[\Pi^3_1,\Pi^3_2]}(f,\pi_2\anno{\Pi^3_1,\Pi^3_2}) 
\hfill \cr
\hfill : {\rm prod}[\Pi^3_1,\Pi^3_2] \rightarrow \Pi^3_3 \qquad \cr}
\end{displaymath}
where $\rho_5$ maps $f$ to ${\rm prod}[\Pi^3_1,\Pi^3_2] \rightarrow
{\rm exp}[\Pi^3_2,\Pi^3_3]$.
\end{enumerate}
From definition~\ref{def-anno-deno}, the denotation of this annotated
expression is a set of morphisms for objects $A$,
$B$ and $C$ and a morphism
variable assignment
\begin{displaymath}
\omega(f,A,B,C): \xi{\rm prod}(A,B) \rightarrow \xi{\rm exp}(B,C).
\end{displaymath}
\begin{enumerate}
\item $(\xi {\rm ev}[\Pi^3_2,\Pi^3_3])_{A,B,C} = \xi{\rm ev}_{B,C}$
\item $(\xi f)_{A,B,C} = \omega(f,A,B,C)$
\item $(\xi \pi_2[\Pi^3_1,\Pi^3_2])_{A,B,C} = {\xi\pi_2}_{A,B}$
\item $(\xi {\rm pair}[\ldots](f,\pi_2[\ldots]))_{A,B,C} = {}$ \\
${} \qquad \xi{\rm pair}_{\xi{\rm exp}(B,C),B,\xi{\rm
prod}(A,B)}(\omega(f,A,B,C),{\xi\pi_2}_{A,B})$
\item $(\xi {\rm ev}[\ldots]\circ{\rm
pair}[\ldots](f,\pi_2[\ldots]))_{A,B,C} = {}$ \\
${} \qquad \xi{\rm ev}_{B,C} \circ \xi{\rm pair}_{\xi{\rm
exp}(B,C),B,\xi{\rm prod}(A,B)}(\omega(f,A,B,C),{\xi\pi_2}_{A,B})$
\end{enumerate}
Therefore, the denotation of ${\rm ev}\circ{\rm pair}(f,\pi_2)$ is
\begin{displaymath}
\xi{\rm ev}_{B,C} \circ \xi{\rm pair}_{\xi{\rm exp}(B,C),B,\xi{\rm
prod}(A,B)}(\omega(f,A,B,C),{\xi\pi_2}_{A,B}) \rlap{\qquad\qquad \qed}
\end{displaymath}
\end{example}

\section{Sentences and Satisfaction Relation of Categorical
Specification Language}
\label{sec-csl-sen}

In this section, we will finish defining the specification language CSL
at last.  First, we define what a CSL sentence is.
\begin{definition}
A {\it CSL conditional equation} is a sequence of CSL expression pairs and a
CSL expression pair.  We usually write it as
\begin{displaymath}
e_1 = e'_1 \land \ldots \land e_n = e'_n \implies e = e',
\end{displaymath}
or simply $e = e'$ if the preceding sequence is empty.  To be typed, it
needs to share the same morphism variable environment, $e_i$ and $e'_i$
have to have the same type and $e$ and $e'$ have to have the same type.
We may write the types as follows:
\begin{displaymath}
\rho \vdash e_1 = e'_1: K_1 \rightarrow L_1 \land\ldots\land
e_n = e'_n: K_n \rightarrow L_n \implies e = e': K \rightarrow L
\end{displaymath}
We write ${\bf CEq}(\Gamma,\Delta,\Psi)$ for the set of all the CSL
conditional equations which can be typed.
\end{definition}

The CSL conditional equations are the CSL sentences.  We now have to
define the {\it satisfaction relation} for CSL.  We have separately
defined what CSL structures are and what CSL conditional equations are.  The
satisfaction relation connects these two together so that we can say a
CSL conditional equation holds or not in a particular CSL structure.
\begin{definition}
\label{def-csl-sat}
Let $\pair{\Gamma,\Delta,\Psi}$ be a CSL signature.  A CSL structure
$\pair{\cC,\xi}$ satisfies a CSL conditional equation
\begin{displaymath}
e_1 = e'_1 \land \ldots \land e_n = e'_n \implies e = e'
\end{displaymath}
having a type
\begin{displaymath}
\rho \vdash e_1 = e'_1: K_1 \rightarrow L_1 \land\ldots\land e_n = e'_n:
K_n \rightarrow L_n \implies e = e': K \rightarrow L,
\end{displaymath}
if and only if for any objects $A_1,\ldots,A_l$ and any morphism
variable assignment $\omega$ we have either
\begin{enumerate}
\item a CSL equation $e_i = e'_i$ such that
$(\xi e_i)_{A_1,\ldots,A_l} \not= (\xi e'_i)_{A_1,\ldots,A_l}$, or
\item $(\xi e)_{A_1,\ldots,A_l} = (\xi e')_{A_1,\ldots,A_l}$.
\end{enumerate}
We will write
\begin{displaymath}
\pair{\cC,\xi} \models e_1 = e'_1 \land \ldots \land e_n = e'_n \implies
e = e'
\end{displaymath}
when $\pair{\cC,\xi}$ satisfies this CSL conditional equation.
\end{definition}

We have defined the specification language, CSL: CSL signatures, CSL
structures, CSL conditional equations and CSL satisfaction relation.  We
could have defined it as an institution (see~\cite{goguen-burstall-83})
by defining {\bf CMod} as a contravariant functor and showing CSL
satisfaction condition.

Finally, let us finish presenting the CSL theory (i.e.\ a pair of CSL
signature and CSL conditional equations) of cartesian
closed categories.
\begin{example}
\label{ex-ccc-th}
We have presented the signature for cartesian closed categories in
example~\ref{ex-ccc-sig} (or figure~\ref{fig-ccc-sig}), so all we have
to do is to list the CSL conditional equations. (Note that they are not
conditional for this example.)
\begin{displaymath}
\displaylines{
\qquad\llap{1.} f = {\rm !} \hfill \cr
\qquad\llap{2.} {\rm 1} = {\bf I} \hfill \cr
\qquad\llap{3.} {\rm pair}(\pi_1,\pi_2) = {\bf I} \hfill \cr
\qquad\llap{4.} \pi_1\circ{\rm pair}(f,g) = f \hfill \cr
\qquad\llap{5.} \pi_2\circ{\rm pair}(f,g) = g \hfill \cr
\qquad\llap{6.} {\rm pair}(f,g)\circ h = {\rm pair}(f\circ h,g\circ h)
\hfill \cr
\qquad\llap{7.} {\rm prod}(f,g) = {\rm pair}(f\circ\pi_1,g\circ\pi_2)
\hfill \cr
\qquad\llap{8.} {\rm curry}({\rm ev}) = {\bf I} \hfill \cr
\qquad\llap{9.} {\rm ev}\circ{\rm curry}({\rm prod}(f,{\bf I})) = f
\hfill \cr
\qquad\llap{10.} {\rm curry}(f)\circ g = {\rm curry}(f\circ{\rm
prod}(g,{\bf I})) \hfill \cr
\qquad\llap{11.} {\rm exp}(f,g) = {\rm curry}(g\circ{\rm eval}\circ{\rm
prod}(f,{\bf I})) \hfill \cr}
\end{displaymath}
The naturality of $\pi_1$, $\pi_2$ and ${\rm ev}$ can be derived from
these equations.  For example, the naturality of $\pi_1$ is shown by
\begin{displaymath}
\pi_1\circ{\rm prod}(f,g) = \pi_1\circ{\rm
pair}(f\circ\pi_1,g\circ\pi_2) = f\circ\pi_1. \rlap{\qquad \qquad \qed}
\end{displaymath}
\end{example}

\section{Free Categories}
\label{sec-free-cat}

One of the major advantages of algebraic specification methods using
equations or conditional equations over other specification methods is
that any theory has an initial model (i.e.\ the initial object in the
category of models which satisfy the theory).  This also holds for CSL,
and in this section, we will construct an initial structure for a CSL
theory.  Remember that a CSL structure is a pair of a category and an
interpretation.  The category of an initial CSL structure corresponds to a
so-called {\it free category}.

Given a CSL signature $\pair{\Gamma,\Delta,\Psi}$ and a set $\Theta$ of
CSL conditional equations, we are going to define a special category
$\cC$ and an interpretation $\xi$.  For simplicity, we assume that
$\Gamma$ does not contain any free-variant functors.  (We can always get
such a signature by discarding free-variant arguments.  This does not
affect its semantics at all.)

\begin{definition}
\label{def-free-cat}
We say that a closed functorial expression is {\it ground} if it has no
variables, that is, its basic form is $\lambda().E$.  We take ground
closed functorial expressions as the objects of $\cC$.
\end{definition}

The definition of morphisms of $\cC$ is a little bit more complicated, so we
define them step by step.
\begin{definition}
A ground annotated expression is an annotated expression such that
\begin{enumerate}
\item all the annotation, $\anno{K_1,\ldots,K_n}$ consists of ground
closed functorial expressions, and
\item it does not contain any morphism variables. \qed
\end{enumerate}
\end{definition}

\begin{proposition}
If a ground annotated expression $e$ has a type $\rho \vdash e: K
\rightarrow L$, and both $K$ and $L$ are ground
functorial expressions. \\
{\bf Proof:} We can easily prove it from
definition~\ref{def-anno-exp-type} by structural induction.
\end{proposition}

The following will give us the basis of the morphisms in $\cC$.
\begin{definition}
For ground closed functorial expressions $K$ and $L$, we define
\begin{displaymath}
{\bf GExp}(K,L) \defeq \{\; e \mid \mbox{$e$ is ground and $\emptyset
\vdash e: K \rightarrow L$} \;\} \qed
\end{displaymath}
\end{definition}

To make ${\bf GExp}$ proper morphisms, we introduce a family of equivalence
relations $\equiv$ indexed by a pair of ground functorial expressions.  Each
$\equiv_{K,L}$ is an equivalence relation on ${\bf GExp}(K,L)$.
\begin{definition}
\label{def-eqv-gfe}
We define $\equiv$ to be the smallest relation satisfying the following
conditions.  (In the following, to simplify the presentation, we omit
indexes of $\equiv$ if there is no ambiguity.)
\begin{enumerate}
\item $\equiv$ is an equivalence relation, that is, reflexive, symmetric
and transitive.
\item If $e_1 \equiv_{K,L} e'_1$ and $e_2 \equiv_{K',K} e'_2$, then $e_1
\circ e_2 \equiv_{K',L} e'_1 \circ e'_2$.
\item If $e \in {\bf GExp}(K,L)$, then ${\bf I}\anno{L}\circ e \equiv e$
and $e \circ {\bf I}\anno{K} \equiv e$.
\item For a functor symbol $F \in \Gamma_{v_1\ldots v_n}$, if $e_1
\equiv e'_1$, \ldots and $e_n \equiv e'_n$, then $F(e_1,\ldots,e_n)
\equiv F(e'_1,\ldots,e'_n)$.
\item For a factorizer symbol $\psi \in
\Psi_{\pair{\pair{K'_1,L'_1}\ldots\pair{K'_l,L'_l},\pair{K',L'}}}$ and
ground functorial expressions $K_1,\ldots,K_n$, if $e_1 \equiv e'_1$, \ldots
and $e_l \equiv e'_l$, then
\begin{displaymath}
\psi\anno{K_1,\ldots,K_n}(e_1,\ldots,e_l) \equiv
\psi\anno{K_1,\ldots,K_n}(e'_1,\ldots,e'_l).
\end{displaymath}
\item Finally, for a conditional CSL equation $e_1 = e'_1 \land \ldots
\land e_n = e'_n \implies e = e' \in \Theta$ whose type is
\begin{displaymath}
\rho \vdash e_1 = e'_1: K'_1 \rightarrow L'_1 \land\ldots\land
e_n = e'_n: K'_n \rightarrow L'_n \implies e = e': K' \rightarrow L'
\end{displaymath}
with $\rho(f_i)$ is $K''_i \rightarrow L''_i$, ground functorial
expressions $K_1,\ldots,K_m$, and ground annotated expressions
\begin{displaymath}
e''_i \in {\bf GExp}(K''_i[K_1,\ldots,K_m],L''_i[K_1,\ldots,K_m]),
\end{displaymath}
if for all $j = 1,\ldots,n$
\begin{displaymath}
\displaylines{
\qquad (e_j[K_1,\ldots,K_m])[e''_1,\ldots,e''_l/f_1,\ldots,f_l] \equiv
{} \hfill \cr
\hfill (e'_j[K_1,\ldots,K_m])[e''_1,\ldots,e''_l/f_1,\ldots,f_l], \qquad
\cr}
\end{displaymath}
then
\begin{displaymath}
\displaylines{
\qquad (e[K_1,\ldots,K_m])[e''_1,\ldots,e''_l/f_1,\ldots,f_l] \equiv {}
\hfill \cr
\hfill (e'[K_1,\ldots,K_m])[e''_1,\ldots,e''_l/f_1,\ldots,f_l]. \qquad
\qed \cr}
\end{displaymath}
\end{enumerate}
\end{definition}

We can now define the morphisms of $\cC$.
\begin{contdefinition}{\ref{def-free-cat}}
The $\cC$ morphisms from $K$ to $L$ are the equivalence
classes of ${\bf GExp}(K,L)$ by $\equiv_{K,L}$, or simply,
\begin{displaymath}
\Hom{\cC}{K}{L} \defeq {\bf GExp}(K,L)/{\equiv_{K,L}}.
\end{displaymath}
We write $\eqvelm{e}$ for the equivalence class to which $e$ belongs.
\end{contdefinition}

\begin{proposition}
$\cC$ is a category. \\
{\bf Proof:} The identity morphism of a $\cC$ object $K$ is given
by $\eqvelm{{\bf I}\anno{K}}$.  The composition of $\eqvelm{e}: K
\rightarrow L$ and $\eqvelm{e'}: K' \rightarrow K$ is defined by
\begin{displaymath}
\eqvelm{e} \circ \eqvelm{e'} \defeq \eqvelm{e \circ e'}.
\end{displaymath}
The composition is trivially associative, and the satisfiability of the
absorption rules of the identities is guaranteed by the third condition of
the equivalence relation $\equiv$ defined in definition~\ref{def-eqv-gfe}.
\end{proposition}

Let us define an interpretation $\xi$ so that $\pair{\cC,\xi}$ is a CSL
structure of the CSL signature $\pair{\Gamma,\Delta,\Psi}$.
\begin{definition}
\label{def-free-xi}
The definition of $\xi$ is divided into three.
\begin{enumerate}
\item For a functor symbol $F \in \Gamma_{v_1\ldots v_n}$, $\xi F$ is a
functor $\cC^{v_1\ldots v_n} \rightarrow \cC$ defined by
\begin{displaymath}
\begin{array}{l}
\xi F(K_1,\ldots,K_n) \defeq F[K_1,\ldots,K_n)], \quad \mbox{and} \\
\xi F(\eqvelm{e_1},\ldots,\eqvelm{e_n}) \defeq
\eqvelm{F(e_1,\ldots,e_n)}. \\
\end{array}
\end{displaymath}
\item For a natural transformation symbol $\alpha \in \Delta_{K,L}$,
$\xi\alpha$ is
\begin{displaymath}
\xi\alpha_{K_1,\ldots,K_n} \defeq \eqvelm{\alpha\anno{K_1,\ldots,K_n}}.
\end{displaymath}
\item Finally, for a factorizer symbol $\psi \in
\Psi_{\pair{\pair{K_1,L_1}\ldots\pair{K_n,L_n},\pair{K,L}}}$, $\xi\psi$
is
\begin{displaymath}
\xi\psi_{K'_1,\ldots,K'_n}(\eqvelm{e_1},\ldots,
\eqvelm{e_n}) \defeq
\eqvelm{\psi\anno{K'_1,\ldots,K'_n}(e_1,\ldots,e_n)}. \qed
\end{displaymath}
\end{enumerate}
\end{definition}

\begin{proposition}
$\pair{\cC,\xi}$ is a CSL structure of $\pair{\Gamma,\Delta,\Psi}$.
Moreover, it is a theory structure of the CSL theory given by the set
$\Theta$ of CSL conditional equations. \\
{\bf Proof:} The condition 6 in definition~\ref{def-eqv-gfe} makes it
satisfy the conditional equations.
\end{proposition}

We have constructed a special CSL structure $\pair{\cC,\xi}$, and we will
next show that it is the initial object in the full-subcategory of ${\bf
CMod}(\pair{\Gamma,\Delta,\Psi})$ of all the structures satisfying $\Theta$.
\begin{theorem}
For any CSL structure satisfying $\Theta$, there is a unique CSL
structure morphism from $\pair{\cC,\xi}$. \\ {\bf Proof:} Let
$\pair{\cD,\zeta}$ be an arbitrary CSL structure satisfying $\Theta$.  Using
the denotation of annotated expressions on this structure (see
definition~\ref{def-anno-deno}), we define a functor $T$ from $\cC$ to
$\cD$ as follows:
\begin{displaymath}
T(K) \defeq \zeta K \quad \mbox{and} \quad T(\eqvelm{e}) \defeq
\zeta e.
\end{displaymath}
Note that, since $K$ is a ground functorial expression $\zeta K$ is an
$\cD$ object, and that, since $e$ is a ground annotated expression,
$\zeta e$ is a $\cD$ morphism (it does not need objects $A_1,\ldots,A_m$
or a morphism variable assignment $\omega$).  It is easy to see that $T$
is a (covariant) functor (note that we have to show the well-definedness
first).  It is also not so difficult to show that $T$ is a CSL structure
morphism and that it is the unique one from $\pair{\cC,\xi}$ to
$\pair{\cD,\zeta}$ (simply extending the result on algebraic
specifications).
\end{theorem}

The advantage of working in an initial CSL structure is that, if we show that
a conditional equation holds in the initial one, then we automatically know
that it holds in any CSL structure.  In chapter~\ref{ch-cpl}, we will define
the symbolic computation in categories and it can be regarded as the
computation in free categories.

\chapter{Categorical Data Types}
\label{ch-cdt}

In chapter~\ref{ch-csl}, we introduced categorical data types from a
point of view of a specification language defining categories.  Although
the specification language CSL is a rigorous language, it is rather
tedious and categoritians may never define categories in that way.  In
this chapter, we will give another presentation of categorical data
types, which will be simpler and more intuitive.  This is also the way
Categorical Data Types originated.  Note that we are not discarding CSL
completely and that the semantics of categorical data types will be
given in terms of CSL.

Section~\ref{sec-what-cdt} is an extended introduction to categorical
data types.  We will investigate some conventional data types and
introduce a new uniform categorical way of defining data types.  In
section~\ref{sec-decl-CDT}, we will make this new way into the CDT
declaration mechanism.  Section~\ref{sec-ex-CDT} will give various
examples of CDT declarations.  In section~\ref{sec-CDT-as-CSL}, CDT
declarations will be connected to CSL theories, and finally in
section~\ref{sec-exist-left-right} we will give a construction of CDT
data types.

\section{What are Categorical Data Types?}
\label{sec-what-cdt}

The need for pairs of two (or more) items of data often arises when we
write programs.  It is often the case that a function or procedure
takes more than one argument and this means that a pair (or a record)
of data needs to be passed to the function or procedure.  In another
situation, we may want to declare a new data type whose element is a
pair of elements of other data types.  In PASCAL, we can define such
data types using its {\tt record}~\ldots~{\tt end} construct.  For
example,
{\tt intchar} whose element is a pair of an integer and a character can
be declared as:
\begin{quote}
\begin{verbatim}
type intchar = record
                   first: integer;
                   second: char
               end;
\end{verbatim}
\end{quote}
In ML, this can be done by
\begin{quote}
\begin{verbatim}
type intchar = int * string;
\end{verbatim}
\end{quote}
(where since ML does not have a type for representing characters, we
have to use `{\tt string}' type whose element is a sequence of
characters).  These two languages have the means of constructing data
types of pairs from already-existing data types.  Let us call the
constructors {\it product type constructors}.  Most of the current
programming languages have product type constructors in one way or
another as their built-in primitives because they are so essential that
we can even say that no programming language is complete without them.

In order to understand the nature of product type constructors, let us
suppose a programming language which does not have them as primitives
and that we have to define them in terms of others.  This might mean
that we need in the language some kind of one level higher operations
which can define not types but type constructors.  Let us refer to an
algebraic specification language CLEAR,\footnote{CLEAR can be
institution independent, but here we refer the one that uses the
equational algebraic institution.} and see its ability to define a
product type constructor.
\begin{quote}
\begin{verbatim}
constant Triv =
    theory
        sort element
    endth

procedure Prod(A:Triv,B:Triv) =
    theory
        data sort prod
        opns pair: element of A,element of B -> prod
             pi1 : prod -> element of A
             pi2 : prod -> element of B
        eqns all a:element of A,b:element of B,
                                   pi1(pair(a,b)) = a
             all a:element of A,b:element of B,
                                   pi2(pair(a,b)) = b
    endth
\end{verbatim}
\end{quote}
This defines an algebra $P$ of three sorts: two `{\tt element}' sorts
(let us call them `{\tt A-element}' and `{\tt B-element}' to distinguish
them) and a `{\tt prod}' sort.  The underlying set $|P|_{\tt prod}$ is
the (set) product of two sets $|P|_{\tt A-element}$ and $|P|_{\tt
B-element}$.  The declaration also defines three operations (or
functions), `{\tt pair}', `{\tt pi1}' and `{\tt pi2}', satisfying the
two equations listed.  Note that the following equation can be proved
using induction on `{\tt prod}' sort (we have the induction principle on
this sort because of the initial data constraint).
\begin{center}
\verb"all x: prod, pair(pi1(x),pi2(x)) = x"
\end{center}

An algebraic specification language like CLEAR is powerful enough to
allow us to define all kinds of type constructors (including ordinary
data types as constant type constructors) in a uniform way without
having any particular primitives.  However, because
of its use of equations, we cannot adopt its declaration mechanism in
ordinary programming languages.\footnote{There is a programming language
OBJ \cite{goguen-tardo-79} which treats equations as a kind of
programs (as rewrite rules).} We might be puzzled that we need equations
even to define such very basic data types as products.

Let us find whether there is any other ways of defining product type
constructors by examining the foundation, namely mathematics.  Modern
mathematics uses set theory extensively because of its power.  Set
theory does not have a product type constructor as its primitive
construct either, so it is defined by means of other constructs.
For sets $A$ and $B$, their (set) product is defined as:
\begin{displaymath}
A\times B \defeq \{\; (x,y) \mid x \in A,\; y \in B \;\}
\end{displaymath}
where $(x,y)$ is really an abbreviation of $\{ \{ x \} \; \{ x \; y \}
\}$.  Although this looks very simple, it actually needs some
work to show from the axioms of set theory that this is actually a set.
Set theory uses the power of first order logic so heavily that it is
much harder to put the set theory formalism into a programming language
than to put the algebraic specification formalism.  As an example, let
$\phi(x)$ be a first order formula.  Then, from the comprehension axiom
(or the replacement axiom), we have the set
\begin{displaymath}
\{\; x \in A \mid \phi(x) \;\}
\end{displaymath}
where $A$ is a set.  $\phi(x)$ can be anything expressible by first
order logic (using quantifiers and negations), and this is too much
powerful to investigate the basic property of data types.  It disfigures
the beauty behind this powerful definition mechanism and we cannot see
through it easily.

Set theory has achieved a firm position as the foundation of
mathematics, but there are some alternatives.  Category theory is one of
them.  It has been proved that category theory has a remarkable ability
to disclose true nature of mathematical objects.  For example, a product
constructor (or categorically a functor $\cC\times\cC \rightarrow \cC$)
is beautifully characterized as the right adjoint of the diagonal
functor $\cC \rightarrow \cC\times\cC$.\footnote{There may be more than
one right adjoint of the diagonal functor, but they are isomorphic.
Therefore, we say `{\it the\/}' right adjoint rather than `{\it a\/}'
right adjoint.} Expanding the definition of adjunctions, this means that
we have the following natural isomorphism:
\begin{displaymath}
\Hom{\cC}{C}{A}\times\Hom{\cC}{C}{B} \simeq \Hom{\cC}{C}{{\rm
prod}(A,B)}\eqno(*)
\end{displaymath}
(natural in $A$, $B$ and $C$).  We use `prod' for the product functor to
follow the notation we use later.  The isomorphic function from the
left-hand side to the right-hand side  is the factorizer of this
adjunction and we write `pair' for it.  We can rewrite this adjoint
situation as the following rule:
\begin{displaymath}
\logicrule{C \rubberrightarrow{f} A \qquad C \rubberrightarrow{g}
B}{C \rubberrightarrow{{\rm pair}(f,g)} {\rm
prod}(A,B)}\eqno(**)
\end{displaymath}
Given two morphisms $f: C \rightarrow A$ and $g: C \rightarrow B$, ${\rm
pair}(f,g)$ give a morphism of $C \rightarrow {\rm prod}(A,B)$.  This
correspondence is one-to-one and is natural in $A$, $B$ and $C$.

Comparing with the product type constructor `{\tt Prod}' defined by
CLEAR, we have the same `pair' though the previous one takes two
elements as the arguments and this one takes two morphisms instead, and
now we can find the things corresponding to two projections `{\tt pi1}'
and `{\tt pi2}' as well.  Replacing $C$ by ${\rm prod}(A,B)$ in $(*)$,
we get:
\begin{displaymath}
\displaylines{
\qquad \Hom{\cC}{{\rm prod}(A,B)}{A} \times \Hom{\cC}{{\rm prod}(A,B)}{B}
\simeq {} \hfill \cr
\hfill \Hom{\cC}{{\rm prod}(A,B)}{{\rm prod}(A,B)}. \qquad \cr}
\end{displaymath}
We have a very special morphism in $\Hom{\cC}{{\rm prod}(A,B)}{{\rm
prod}(A,B)}$, namely the identity.  Because of the isomorphism, there
exist unique morphisms of ${\rm prod}(A,B) \rightarrow A$ and ${\rm
prod}(A,B) \rightarrow B$ which are mapped to the identity by `pair',
and these are the projections.  We name them `pi1' and `pi2' as well.
Because of the very way they are defined, it is trivial that
\begin{displaymath}
{\rm pair}({\rm pi1},{\rm pi2}) = {\bf I}.\eqno(+)
\end{displaymath}
Furthermore, from the naturality in $C$ of $(*)$, we have the rule
\begin{displaymath}
\logicrule{C \rubberrightarrow{{\rm pair}(f,g)} {\rm prod}(A,B)
\rubberrightarrow{{\rm pi1}} A \qquad C \rubberrightarrow{{\rm
pair}(f,g)} {\rm prod}(A,B) \rubberrightarrow{{\rm pi2}}
B}{C \rubberrightarrow{{\rm pair}(f,g)} {\rm prod}(A,B)
\rubberrightarrow{{\rm pair}({\rm pi1},{\rm pi2})} {\rm prod}(A,B)},
\end{displaymath}
and, if we express it by equations and use $(+)$,
\begin{displaymath}
{\rm pair}({\rm pi1}\circ{\rm pair}(f,g),{\rm pi2}\circ{\rm
pair}(f,g)) = {\rm pair}({\rm pi1},{\rm pi2})\circ{\rm pair}(f,g) =
{\rm pair}(f,g).
\end{displaymath}
Since `pair' is isomorphic, we can conclude that the following
equations hold:
\begin{displaymath}
{\rm pi1}\circ{\rm pair}(f,g) = f \quad{\rm and}\quad {\rm pi2}\circ{\rm
pair}(f,g) = g.
\end{displaymath}
These are exactly the ones which we listed when defining `{\tt Prod}' in
CLEAR.  Note that this time they are derived equations.  By saying that
`prod' is the right adjoint to the diagonal functor, we get these
equations automatically.  This shows how neat the categorical definition
is.

Another advantage of categorical definition is that we can form the dual
definition easily.  We defined the product functor as the right adjoint
of the diagonal functor.  Then, it is natural to ask what is the left
adjoint of the diagonal functor.  It is the coproduct functor
$\cC\times\cC \rightarrow \cC$.  In the category of sets, the copoduct
of two sets $A$ and $B$ is their disjoint sum
\begin{displaymath}
A + B \defeq \{\; 0 \;\} \times A \bigcup \{\; 1 \;\} \times B.
\end{displaymath}
It is not easy to see in set theory that this is the dual of $A \times
B$.  In PASCAL, we can define coproducts by means of variant record.  In
ML, we used to have a built-in coproduct type constructor `{\tt +}', but
the new Standard ML does not.  Instead, `{\tt +}' can be defined by the
following `{\tt datatype}' declaration.
\begin{center}
\verb"datatype 'a + 'b = in1 of 'a | in2 of 'b;"
\end{center}
We cannot define the product type constructor by a `{\tt datatype}'
declaration in ML, but we can define its dual.  ML looks non-symmetric
from this.  In CLEAR, we can define a coproduct type constructor as
follows:
\begin{quote}
\begin{verbatim}
Procedure CoProd(A:Triv,B:Triv) =
    theory
        data sort coprod
        opns in1: element of A -> coprod
             in2: element of B -> coprod
    endth
\end{verbatim}
\end{quote}
Again, this cannot be seen as the dual of `{\tt Prod}' we defined
earlier; here we do not use equations; we have only two operations
whereas we had three.  This shows that CLEAR is not symmetric either.

Now, in category theory, the coproduct functor $\cC\times\cC \rightarrow
\cC$ which we call `coprod' is defined by the dual of $(*)$, by just
changing the direction of arrows.
\begin{displaymath}
\Hom{\cC}{A}{C}\times\Hom{\cC}{B}{C} \simeq \Hom{\cC}{{\rm coprod}(A,B)}{C}
\end{displaymath}
We name the isomorphic function going from the left-hand side to the
right-hand side `case' (we could call it `copair' to emphasize the
duality to `pair', but, since it plays a role of `{\tt case}' statements
of ML or C (or PASCAL), we call it `case').  Writing the adjunction as a
rule,
\begin{displaymath}
\logicrule{A \rubberrightarrow{f} C \qquad B \rubberrightarrow{g}
C}{{\rm coprod}(A,B) \rubberrightarrow{{\rm case}(f,g)} C}.
\end{displaymath}
Two injections ${\rm in1}: A \rightarrow {\rm coprod}(A,B)$ and ${\rm
in2}: B \rightarrow {\rm coprod}(A,B)$ are defined as the morphisms
which `case' maps to the identity of ${\rm coprod}(A,B)$.  As before, we
can derive some equations easily.

From what we have looked at, it seems a good idea to design a category
theory based (programming or specification) language which has the
ability to define functors by means of adjunctions.  Since it is
convenient to introduce names for unit natural transformations and
factorizers at the same time, we regard an adjunction as a triple of a
functor, a unit natural transformation and a factorizer (see, for
example, \cite{maclane-71} for many equivalent ways of defining
adjunctions).  Therefore, a category theory based language may have the
following two forms of declaring new functors:
\begin{center}
\begin{tabular}{l}
let $\pair{F,\alpha,\psi}$ be right adjoint of $G$ \\
let $\pair{F,\alpha,\psi}$ be left adjoint of $G$
\end{tabular}
\end{center}
where $G$ is a functor we already have, $F$ is the new functor we are
defining, $\alpha$ is the associated unit natural transformation and
$\psi$ is the associated factorizer.  One problem is that we need to
have some primitive functors with which we start.  We definitely need
diagonal functors for we want to define product and coproduct functors.  In
order to define the natural number object (which is a constant functor),
we need a pretty complicated functor $G$.  The problem is how to
represent such $G$.

Let us investigate how other languages and theories define the data type
of natural numbers.  In set theory, it has the axiom of infinity which
says the existence of natural numbers.  This may look rather artificial.
In PASCAL, there is no intuitively easy way to define it.  In ML, though
it is a built-in data type for efficiency, we could define it as:
\begin{center}
\verb"datatype nat = zero | succ of nat;"
\end{center}
Note the recursiveness in this definition.  Essentially, we need some
kind of recursiveness to define a data type of natural numbers.  In
CLEAR, one can define it as
\begin{quote}
\begin{verbatim}
constant Nat =
  theory
    data sort nat
    opns zero: nat
         succ: nat -> nat
  endth
\end{verbatim}
\end{quote}
This is very much similar to the one in ML, though we often think that
the CLEAR definition is based on the initial algebra semantics whereas
the ML definition is based on domain theory.  In domain theory, a data
type of natural numbers can be defined as the solution of the following
domain equation
\begin{displaymath}
N \iso 1 + N.\eqno(*)
\end{displaymath}
The initial solution of $(*)$ can be calculated as a colimit of a
sequence of domains, but we do not go into its detail here.  As a
connection to the initial algebra semantics, the initial solution can be
characterized as the initial $T$-algebra, where $T$ is a functor $T(X)
\defeq 1+X$ in this case.  In general, given a category and an
endo-functor $T$, we can form a category of $T$-algebras.
\begin{definition}
\label{def-T-alg}
For a category $\cC$ and an endo-functor $T: \cC \rightarrow \cC$, the
category of $T$-algebras is defined
\begin{enumerate}
\item its objects are pairs $\pair{A,f}$ where $A$ is a $\cC$ object and
$f$ is a $\cC$ morphism $T(A) \rightarrow A$, and
\item its morphisms $h: \pair{A,f} \rightarrow \pair{B,g}$ are $\cC$
morphisms $h: A \rightarrow B$ which make the following diagram commute.
\begin{displaymath}
\sqdiagram{T(A)}{f}{A}{T(h)}{h}{T(B)}{g}{B}{}
\end{displaymath}
\end{enumerate}
Note that this is a weaker version of the category of $T$-algebras
defined in many category theory books (e.g.\ \cite{maclane-71}) where
$T$ needs to be a monad.
\end{definition}

We can dualize definition~\ref{def-T-alg} to define $T$-coalgebras.
However, in the theory of categorical data types (`CDT theory' for
short), we combine the two definitions together.
\begin{definition}
\label{def-dbl-alg}
Let $\cC$ and $\cD$ be categories and both $F$ and $G$ be functors from
$\cC$ to $\cD$.  We define an $F,G$-dialgebra\footnote{The name {\it
dialgebra} was suggested by Bob McKay.} as
\begin{enumerate}
\item its objects are pairs $\pair{A,f}$ where $A$ is a $\cC$ object and
$f$ is a $\cD$ morphism of $F(A) \rightarrow G(A)$, and
\item its morphisms $h: \pair{A,f} \rightarrow \pair{B,g}$ are $\cC$
morphisms $h: A \rightarrow B$ such that the following diagram commutes.
\begin{displaymath}
\sqdiagram{F(A)}{f}{G(A)}{F(h)}{G(h)}{F(B)}{g}{G(B)}{}
\end{displaymath}
In the case where $F$ or $G$ is contravariant, we have to modify the
direction of some arrows.
\end{enumerate}
It is easy to show that it is a category; let us write ${\bf
DAlg}(F,G)$ for it.  Note that ${\bf DAlg}(T,{\bf I})$ is the category
of $T$-algebras and ${\bf DAlg}({\bf I},T)$ is the category of
$T$-coalgebras.
\end{definition}
This is a very simple extension of definition~\ref{def-T-alg}, yet its
symmetry and dividing the source category from the target one give us
greater freedom.  With $T$-algebras, we need to use the coproduct
functor to define the domain of natural numbers, but by $F,G$-dialgebra we
do not.  Let $\cC$ be any category and $\cD$ be its product
$\cC\times\cC$.  We define the functors $F$ and $G$ as
\begin{displaymath}
F(A) \defeq \pair{1,A} \quad\mbox{and}\quad G(A) \defeq \pair{A,A}.
\end{displaymath}
Let $\pair{{\rm nat},\pair{{\rm zero},{\rm succ}}}$ be the initial
$F,G$-dialgebra.  From the definition, `nat' is a $\cC$ object, `zero'
is a $\cC$ morphism of $1 \rightarrow {\rm nat}$ and `succ' is a $\cC$
morphism of ${\rm nat} \rightarrow {\rm nat}$.  The initiality means
that for any ${\bf DAlg}(F,G)$ object $\pair{A,\pair{f,g}}$ there exists
a unique ${\bf DAlg}(F,G)$ morphism $h: \pair{{\rm nat},\pair{{\rm
zero},{\rm succ}}} \rightarrow \pair{A,\pair{f,g}}$.  If we spell out
the definition, this means that for any $\cC$ object $A$ and any $\cC$
morphisms $f: 1
\rightarrow A$ and $g: A \rightarrow A$ there exists a unique $\cC$
morphism $h: {\rm nat} \rightarrow A$ which makes the following diagram
commute.
\begin{displaymath}
\setlength{\unitlength}{1mm}
\begin{picture}(75,45)(0,2.5)
\put(5,40){\makebox(0,0){1}}
\put(20,42){\makebox(0,0)[b]{zero}}
\put(35,40){\makebox(0,0){nat}}
\put(50,42){\makebox(0,0)[b]{succ}}
\put(65,40){\makebox(0,0){nat}}
\put(15,22){\makebox(0,0){$f$}}
\put(25,30){\makebox(0,0){\commute}}
\put(37,25){\makebox(0,0)[l]{$h$}}
\put(50,25){\makebox(0,0){\commute}}
\put(67,25){\makebox(0,0)[l]{$h$}}
\put(35,10){\makebox(0,0){$A$}}
\put(65,10){\makebox(0,0){$A$}}
\put(50,8){\makebox(0,0)[t]{$g$}}
\put(7.5,40){\vector(1,0){22.5}}
\put(40,40){\vector(1,0){20}}
\put(7.5,37.5){\vector(1,-1){25}}
\put(40,10){\vector(1,0){20}}
\multiput(35,37.5)(0,-5){4}{\line(0,-1){3}}
\put(35,17.5){\vector(0,-1){5}}
\multiput(65,37.5)(0,-5){4}{\line(0,-1){3}}
\put(65,17.5){\vector(0,-1){5}}
\end{picture}
\end{displaymath}
This is exactly the definition of `nat' being a natural number object
in category theory.

To get further generality of $F,G$-dialgebras, we parametrize $F$ and $G$.
\begin{definition}
\label{def-Right-Left}
Let $\cC$, $\cD$ and $\cE$ be categories, and let $F: \cC\times\cD \rightarrow
\cE$ and $G: \cC\times\cD^- \rightarrow \cE$ be functors.  We define
$\pair{{\bf Left}[F,G](A),\eta_A}$ for a $\cD$ object $A$ to be the
initial object in the category ${\bf
DAlg}(F(\;\cdot\;,A),G(\;\cdot\;,A))$.  Dually, we define
$\pair{{\bf Right}[F,G](A),\epsilon_A}$ to be the final object.  We may
write ${\bf Right}(A)$ for ${\bf Right}[F,G](A)$ and ${\bf Left}(A)$ for
${\bf Left}[F,G](A)$ if the context makes $F$ and $G$ clear.
\end{definition}

\begin{proposition}
\label{prop-Right-Left}
If ${\bf Left}[F,G](A)$ exists for every object $A \in |\cD|$, ${\bf
Left}[F,G]$ denotes a functor $\cD \rightarrow \cC$ (i.e.\ we can extend
it to $\cD$ morphisms).  Dually, if ${\bf Right}[F,G](A)$ exists for
every $A \in |\cD|$, ${\bf Right}[F,G]$ denotes a functor $\cD^-
\rightarrow \cC$. \\
{\bf Proof:} We first need to define what ${\bf Left}[F,G](f)$ is for a
$\cD$ morphism $f: A \rightarrow B$.  We define it as the morphism $h: {\bf
Left}(A) \rightarrow {\bf Left}(B)$ which fills in the following
diagram.
\begin{displaymath}
\setlength{\unitlength}{1mm}
\begin{picture}(140,75)(0,2.5)
\put(15,70){\makebox(0,0){$F({\bf Left}(A),A)$}}
\put(35,72){\makebox(0,0)[b]{$\eta_A$}}
\put(55,70){\makebox(0,0){$G({\bf Left}(A),A)$}}
\put(105,70){\makebox(0,0){$\pair{{\bf Left}(A),\eta_A}$}}
\put(13,55){\makebox(0,0)[r]{$F(h,{\bf I})$}}
\put(57,55){\makebox(0,0)[l]{$G(h,{\bf I})$}}
\put(107,55){\makebox(0,0)[l]{$h$}}
\put(15,40){\makebox(0,0){$F({\bf Left}(B),A)$}}
\put(55,40){\makebox(0,0){$G({\bf Left}(B),A)$}}
\put(105,40){\makebox(0,0){$\pair{{\bf Left}(B),G({\bf I},f) \circ
\eta_B \circ F({\bf I},f)}$}}
\put(13,25){\makebox(0,0)[r]{$F({\bf I},f)$}}
\put(57,25){\makebox(0,0)[l]{$G({\bf I},f)$}}
\put(15,10){\makebox(0,0){$F({\bf Left}(B),B)$}}
\put(35,8){\makebox(0,0)[t]{$\eta_B$}}
\put(55,10){\makebox(0,0){$G({\bf Left}(B),B)$}}
\put(30,70){\vector(1,0){10}}
\multiput(15,67.5)(0,-5){4}{\line(0,-1){3}}
\put(15,47.5){\vector(0,-1){5}}
\multiput(55,67.5)(0,-5){4}{\line(0,-1){3}}
\put(55,47.5){\vector(0,-1){5}}
\multiput(105,67.5)(0,-5){4}{\line(0,-1){3}}
\put(105,47.5){\vector(0,-1){5}}
\put(15,37.5){\vector(0,-1){25}}
\put(55,12.5){\vector(0,1){25}}
\put(30,10){\vector(1,0){10}}
\end{picture}
\end{displaymath}
The unique existence of the morphism is provided because $\pair{{\bf
Left}(A),\eta_A}$ is the initial object of ${\bf
DAlg}(F(\;\cdot\;,A),G(\;\cdot\;,A))$.  In other words, ${\bf
Left}(f)$ is the unique morphism which satisfies
\begin{displaymath}
G({\bf Left}(f),f) \circ \eta_B \circ F({\bf Left}(f),f) = \eta_A.
\end{displaymath}
Let us check that {\bf Left} is in fact a functor.  Trivially,
\begin{displaymath}
G({\bf I},{\bf I}) \circ \eta_A \circ F({\bf I},{\bf I}) = \eta_A.
\end{displaymath}
Therefore, ${\bf Left}({\bf I}_A) = {\bf I}_{{\bf Left}(A)}$.  For
morphisms $f: A \rightarrow B$ and $g: B \rightarrow C$,
\begin{displaymath}
\begin{array}{rl}
& G({\bf Left}(g)\circ{\bf Left}(f),g\circ f) \circ \eta_C \circ F({\bf
Left}(g)\circ{\bf Left}(f),g\circ f) \\
=\;& G({\bf Left}(f),f) \circ G({\bf Left}(g),g)\circ \eta_C \circ F({\bf
Left}(g),g) \circ F({\bf Left}(f),f) \\
=\;& G({\bf Left}(f),f) \circ \eta_B \circ F({\bf Left}(f),f) \\
=\;& \eta_A \\
\end{array}
\end{displaymath}
Therefore, ${\bf Left}(g)\circ{\bf Left}(f) = {\bf Left}(g\circ f)$.
\begin{displaymath}
\setlength{\unitlength}{1mm}
\begin{picture}(115,70)(0,2.5)
\put(40,65){\makebox(0,0){$F({\bf Left}(A),A)$}}
\put(80,65){\makebox(0,0){$G({\bf Left}(A),A)$}}
\put(40,35){\makebox(0,0){$F({\bf Left}(B),B)$}}
\put(80,35){\makebox(0,0){$G({\bf Left}(B),B)$}}
\put(40,5){\makebox(0,0){$F({\bf Left}(C),C)$}}
\put(80,5){\makebox(0,0){$G({\bf Left}(C),C)$}}
\put(60,67){\makebox(0,0)[b]{$\eta_A$}}
\put(60,37){\makebox(0,0)[b]{$\eta_B$}}
\put(60,7){\makebox(0,0)[b]{$\eta_C$}}
\put(38,50){\makebox(0,0)[r]{$F({\bf Left}(f),f)$}}
\put(82,50){\makebox(0,0)[l]{$G({\bf Left}(f),f)$}}
\put(38,20){\makebox(0,0)[r]{$F({\bf Left}(g),g)$}}
\put(82,20){\makebox(0,0)[l]{$G({\bf Left}(g),g)$}}
\put(25,35){\makebox(0,0)[r]{$F({\bf Left}(g\circ f),g\circ f)$}}
\put(95,35){\makebox(0,0)[l]{$G({\bf Left}(g\circ f),g\circ f)$}}
\put(55,65){\vector(1,0){10}}
\put(55,35){\vector(1,0){10}}
\put(55,5){\vector(1,0){10}}
\put(40,62.5){\vector(0,-1){25}}
\put(80,62.5){\vector(0,-1){25}}
\put(40,32.5){\vector(0,-1){25}}
\put(80,32.5){\vector(0,-1){25}}
\put(25,65){\line(-1,0){20}}
\put(5,65){\line(0,-1){27.5}}
\put(5,32.5){\line(0,-1){27.5}}
\put(5,5){\vector(1,0){20}}
\put(95,65){\line(1,0){20}}
\put(115,65){\line(0,-1){27.5}}
\put(115,32.5){\line(0,-1){27.5}}
\put(115,5){\vector(-1,0){20}}
\end{picture}
\end{displaymath}
Dually, we can prove that ${\bf Right}$ is a functor.
\end{proposition}

`{\bf Left}' and `{\bf Right}' may suggest a connection with left and
right adjoint functors.  In fact,
\begin{proposition}
For a functor $F: \cC \rightarrow \cD$, its left adjoint functor can be
denoted by
\begin{displaymath}
{\bf Left}[\lambda(X,Y).Y,\lambda(X,Y).F(X)]
\end{displaymath}
and, dually, its right adjoint functor can be denoted by
\begin{displaymath}
{\bf Right}[\lambda(X,Y).F(X),\lambda(X,Y).Y].
\end{displaymath}
{\bf Proof:} Let us only check the left adjoint case.  We see
$\lambda(X,Y).Y$ as a functor $\cC\times\cD \rightarrow \cD$ and
$\lambda(X,Y).F(X)$ as a functor $\cC\times\cD^- \rightarrow \cD$.  From
definition~\ref{def-Right-Left}, {\bf Left} is a functor $\cD
\rightarrow \cC$.  If we spell out the condition of $\pair{{\bf
Left}(A),\eta_A}$ being the initial algebra, it means that for any
$\cC$ object $B$ and a $\cC$ morphism $f: A \rightarrow F(B)$ there
exists a unique $\cC$ morphism $h: {\bf Left}(A) \rightarrow B$ such that
the following diagram commutes.
\begin{displaymath}
\setlength{\unitlength}{1mm}
\begin{picture}(70,40)(0,2.5)
\put(5,35){\makebox(0,0){$A$}}
\put(15,37){\makebox(0,0)[b]{$\eta_A$}}
\put(35,35){\makebox(0,0){$F({\bf Left}(A))$}}
\put(65,35){\makebox(0,0){${\bf Left}(A)$}}
\put(5,5){\makebox(0,0){$A$}}
\put(35,5){\makebox(0,0){$F(B)$}}
\put(65,5){\makebox(0,0){$B$}}
\put(20,3){\makebox(0,0)[t]{$f$}}
\put(20,20){\makebox(0,0){\commute}}
\put(37,20){\makebox(0,0)[l]{$F(h)$}}
\put(67,20){\makebox(0,0)[l]{$h$}}
\put(7.5,35){\vector(1,0){17.5}}
\put(4.5,32.5){\line(0,-1){25}}
\put(5.5,32.5){\line(0,-1){25}}
\put(7.5,5){\vector(1,0){22.5}}
\multiput(35,33.5)(0,-5){4}{\line(0,-1){3}}
\put(35,12.5){\vector(0,-1){5}}
\multiput(65,33.5)(0,-5){4}{\line(0,-1){3}}
\put(65,12.5){\vector(0,-1){5}}
\end{picture}
\end{displaymath}
This is exactly the condition of {\bf Left} being the left adjoint
functor of $F$.  Dually, we can prove that ${\bf
Right}[\lambda(X,Y).F(X),\lambda(X,Y).Y]$ is the right adjoint.
\end{proposition}

Hence, definition~\ref{def-dbl-alg} of $F,G$-dialgebras covers both
$T$-algebras and adjoints so that it enables us to define products,
coproducts, natural number object, and so on in a uniform way.

\section[Data Type Declarations in Categorical Data Types]{Data Type
Declarations in\\ Categorical Data Types}
\label{sec-decl-CDT}

In the previous section, we have looked at some ways of defining data
types in some languages.  In this section, we will introduce
how to define data types in CDT.

If we were only interested in functors, only ${\bf Left}[F,G]$ and ${\bf
Right}[F,G]$ defined in the previous section would be needed, but we do
want morphisms (or natural transformations) and factorizers which will
make up some kind of programs, the meaning of which we will examine in
chapter~\ref{ch-cpl} (e.g.\ how to execute them).

${\bf Left}[F,G]$ and ${\bf Right}[F,G]$ have been defined for functors
$F: \cC\times\cD \rightarrow \cE$ and $G: \cC\times\cD^- \rightarrow
\cE$, where $\cC$, $\cD$ and $\cE$ are some categories.  Remember the
aim of CDT; we would like to define (or specify, or study) a category of
data types.  Therefore, $\cC$, $\cD$ and $\cE$ should somehow be related
to this category.  The simplest we can think of is that they are the
product categories of this category and all the functors are in the form
of $\cC^s \rightarrow \cC$, where $s$ is a sequence of variances (i.e.\
$s \in {\bf Var}^\ast$) and $\cC^{v_1\ldots v_n} \defeq
\cC^{v_1}\times\cdots\times \cC^{v_n}$.

In order to simplify the presentation, let us use the vector notation
and write, for example, $\vec F$ for a sequence of functors
$\pair{F_1,\ldots,F_n}$ where all of them have the same type $\cC^s
\rightarrow \cC$, that is, $\vec F$ is a functor of $\cC^s \rightarrow
\cC^n$.

From definition~\ref{def-Right-Left}, for $\vec F: \cC\times\cC^s
\rightarrow \cC^n$ and $\vec G: \cC\times\cC^{-\bullet s} \rightarrow
\cC^n$, ${\bf Left}[F,G]$ is a functor $\cC^s \rightarrow \cC$ and ${\bf
Right}[F,G]$ is a functor $\cC^{-\bullet s} \rightarrow \cC$, where $u
\bullet {v_1\ldots v_n} \defeq u\bullet v_1 \ldots u\bullet v_n$.

Hence, we come to the definition of CDT declarations.
\begin{definition}
\label{def-CDT-decl}
In CDT theory, there are two forms of declaring new functors.  One is to
define a functor $L: \cC^s \rightarrow \cC$ by
\begin{displaymath}
\begin{tabular}{l}
left object $L(\vec X)$ with $\psi$ is \\
$\qquad \vec\alpha: \vec F(L,\vec X) \rightarrow \vec G(L,\vec X)$ \\
end object \\
\end{tabular}
\eqno(*)
\end{displaymath}
and the other is to define a functor $R: \cC^{-\bullet s} \rightarrow
\cC$ by
\begin{center}
\begin{tabular}{l}
right object $R(\vec X)$ with $\psi$ is \\
$\qquad \vec\alpha: \vec F(R,\vec X) \rightarrow \vec G(R,\vec X)$ \\
end object \\
\end{tabular}
\end{center}
where $\vec X$ is a sequence $\pair{X_1,\ldots,X_n}$ of variables,
$\psi$ is the associated factorizer, $\vec\alpha$ is a sequence
$\pair{\alpha_1,\ldots,\alpha_m}$ of the associated natural
transformations, and $\vec F$ and $\vec G$ are sequences
$\pair{F_1,\ldots,F_m}$ and $\pair{G_1,\ldots,G_m}$, respectively, of
functors which we have as primitives or we have already defined and
whose type is $F_i: \cC\times\cC^s \rightarrow \cC$ and $G_i:
\cC\times\cC^{-\bullet s} \rightarrow \cC$, respectively.  Semantically,
$L$ is ${\bf Left}[\vec F,\vec G]$ and $R$ is ${\bf Right}[\vec F,\vec G]$.
We may call $L$ left functor or left object and $R$ right functor or
right object.
\end{definition}

If we expand the definition of {\bf Left}, $(*)$ defines for any $\cC$
objects $\vec A = \pair{A_1,\ldots,A_n}$ an object $L(\vec A)$ and a
morphism
\begin{displaymath}
\ardiagram{\vec F(L(\vec A),\vec A)}{\vec\alpha_{\vec A}}{\vec G(L(\vec
A),\vec A),}
\end{displaymath}
and for any object $B$ and a sequence of morphisms $\vec f: \vec
F(B,\vec A) \rightarrow \vec G(B,\vec A)$, $\psi(\vec f)$ denotes the
unique morphism which makes the following diagram commute.
\begin{displaymath}
\setlength{\unitlength}{1mm}
\begin{picture}(95,40)(0,2.5)
\put(10,35){\makebox(0,0){$\vec F(L(\vec A),\vec A)$}}
\put(32.5,37){\makebox(0,0)[b]{$\vec\alpha_{\vec A}$}}
\put(55,35){\makebox(0,0){$\vec G(L(\vec A),\vec A)$}}
\put(90,35){\makebox(0,0){$L(\vec A)$}}
\put(8,20){\makebox(0,0)[r]{$\vec F(\psi(\vec f),\vec A)$}}
\put(32.5,20){\makebox(0,0){\commute}}
\put(57,20){\makebox(0,0)[l]{$\vec G(\psi(\vec f),\vec A)$}}
\put(92,20){\makebox(0,0)[l]{$\psi(\vec f)$}}
\put(10,5){\makebox(0,0){$\vec F(B,\vec A)$}}
\put(32.5,3){\makebox(0,0)[t]{$\vec f$}}
\put(55,5){\makebox(0,0){$\vec G(B,\vec A)$}}
\put(90,5){\makebox(0,0){$B$}}
\put(22.5,35){\vector(1,0){20}}
\put(17.5,5){\vector(1,0){30}}
\multiput(10,32.5)(0,-5){4}{\line(0,-1){3}}
\put(10,12.5){\vector(0,-1){5}}
\multiput(55,32.5)(0,-5){4}{\line(0,-1){3}}
\put(55,12.5){\vector(0,-1){5}}
\multiput(90,32.5)(0,-5){4}{\line(0,-1){3}}
\put(90,12.5){\vector(0,-1){5}}
\end{picture}
\end{displaymath}

In definition~\ref{def-CDT-decl}, it is not immediately clear what kind
of $\vec F$ and $\vec G$ is allowed.  We vaguely stated that they are
primitive or have been defined already.  In order to clarify this point,
we go back to CSL and regard a CDT declaration as an extension of a
given CSL signature.
\begin{definition}
\label{def-CDT-CSL}
Let $\pair{\Gamma,\Delta,\Psi}$ be a CSL signature.  A CDT declaration
$D \in {\bf Decl}$ is given by the following BNF expression.
\begin{quote}
\begin{tabbing}
$D \coloneq {}$ \= $\{$ left \=$\mid$ right $\}$ object
$F(X_1,\ldots,X_n)$ with $\psi$ is \\
\>\> $\alpha_1: E_1 \rightarrow E'_1$ \\
\>\> $\qquad \ldots$ \\
\>\> $\alpha_m: E_m \rightarrow E'_m$ \\
\> end object \\
\end{tabbing}
\end{quote}
where $F$ is a new functor symbol, $\psi$ is a new factorizer symbol,
$\alpha_1,\ldots,a_m$ are new natural transformation symbols, and  $E_i$
and $E'_i$ ($i=1,\ldots,m$) are well-formed functorial expressions
(under this signature $\pair{\Gamma,\Delta,\Psi}$) whose variables are
$X_1,\ldots,X_n$ and $F$ (here we use $F$ as a formal parameter like we
use its function name inside a function body in PASCAL).
\end{definition}

We need to put restriction on the variance of $F$ in the functorial
expressions such that for each $i = 1,\ldots,m$
\begin{enumerate}
\item the variance of $F$ in $E_i$ should be either covariant or free,
\item the variance of $F$ in $E'_i$ should also be either covariant or
free, and
\item either the variance of $F$ in $E_i$ should be covariant or that in
$E'_i$ should be covariant.
\end{enumerate}
We could have allowed $F$ to be contravariant (as indeed the original
definition of CDT did), but it turned out that the generality by
contravariance was of very little use, so, because it simplifies the
following argument, we restrict ourselves only to covariant functors.
The third condition above is to make the extension consistent (each
$\alpha_i$ should somehow relate to the functor we are declaring).

Let us calculate the variance of $F$. If the variance of
$\lambda(F,X_1,\ldots,X_n).E_i$ is $v_i s_i$ ($v_i \in {\bf Var}$ for
the variance of $F$ and $s_i \in {\bf Var}^\ast$ for the variance of
$X_1,\ldots,X_n$) and that of $\lambda(F,X_1,\ldots,X_n).E'_i$ is $v'_i
s'_i$, $\lambda(F,X_1,\ldots,X_n).E_i$ denotes a functor
$\cC^{v_i}\times\cC^{s_i} \rightarrow \cC$ and
$\lambda(F,X_1,\ldots,X_n).E'_i$ denotes a functor
$\cC^{v'_i}\times\cC^{s'_i} \rightarrow \cC$.  The restriction above
states that $v_i\lub v'_i = {+}$.  From
proposition~\ref{prop-Right-Left}, the variance of $F$ in case that it
is declared by a left CDT declaration should be
\begin{displaymath}
\Lub\limits_{i=1}^{m} s_i\lub -\bullet s'_i,\eqno(*)
\end{displaymath}
and in case that it is declared by a right one, the variance should be
\begin{displaymath}
\Lub\limits_{i=1}^{m} -\bullet s_i\lub s'_i,\eqno(**)
\end{displaymath}
where ${\bf Var}^\ast$ is a partially ordered set with the ordering
given by $u_1\ldots u_n \lleq v_1\ldots v_n$ if and only if $u_1\lleq
v_1$, \ldots and $u_n \lleq v_n$.

\begin{contdefinition}{\ref{def-CDT-CSL}}
A CDT declaration gives an extension of CSL signature.
\begin{displaymath}
\pair{\Gamma,\Delta,\Psi} \hookrightarrow \pair{\Gamma \cup \{\; F \;\},
\Delta \cup \{\; \alpha_1,\ldots,\alpha_m \;\}, \Psi \cup \{\; \psi \;\}}
\end{displaymath}
where the variance of $F$ is given by $(*)$ or $(**)$, the type of
$\alpha_i$ is
\begin{displaymath}
\lambda(X_1,\ldots,X_n).E_i[F(X_1,\ldots,X_n)/F] \natrightarrow
\lambda(X_1,\ldots,X_n).E'_i[F(X_1,\ldots,X_n)/F],
\end{displaymath}
and the type of $\psi$ is
\begin{displaymath}
\logicrule{f_i: \lambda(X,X_1,\ldots,X_n).E_i[X/F] \rightarrow
\lambda(X,X_1,\ldots,X_n).E'_i[X/F] \quad (i =
1,\ldots,m)}{\psi(f_1,\ldots,f_m):
\lambda(X,X_1,\ldots,X_n).F(X_1,\ldots,X_n) \rightarrow
\lambda(X,X_1,\ldots,X_n).X}
\end{displaymath}
by a left CDT declaration and
\begin{displaymath}
\logicrule{f_i: \lambda(X,X_1,\ldots,X_n).E_i[X/F] \rightarrow
\lambda(X,X_1,\ldots,X_n).E'_i[X/F] \quad (i =
1,\ldots,m)}{\psi(f_1,\ldots,f_m):
\lambda(X,X_1,\ldots,X_n).X \rightarrow
\lambda(X,X_1,\ldots,X_n).F(X_1,\ldots,X_n)}
\end{displaymath}
by a right one.
\end{contdefinition}

We will see in section~\ref{sec-CDT-as-CSL} a CDT declaration
as a CSL theory extension so that the semantics of a CDT declaration can
be given by a CSL structure.

\section{Examples of Categorical Data Types}
\label{sec-ex-CDT}

In this section, we will present several examples of categorical data
types declared by
\begin{displaymath}
\begin{tabular}{l}
left object $F(X_1,\ldots,X_n)$ with $\psi$ is \\
$\qquad \alpha_1: E_1 \rightarrow E'_1$ \\
$\qquad \qquad \ldots$ \\
$\qquad \alpha_m: E_m \rightarrow E'_m$ \\
end object \\
\end{tabular}
\eqno(*)
\end{displaymath}
for left objects and by
\begin{displaymath}
\begin{tabular}{l}
right object $F(X_1,\ldots,X_n)$ with $\psi$ is \\
$\qquad \alpha_1: E_1 \rightarrow E'_1$ \\
$\qquad \qquad \ldots$ \\
$\qquad \alpha_m: E_m \rightarrow E'_m$ \\
end object \\
\end{tabular}
\eqno(**)
\end{displaymath}
for right objects.

\subsection{Terminal and Initial Objects}
\label{ssec-terminal-initial}

Let us start with an empty CSL signature
$\pair{\emptyset,\emptyset,\emptyset}$.  The simplest case of $(*)$ and
$(**)$ is when $n = m = 0$.  If we consider the case when $n = m = 0$ in
$(**)$, we get the declaration of the terminal object.
\begin{center}
\begin{tabular}{l}
right object 1 with ! \\
end object \\
\end{tabular}
\end{center}
(We omitted the keyword `is' to make the declaration look nicer.)  From
the definition, this defines an object `1' and for any object $A$ there
is a unique morphism `!' from $A$ to `1'.
\begin{displaymath}
\ardiagram{A}{\rm !}{\rm 1}
\end{displaymath}
Therefore, it really is the terminal object.

Dually, if we change the keyword `right' to `left' in the definition of
the terminal object, we get the definition of the initial object.
\begin{center}
\begin{tabular}{l}
left object 0 with !! \\
end object \\
\end{tabular}
\end{center}
The factorizer `!!' gives a unique morphism from `0' to any object $A$.
\begin{displaymath}
\ardiagram{\rm 0}{\rm !!}{A}
\end{displaymath}

\subsection{Products and CoProducts}
\label{ssec-product-coproduct}

Next, we define products and coproducts.  The binary product functor can
be declared as the following right object.
\begin{displaymath}
\begin{tabular}{l}
right object ${\rm prod}(X,Y)$ with pair is \\
$\qquad {\rm pi1}: {\rm prod} \rightarrow X$ \\
$\qquad {\rm pi2}: {\rm prod} \rightarrow Y$ \\
end object \\
\end{tabular}
\end{displaymath}
From definition~\ref{def-CDT-CSL}, this defines a functor symbol `prod'
whose variance is `${+}{+}$' (i.e.\ covariant in both arguments), two
natural transformation symbols `pi1' and `pi2' whose types are
\begin{displaymath}
{\rm pi1}: \lambda(X,Y).{\rm prod}(X,Y) \natrightarrow \lambda(X,Y).X
\end{displaymath}
\begin{displaymath}
{\rm pi2}: \lambda(X,Y).{\rm prod}(X,Y) \natrightarrow \lambda(X,Y).Y
\end{displaymath}
and a factorizer symbol `pair' whose type is
\begin{displaymath}
\logicrule{f: \lambda(Z,X,Y).Z \rightarrow \lambda(Z,X,Y).X \qquad
g: \lambda(Z,X,Y).Z \rightarrow \lambda(Z,X,Y).Y}{{\rm pair}(f,g):
\lambda(Z,X,Y).Z \rightarrow \lambda(Z,X,Y).{\rm prod}(X,Y)}.
\end{displaymath}
If we write this down in a more understandable way, it becomes the
familiar definition of the binary product, that is `prod' has two unit
morphisms
\begin{displaymath}
\setlength{\unitlength}{1mm}
\begin{picture}(70,10)(0,2.5)
\put(5,5){\makebox(0,0){$A$}}
\put(35,5){\makebox(0,0){${\rm prod}(A,B)$}}
\put(65,5){\makebox(0,0){$B$}}
\put(17,7){\makebox(0,0)[b]{pi1}}
\put(53,7){\makebox(0,0)[b]{pi2}}
\put(25,5){\vector(-1,0){17.5}}
\put(45,5){\vector(1,0){17.5}}
\end{picture}
\end{displaymath}
and ${\rm pair}(f,g)$ gives the unique morphism for any morphisms $f: C
\rightarrow A$ and $g: C \rightarrow B$ such that the following diagram
commutes.
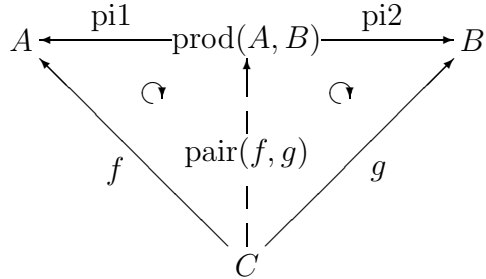
\begin{displaymath}
\setlength{\unitlength}{1mm}
\begin{picture}(70,40)(0,2.5)
\put(5,35){\makebox(0,0){$A$}}
\put(35,35){\makebox(0,0){${\rm prod}(A,B)$}}
\put(65,35){\makebox(0,0){$B$}}
\put(17,37){\makebox(0,0)[b]{pi1}}
\put(53,37){\makebox(0,0)[b]{pi2}}
\put(25,35){\vector(-1,0){17.5}}
\put(45,35){\vector(1,0){17.5}}
\put(35,5){\makebox(0,0){$C$}}
\put(17.5,17.5){\makebox(0,0){$f$}}
\put(52.5,17.5){\makebox(0,0){$g$}}
\put(35,20){\makebox(0,0){${\rm pair}(f,g)$}}
\put(22.5,27.5){\makebox(0,0){\commute}}
\put(47.5,27.5){\makebox(0,0){\commute}}
\put(32.5,7.5){\vector(-1,1){25}}
\put(37.5,7.5){\vector(1,1){25}}
\multiput(35,7.5)(0,5){2}{\line(0,1){3}}
\put(35,22.5){\line(0,1){3}}
\put(35,27.5){\vector(0,1){5}}
\end{picture}
\end{displaymath}

Note that the CDT declaration of the binary product functor is very
similar to the `Prod' theory in CLEAR defined in~\ref{sec-what-cdt}.
One of the differences is that in CLEAR `pair' is treated as a function
in the same class as `pi1' and `pi2' but in CDT `pair' is quite
different from `pi1' and `pi2'.  Another one is that in CLEAR `Prod' is
declared as the initial algebra so it is close to CDT's left object but
in CDT `prod' is naturally a right object because the product functor is
the {\it right} adjoint of the diagonal functor.

Dually, we can define the binary coproduct functor as a left object.
\begin{displaymath}
\begin{tabular}{l}
left object ${\rm coprod}(X,Y)$ with case is \\
$\qquad {\rm in1}: X \rightarrow {\rm coprod}$ \\
$\qquad {\rm in2}: Y \rightarrow {\rm coprod}$ \\
end object \\
\end{tabular}
\end{displaymath}
Again, this declaration looks very close to the one in CLEAR (defined in
section~\ref{sec-what-cdt}), but note that we have `case' in CDT so that
we can use it to write programs or to specify some properties.

Just writing the situation as a diagram,
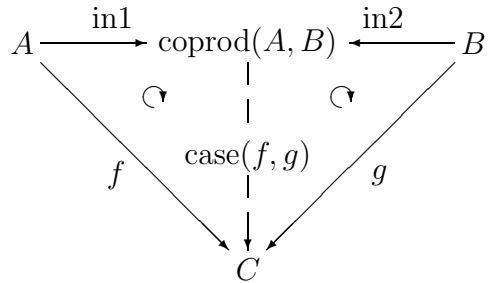
\begin{displaymath}
\setlength{\unitlength}{1mm}
\begin{picture}(70,40)(0,2.5)
\put(5,35){\makebox(0,0){$A$}}
\put(35,35){\makebox(0,0){${\rm coprod}(A,B)$}}
\put(65,35){\makebox(0,0){$B$}}
\put(17,37){\makebox(0,0)[b]{in1}}
\put(53,37){\makebox(0,0)[b]{in2}}
\put(7.5,35){\vector(1,0){14}}
\put(62.5,35){\vector(-1,0){14}}
\put(35,5){\makebox(0,0){$C$}}
\put(17.5,17.5){\makebox(0,0){$f$}}
\put(52.5,17.5){\makebox(0,0){$g$}}
\put(35,20){\makebox(0,0){${\rm case}(f,g)$}}
\put(22.5,27.5){\makebox(0,0){\commute}}
\put(47.5,27.5){\makebox(0,0){\commute}}
\put(7.5,32.5){\vector(1,-1){25}}
\put(62.5,32.5){\vector(-1,-1){25}}
\multiput(35,32.5)(0,-5){2}{\line(0,-1){3}}
\put(35,17.5){\line(0,-1){3}}
\put(35,12.5){\vector(0,-1){5}}
\end{picture}
\end{displaymath}

\subsection{Exponentials}
\label{ssec-exponential}

One of the objections against algebraic specification methods is that it
cannot handle function spaces.  CDT's declaration mechanism
looks very close to that of algebraic specification methods, but CDT is
based on category theory not on many-sorted algebras, and in category
theory function spaces can be defined as exponentials.  For objects $A$
and $B$, the exponential of $B$ by $A$ is written as ${\rm
exp}(A,B)$\footnote{Many category theory books use the notation $B^A$
for the exponential of $B$ by $A$.} satisfies the following natural
isomorphism.
\begin{displaymath}
\Hom{\cC}{{\rm prod}(C,A)}{B} \simeq \Hom{\cC}{C}{{\rm exp}(A,B)}
\end{displaymath}
In other words, The functor ${\rm exp}(A,\;\cdot\;)$ is the right
adjoint of ${\rm prod}(\;\cdot\;,A)$.  We write `curry' for the
factorizer and `eval' for the counit natural transformation.  Then, for
any object $C$ and any morphism $f: {\rm prod}(C,A) \rightarrow B$,
${\rm curry}(f)$ is the unique morphism from ${\rm exp}(A,B)$ to $C$ such that
the following diagram commutes.
\begin{displaymath}
\setlength{\unitlength}{1mm}
\begin{picture}(90,40)(0,2.5)
\put(10,35){\makebox(0,0){$C$}}
\put(10,5){\makebox(0,0){${\rm exp}(A,B)$}}
\put(8,20){\makebox(0,0)[r]{${\rm curry}(f)$}}
\multiput(10,32.5)(0,-5){4}{\line(0,-1){3}}
\put(10,12.5){\vector(0,-1){5}}
\put(50,35){\makebox(0,0){${\rm prod}(C,A)$}}
\put(50,5){\makebox(0,0){${\rm prod}({\rm exp}(A,B),A)$}}
\put(90,5){\makebox(0,0){$B$}}
\put(48,20){\makebox(0,0)[r]{${\rm prod}({\rm curry}(f),{\bf I})$}}
\put(72,22){\makebox(0,0)[bl]{$f$}}
\put(62.5,17.5){\makebox(0,0){\commute}}
\put(75,3){\makebox(0,0)[t]{eval}}
\put(55,31){\vector(4,-3){32.5}}
\put(70,5){\vector(1,0){17.5}}
\multiput(50,32.5)(0,-5){4}{\line(0,-1){3}}
\put(50,12.5){\vector(0,-1){5}}
\end{picture}
\end{displaymath}
The reason why the exponentials are function spaces is that their global
elements are just morphisms.\footnote{A {\it global element} of an
object $A$ is a morphism from the terminal object to $A$.}
\begin{displaymath}
\Hom{\cC}{1}{{\rm exp}(A,B)} \simeq \Hom{\cC}{{\rm prod}(1,A)}{B} \simeq
\Hom{\cC}{A}{B}
\end{displaymath}

Let us write down the definition as a CDT declaration.  Assume a CSL
signature $\pair{\Gamma,\Delta,\Psi}$ which contains the definition of
the binary product functor as we defined in
subsection~\ref{ssec-product-coproduct}.  Then, the exponential functor
can be declared as follows.
\begin{displaymath}
\begin{tabular}{l}
right object ${\rm exp}(X,Y)$ with curry is \\
$\qquad {\rm eval}: {\rm prod}({\rm exp},X) \rightarrow Y$ \\
end object \\
\end{tabular}
\end{displaymath}
This is so far the most complicated CDT declaration.  In the previous
examples, functorial expressions $E_i$ and $E'_i$ in $(*)$ and $(**)$
are all simply variables.  From definition~\ref{def-CDT-CSL}, the CDT
declaration above defines a functor symbol `exp' of type ${-}{+}$, a
natural transformation `eval' of type
\begin{displaymath}
\lambda(X,Y).{\rm prod}({\rm exp}(X,Y),X) \natrightarrow \lambda(X,Y).Y
\end{displaymath}
and a factorizer `curry' of type
\begin{displaymath}
\logicrule{f: \lambda(Z,X,Y).{\rm prod}(Z,X) \rightarrow
\lambda(Z,X,Y).Y}{{\rm curry}(f): \lambda(Z,X,Y).Z \rightarrow
\lambda(Z,X,Y).{\rm exp}(X,Y)}.
\end{displaymath}
These types are what we expect them to be from the exponential
adjunction.  Let us once more convince ourselves that the semantics by
$F,G$-dialgebras really defines the exponentials.  $\pair{{\rm
exp}(A,B),{\rm eval}_{A,B}}$ is the final $F,G$-dialgebra where $F(C)
\defeq {\rm prod}(C,A)$ and $G(C) \defeq B$.  This means that, for any
$\pair{C,f}$ where $C$ is an object and $f$ is a morphism of $F(C)
\rightarrow G(C)$, ${\rm curry}(f)$ is the unique morphism of
\begin{displaymath}
\pair{{\rm exp}(A,B),{\rm eval}_{A,B}} \rightarrow \pair{C,f}.
\end{displaymath}
From definition~\ref{def-dbl-alg} of $F,G$-dialgebras, ${\rm curry}(f)$ is
the unique morphism $C \rightarrow {\rm exp}(A,B)$ which makes the
following diagram commute.
\begin{displaymath}
\setlength{\unitlength}{1mm}
\begin{picture}(90,40)(-20,2.5)
\put(12.5,35){\makebox(0,0){$\llap{$F(C)={}$}{\rm prod}(C,A)$}}
\put(60,35){\makebox(0,0){$B\rlap{${}=G(C)$}$}}
\put(37.5,37){\makebox(0,0)[b]{$f$}}
\put(12.5,5){\makebox(0,0){$\llap{$F({\rm exp}(A,B))={}$}{\rm prod}({\rm
exp}(A,B),A)$}}
\put(60,5){\makebox(0,0){$B\rlap{${}=G({\rm exp}(A,B))$}$}}
\put(42.5,3){\makebox(0,0)[t]{${\rm eval}_{A,B}$}}
\put(10.5,20){\makebox(0,0)[r]{$F({\rm curry}(f)) = {\rm prod}({\rm
curry}(f),{\bf I})$}}
\put(40,20){\makebox(0,0){\commute}}
\put(62.5,20){\makebox(0,0)[l]{${\bf I} = G({\rm curry}(f))$}}
\put(22.5,35){\vector(1,0){35}}
\put(30,5){\vector(1,0){27.5}}
\put(12.5,32.5){\vector(0,-1){25}}
\put(59.5,32.5){\line(0,-1){25}}
\put(60.5,32.5){\line(0,-1){25}}
\end{picture}
\end{displaymath}
This is exactly the condition of `exp' being the exponential functor.

The declaration of the exponential functor in CDT very much looks like
a declaration in a algebraic specification
language (e.g.\ in CLEAR), but, as is well-known, we cannot define function
spaces as algebras.  The essential difference lies in that `exp' is a right
object, in other words, defined by the terminal data constraint rather
than the initial one which CLEAR uses and in the availability of the
factorizer `curry'.  If we define `curry' as an ordinary function (or an
ML function), its type is
\begin{displaymath}
(C\times A \rightarrow B) \rightarrow (C \rightarrow (A \rightarrow B))
\end{displaymath}
and this could never be a type of algebraic functions (i.e.\ functions
defined by algebraic specification methods).

\subsection{Natural Number Object}
\label{ssec-NNO}

As we have already shown that the natural number object can be given by
`{\bf Left}', let us write it down as a CDT declaration.  Although we
can define the natural number object if we have only the terminal
object, it is often very convenient to assume that a CSL signature
$\pair{\Gamma,\Delta,\Psi}$ contains not only the terminal object but
also the product functor and the exponential functor.  The declaration
of the natural number object as a CDT is
\begin{displaymath}
\begin{tabular}{l}
left object nat with pr is \\
$\qquad {\rm zero}: 1 \rightarrow {\rm nat}$\\
$\qquad {\rm succ}: {\rm nat} \rightarrow {\rm nat}$\\
end object \\
\end{tabular}
\end{displaymath}
which defines a constant functor (i.e.\ an object) `nat' with two
morphisms `zero' and `succ'.
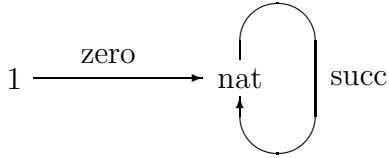
\begin{displaymath}
\setlength{\unitlength}{1mm}
\begin{picture}(50,20)(0,2.5)
\put(5,10){\makebox(0,0){1}}
\put(35,10){\makebox(0,0){nat}}
\put(17.5,12){\makebox(0,0)[b]{zero}}
\put(47,10){\makebox(0,0)[l]{succ}}
\put(7.5,10){\vector(1,0){22.5}}
\put(35,12.5){\line(0,1){2.5}}
\put(40,15){\oval(10,10)[t]}
\put(45,15){\line(0,-1){10}}
\put(40,5){\oval(10,10)[b]}
\put(35,5){\vector(0,1){2.5}}
\end{picture}
\end{displaymath}
In addition, the factorizer `pr' (standing for {\it p\/}rimitive {\it
r\/}ecursion) gives for any morphisms $f: 1 \rightarrow A$ and $g: A
\rightarrow A$ a unique morphism ${\rm pr}(f,g): {\rm nat} \rightarrow A$
such that the following diagram commutes.
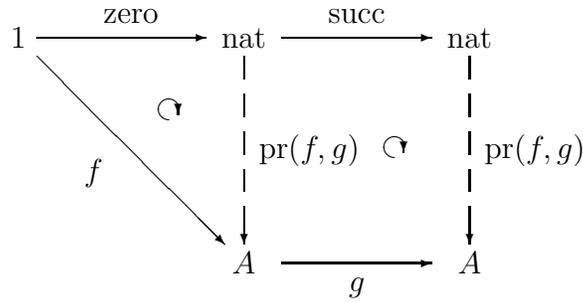
\begin{displaymath}
\setlength{\unitlength}{1mm}
\begin{picture}(75,45)(0,2.5)
\put(5,40){\makebox(0,0){1}}
\put(20,42){\makebox(0,0)[b]{zero}}
\put(35,40){\makebox(0,0){nat}}
\put(50,42){\makebox(0,0)[b]{succ}}
\put(65,40){\makebox(0,0){nat}}
\put(15,22){\makebox(0,0){$f$}}
\put(25,30){\makebox(0,0){\commute}}
\put(37,25){\makebox(0,0)[l]{${\rm pr}(f,g)$}}
\put(55,25){\makebox(0,0){\commute}}
\put(67,25){\makebox(0,0)[l]{${\rm pr}(f,g)$}}
\put(35,10){\makebox(0,0){$A$}}
\put(65,10){\makebox(0,0){$A$}}
\put(50,8){\makebox(0,0)[t]{$g$}}
\put(7.5,40){\vector(1,0){22.5}}
\put(40,40){\vector(1,0){20}}
\put(7.5,37.5){\vector(1,-1){25}}
\put(40,10){\vector(1,0){20}}
\multiput(35,37.5)(0,-5){4}{\line(0,-1){3}}
\put(35,17.5){\vector(0,-1){5}}
\multiput(65,37.5)(0,-5){4}{\line(0,-1){3}}
\put(65,17.5){\vector(0,-1){5}}
\end{picture}
\end{displaymath}
As is well-known (e.g.~\cite{goldblatt-79} chapter 13), `pr' provides us
to define any primitive recursive function.
\begin{definition}
\label{def-pr-rec}
For a category $\cC$ with the natural number object `nat', the terminal
object `1' and the binary product functor `prod', a morphism $f$ is {\it
primitive recursive} (on natural numbers) if it can be generated after
finitely many steps by means of the following rules:
\begin{enumerate}
\item $f = {\bf I}_{\rm nat}: {\rm nat} \rightarrow {\rm nat}$,
\item $f = {\rm zero}: 1 \rightarrow {\rm nat}$,
\item $f = {\rm succ}: {\rm nat} \rightarrow {\rm nat}$,
\item $f = {\rm pi1}: {\rm prod}({\rm nat},{\rm nat}) \rightarrow {\rm
nat}$,
\item $f = {\rm pi2}: {\rm prod}({\rm nat},{\rm nat}) \rightarrow {\rm
nat}$,
\item $f = g\circ{\rm pair}(h,k): A \rightarrow {\rm nat}$ for primitive
recursive morphisms\\
$g: {\rm prod}({\rm nat},{\rm nat}) \rightarrow {\rm nat}$, $h: A
\rightarrow {\rm nat}$ and $k: A \rightarrow {\rm nat}$,
\item $f = g\circ{\rm prod}(h,k): {\rm prod}(A,B) \rightarrow {\rm nat}$
for primitive recursive morphisms\\
$g: {\rm prod}({\rm nat},{\rm nat}) \rightarrow {\rm nat}$, $h: A
\rightarrow {\rm nat}$ and $k: B \rightarrow {\rm nat}$, and
\item $f: {\rm prod}({\rm nat},A) \rightarrow {\rm nat}$ satisfying
\begin{enumerate}
\item $f\circ{\rm pair}({\rm zero}\circ{\rm !},{\bf I}) = g$, and
\item $f\circ{\rm pair}({\rm succ}\circ{\rm pi1},{\rm pi2}) =
h\circ{\rm pair}(f,{\bf I})$
\end{enumerate}
for primitive recursive $g: A \rightarrow {\rm nat}$ and $h:
{\rm prod}({\rm nat},{\rm prod}({\rm nat},A)) \rightarrow {\rm nat}$. \qed
\end{enumerate}
\end{definition}
This is a straight copy of the standard definition of primitive
recursive functions on natural numbers.
\begin{proposition}
If a cartesian closed category $\cC$ has the natural number object, it
has all the primitive recursive morphisms. \\
{\bf Proof:} It is suffice to show that there exists a morphism $f: {\rm
prod}({\rm nat},A) \rightarrow {\rm nat}$ for any morphisms $g: A
 \rightarrow {\rm nat}$ and $h: {\rm  prod}({\rm nat},{\rm prod}({\rm
nat},A)) \rightarrow {\rm nat}$ such that
\begin{enumerate}
\item $f\circ{\rm pair}({\rm zero}\circ{\rm !},{\bf I}) = g$, and
\item $f\circ{\rm pair}({\rm succ}\circ{\rm pi1},{\rm pi2}) =
h\circ{\rm pair}(f,{\bf I})$.
\end{enumerate}
There is a morphism $k: {\rm nat} \rightarrow {\rm prod}({\rm
exp}(A,{\rm nat}),{\rm nat})$ such that the following diagram commutes.
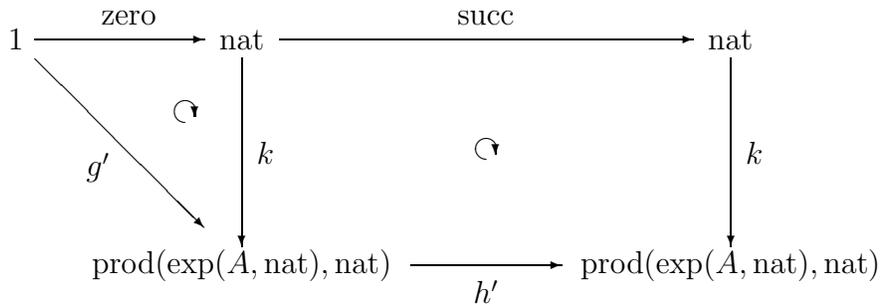
\begin{displaymath}
\setlength{\unitlength}{1mm}
\begin{picture}(115,40)(0,2.5)
\put(2.5,35){\makebox(0,0){1}}
\put(32.5,35){\makebox(0,0){nat}}
\put(97.5,35){\makebox(0,0){nat}}
\put(32.5,5){\makebox(0,0){${\rm prod}({\rm exp}(A,{\rm nat}),{\rm nat})$}}
\put(97.5,5){\makebox(0,0){${\rm prod}({\rm exp}(A,{\rm nat}),{\rm nat})$}}
\put(17.5,37){\makebox(0,0)[b]{zero}}
\put(65,37){\makebox(0,0)[b]{succ}}
\put(34.5,20){\makebox(0,0)[l]{$k$}}
\put(99.5,20){\makebox(0,0)[l]{$k$}}
\put(15,20){\makebox(0,0)[tr]{$g'$}}
\put(65,3){\makebox(0,0)[t]{$h'$}}
\put(25,25){\makebox(0,0){\commute}}
\put(65,20){\makebox(0,0){\commute}}
\put(5,35){\vector(1,0){22.5}}
\put(37.5,35){\vector(1,0){55}}
\put(5,32.5){\vector(1,-1){22.5}}
\put(32.5,32.5){\vector(0,-1){25}}
\put(97.5,32.5){\vector(0,-1){25}}
\put(55,5){\vector(1,0){20}}
\end{picture}
\end{displaymath}
where $g'$ and $h'$ are
\begin{displaymath}
\end{displaymath}
\begin{displaymath}
\begin{array}{llll}
\multicolumn{4}{l}{g' \defeq {\rm pair}({\rm curry}(g\circ{\rm pi2}),{\rm
zero})} \\
h' \defeq {\rm pair}( & {\rm curry}(h\circ{\rm pair}( & {\rm eval}\circ{\rm
pair}( & {\rm pi2}, \\
& & & {\rm pi1}\circ{\rm pi1}), \\
& & \multicolumn{2}{l}{{\rm pi2}\circ{\rm pi1})),} \\
& \multicolumn{3}{l}{{\rm succ}\circ{\rm pi2}).} \\
\end{array}
\end{displaymath}
Therefore, $k$ is ${\rm pr}(g',h')$.  Then, $f \defeq {\rm eval}\circ{\rm
prod}({\rm pi1}\circ k,{\bf I})$.  We can easily show that this is what
we wanted.
\end{proposition}
For example, the morphism `add' corresponding to the addition function
of natural numbers can be given as
\begin{displaymath}
{\rm add} \defeq {\rm eval}\circ{\rm prod}({\rm pr}({\rm curry}({\rm
pi2}),{\rm curry}({\rm succ}\circ{\rm eval})),{\bf I}).
\end{displaymath}
It corresponds to the following usual definition of `add'.
\begin{displaymath}
\begin{array}{l}
{\rm add}(0,y) = y \\
{\rm add}(x+1,y) = {\rm add}(x,y)+1 \\
\end{array}
\end{displaymath}
Furthermore, we can easily prove categorically that `add' satisfies
familiar laws like commutativity (categorically ${\rm add} \circ
{\rm pair}({\rm pi2},{\rm pi1}) = {\rm add}$) and so on.

\subsection{Lists}
\label{ssec-list}

We have defined the data type of natural numbers in the previous
subsection.  Another algebraic data type which is often used in
programming is the data type of lists.  In CDT, the data type of lists
is defined as follows:
\begin{displaymath}
\begin{tabular}{l}
left object ${\rm list}(X)$ with prl is \\
$\qquad{\rm nil}: 1 \rightarrow {\rm list}$ \\
$\qquad{\rm cons}: {\rm prod}(X,{\rm list}) \rightarrow {\rm list}$ \\
end object \\
\end{tabular}
\end{displaymath}
We needed to assume a CSL signature having the terminal object and the
product functor.  The declaration above defines a one argument covariant
functor `list', two natural transformations
\begin{displaymath}
{\rm nil}: \lambda(X).1 \natrightarrow \lambda(X).{\rm list}(X)
\quad\mbox{and}\quad
{\rm cons}: \lambda(X).{\rm prod}(X,{\rm list}(X)) \natrightarrow
\lambda(X).{\rm list}(X)
\end{displaymath}
and a factorizer `prl' (standing for {\it p\/}rimitive {\it r\/}ecursion
on {\it l\/}ist) whose type is
\begin{displaymath}
\logicrule{f: \lambda(Y,X).1 \rightarrow \lambda(Y,X).Y \qquad g:
\lambda(Y,X).{\rm prod}(X,Y) \rightarrow \lambda(Y,X).Y}{{\rm prl}(f,g):
\lambda(Y,X).{\rm list}(X) \rightarrow \lambda(Y,X).Y}.
\end{displaymath}
As usual, we can express the situation as a diagram.
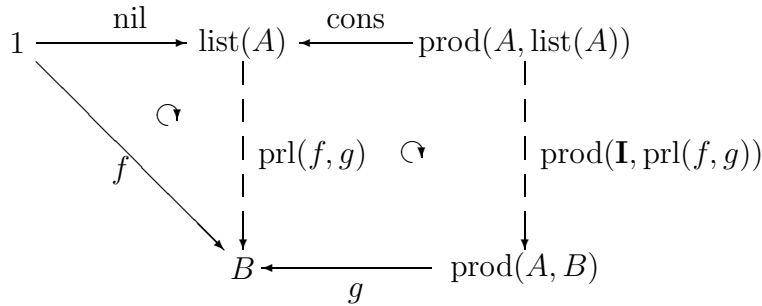
\begin{displaymath}
\setlength{\unitlength}{1mm}
\begin{picture}(100,35)(0,2.5)
\put(5,35){\makebox(0,0){1}}
\put(35,35){\makebox(0,0){${\rm list}(A)$}}
\put(72.5,35){\makebox(0,0){${\rm prod}(A,{\rm list}(A))$}}
\put(35,5){\makebox(0,0){$B$}}
\put(72.5,5){\makebox(0,0){${\rm prod}(A,B)$}}
\put(20,37){\makebox(0,0)[b]{nil}}
\put(50,37){\makebox(0,0)[b]{cons}}
\put(20,20){\makebox(0,0)[tr]{$f$}}
\put(25,25){\makebox(0,0){\commute}}
\put(37,20){\makebox(0,0)[l]{${\rm prl}(f,g)$}}
\put(57.5,20){\makebox(0,0){\commute}}
\put(74.5,20){\makebox(0,0)[l]{${\rm prod}({\bf I},{\rm prl}(f,g))$}}
\put(50,3){\makebox(0,0)[t]{$g$}}
\put(7.5,35){\vector(1,0){20}}
\put(7.5,32.5){\vector(1,-1){25}}
\put(57.5,35){\vector(-1,0){15}}
\put(60,5){\vector(-1,0){22.5}}
\multiput(35,32.5)(0,-5){4}{\line(0,-1){3}}
\put(35,12.5){\vector(0,-1){5}}
\multiput(72.5,32.5)(0,-5){4}{\line(0,-1){3}}
\put(72.5,12.5){\vector(0,-1){5}}
\end{picture}
\end{displaymath}
A global element of ${\rm list}(A)$ is normally constructed from
`nil' and `cons'.  For example, ${\rm list}({\rm nat})$ has
\begin{displaymath}
{\rm cons}\circ{\rm pair}({\rm succ}\circ{\rm zero},{\rm cons}\circ{\rm
pair}({\rm succ}\circ{\rm succ}\circ{\rm zero},{\rm nil}))
\end{displaymath}
as a global element (in plain words, this element is the list of 1 and
2).  `nil' and `cons' are usually called constructors of `list'.  We can
see in general to define an algebraic CDT by listing its constructors.
Destructors are defined using factorizers.  In the case of `list', `hd'
({\it h\/}ea{\it d\/}) and `tl' ({\it t\/}ai{\it l\/}) can be defined as
follows.
\begin{displaymath}
\begin{array}{l}
{\rm hd} \defeq {\rm prl}({\rm in2},{\rm in1}\circ{\rm pi1}) \\
{\rm tl} \defeq {\rm coprod}({\rm pi2},{\bf I})\circ{\rm prl}({\rm
in2},{\rm in1}\circ{\rm prod}({\bf I},{\rm case}({\rm cons},{\rm nil}))) \\
\end{array}
\end{displaymath}
Note that we have to define `hd' and `tl' as total functions (in a
sense).  The type of `hd' is ${\rm list}(A) \rightarrow {\rm
coprod}(A,1)$ and is not ${\rm list}(A) \rightarrow A$.  The type of
`tl' is also ${\rm list}(A) \rightarrow {\rm coprod}({\rm list}(A),1)$.
The type `1' is for error (like $\bot$ in a domain) and, for example,
${\rm hd} \circ {\rm nil} = {\rm in2}$.

As `list' is a covariant functor, for a morphism $f: A \rightarrow B$ ${\rm
list}(f): {\rm list}(A) \rightarrow {\rm list}(B)$ transforms a list of
$A$ elements to a list of $B$ elements by applying $f$ to each element.
For example, ${\rm list}({\rm succ}): {\rm list}({\rm nat}) \rightarrow
{\rm list}({\rm nat})$ increments every element in a given list by one.
In general, we have the following equations:
\begin{displaymath}
\begin{array}{l}
{\rm list}(f) = {\rm prl}({\rm nil},{\rm cons}\circ{\rm prod}(f,{\bf
I})) \\
{\rm list}(f)\circ{\rm nil} = {\rm nil} \\
{\rm list}(f)\circ{\rm cons}\circ{\rm pair}(x,l) = {\rm cons}\circ{\rm
pair}(f\circ x,{\rm list}(f)\circ l)
\end{array}
\end{displaymath}
`list' corresponds to `{\tt map}' function in ML and `{\tt MAPCAR}' in
LISP.

\subsection{Final Co-Algebras (Infinite Lists and Co-Natural Number
Object)}
\label{ssec-final-coalg}

The objects we defined in the preceding subsections are all familiar
either in category theory or in programming languages.  Particularly, we
have seen in subsection~\ref{ssec-NNO} and~\ref{ssec-list} the natural
number object and the data type of lists, which are typical initial
algebras.  Recently, several works have been done about final
coalgebras, which are the dual of initial algebras (see
\cite{arbib-manes-80}).  From their symmetry of CDT declarations, we can
easily define final coalgebras in CDT as well as initial algebras.

Let us dualize the declaration of the natural number object defined in
subsection~\ref{ssec-NNO} by
\begin{displaymath}
\begin{tabular}{l}
left object nat with pr is \\
$\qquad {\rm zero}: 1 \rightarrow {\rm nat}$\\
$\qquad {\rm succ}: {\rm nat} \rightarrow {\rm nat}$\\
end object \\
\end{tabular}
\end{displaymath}
If we simply replace `left' by `right' and change the direction of
arrows, we get
\begin{displaymath}
\begin{tabular}{l}
right object conat with copr is \\
$\qquad {\rm cozero}: {\rm nat} \rightarrow 1$\\
$\qquad {\rm cosucc}: {\rm nat} \rightarrow {\rm nat}$\\
end object \\
\end{tabular}
\end{displaymath}
Unfortunately, this is not an exciting object.  We can prove that
`conat' is isomorphic to the terminal object as follows: from the
uniqueness of terminal objects up to isomorphism in any category, we
simply need to show that `1' is the terminal $F,G$-dialgebra for these
particular $F$ and $G$, that is, for any object $A$ and morphisms $f: A
\rightarrow 1$ and $g: A \rightarrow A$, there exists a unique morphism
$h: A \rightarrow 1$ such that the following diagram commutes.
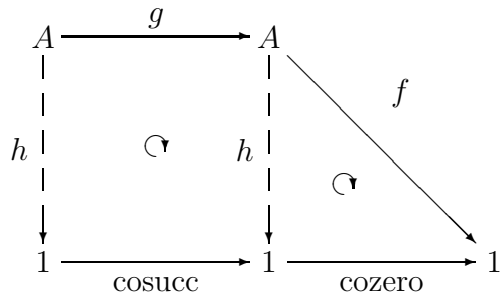
\begin{displaymath}
\setlength{\unitlength}{1mm}
\begin{picture}(70,40)(0,2.5)
\put(5,35){\makebox(0,0){$A$}}
\put(20,37){\makebox(0,0)[b]{$g$}}
\put(35,35){\makebox(0,0){$A$}}
\put(3,20){\makebox(0,0)[r]{$h$}}
\put(20,20){\makebox(0,0){\commute}}
\put(33,20){\makebox(0,0)[r]{$h$}}
\put(45,15){\makebox(0,0){\commute}}
\put(52.5,27.5){\makebox(0,0){$f$}}
\put(5,5){\makebox(0,0){1}}
\put(35,5){\makebox(0,0){1}}
\put(65,5){\makebox(0,0){1}}
\put(20,3){\makebox(0,0)[t]{cosucc}}
\put(50,3){\makebox(0,0)[t]{cozero}}
\put(7.5,35){\vector(1,0){25}}
\put(7.5,5){\vector(1,0){25}}
\put(37.5,5){\vector(1,0){25}}
\put(37.5,32.5){\vector(1,-1){25}}
\multiput(5,32.5)(0,-5){4}{\line(0,-1){3}}
\put(5,12.5){\vector(0,-1){5}}
\multiput(35,32.5)(0,-5){4}{\line(0,-1){3}}
\put(35,12.5){\vector(0,-1){5}}
\end{picture}
\end{displaymath}
Indeed, we have this unique morphism $A \rightarrow 1$ because `1' is
the terminal object and the above diagram trivially commutes.

Although the exact dual of the natural number object is not an
interesting thing, we can modify it to get a CDT data type of infinite
lists.
\begin{displaymath}
\begin{tabular}{l}
right object ${\rm inflist}(X)$ with fold is \\
$\qquad {\rm head}: {\rm inflist} \rightarrow X$ \\
$\qquad {\rm tail}: {\rm inflist} \rightarrow {\rm inflist}$ \\
end object \\
\end{tabular}
\end{displaymath}
The diagram of explaining `inflist' is
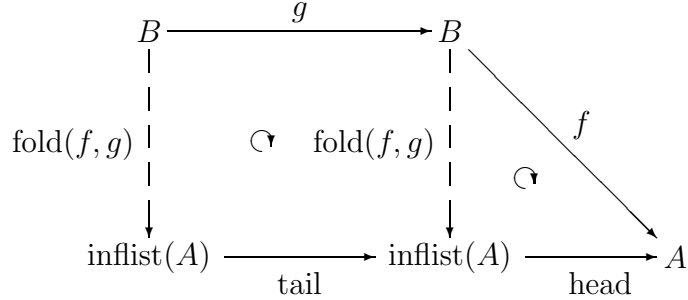
\begin{displaymath}
\setlength{\unitlength}{1mm}
\begin{picture}(90,40)(0,2.5)
\put(15,35){\makebox(0,0){$B$}}
\put(55,35){\makebox(0,0){$B$}}
\put(13,20){\makebox(0,0)[r]{${\rm fold}(f,g)$}}
\put(53,20){\makebox(0,0)[r]{${\rm fold}(f,g)$}}
\put(72.5,22.5){\makebox(0,0){$f$}}
\put(30,20){\makebox(0,0){\commute}}
\put(65,15){\makebox(0,0){\commute}}
\put(15,5){\makebox(0,0){${\rm inflist}(A)$}}
\put(55,5){\makebox(0,0){${\rm inflist}(A)$}}
\put(85,5){\makebox(0,0){$A$}}
\put(35,3){\makebox(0,0)[t]{tail}}
\put(75,3){\makebox(0,0)[t]{head}}
\put(17.5,35){\vector(1,0){35}}
\put(25,5){\vector(1,0){20}}
\put(65,5){\vector(1,0){17.5}}
\put(57.5,32.5){\vector(1,-1){25}}
\put(35,37){\makebox(0,0)[b]{$g$}}
\multiput(15,32.5)(0,-5){4}{\line(0,-1){3}}
\put(15,12.5){\vector(0,-1){5}}
\multiput(55,32.5)(0,-5){4}{\line(0,-1){3}}
\put(55,12.5){\vector(0,-1){5}}
\end{picture}
\end{displaymath}

Since the functor `inflist' is not so familiar in category theory or in
conventional programming languages, let us find out what it is in the
category of sets.  We expect it to be a set of {\it infinite lists} in some
sense.
\begin{proposition}
\label{prop-inflist-set}
In the category of sets, for a set $A$, ${\rm inflist}(A)$ is the
following set of $\omega$-infinite sequences of elements in $A$.
\begin{displaymath}
\{\; (x_0,x_1,\ldots,x_n,\ldots) \mid x_i \in A \;\}
\end{displaymath}
{\bf Proof:} We define `head' and `tail' as follows:
\begin{displaymath}
\begin{array}{l}
{\rm head}((x_0,x_1,\ldots,x_n,\ldots)) \defeq x_0 \\
{\rm tail}((x_0,x_1,\ldots,x_n,\ldots)) \defeq
(x_1,x_2,\ldots,x_{n+1},\ldots) \\
\end{array}
\end{displaymath}
Let ${\rm fold}(f,g)(x)$ be a sequence
$(h_0(x),h_1(x),\ldots,h_n(x),\ldots)$ for functions $f: B \rightarrow
A$ and $g: B \rightarrow B$.  The commutativity of the diagram above
forces the following equations.
\begin{displaymath}
\begin{array}{l}
h_0(x) = f(x) \\
(h_1(x),h_2(x),\ldots,h_{n+1}(x),\ldots) =
(h_0(x),h_1(x),\ldots,h_n(x),\ldots) \\
\end{array}
\end{displaymath}
Therefore, ${\rm fold}(f,g)(x)$ is uniquely determined as
\begin{displaymath}
{\rm fold}(f,g)(x) \defeq (f(x),f(g(x)),\ldots,f(g^n(x)),\ldots)
\rlap{\qquad \qquad \qed}
\end{displaymath}
\end{proposition}
Hence, at least in the category of sets, ${\rm inflist}(A)$ is really the
data type of infinite lists of $A$ elements.

More generally,
\begin{proposition}
In a cartesian closed category $\cC$ with the natural number object,
${\rm inflist}(A)$ is isomorphic to ${\rm exp}({\rm nat},A)$. \\
{\bf Proof:} Let us define $h: {\rm exp}({\rm nat},A) \rightarrow {\rm
inflist}(A)$ to be the fill-in morphism of the following diagram.
\begin{displaymath}
\setlength{\unitlength}{1mm}
\begin{picture}(90,40)(0,2.5)
\put(10,35){\makebox(0,0){${\rm exp}({\rm nat},A)$}}
\put(35,37){\makebox(0,0)[b]{${\rm tail}'$}}
\put(60,35){\makebox(0,0){${\rm exp}({\rm nat},A)$}}
\put(8,20){\makebox(0,0)[r]{$h$}}
\put(35,20){\makebox(0,0){\commute}}
\put(58,20){\makebox(0,0)[r]{$h$}}
\put(77,22){\makebox(0,0)[bl]{${\rm head}'$}}
\put(70,15){\makebox(0,0){\commute}}
\put(10,5){\makebox(0,0){${\rm inflist}(A)$}}
\put(35,3){\makebox(0,0)[t]{tail}}
\put(60,5){\makebox(0,0){${\rm inflist}(A)$}}
\put(77.5,3){\makebox(0,0)[t]{head}}
\put(90,5){\makebox(0,0){$A$}}
\put(22.5,35){\vector(1,0){25}}
\put(20,5){\vector(1,0){30}}
\put(70,5){\vector(1,0){17.5}}
\put(62.5,32.5){\vector(1,-1){25}}
\multiput(10,32.5)(0,-5){4}{\line(0,-1){3}}
\put(10,12.5){\vector(0,-1){5}}
\multiput(60,32.5)(0,-5){4}{\line(0,-1){3}}
\put(60,12.5){\vector(0,-1){5}}
\end{picture}
\end{displaymath}
where `${\rm head}'$' and `${\rm tail}'$' are
\begin{displaymath}
\begin{array}{l}
{\rm head}' \defeq {\rm eval}\circ{\rm pair}({\bf I},{\rm zero}\circ{\rm
!}), \\
{\rm tail}' \defeq {\rm curry}({\rm eval}\circ{\rm prod}({\bf I},{\rm
succ})). \\
\end{array}
\end{displaymath}
We define $h': {\rm inflist}(A) \rightarrow {\rm exp}({\rm nat},A)$ to
be
\begin{displaymath}
h' \defeq {\rm curry}({\rm eval}\circ{\rm pair}({\rm pr}({\rm
curry}({\rm head}\circ{\rm pi2}),{\rm exp}({\rm tail},{\bf I}))\circ{\rm
pi2},{\rm pi1})).
\end{displaymath}
After some calculation, we can show that $h\circ h' = {\bf I}$ and
$h'\circ h = {\bf I}$, so ${\rm inflist}(A)$ is isomorphic to ${\rm
exp}({\rm nat},A)$.
\end{proposition}
This proposition tells us that
\begin{displaymath}
{\rm inflist}(A) \iso A\times A\times\cdots\times A\times\cdots \iso
\prod_{i=1}^{\infty} A.
\end{displaymath}
Indeed, this is the dual of $\sum_{i=1}^{\infty} A$, the special case of
which is the natural number object, ${\rm nat} \iso \sum_{i=1}^{\infty}
1$.

We started this subsection by considering the dual of the natural number
object.  It led us to the CDT data type of infinite lists.  We still
have a different question whether there is a final coalgebra which
resembles a data type of natural numbers.  The answer is yes.  The
following right object defines a CDT data type of natural numbers plus
alpha.
\begin{displaymath}
\begin{tabular}{l}
right object conat with copr is \\
$\qquad {\rm pred}: {\rm conat} \rightarrow {\rm coprod}(1,{\rm conat})$
\\
end object \\
\end{tabular}
\end{displaymath}
The situation can be written as a diagram
\begin{displaymath}
\setlength{\unitlength}{1mm}
\begin{picture}(110,40)(0,2.5)
\put(20,35){\makebox(0,0){$A$}}
\put(40,37){\makebox(0,0)[b]{$f$}}
\put(70,35){\makebox(0,0){${\rm coprod}(1,A)$}}
\put(18,20){\makebox(0,0)[r]{${\rm copr}(f)$}}
\put(45,20){\makebox(0,0){\commute}}
\put(72,20){\makebox(0,0)[l]{${\rm coprod}(1,{\rm copr}(f))$}}
\put(20,5){\makebox(0,0){conat}}
\put(40,3){\makebox(0,0)[t]{pred}}
\put(70,5){\makebox(0,0){${\rm coprod}(1,{\rm conat})$}}
\put(22.5,35){\vector(1,0){35}}
\put(27.5,5){\vector(1,0){25}}
\multiput(20,32.5)(0,-5){4}{\line(0,-1){5}}
\put(20,12.5){\vector(0,-1){5}}
\multiput(70,32.5)(0,-5){4}{\line(0,-1){5}}
\put(70,12.5){\vector(0,-1){5}}
\end{picture}
\end{displaymath}
The natural transformation `pred' is the predecessor function and there
is a morphism from `nat' to `conat' given by
\begin{displaymath}
{\rm copr}({\rm pr}({\rm in1},{\rm in2}\circ{\rm case}({\rm zero},{\rm
succ})))
\end{displaymath}
which we expect to be injective, but so far the author has been able
neither to prove it nor to give a counter example.  Note that there is
always a morphism from the left object to its corresponding right
object.  `Conat' has an interesting extra element, namely the infinity
($\infty$).  The ground element to denote it is
\begin{displaymath}
{\rm infinity} \defeq {\rm copr}({\rm in2})
\end{displaymath}
It is easy to prove that the predecessor of the infinity is itself (i.e.
${\rm pred}\circ{\rm infinity} = {\rm in2}\circ{\rm infinity}$).

In the category of sets, `conat' is really the set of natural numbers
and the infinity.
\begin{proposition}
In {\bf Set}, `conat' is `${\rm nat} \cup \{\; \infty \;\}$'. \\
{\bf Proof:} The predecessor function is defined as usual.  Roughly
speaking, for any function $f : A \rightarrow {\rm coprod}(1,A)$, ${\rm
copr}(f)(x)$ is the number of applications of $f$ to $x$ to get the
element of 1
\begin{displaymath}
{\rm copr}(f)(x) = n \qquad\mbox{where}\quad f^n(x) \in 1,
\end{displaymath}
and, if it never results in the element of 1, ${\rm copr}(f)(x) =
\infty$.  We can easily show that this is the unique function which makes
the `conat' diagram commute.
\end{proposition}
Therefore, in the category of sets, `conat' is isomorphic to `nat', but
this is not the case for all the categories.  There are some categories
where `conat' and `nat' are not isomorphic.
\begin{proposition}
In category {\bf TRF} of subsets of natural numbers as objects and total
recursive functions as morphisms, there exists the natural number object
but does not exist the co-natural number object. \\ {\bf Proof:} {\bf
TRF}'s terminal object, initial object, product functor and coproduct
functor are all the same as those of {\bf Set}.  For example, injections
`in1' and `in2' for ${\rm coprod}(A,B)$ are trivially total and
recursive, and for any two total recursive functions $f: A \rightarrow
C$ and $g: B \rightarrow C$ ${\rm case}(f,g)$ is also total recursive
function.  We can write it down as a kind of program.
\begin{displaymath}
\setlength{\tabcolsep}{0pt}
\begin{tabular}{rl}
${\rm case}(f,g)(x) \defeq {}$ & if $x \in A$ then $f(x)$ \\
& else $g(x)$ \\
\end{tabular}
\end{displaymath}
The natural number object `nat' in {\bf TRF} is also the ordinary set of
natural numbers.  `Zero' and `succ' are total recursive functions from
the very definition of recursive functions and for $f: 1 \rightarrow A$
and $g: A \rightarrow A$ of total recursive functions, ${\rm pr}(f,g)$
can be written as the following program.
\begin{displaymath}
\setlength{\tabcolsep}{0pt}
\begin{tabular}{rl}
${\rm pr}(f,g)(n) \defeq {}$ & if $n=0$ then $f()$ \\
& else $g({\rm pr}(f,g)(n-1))$ \\
\end{tabular}
\end{displaymath}
which defines a total recursive function.

However, we cannot have the co-natural number object in {\bf TRF}.  The
program of ${\rm copr}(f)$ for a total recursive function $f: A \rightarrow
{\rm coprod}(1,A)$ can only be
\begin{displaymath}
\setlength{\tabcolsep}{0pt}
\begin{tabular}{rl}
${\rm copr}(f)(x) \defeq {}$ & if $f(x) \in 1$ then 0 \\
& else ${\rm copr}(f)(f(x))+1$ \\
\end{tabular}
\end{displaymath}
which is recursive but not total.
\end{proposition}
There is also a category which has both `nat' and `conat' and in which
they are non-isomorphic.  We will show this in chapter~\ref{ch-cpl}.

From the point of view of finding fixed points of functors, `nat' is
the initial fixed point of $F(X) \defeq 1+X$ and `conat' is the final
fixed point of the same functor.

\subsection{Automata}
\label{ssec-automata}

The declarations of initial algebras and final coalgebras do not use the
full power of the CDT declaration mechanism.  Their unit and counit
natural transformations always have the form
\begin{displaymath}
\alpha: E \rightarrow F
\end{displaymath}
for initial algebras and have the form
\begin{displaymath}
\alpha: F \rightarrow E
\end{displaymath}
for final coalgebras, where $E$ is any functorial expression but F is a
variable (more specially, the variable which denotes the object we
declare).  Therefore, all we are doing is just listing constructors for
initial algebras and listing destructors for final coalgebras.  We will
see what kinds of form define sensible functors in
section~\ref{sec-exist-left-right}.  So far, the only exception was the
exponentials.  We will see another example in this subsection.

One of the interesting applications of category theory to computer
science is to automata theory.  \cite{arbib-manes-75} presents the
category ${\bf Dyn}(I)$ of $I$-dynamics whose object is
\begin{displaymath}
\ardiagram{Q \times I}{\delta}{Q,}
\end{displaymath}
where $Q$ is the set of states, $I$ is the set of inputs and $\delta$ is
a {\it dynamics} which is a function determining the next state of the
automaton according to the current state and input.

From this, we can construct a categorical data type of automata.
\begin{displaymath}
\begin{tabular}{l}
right object ${\rm dyn}(I)$ with univ is \\
$\qquad {\rm next}: {\rm prod}({\rm dyn},I) \rightarrow {\rm dyn}$ \\
end object \\
\end{tabular}
\end{displaymath}
Note that our ${\rm dyn}(I)$ for an object $I$ is just an object; it is not a
category like ${\bf Dyn}(I)$ is.  The diagram which explains this right
object is
\begin{displaymath}
\setlength{\unitlength}{1mm}
\begin{picture}(70,40)(0,2.5)
\put(20,35){\makebox(0,0){${\rm prod}(Q,I)$}}
\put(42.5,37){\makebox(0,0)[b]{$\delta$}}
\put(60,35){\makebox(0,0){$Q$}}
\put(18,20){\makebox(0,0)[r]{${\rm prod}({\rm univ}(\delta),I)$}}
\put(42.5,20){\makebox(0,0){\commute}}
\put(62,20){\makebox(0,0)[l]{${\rm univ}(\delta)$}}
\put(20,5){\makebox(0,0){${\rm prod}({\rm dyn}(I),I)$}}
\put(42.5,3){\makebox(0,0)[t]{next}}
\put(60,5){\makebox(0,0){dyn}}
\put(30,35){\vector(1,0){27.5}}
\put(32.5,5){\vector(1,0){22.5}}
\multiput(20,32.5)(0,-5){4}{\line(0,-1){3}}
\put(20,12.5){\vector(0,-1){5}}
\multiput(60,32.5)(0,-5){4}{\line(0,-1){3}}
\put(60,12.5){\vector(0,-1){5}}
\end{picture}
\end{displaymath}
For any dynamics $\delta: {\rm prod}(Q,I) \rightarrow Q$ and an
initial state $q_0: 1 \rightarrow Q$, we get an automaton
\begin{displaymath}
\ardiagram{\rm 1}{{\rm univ}(\delta)\circ q_0}{{\rm dyn}(I)}
\end{displaymath}
as a global element of ${\rm dyn}(I)$.  Though we can put this
automaton into the next state by applying `next', we are never ever able
to see its behaviour from the outside.  Moreover, because of this
non-observability, we can easily prove that ${\rm dyn}(I)$ is in fact
isomorphic to the terminal object.  In order to make `dyn' a more
sensible object, we need to add an output function.  The new categorical data
type of automata is\footnote{The original definition we used was
\begin{displaymath}
\begin{tabular}{l}
right object ${\rm dyn}'(I,O)$ with ${\rm univ}'$ is \\
$\qquad{\rm next}': {\rm prod}({\rm dyn}',I) \rightarrow
{\rm prod}({\rm dyn}',O)$ \\
end object \\
\end{tabular}
\end{displaymath}
which gave us the categorical data type of Mealy automata.
The current definition gives us the data type of Moore automata.}
\begin{displaymath}
\begin{tabular}{l}
right object ${\rm dyn}'(I,O)$ with ${\rm univ}'$ is \\
$\qquad {\rm next}': {\rm prod}({\rm dyn}',I) \rightarrow {\rm dyn}'$ \\
$\qquad {\rm output}': {\rm dyn}' \rightarrow O$ \\
end object \\
\end{tabular}
\end{displaymath}
For any dynamics $\delta: {\rm prod}(Q,I) \rightarrow Q$, any
output function $\beta: Q \rightarrow O$ and an initial state
$q_0: 1 \rightarrow Q$, we have a global element in ${\rm dyn}'(I,O)$
\begin{displaymath}
\ardiagram{1}{{\rm univ}'(\delta,\beta)\circ q_0}{{\rm dyn}'(I,O).}
\end{displaymath}
We can obtain its next state by applying `${\rm next}'$' and its output
by `$\rm{output}'$'.  In addition, the following proposition holds.
\begin{proposition}
In a cartesian closed category, the categorical data type of Moore automata,
${\rm dyn}'(I,O)$, is isomorphic to ${\rm exp}({\rm list}(I),O)$. \\
{\bf Proof:} By defining two morphisms between them and proving
that they form an isomorphism.
\end{proposition}

\subsection{Obscure Categorical Data Types}
\label{ssec-obscure}

We have defined more or less familiar data types as categorical data
types in the preceding subsections.  One might ask whether CDT can
define any data types which are unable to be defined in other languages
or methods.  The data type of automata is such an example and we can
invent similar examples more, but still they are familiar (or we are
just trying to express our familiar data types in CDT).  In fact, CDT
allows us very obscure data types, some of which may not be
conceptualized in the human brain (at least not in the author's brain).

From the prime requirement of CDT, it can define right and left adjoint
functors of any existing functors, and in subsection~\ref{ssec-list}, we
defined `list' as a covariant functor, so that we can declare its left
and right adjoint functors in CDT as follows.
\begin{displaymath}
\begin{tabular}{l}
left object ${\rm ladjlist}(X)$ with $\psi$ is \\
$\qquad \alpha: X \rightarrow {\rm list}({\rm ladjlist})$ \\
end object \\
\end{tabular}
\end{displaymath}
\begin{displaymath}
\begin{tabular}{l}
right object ${\rm radjlist}(X)$ with $\psi'$ is \\
$\qquad \alpha': {\rm list}({\rm radjlist}) \rightarrow X$ \\
end object \\
\end{tabular}
\end{displaymath}
Some questions arise immediately after defining these data types: are
they familiar data types, and are they in any way useful?  The answers
to the both questions are unfortunately negative.  For the left adjoint,
\begin{proposition}
In a cartesian closed category, `${\rm ladjlist}(A)$' for any object $A$
is isomorphic to the initial object. \\
{\bf Proof:} It is easy to show that the initial object makes the
characteristic diagram of `ladjlist' commute.  Note that in a cartesian
closed category `${\rm list}(0)$' is isomorphic to the terminal object so
that the unit morphism of `ladjlist' is the unique morphism to the
terminal object.
\end{proposition}

The right adjoint functor is more harmful than the left one.
\begin{proposition}
A cartesian closed category with `radjlist' degenerates (i.e.\ all the
objects are isomorphic). \\
{\bf Proof:} We have the following morphism from the initial object.
\begin{displaymath}
1 \rubberrightarrow{{\rm nil}} {\rm list}({\rm radjlist}(0))
\rubberrightarrow{\alpha'} 0
\end{displaymath}
Then, it is easy to show that the terminal object is isomorphic to the
initial one.  This further implies that any object in the category is
isomorphic to the initial object.
\begin{displaymath}
A \iso {\rm prod}(A,1) \iso {\rm prod}(A,0) \iso 0 \rlap{\qquad \qquad \qed}
\end{displaymath}
\end{proposition}

Most of the left and right adjoint functors of conventional data types
follow the same pattern as `list', that is, they are either trivial or
destructive, so they are useless.

Hence, a natural question to ask ourselves is that what kind of
categorical data types are useful.  But what is the formal criteria of
{\it useful} data types?  We have not yet defined this.  We will come
back to this in chapter~\ref{ch-cpl} and see it from a point of view of
computability of categorical data types.

\section{Semantics of Categorical Data Types}
\label{sec-CDT-as-CSL}

In definition~\ref{def-CDT-CSL}, we associated a CDT declaration to a CSL
signature extension (an injective morphism in {\bf CSig}).  In this
section, we will see it as a CSL theory extension and give the precise
semantics of CDT declarations.

First, from our informal intention of CDT declarations we have to figure
out the CSL statements which characterize them.  A CDT
declaration
\begin{displaymath}
\begin{tabular}{l}
left object $L(X_1,\ldots,X_n)$ with $\psi$ is \\
$\qquad \alpha_1: E_1 \rightarrow E'_1$ \\
$\qquad\qquad \cdots$ \\
$\qquad \alpha_m: E_m \rightarrow E'_m$ \\
end object \\
\end{tabular}
\end{displaymath}
is the syntactic form of defining the functor $L = {\bf Left}[\vec
F,\vec G]$, where $\vec F$ and $\vec G$ are corresponding functors for
$E_1,\ldots,E_m$ and $E'_1,\ldots,E'_m$, respectively.  $\pair{L(\vec
A),\vec \alpha}$ is the initial object of ${\bf DAlg}(\vec F,\vec G)$
and $\psi$ is its mediating morphism, that is, for any morphisms $\vec
f: \vec F(B,\vec A) \rightarrow \vec G(B,\vec A)$, $\psi(\vec f)$ gives
a unique morphism from $L(\vec A)$ to $B$ such that the following
diagram commutes.
\begin{displaymath}
\setlength{\unitlength}{1mm}
\begin{picture}(95,40)(0,2.5)
\put(10,35){\makebox(0,0){$\vec F(L(\vec A),\vec A)$}}
\put(32.5,37){\makebox(0,0)[b]{$\vec\alpha_{\vec A}$}}
\put(55,35){\makebox(0,0){$\vec G(L(\vec A),\vec A)$}}
\put(90,35){\makebox(0,0){$L(\vec A)$}}
\put(8,20){\makebox(0,0)[r]{$\vec F(\psi(\vec f),\vec A)$}}
\put(32.5,20){\makebox(0,0){\commute}}
\put(57,20){\makebox(0,0)[l]{$\vec G(\psi(\vec f),\vec A)$}}
\put(92,20){\makebox(0,0)[l]{$\psi(\vec f)$}}
\put(10,5){\makebox(0,0){$\vec F(B,\vec A)$}}
\put(32.5,3){\makebox(0,0)[t]{$\vec f$}}
\put(55,5){\makebox(0,0){$\vec G(B,\vec A)$}}
\put(90,5){\makebox(0,0){$B$}}
\put(22.5,35){\vector(1,0){20}}
\put(17.5,5){\vector(1,0){30}}
\multiput(10,32.5)(0,-5){4}{\line(0,-1){3}}
\put(10,12.5){\vector(0,-1){5}}
\multiput(55,32.5)(0,-5){4}{\line(0,-1){3}}
\put(55,12.5){\vector(0,-1){5}}
\multiput(90,32.5)(0,-5){4}{\line(0,-1){3}}
\put(90,12.5){\vector(0,-1){5}}
\end{picture}
\end{displaymath}
The commutativity of this diagram can be expressed as the following equation:
\begin{displaymath}
\vec G(\psi(\vec f),\vec A) \circ \vec\alpha = \vec f \circ \vec
F(\psi(\vec f),\vec A),
\eqno(*)
\end{displaymath}
and the uniqueness can be expressed as the following conditional equation:
\begin{displaymath}
\vec G(h,\vec A) \circ \vec\alpha = \vec f \circ \vec
F(h,\vec A) \implies h = \psi(\vec f)
\eqno(**)
\end{displaymath}
The two equations say everything about $L$, $\vec \alpha$ and $\psi$.
Let us now translate them to CSL statements in $\pair{\Gamma\cup\{L\},
\Delta\cup\{\alpha_1,\ldots,\alpha_m\},
\Psi\cup\{\psi\}}$ so as to give complete description of the CDT declaration
above in CSL.  $(*)$ corresponds to the following $m$ CSL equations:
\begin{displaymath}
E'_i[\psi(f_1,\ldots,f_m)/L] \circ \alpha_i = f_i \circ
E_i[\psi(f_1,\ldots,f_m)/L]\footnote{$E_i[\psi(f_1,\ldots,f_m)/L]$ means
replacing the variable $L$ by $\psi(f_1,\ldots,f_m)$ and replacing the
other variables $X_1,\ldots,X_n$ by the identities, that is it is a
shorthand for
\begin{displaymath}
E_i[\psi(f_1,\ldots,f_m)/L,{\bf I}/X_1,\ldots,{\bf I}/X_n].
\end{displaymath}
}
\eqno ({\rm LEQ}_i) 
\end{displaymath}
($i=1,\ldots,m$) and $(**)$ corresponds to the following conditional CSL
equation.
\begin{displaymath}
\displaylines{
\qquad E'_1[g/L] \circ \alpha_1 = f_1 \circ E_1[g/L] \land \ldots {} \hfill \cr
\hfill {} \land E'_m[g/L] \circ \alpha_m = f_m \circ E_m[g/L] \implies
g = \psi(f_1,\ldots,f_m) \quad ({\rm LCEQ}) \cr}
\end{displaymath}

In addition, we should have a CSL equation expressing functors by
factorizers and natural transformations.  We can extract such an equation
from proposition~\ref{prop-Right-Left}.
\begin{displaymath}
\displaylines{
\qquad L(h_1,\ldots,h_n) = {} \hfill \cr
\hfill \psi(E'_1[h_i/X_i]\circ \alpha_1\circ
E_1[h_i/X_i],\ldots,E'_m[h_i/X_i]\circ \alpha_m\circ
E_m[h_i/X_i])\footnotemark
\quad ({\rm LFEQ}) \cr}
\end{displaymath}
\footnotetext{$E_1[h_i/X_i]$ is a shorthand for $E_1[{\bf
I}/L,h_1/X_1,\ldots,h_n/X_n]$.}

Therefore, the semantics of CDT declaration can be given as a CSL theory
extension as follows.
\begin{definition}
\label{def-CDT-sem-eq}
Given a CSL theory $\pair{\Gamma,\Delta,\Psi,\Theta}$, a CDT declaration
\begin{displaymath}
\begin{tabular}{l}
left object $L(X_1,\ldots,X_n)$ with $\psi$ is \\
$\qquad \alpha_1: E_1 \rightarrow E'_1$ \\
$\qquad\qquad \cdots$ \\
$\qquad \alpha_m: E_m \rightarrow E'_m$ \\
end object, \\
\end{tabular}
\end{displaymath}
where $E_i$ and $E'_i$ ($i=1,\ldots,m$) are CSL functorial expressions
over $\pair{\Gamma,\Delta,\Psi}$ whose variables are $X_1,\ldots,X_n$
and $L$, is associated with a CSL theory morphism
\begin{displaymath}
\displaylines{
\qquad \sigma_L: \pair{\Gamma,\Delta,\Psi,\Theta} \rightarrow
\langle \Gamma\cup\{L\}, \Delta\cup\{\alpha_1,\ldots,\alpha_m\},
\Psi\cup\{\psi\}, \hfill \cr
\hfill \Theta\cup\{{\rm LEQ}_1,\ldots,{\rm LEQ}_m,{\rm LCEQ},{\rm
LFEQ}\} \rangle \qquad \cr}
\end{displaymath}
where the types of $L$, $\alpha_1,\ldots,\alpha_m$ and $\psi$ are as
given in definition~\ref{def-CDT-CSL}.  Dually, we can associate to a
right object $R$
\begin{displaymath}
\begin{tabular}{l}
right object $R(X_1,\ldots,X_n)$ with $\psi$ is \\
$\qquad \alpha_1: E_1 \rightarrow E'_1$ \\
$\qquad\qquad \cdots$ \\
$\qquad \alpha_m: E_m \rightarrow E'_m$ \\
end object, \\
\end{tabular}
\end{displaymath}
a CSL theory morphism $\sigma_R$.
\begin{displaymath}
\displaylines{
\qquad \sigma_R: \pair{\Gamma,\Delta,\Psi,\Theta} \rightarrow
\langle \Gamma\cup\{R\}, \Delta\cup\{\alpha_1,\ldots,\alpha_m\},
\Psi\cup\{\psi\}, \hfill \cr
\hfill \Theta\cup\{{\rm REQ}_1,\ldots,{\rm REQ}_m,{\rm RCEQ},{\rm
RFEQ}\} \rangle \qquad \cr}
\end{displaymath}
where ${\rm REQ}_i$, ${\rm RCEQ}$ and ${\rm RFEQ}$ are
\begin{displaymath}
\alpha_i \circ E_i[\psi(f_1,\ldots,f_m)/R] =
E'_i[\psi(f_1,\ldots,f_m)/R] \circ f_i \eqno({\rm REQ}_i)
\end{displaymath}
\begin{displaymath}
\displaylines{
\qquad \alpha_1 \circ E_1[g/R] = E'_1[g/R] \circ f_1 \land \ldots {}
\hfill \cr
\hfill {} \land 
\alpha_m \circ E_m[g/R] = E'_m[g/R] \circ f_m \implies g =
\psi(f_1,\ldots,f_m) \quad ({\rm RCEQ}) \cr}
\end{displaymath}
\begin{displaymath}
\displaylines{
\qquad R(h_1,\ldots,h_n) = \hfill \cr
\hfill \psi(E'_1[h_1/X_i]\circ \alpha_1\circ
E_1[h_i/X_i],\ldots,E'_m[h_i/X_i]\circ \alpha_m \circ E_m[h_i/X_i])
\quad ({\rm RFEQ}) \cr}
\end{displaymath}
\end{definition}

\begin{example}
Let $\pair{\Gamma,\Delta,\Psi,\Theta}$ be the CSL theory of cartesian
closed categories (see examples~\ref{ex-ccc-sig} and~\ref{ex-ccc-th}).
On top of this, as we have seen in section~\ref{ssec-NNO}, we can define
a natural number object by
\begin{displaymath}
\begin{tabular}{l}
left object nat with pr is \\
$\qquad {\rm zero}: 1 \rightarrow {\rm nat}$\\
$\qquad {\rm succ}: {\rm nat} \rightarrow {\rm nat}$\\
end object. \\
\end{tabular}
\end{displaymath}
The CSL statements characterizing this object are
\begin{displaymath}
\displaylines{
\qquad {\rm pr}(f,g)\circ {\rm zero} = f \hfill ({\rm LEQ}_{{\rm
nat},1}) \cr
\qquad {\rm pr}(f,g)\circ {\rm succ} = g\circ {\rm pr}(f,g) \hfill ({\rm
LEQ}_{{\rm nat},2}) \cr
\qquad h \circ {\rm zero} = f \land h \circ {\rm succ} = g \circ h \implies h = {\rm
pr}(f,g) \hfill ({\rm LCEQ}_{\rm nat}) \cr}
\end{displaymath}
The CSL extension $\sigma_{\rm nat}$ associated with the declaration
above is:
\begin{displaymath}
\displaylines{
\qquad \pair{\Gamma,\Delta,\Psi} \rubberrightarrow{\sigma_{\rm nat}}
\langle \Gamma\cup\{{\rm nat}\}, \Delta\cup\{{\rm zero},{\rm succ}\},
\Psi\cup\{{\rm pr}\}, \hfill \cr
\hfill \Theta\cup\{{\rm LEQ}_{{\rm nat},1},{\rm LEQ}_{{\rm nat},2},{\rm
LCEQ}_{\rm nat}\} \rangle \qed \qquad \cr}
\end{displaymath}
\end{example}

Thus, each CDT declaration can be associated with a CSL theory
extension.  This can be thought as a semantics of CDT declarations.
However, it is sometimes convenient to regard their semantics to be real
categories.
\begin{definition}
\label{def-seq-CDT-decl}
A sequence of CDT declarations $D_1,\ldots,D_l$ defines a sequence of
CSL theory extensions starting from the empty CSL theory.
\begin{displaymath}
\pair{\emptyset,\emptyset,\emptyset,\emptyset}
\rubberrightarrow{\sigma_{D_1}} \pair{\Gamma_1,\Delta_1,\Psi_1,\Theta_1}
\rubberrightarrow{\sigma_{D_2}} \cdots \rubberrightarrow{\sigma_{D_l}}
\pair{\Gamma_l,\Delta_l,\Psi_l,\Theta_l}
\end{displaymath}
We define a CSL structure of the CDT declaration sequence
$D_1,\ldots,D_l$ to be a category which is a CSL theory structure of
$\pair{\Gamma_l,\Delta_l,\Psi_l,\Theta_l}$, and the free structure of
$D_1,\ldots,D_l$ to be the free category of this CSL theory (see
section~\ref{sec-free-cat}).
\end{definition}
If we do not rely on any ways of defining functors other than CDT
declarations and if we do not accept any pre-defined functors, it is
inevitable to start with the empty CSL theory.  We have defined cartesian
closed categories as a CSL theory in examples~\ref{ex-ccc-sig}
and~\ref{ex-ccc-th}, but we can do so in CDT starting from the empty
theory by just declaring the terminal object (see
subsection~\ref{ssec-terminal-initial}), products (see
subsection~\ref{ssec-product-coproduct}) and exponentials (see
subsection~\ref{ssec-exponential}).  The advantage of the latter is that
we neither need to think about equations nor need to do tedious typing of
functors, natural transformations or factorizers.  These things come out
automatically, so it is easy to define categories and there is less
chance to make mistakes.

We have introduced CDT declarations from {\bf Left} and {\bf Right}, but
we could not connect them formally.  Now, after having models of CDT as
categories, we can do so.
\begin{proposition}
Let $\pair{\cC,\xi}$ be a CSL structure of a CDT declaration sequence
$D_1$, \ldots, $D_l$.  If $D_i$ is
\begin{displaymath}
\begin{tabular}{l}
left object $L(X_1,\ldots,X_n)$ with $\psi$ is \\
$\qquad \alpha_1: E_1 \rightarrow E'_1$ \\
$\qquad\qquad \cdots$ \\
$\qquad \alpha_m: E_m \rightarrow E'_m$ \\
end object \\
\end{tabular}
\end{displaymath}
then $\xi L = {\bf Left}(\vec F,\vec G)$ where
\begin{displaymath}
\begin{array}{l}
\vec F \defeq \pair{\xi \lambda (L,X_1,\ldots ,X_n).E_1,\ldots
,\xi \lambda (L,X_1,\ldots,X_n).E_m} \\
\vec G \defeq \pair{\lambda (L,X_1,\ldots ,X_n).E'_1,\ldots
,\lambda (L,X_1,\ldots,X_n).E'_m}. \\
\end{array}
\end{displaymath}
The similar thing holds for right CDT declarations. \\
{\bf Proof:} Trivial, because we set the CSL statements so that this
proposition holds.
\end{proposition}

Finally, in this section, let us summarize the objects we have defined
in this chapter and their characteristic CSL statements in
figure~\ref{fig-objects-eqns}.
\begin{figure}
\scriptsize
\begin{center}
\begin{tabular}{|r|l|l|}
\hline
\multicolumn{1}{|c|}{Name} &
\multicolumn{1}{c|}{CDT Declaration} &
\multicolumn{1}{c|}{CSL Statements} \\
\hline
Terminal &
\begin{tabular}[t]{l}
right object 1 with ! \\
end object \\
\end{tabular} &
$f = {\rm !}$ \\
\hline
Initial &
\begin{tabular}[t]{l}
left object 0 with !! \\
end object \\
\end{tabular} &
$f = {\rm !!}$ \\
\hline
Products &
\begin{tabular}[t]{l}
right object ${\rm prod}(X,Y)$ with pair is \\
$\qquad {\rm pi1}: {\rm prod} \rightarrow X$ \\
$\qquad {\rm pi2}: {\rm prod} \rightarrow Y$ \\
end object \\
\end{tabular} &
\begin{array}[t]{l}
{\rm pi1} \circ {\rm pair}(f,g) = f \\
{\rm pi2} \circ {\rm pair}(f,g) = g \\
h = {\rm pair}({\rm pi1}\circ h,{\rm pi2}\circ h) \\
{\rm prod}(f,g) = {\rm pair}(f\circ {\rm pi1},g\circ {\rm pi2}) \\
\end{array} \\
\hline
Coproducts &
\begin{tabular}[t]{l}
left object ${\rm coprod}(X,Y)$ with case is \\
$\qquad {\rm in1}: X \rightarrow {\rm coprod}$ \\
$\qquad {\rm in2}: Y \rightarrow {\rm coprod}$ \\
end object \\
\end{tabular} &
\begin{array}[t]{l}
{\rm case}(f,g)\circ {\rm in1} = f \\
{\rm case}(f,g)\circ {\rm in2} = g \\
h = {\rm case}(h\circ {\rm in1}, h\circ {\rm in2})\\
{\rm coprod}(f,g) = {\rm case}({\rm in1}\circ f,{\rm in2}\circ g) \\
\end{array} \\
\hline
Exponentials &
\begin{tabular}[t]{l}
right object ${\rm exp}(X,Y)$ with curry is \\
$\qquad {\rm eval}: {\rm prod}({\rm exp},X) \rightarrow Y$ \\
end object \\
\end{tabular} &
\begin{array}[t]{l}
{\rm eval}\circ {\rm prod}({\rm curry}(f),{\bf I}) = f \\
h = {\rm curry}({\rm eval}\circ {\rm prod}(h,{\bf I})) \\
{\rm exp}(f,g) = {\rm curry}(g\circ {\rm eval} \circ {\rm prod}({\bf
I},f)) \\
\end{array} \\
\hline
NNO &
\begin{tabular}[t]{l}
left object nat with pr is \\
$\qquad {\rm zero}: 1 \rightarrow {\rm nat}$\\
$\qquad {\rm succ}: {\rm nat} \rightarrow {\rm nat}$\\
end object \\
\end{tabular} &
\begin{array}[t]{l}
{\rm pr}(f,g)\circ {\rm zero} = f \\
{\rm pr}(f,g)\circ {\rm succ} = g\circ {\rm pr}(f,g) \\
h\circ {\rm succ} = g\circ h \implies h = {\rm pr}(h\circ {\rm zero},g) \\
\end{array} \\
\hline
Lists & 
\begin{tabular}[t]{l}
left object ${\rm list}(X)$ with prl is \\
$\qquad{\rm nil}: 1 \rightarrow {\rm list}$ \\
$\qquad{\rm cons}: {\rm prod}(X,{\rm list}) \rightarrow {\rm list}$ \\
end object \\
\end{tabular} &
\begin{array}[t]{l}
{\rm prl}(f,g)\circ {\rm nil} = f \\
{\rm prl}(f,g)\circ {\rm cons} = {\rm prod}({\bf I},{\rm prl}(f,g))\circ
g \\
h\circ {\rm cons} = {\rm prod}({\bf I},h)\circ g \implies \\
\qquad h = {\rm prl}(h\circ {\rm nil},g) \\
{\rm list}(f) = {\rm prl}({\rm nil},{\rm cons}\circ {\rm prod}(f,{\bf
I})) \\
\end{array} \\
\hline
Infinite Lists &
\begin{tabular}[t]{l}
right object ${\rm inflist}(X)$ with fold is \\
$\qquad {\rm head}: {\rm inflist} \rightarrow X$ \\
$\qquad {\rm tail}: {\rm inflist} \rightarrow {\rm inflist}$ \\
end object \\
\end{tabular} &
\begin{array}[t]{l}
{\rm head}\circ {\rm fold}(f,g) = f \\
{\rm tail}\circ {\rm fold}(f,g) = {\rm fold}(f,g)\circ g \\
{\rm tail}\circ h = g\circ h \implies h = {\rm fold}({\rm head}\circ h,g)
\\
{\rm inflist}(f) = {\rm fold}(f\circ{\rm head},{\rm tail}) \\
\end{array} \\
\hline
Co-NNO &
\begin{tabular}[t]{l}
right object conat with copr is \\
$\qquad {\rm pred}: {\rm conat} \rightarrow {\rm coprod}(1,{\rm conat})$
\\
end object \\
\end{tabular} &
\begin{array}[t]{l}
{\rm pred}\circ {\rm copr}(f) = {\rm coprod}({\bf I},{\rm
copr}(f))\circ f \\
h\circ {\rm pred} = f\circ {\rm coprod}({\bf I},h) \implies \\
\qquad h = {\rm copr}(f) \\
\end{array} \\
\hline
Automata &
\begin{tabular}[t]{l}
right object ${\rm dyn}'(I,O)$ with ${\rm univ}'$ is \\
$\qquad {\rm next}': {\rm prod}({\rm dyn}',I) \rightarrow {\rm dyn}'$ \\
$\qquad {\rm output}': {\rm dyn}' \rightarrow O$ \\
end object \\
\end{tabular} &
\begin{array}[t]{l}
{\rm next}'\circ {\rm prod}({\rm univ}'(f,g),{\bf I}) = {\rm
univ}'(f,g)\circ f \\
{\rm output}'\circ {\rm univ}'(f,g) = g \\
{\rm next}'\circ {\rm prod}(h,{\bf I}) = h\circ f \implies \\
\qquad h = {\rm univ}'(f,{\rm output}'\circ h) \\
\end{array} \\
\hline
\end{tabular}
\end{center}
\caption{CDT Objects}\label{fig-objects-eqns}
\end{figure}

\section{Existence of {\bf Left} and {\bf Right}}
\label{sec-exist-left-right}

In section~\ref{sec-what-cdt}, we have introduced functors ${\bf
Left}[F,G]$ and ${\bf Right}[F,G]$ with the condition which
characterizes them, but we did not consider whether such functors exist
or not.  In this section, we study them mathematically and
present a condition of the existence.

Let us recall the standard construction of initial $T$-algebras (see,
for example, \cite{scott-76,lehmann-smyth-81}).
\begin{proposition}
\label{prop-init-T-alg}
For a $\omega$-cocomplete category $\cC$ (i.e.\ it has colimit of any
$\omega$-chain diagram) and an endo-functor $T: \cC \rightarrow \cC$
which is $\omega$-cocontinuous (i.e.\ it preserves colimit of any
$\omega$-chain diagram), its initial $T$-algebra is given by the colimit
of the following $\omega$-chain diagram.\footnote{In general, this
sequence might not converge at $\omega$.  In such a case, we may extend
the sequence up to any ordinal such that
\begin{quote}
\begin{enumerate}
\item $T^{\alpha+1}(0) = T(T^\alpha(0))$, and
\item $T^\beta(0) = \mathop{\rm colimit}\limits_{\alpha < \beta}
T^\alpha(0)$ for a limit ordinal $\beta$ (treating 0 as a limit ordinal).
\end{enumerate}
\end{quote}}
\begin{displaymath}
0 \rubberrightarrow{\rm !!} T(0) \rubberrightarrow{T({\rm !!})} T^2(0)
\rubberrightarrow{T^2({\rm !!})} \cdots \rubberrightarrow{T^{n-1}({\rm !!})}
T^n(0) \rubberrightarrow{T^n({\rm !!})} \cdots {} \rlap{\qquad \qed}
\end{displaymath}
\end{proposition}

As we presented in section~\ref{sec-what-cdt}, ${\bf Left}[F,G]$ is a
generalization of initial $T$-algebras, where $F$ is a functor of $\cC
\times \cD \rightarrow \cE$ and $G$ is of $\cC \times \cD^- \rightarrow
\cE$.  We will reduce the existence problem of ${\bf Left}$ to that of
corresponding $T$-algebras.  For a $\cD$ object  $A$, from its
definition $\pair{{\bf Left}[F,G](A),\eta_A}$ is the initial object in
the category ${\bf DAlg}(F(\;\cdot\;,A),G(\;\cdot\;,A))$, so
\begin{displaymath}
\ardiagram{F({\bf Left}[F,G](A),A)}{\eta_A}{G({\bf Left}[F,G](A),A)}
\end{displaymath}
Now, if $G(\;\cdot\;,A)$ has a left adjoint functor, say
$H(\;\cdot\;,A): \cE \rightarrow \cC$, this morphism $\eta_A$ has its
one-to-one corresponding morphism
\begin{displaymath}
\ardiagram{H(F({\bf Left}[F,G](A),A),A)}{}{{\bf Left}[F,G](A).}
\end{displaymath}
This means that ${\bf Left}[F,G](A)$ is a $T$-algebra, where $T(B)
\defeq H(F(B,A),A)$, and we naturally expect this $T$-algebra to be
special.  It really is the initial $T$-algebra, so we can formulate the
following theorem.
\begin{theorem}
\label{th-ex-left}
Let $F: \cC \times \cD \rightarrow \cE$ and $G: \cC \times \cD^-
\rightarrow \cE$ be functors.  If
\begin{enumerate}
\item $\cC$ is $\omega$-cocomplete,
\item for each $\cD$ object $A$, $G_A \defeq G(\;\cdot\;,A): \cC
\rightarrow \cE$ has a left adjoint $H_A: \cE \rightarrow \cC$, and
\item for each $\cD$ object $A$, $F_A \defeq F(\;\cdot\;,A): \cC
\rightarrow \cE$ is $\omega$-cocontinuous,
\end{enumerate}
then ${\bf Left}[F,G](A)$ exists in $\cC$ and
\begin{displaymath}
{\bf Left}[F,G](A) = \mathop{\rm colimit}\limits_n (H_A \circ
F_A)^n(0)
\end{displaymath}
{\bf Proof:} Since a left adjoint is cocontinuous, $T_A
\defeq H_A \circ F_A$ is $\omega$-cocontinuous if $F_A$ is so.  All we
have to show is that the initial $T_A$-algebra gives the initial object
of ${\bf DAlg}(F_A,G_A)$, and the rest follows from
proposition~\ref{prop-init-T-alg}.

Let the initial $T_A$-algebra be $I$ paired with a morphism
\begin{displaymath}
\ardiagram{H_A(F_A(I))}{\iota}{I.}
\end{displaymath}
For an object $\pair{B,f}$ in ${\bf DAlg}(F_A,G_A)$
\begin{displaymath}
\ardiagram{F_A(B)}{f}{G_A(B)}
\end{displaymath}
we have to construct the unique morphism from $I$ to $B$ which makes a
certain diagram commute.  To do so, let us name the factorizer of the
adjunction $H_A \adjoint G_A$ $\psi$, that is, $\psi$ is the natural
isomorphism $\Hom{\cC}{H_A(C)}{D} \stackrel{\iso}{\longrightarrow}
\Hom{\cE}{C}{G_A(D)}$.  Then, $\psi^{-1}(f)$ gives a $T_A$-algebra
\begin{displaymath}
\ardiagram{H_A(F_A(B))}{\psi^{-1}(f)}{B}
\end{displaymath}
and since $\pair{I,\iota}$ is the initial $T_A$-algebra, there exists a
unique morphism $g: I \rightarrow B$ such that the following diagram
commutes.
\begin{displaymath}
\raisebox{-22.5mm}[25mm][22.5mm]{
\setlength{\unitlength}{1mm}
\begin{picture}(50,45)(0,2.5)
\put(15,40){\makebox(0,0){$H_A(F_A(I))$}}
\put(32.5,42){\makebox(0,0)[b]{$\iota$}}
\put(45,40){\makebox(0,0){$I$}}
\put(13,25){\makebox(0,0)[r]{$H_A(F_A(g))$}}
\put(30,25){\makebox(0,0){\commute}}
\put(47,25){\makebox(0,0)[l]{$g$}}
\put(15,10){\makebox(0,0){$H_A(F_A(B))$}}
\put(32.5,8){\makebox(0,0)[t]{$\psi^{-1}(f)$}}
\put(45,10){\makebox(0,0){$B$}}
\put(25,40){\vector(1,0){17.5}}
\put(15,37.5){\vector(0,-1){25}}
\put(45,37.5){\vector(0,-1){25}}
\put(25,10){\vector(1,0){17.5}}
\end{picture}}
\eqno(*)
\end{displaymath}
The naturality of $\psi$ converts this diagram to the following
commutative diagram.
\begin{displaymath}
\raisebox{-22.5mm}[22.5mm][22.5mm]{
\setlength{\unitlength}{1mm}
\begin{picture}(60,45)(0,2.5)
\put(10,40){\makebox(0,0){$F_A(I)$}}
\put(25,42){\makebox(0,0)[b]{$\psi(\iota)$}}
\put(40,40){\makebox(0,0){$G_A(I)$}}
\put(17.5,40){\vector(1,0){15}}
\put(8,25){\makebox(0,0)[r]{$F_A(g)$}}
\put(25,25){\makebox(0,0){\commute}}
\put(42,25){\makebox(0,0)[l]{$G_A(g)$}}
\put(10,37.5){\vector(0,-1){25}}
\put(40,37.5){\vector(0,-1){25}}
\put(10,10){\makebox(0,0){$F_A(B)$}}
\put(25,8){\makebox(0,0)[t]{$f$}}
\put(17.5,10){\vector(1,0){15}}
\put(40,10){\makebox(0,0){$G_A(B)$}}
\end{picture}}
\eqno(**)
\end{displaymath}
(i.e.\ $\psi(g\circ \iota) = G_A(g)\circ \psi(\iota)$ and
$\psi(\psi^{-1}(f)\circ H_A(F_A(g))) = f \circ F_A(g)$).  That showed
the existence of $g$.  The uniqueness of $g$ is no more difficult.  If
$g: I \rightarrow B$ satisfies $(**)$, then by applying $\psi^{-1}$ we
get $(*)$ back again and so there $g$ should be unique.
\end{theorem}
\begin{example}
Trivially, if $\cE$ is $\cC$ and $G: \cC \times \cD^- \rightarrow \cC$
is the projection functor, $G_A$ is the identity which has the left
adjoint $H_A$ which is also the identity functor.  In this case, the
above theorem is essentially stating the same thing as
proposition~\ref{prop-init-T-alg}.
\end{example}
\begin{example}
Another simple case is that $\cE$ is $\cC \times \cC$ and $G_A$ is the
diagonal functor.  Its left adjoint is the binary coproduct functor.
$F_A: \cC \rightarrow \cC \times \cC$ can be decomposed into
$F'_A$ and $F''_A$ both of which are functors of $\cC \times \cD
\rightarrow \cC$ such that $F_A(B) = \pair{F'_A(B),F''_A(B)}$.  The theorem
states that ${\bf Left}[F,G](A)$ is the initial $T_A$-algebra where $T_A(B)
\defeq F'_A(B) + F''_A(B)$.  This explains that our natural number object
in subsection~\ref{ssec-NNO} is the initial $T$-algebra where $T(B)
\defeq 1+B$.
\end{example}

Theorem~\ref{th-ex-left} has its dual form for ${\bf Right}[F,G]$.
\begin{theorem}
\label{th-ex-right}
Let $F: \cC \times \cD \rightarrow \cE$ and $G: \cC \times \cD^-
\rightarrow \cE$ be functors.  If
\begin{enumerate}
\item $\cC$ is $\omega$-complete,
\item for each $\cD$ object $A$, $F_A \defeq F(\;\cdot\;,A): \cC
\rightarrow \cE$ has a right adjoint $K_A: \cE \rightarrow \cC$, and
\item for each $\cD$ object $A$, $G_A \defeq G(\;\cdot\;,A): \cC
\rightarrow \cE$ is $\omega$-continuous,
\end{enumerate}
then ${\bf Right}[F,G](A)$ exists in $\cC$ and
\begin{displaymath}
{\bf Right}[F,G](A) = \mathop{\rm limit}\limits_n (K_A \circ F_A)^n(1)
\rlap{\qquad \qquad \qed}
\end{displaymath}
\end{theorem}
\begin{example}
As an application of this theorem, let us calculate ${\rm dyn}'(I,O)$.
For this, we should take $\cD$ to be $\cC \times \cC^-$, $\cE$ to be
$\cC \times \cC$, $F: \cC \times \cC \times \cC^- \rightarrow \cC \times
\cC$ to be $F(A,I,O) \defeq \pair{A \times I,A}$ and $G: \cC \times \cC^-
\times \cC \rightarrow \cC \times \cC$ to be $G(A,I,O) \defeq
\pair{A,O}$.  $F_{I,O}$ has a right adjoint $H_{I,O}(D,E) \defeq {\rm
exp}(I,D) \times E$.  Therefore, ${\rm dyn}'(I,O)$ is the final
$T_{I,O}$-coalgebra, where $T_{I,O}(A) \defeq H_{I,O}(G_{I,O}(A)) = {\rm
exp}(I,A) \times O$.  Now, let us calculate $\mathop{\rm
limit}\limits_n T_{I,O}^n(1)$.
\begin{displaymath}
\begin{array}{rl}
T_{I,O}^0(1) & {} \iso 1 \\
T_{I,O}^1(1) & {} \iso {\rm exp}(I,1)\times O \iso O \iso {\rm exp}(1,O) \\
T_{I,O}^2(1) & {} \iso T_{I,O}(O) \iso {\rm exp}(I,O)\times O \iso {\rm
exp}(1+I,O) \\
T_{I,O}^3(1) & {} \iso {\rm exp}(I,{\rm exp}(1+I,O))\times O \\
& {} \iso {\rm exp}(I \times (1+I),O)\times O \\
& {} \iso {\rm exp}(1+I+I^2,O) \\
\multicolumn{2}{c}{\cdots} \\
T_{I,O}^n(1) & {} \iso {\rm exp}(1+I+I^2+\cdots+I^{n-1},O) \\
\multicolumn{2}{c}{\cdots} \\
\end{array}
\end{displaymath}
Therefore, ${\rm dyn}'(I,O) \iso \mathop{\rm limit}\limits_n
T_{I,O}^n(1) \iso {\rm exp}(1+I+I^2+\cdots,O) \iso {\rm exp}({\rm
list}(I),O)$.
\end{example}

\chapter{Computation and Categorical Data Types}
\label{ch-cpl}

In chapter~\ref{ch-csl} section~\ref{sec-fun-calc-2}, we introduced CSL
expressions ${\bf Exp}(\Gamma,\Delta,\Psi)$ which denote polymorphic
functions in CSL structures.  In this chapter, we will see them as
programs and see how they can be executed.  One of the ways to treat
specifications as programs is to regard equations as rewrite rules, but
in our case, CSL statements are in general conditional equations and,
therefore, it is quite difficult to treat them as rewrite rules.
Furthermore general rewriting cannot be regarded as real computation
unless rules are Church-Rosser, otherwise, rewriting is more close to
theorem proving.

There is no other way so long as we are dealing with CSL specifications.
Remember that CSL was introduced in order to give semantics to
categorical data types we have investigated in chapter~\ref{ch-cdt}.  In
CDT theory, we are not dealing with arbitrary CSL specifications, but
some special ones, ones which are associated with CDT declarations.
Therefore, we have much more hope for executing these special CSL
specifications than arbitrary ones.  For example, cartesian closed
categories are, as is well-known, connected to lambda calculus which is
a model of computation, so we can put some evaluation mechanism into
them.  \cite{curien-86}, for example, has developed such a system.  A
difference of our approach from his is that we do not restrict ourselves
only to cartesian closed categories.  One of the main aims of CDT is to
study categories formed by programming languages, and we should not
presume any structure in the categories without proper reasons.  We can
only accept the cartesian closed structure in CDT if this is necessary
for putting the concept of computation to it.  As we will see later in
this chapter, our categories are cartesian closed (with some extra
structure), and by then we should know why the cartesian closed
structure is necessary.

CSL expressions in section~\ref{sec-fun-calc-2} and the CDT
declaration mechanism in chapter~\ref{ch-cdt} give us the basis of
Categorical Programming Language (CPL) to which we will devote this
chapter.  CPL tries to extract the computable part of CDT.  As we have
seen in subsection~\ref{ssec-obscure}, CDT in general allows us to
define very strange objects which have no concept of being computable.
We are going to put restrictions to the form of CDT declarations in CPL.

Therefore, in CPL, we can declare data types by CDT (with restrictions).
As we have seen in section~\ref{sec-CDT-as-CSL}, this will determine a
CSL in which we can have CSL expressions introduced in
section~\ref{sec-fun-calc-2}.  CSL expressions are the programs in CPL.
There is no difference between programs and data.  From the categorical
point of view, there is nothing inside objects, that is, there are no
data inside data types.  CSL expressions whose domain is the terminal
object are called {\it CPL elements}.  The execution in CPL is
essentially a reduction of a CPL element to a canonical irreducible CPL
element.

Following the result of this chapter, CPL will not only exist on paper,
but also can be implemented.  This will be presented in
chapter~\ref{ch-application}.

In section~\ref{sec-reduct-rule} we introduce a restriction to objects
we can declare in CPL and a set of reduction rules for CPL computation.
In section~\ref{sec-reduct-example} we see an example of computation in
CPL.  In section~\ref{sec-reduct-complete} we prove that any computation
in CPL terminates (i.e.\ the reductions are normalizing) by Tait's
computability method.  In section~\ref{sec-prop-comp-object} we show
some properties about objects in CPL and, finally,
section~\ref{sec-reduct-rule-full} gives another set of reduction rules
for CPL computation which reduces CPL elements into intuitively more
canonical elements.

\section{Reduction Rules for Categorical Programming Language}
\label{sec-reduct-rule}

In this section, we will present some basic definitions together with
the reduction rules for CPL.  Let us assume in the following discussion
that we have defined categorical data types by a sequence of CDT
declarations, $D_1,\ldots,D_l$, and that we have obtained the
corresponding CSL, $\pair{\Gamma,\Delta,\Psi,\Theta}$ as in
definition~\ref{def-seq-CDT-decl}.

In ordinary programming languages, we distinguish programs and data.  We
feed data into a program; it processes the data and then outputs the result
data.  We cannot feed a program into another program and data cannot
process other data or programs.  However, it is true that data are a
special kind of programs, very simple ones.  We can write data directly
into programs as initialization statements or as assignments.  For
example, natural numbers like 1, 132, 59, etc.\ are data as well as
constants in programs.  In some languages like LISP, there is no
difference between data and programs at all.  CPL is not as liberated as
LISP, but both data and programs are morphisms and data are just special
morphisms having a special domain object.

As we know, category theory deals with the external structure of objects
rather than their inner structure so it is not proper to think about
data inside objects.  However, we do sometimes need something
similar to elements in set theory.  We say {\it elements} in category
theory are morphisms whose domain is the terminal object.
\begin{displaymath}
1 \rubberrightarrow{e} A
\end{displaymath}
We say $e$ is an element of $A$.  If we think in the category of sets ({\bf
Set}), `1' is the one-point set and a morphism from the one-point set to
a set corresponds to an element in the set.
\begin{displaymath}
\Hom{\bf Set}{1}{A} \iso A
\end{displaymath}
Hence, the definition of elements in CPL is:
\begin{definition}
Given a CSL signature $\pair{\Gamma,\Delta,\Psi}$, a {\it CPL element}
$e$ is a CSL expression with no morphism variables whose domain type is
the terminal object.
\begin{displaymath}
\rho \vdash e: \lambda(X_1,\ldots,X_n).1 \rightarrow K
\end{displaymath}
(see section~\ref{sec-fun-calc-2} for typing).  In case $n=0$ which
is very often the case, we may write $\vdash e: 1 \rightarrow E$ where
$E$ is a functorial expression without variables.  We also say that $e$
is an element of $E$.
\end{definition}
In order for this definition to be sensible, we need to have the
terminal object in our category.  For simplicity, we assume that
$D_1$ is the CDT declaration of the terminal object as is presented in
subsection~\ref{ssec-terminal-initial}.

Though this definition is a natural way of defining elements in category
theory, it introduces non-symmetry in CPL.  Remember that CDT is meant
to be symmetric: e.g.\ if we have an object of natural numbers, we
should have its dual, an object of co-natural numbers, and so on.  Since
we treat `1' as a special object and the elements in CPL are defined in
this way, we destroy the beauty of symmetry.  We will see the
consequence of this shortly.

\begin{example}
Assume that we have defined all the objects in section~\ref{sec-ex-CDT}.
Then
\begin{sdisplaymath}
\displaylines{
{\rm succ}\circ{\rm zero}, \cr
{\rm pi2}\circ{\rm pair}({\rm succ}\circ{\rm zero},{\rm nil}), \cr
{\rm eval}\circ{\rm prod}({\rm pr}({\rm curry}({\rm pi2}),{\rm
curry}({\rm succ}\circ{\rm eval})),{\bf I})\circ{\rm pair}({\rm
succ} \circ {\rm succ}\circ{\rm zero},{\rm succ}\circ{\rm zero}) \cr}
\end{sdisplaymath}
are all CPL elements.
\end{example}

As we can see from this example, we cannot regard all the elements as
data.  `${\rm succ}\circ{\rm zero}$' and `${\rm pair}({\rm
succ}\circ{\rm zero},{\rm nil})$' can be data, but `${\rm pi1}\circ{\rm
pair}({\rm succ}\circ{\rm zero},{\rm nil})$' cannot be.  We call special
data-like CPL elements {\it canonical CPL elements}.  The definition is:
\begin{definition}
\label{def-cpl-canonical}
A {\it canonical CPL element} is a CPL element which wholly consists of
natural transformations introduced by left objects and factorizers by
right objects.  Formally, a canonical CPL element, $c \in {\bf CE}$, is
defined by
\begin{displaymath}
c \coloneq {\bf I} \mid \alpha_L\circ c \mid \psi_R(e_1,\ldots,e_n)\circ
c
\end{displaymath}
where $\alpha_L$ is a natural transformation of a left object and
$\psi_R$ is a factorizer of a right object.  Note we often do not write
{\bf I} at the tail of canonical elements.
\end{definition}
From this definition, `${\rm pair}$' and `${\rm curry}$' can form canonical
elements, but `${\rm pi1}$', `${\rm pi2}$' or `${\rm eval}$' cannot.  So
for the right objects, factorizers are the means of creating canonical
elements.  On the other hand, `${\rm zero}$', `${\rm succ}$' and `${\rm
cons}$' can form canonical elements, but not `${\rm pr}$' or `${\rm
case}$'.  So for the left objects, natural transformations are the means
of creating canonical elements.
\begin{displaymath}
\begin{tabular}{|r|c|c|}
\cline{2-3}
\multicolumn{1}{c}{} & \multicolumn{1}{|c|}{\bf Canonical} &
\multicolumn{1}{c|}{\bf Non-Canonical} \\
\hline
left object &
\begin{tabular}{c}
natural transformations \\
e.g.\ zero, succ, nil, cons \\
\end{tabular}
&
\begin{tabular}{c}
factorizers \\
e.g.\ case, pr, prl \\
\end{tabular}
\\
\hline
right object &
\begin{tabular}{c}
factorzers \\
e.g.\ pair, curry \\
\end{tabular}
&
\begin{tabular}{c}
natural transformations \\
e.g.\ pi1, pi2, eval \\
\end{tabular}
\\
\hline
\end{tabular}
\end{displaymath}
Definition~\ref{def-cpl-canonical} is beautifully symmetrical as we hoped
from the symmetry of CDT.  In addition, if we look at the canonical
elements in other programming languages, we note that `${\rm pair}$'
corresponds to making pairs of data, `${\rm curry}$' corresponds to lambda
abstraction and both create canonical things (i.e.\ we cannot process
them any more as themselves), and also we note that `${\rm pr}$' and
`${\rm case}$' are programming constructs corresponding to primitive
recursive definitions and case statements and they can never be data.
Therefore, definition~\ref{def-cpl-canonical} fits well with the usual
notion of data.

Of course, this is not the only definition of canonical elements. It
defines quite lazy canonical elements.  It does not care what are inside
factorizers.  For example,
\begin{displaymath}
{\rm pair}({\rm pi1}\circ{\rm pair}({\rm succ}\circ{\rm zero},{\rm
nil}),{\rm zero})
\end{displaymath}
is a canonical element of `${\rm prod}({\rm nat},{\rm nat})$' from this
definition.  Some might not want to call it canonical because it is
equal to another canonical element
\begin{displaymath}
{\rm pair}({\rm succ}\circ{\rm zero},{\rm zero})
\end{displaymath}
which looks more {\it canonical}.  However, this is because `prod' is a
rather special object.  In general, expressions inside factorizers may
not be elements, so we cannot demand them to be canonical (or we have to
treat product-like objects special).  One might still accept this
easily since canonical elements in Martin-L\"of's type theory are
similar to ours, but another queer aspect of our definition is that we
accept
\begin{displaymath}
{\rm pair}({\rm pi2},{\rm pi1})\circ{\rm pair}({\rm succ},{\bf
I})\circ{\rm zero}
\end{displaymath}
as a canonical element.  It is equal to the following element
\begin{displaymath}
{\rm pair}({\rm pi2}\circ{\rm pair}({\rm succ},{\bf
I})\circ{\rm zero},{\rm pi1}\circ{\rm pair}({\rm succ},{\bf
I})\circ{\rm zero}),
\end{displaymath}
but this is again because of the special property of `prod'.  In
general
\begin{displaymath}
\psi_R(e_1,\ldots,e_n) \circ c
\end{displaymath}
is not equal to something in the form of $\psi_R(e'_1,\ldots,e'_n)$.
For example, we cannot always find other simpler canonical elements
which is equal to `${\rm fold}(f,g)\circ c$' (where `fold' is the
factorizer of `inflist').

There are many views of computation, but in CPL computation is the
reduction of an element to a canonical element equal to the given
element.
\begin{displaymath}
e \implies c
\end{displaymath}
Since the reduction is not straightforward, we want to do step-by-step
reduction.  Therefore, we modify the above form of reduction to the
following one.
\begin{displaymath}
\pair{e,c} \implies c'   %\eqno(*)
\end{displaymath}
This means that a CSL expression $e$ applied to a CPL canonical element
$c$ is reduced to a CPL canonical element $c'$.  It is like calculating
the result of applying an element $c$ to a function $e$. Obviously,
$e\circ c$ should be semantically equal to $c'$.  Since a CPL element is
a morphism from the terminal object and `{\bf I}' is its canonical element, if
we want to reduce an arbitrary CPL element $e$ we can ask for the following
reduction.
\begin{displaymath}
\pair{e,{\bf I}} \implies c
\end{displaymath}

In the following, we assume that the associativity of `$\circ$' (the
composition of morphisms) has been taken care by some means so that
in our rules we do not consider it.  We also assume that there are no
functors in CPL elements because we can always replace them with
factorizers and natural transformations.  For example,
\begin{displaymath}
{\rm eval} \circ {\rm prod}({\rm curry}({\rm succ}),{\rm nil})
\end{displaymath}
is equivalent to
\begin{displaymath}
{\rm eval} \circ {\rm pair}({\rm curry}({\rm succ})\circ{\rm pi1},{\rm
nil}\circ{\rm pi2}).
\end{displaymath}

Let us now define rules for the reductions.  The simplest one is for
identities.  Since ${\bf I}\circ c$ is equal to $c$, we should have the
following rule.
\begin{displaymath}
\pair{{\bf I},c} \implies c \eqno(\mbox{IDENT})
\end{displaymath}

Next, for the composition, we naturally have
\begin{displaymath}
\logicrule{\pair{e_2,c} \implies c'' \qquad \pair{e_1,c''} \implies
c'}{\pair{e_1\circ e_2,c} \implies c'} \eqno(\mbox{COMP})
\end{displaymath}

In case that $e$ is a natural transformation introduced by a left object
or a factorizer introduced by a right object, the rule is easy because
the composition of $e$ with a canonical element is canonical by
definition~\ref{def-cpl-canonical}.
\begin{displaymath}
\pair{\alpha_L,c} \implies \alpha_L\circ c \eqno(\mbox{L-NAT})
\end{displaymath}
\begin{displaymath}
\pair{\psi_R(e_1,\ldots,e_n),c} \implies \psi_R(e_1,\ldots,e_n)\circ c
\eqno(\mbox{R-FACT})
\end{displaymath}
For example, $\pair{{\rm succ},{\rm zero}} \implies {\rm succ}\circ {\rm
zero}$ and
\begin{displaymath}
\pair{{\rm curry}({\rm pi2}),{\rm pair}({\rm zero},{\rm
succ}\circ {\rm zero})} \implies {\rm curry}({\rm pi2})\circ {\rm
pair}({\rm zero},{\rm succ}\circ {\rm zero}).
\end{displaymath}

Difficulty comes in for the other cases, i.e.\ when $e$ is a natural
transformation of a right object or a factorizer of a left object.
Let us first consider the case for a factorizer $\psi_L$ introduced by
the following left object.
\begin{displaymath}
\begin{tabular}{l}
left object $L(X)$ with $\psi_L$ is \\
$\qquad \alpha_L: E(L,X) \rightarrow E'(L,X)$ \\
end object \\
\end{tabular}
\end{displaymath}
From the property of this object, we have a CSL equation
\begin{displaymath}
E'(\psi_L(e),{\bf I}) \circ \alpha_L = e \circ E(\psi_L(e),{\bf I}).
\end{displaymath}
An instance of this equation is `${\rm pr}(f,g)\circ {\rm succ} = g\circ
{\rm pr}(f,g)$', and in this case we should have a rewrite rule from
`${\rm pr}(f,g)\circ {\rm succ}$' to `$g\circ {\rm pr}(f,g)$'.
Therefore, in general we might have the following rule.
\begin{displaymath}
\logicrule{\pair{e\circ E(\psi_L(e),{\bf I}),c} \implies
c'}{\pair{E'(\psi_L(e),{\bf I}),\alpha_L\circ c} \implies c'}
\end{displaymath}
However, this rule is not what we wanted.  We wanted a rule for
$\pair{\psi_L(e),c''} \implies c'''$.  In order that the above rule to be
a one we want, $E'(L,X)$ should be simply $L$, and we get
\begin{displaymath}
\logicrule{\pair{e\circ E(\psi_L(e),{\bf I}),c} \implies
c'}{\pair{\psi_L(e),\alpha_L\circ c} \implies c'}. \eqno(\mbox{L-FACT})
\end{displaymath}
The restriction that demands $E'(L,X)$ should be $L$ is the first
restriction we put onto objects in order to obtain the CPL computation
rules.  The left objects introduced in chapter~\ref{ch-cdt}, initial
object, coproducts, natural number object and lists, all satisfy this
restriction except the left adjoint functor of `list' which we do not
expect to be in the world of computation.  If $E'(L,X)$ is something
other than $L$, we are allowing to have the left adjoint $L'$ of
$E'(\;\cdot\;,X)$ by
\begin{displaymath}
\begin{tabular}{l}
left object $L'(X,Y)$ with $\psi_{L'}$ is \\
$\qquad \alpha_{L'}: Y \rightarrow E'(LL',X)$ \\
end object. \\
\end{tabular}
\end{displaymath}
In familiar categories (e.g.\ the category of sets), a functor $F: \cC
\rightarrow \cC$ hardly has a left adjoint: e.g.\ the product functor
$\bullet \times A$ does not have one, nor does the coproduct functor
$\bullet + A$.

As an example of L-FACT, let us write the rules for the factorizer
associated with the natural number object.  There are two rules for two
natural transformations, `zero' and `succ'.
\begin{displaymath}
\logicrule{\pair{e,c} \implies c'}{\pair{{\rm pr}(e,e'),{\rm zero}\circ
c} \implies c'}
\qquad
\logicrule{\pair{e'\circ {\rm pr}(e,e'),c} \implies c'}{\pair{{\rm
pr}(e,e'),{\rm succ}\circ c} \implies c'}
\end{displaymath}

Let us now consider the last case, the case for the reduction rule of a
natural transformation $\alpha_R$ introduced by the following right
object declaration.
\begin{displaymath}
\begin{tabular}{l}
right object $R(X)$ with $\psi_R$ is \\
$\qquad \alpha_R: E(R,X) \rightarrow E'(R,X)$ \\
end object \\
\end{tabular}
\end{displaymath}
From the property of this object, we have a CSL equation
\begin{displaymath}
\alpha_R \circ E(\psi_R(e),{\bf I}) = E'(\psi_R(e),{\bf I}) \circ e.
\end{displaymath}
An instance of this equation is `${\rm pi1}\circ{\rm pair}(f,g) = f$', and
in this case we should have a rule to rewrite `${\rm pi1}\circ{\rm
pair}(f,g)$' to `$f$'.  Therefore, in general we have a rule rewriting
from the left-hand side to the right-hand side as the following rule.
\begin{displaymath}
\logicrule{\pair{E'(\psi_R(e),{\bf I})\circ e,c} \implies
c'}{\pair{\alpha_R,E(\psi_R(e),{\bf I})\circ c} \implies c'}
\end{displaymath}
However, this rule is not quite right because $E(\psi_R(e),{\bf I})\circ
c$ is not a canonical element.  We cannot have functors in canonical
elements.  Therefore, what the rule should really look like
is
\begin{displaymath}
\logicrule{c = E(\psi_R(e),{\bf I})\circ c'' \qquad
\pair{E'(\psi_R(e),{\bf I})\circ e,c''} \implies c'}{\pair{\alpha_R,c}
\implies c'}. \eqno(+)
\end{displaymath}
We now have a different problem of finding out an expression $e$ and a
canonical element $c''$ from a given canonical element $c$ such that
\begin{displaymath}
c = E(\psi_R(e),{\bf I})\circ c''.
\end{displaymath}
Since we are dealing with computation, we need a mechanical way of
solving this problem.  Let us introduce another form of reduction rules.
\begin{displaymath}
\pair{c,E,R} \leadsto \pair{\psi_R(e),c''} \eqno(*)
\end{displaymath}
such that $c = E(\psi_R(e),{\bf I})\circ c''$ where $E$ is a functorial
expression in which $R$ is a variable.  By these rules, we can rewrite
the rule $(+)$ to
\begin{displaymath}
\logicrule{\pair{c,E(R,X),R} \leadsto \pair{\psi_R(e),c''} \qquad
\pair{E'(\psi_R(e),{\bf I})\circ e,c''} \implies c'}{\pair{\alpha_R,c}
\implies c'}. \eqno(\mbox{R-NAT})
\end{displaymath}

We now have to list the rules for $(*)$.  If $E$ is simply $R$ itself,
$c$ should be $E(\psi_R(e),{\bf I})\circ c'' = \psi_R(e)\circ c''$.
Therefore, the rule should be
\begin{displaymath}
\pair{\psi_R(e)\circ c'',R,R} \leadsto \pair{\psi_R(e),c''}.
\eqno(\mbox{R-NAT-V})
\end{displaymath}

Next, if $E(R,X)$ does not depend on $R$, $c$ is $E(\psi_R(e),{\bf
I})\circ c'' = c''$, so the rule may be
\begin{displaymath}
\pair{c,E,R} \leadsto \pair{\psi_R(e),c},
\end{displaymath}
but where does $\psi_R(e)$ come from?  We cannot determine $e$.
Therefore, $E(R,X)$ must not be independent from $R$ (i.e.\ $E(R,X)$
should not be free of $R$).

The case left is when $E(R,X)$ is not a variable but a real functorial
expression.  Let it be
\begin{displaymath}
F(\hat E_1(R,X),\ldots,\hat E_n(R,X)), \eqno(1)
\end{displaymath}
where $F$ is a functor name.  In this case, the equation we are solving
is
\begin{displaymath}
c = F(\hat E_1(\psi_R(e),{\bf I}),\ldots,\hat E_n(\psi_R(e),{\bf
I}))\circ c''. \eqno(2)
\end{displaymath}
Since functors are always represented by their associated factorizers,
the equation looks like
\begin{displaymath}
c = \psi_F(\cdots\circ\hat E_1(\psi_R(e),{\bf I})\circ\cdots\circ\hat
E_n(\psi_R(e),{\bf I})\circ\cdots)\circ c''. \eqno(3)
\end{displaymath}
As we can see, we might have to pick up the same $\psi_R(e)$ from more
than one place and this would be a trouble.  Because $\psi_R(e)$ is a
general CSL expression and we cannot do theorem proving to show two CSL
expressions are equal when doing the CPL computation.  Therefore, we
need to restrict that $\psi_R(e)$ should appear only once in $(3)$.  In
order for this, we have to first restrict that only one $\hat E_i$ in
$(2)$ contains $R$.  Without loss of generality, we can assume it is
$\hat E_1(R,X)$, and since we are only interested in the first argument
of $F$, we assume that $F$ is a unary functor and $E(R,X)$ is $F(\hat
E(R,X))$.  Now (2) looks like
\begin{displaymath}
c = F(\hat E(\psi_R(e),{\bf I}))\circ c'', \eqno(2')
\end{displaymath}
but still $(3)$ might have more than one $\psi_R(e)$ because when we
expand functors by factorizers using $(\mbox{LFEQ})$ or $(\mbox{RFEQ})$
in definition~\ref{def-CDT-sem-eq}, we might duplicate $\psi_R(e)$.
Before stating the restriction to guarantee the no-duplication of
$\psi_R(e)$, let us note that $F$ should be a functor introduced by a
right object declaration.  This is because, if we look at the equation
$(3)$, $c$ is a canonical element consisting of left natural
transformations and right factorizers, so $\psi_F$ should be a right
factorizer.  Let the following be the declaration of $F$.
\begin{displaymath}
\begin{tabular}{l}
right object $F(Y)$ with $\psi_F$ is \\
$\qquad \beta_1: \tilde E_1(F,Y) \rightarrow \tilde E'_1(F,Y)$ \\
\multicolumn{1}{c}{$\cdots$} \\
$\qquad \beta_m: \tilde E_m(F,Y) \rightarrow \tilde E'_m(F,Y)$ \\
end object \\
\end{tabular}
\eqno(4)
\end{displaymath}
From the equation $(\mbox{RFEQ})$ in definition~\ref{def-CDT-sem-eq},
$(3)$ really is
\begin{displaymath}
\begin{array}{rcl}
c = \psi_F( & \tilde E'_1({\bf I},\hat E(\psi_R(e),{\bf I}))\circ \beta_1
\circ \tilde E_1({\bf I},\hat E(\psi_R(e),{\bf I})), & \\
& \cdots & \\
& \tilde E'_m({\bf I},\hat E(\psi_R(e),{\bf I}))\circ \beta_m
\circ \tilde E_m({\bf I},\hat E(\psi_R(e),{\bf I})) & )\circ c'' \\
\end{array}
\eqno(3')
\end{displaymath}
In order that $\psi_R(e)$ should appear only once in this equation, $Y$
should appear only once in one of $\tilde E_i$ and $\tilde E'_i$.  In
order to show that $Y$ should not be in one of $\tilde E'_i$, let $F$ be
simply
\begin{displaymath}
\begin{tabular}{l}
right object $F(Y)$ with $\psi_F$ is \\
$\qquad \beta: Y \rightarrow F$ \\
end object. \\
\end{tabular}
\eqno(4')
\end{displaymath}
$(3)$ becomes
\begin{displaymath}
c = \psi_F(\beta\circ \hat E(\psi_R(e),{\bf I}))\circ c''.
\end{displaymath}
We may demand that $c$ should be $\psi_F(\hat e)\circ \hat c$ and find
$\psi_R(e)$ such that
\begin{displaymath}
\hat e = \beta\circ \hat E(\psi_R(e),{\bf I}),
\end{displaymath}
but how can we solve this equation?  In general, we need theorem proving
for this.  We should have reduced the problem recursively into
\begin{displaymath}
\pair{\check c,\check E,R} \leadsto \pair{\psi_R(e),\check c},
\end{displaymath}
but there is no way to do this if $Y$ appears in one of $\tilde E_i$,
because $\hat e$ should never be an element.  The typing rule in
definition~\ref{def-CDT-CSL} gives us
\begin{displaymath}
\logicrule{\hat e: Y \rightarrow Z}{\psi_F(\hat e): Z \rightarrow F(Y)}
\end{displaymath}
and we cannot choose $Y$ to be the terminal object.  If $Y$ and $F$ were
the other way round in $(4')$, we could choose $Z$ to be the terminal
object to make $\hat e$ an element.  Therefore, if $Y$ is in one of
$\tilde E'_i$ in $(4)$, we have a possibility of reducing the problem of
solving $(3')$ into a smaller problem of the same kind.  Without losing
generality, we can assume that $Y$ only appears in $\tilde E'_1$ in
$(4)$.  Now, $(3')$ is
\begin{displaymath}
c = \psi_F(\tilde E'_1({\bf I},\hat E(\psi_R(e),{\bf I}))\circ
\beta_1,\beta_2,\ldots,\beta_m)\circ c''. \eqno(3'')
\end{displaymath}
We demand $c$ to be $\psi_F(\hat e_1,\hat e_2,\ldots,\hat e_m)\circ \hat
c$, so $(3'')$ is further rewritten to
\begin{displaymath}
\psi_F(\hat e_1,\hat e_2,\ldots,\hat e_m)\circ \hat c =  \psi_F(\tilde
E'_1({\bf I},\hat E(\psi_R(e),{\bf I}))\circ
\beta_1,\beta_2,\ldots,\beta_m)\circ c''. \eqno(3''')
\end{displaymath}
Here, we cannot jump to the conclusion that $\hat e_1$ is $\tilde
E'_1({\bf I},\hat E(\psi_R(e),{\bf I}))\circ \beta_1$, $\hat e_2$ is
$\beta_2$, \ldots, and $\hat c$ is $c''$, because a part of $\hat c$ may
well contribute to form $\tilde E'_1({\bf I},\hat E(\psi_R(e),{\bf
I}))$.  What is desirable is that we could rewrite $\psi_F(\hat e_1,\hat
e_2,\ldots,\hat e_m)\circ \hat c$ to $\psi_F(\hat e'_1,\hat
e'_2,\ldots,\hat e'_m)$ like ${\rm pair}(f,g)\circ h = {\rm pair}(f\circ
h,g\circ h)$.  This is possible when each $\tilde E'_i$ in $(4)$ does
not depend on $F$.
\begin{proposition}
\label{prop-unconditioned}
Let $R$ be a right object defined by
\begin{displaymath}
\begin{tabular}{l}
right object $R(X_1,\ldots,X_n)$ with $\psi_R$ is \\
$\qquad \alpha_1: E_1(R,X_1,\ldots,X_n) \rightarrow
E'_1(X_1,\ldots,X_n)$ \\
\multicolumn{1}{c}{$\cdots$} \\
$\qquad \alpha_m: E_m(R,X_1,\ldots,X_n) \rightarrow
E'_m(X_1,\ldots,X_n)$ \\
end object \\
\end{tabular}
\end{displaymath}
(note that $E'_i$ does not depend on $R$), then
\begin{displaymath}
\psi_R(e_1,\ldots,e_m)\circ e' = \psi_R(e_1\circ E_1(e',{\bf
I},\ldots,{\bf I}),\ldots,e_m\circ E_m(e',{\bf I},\ldots,{\bf I}))
\eqno(\mbox{REQC})
\end{displaymath}
{\bf Proof:} From $(\mbox{RCEQ})$ in definition~\ref{def-CDT-sem-eq},
\begin{displaymath}
\begin{array}{c}
\alpha_1\circ E_1(\psi_R(e_1,\ldots,e_m)\circ e',{\bf I},\ldots,{\bf I})
= f_1 \land {} \\
\cdots \\
\alpha_m\circ E_m(\psi_R(e_1,\ldots,e_m)\circ e',{\bf I},\ldots,{\bf I})
= f_m \land {} \\
{} \implies \psi_R(e_1,\ldots,e_m)\circ e' = \psi_R(f_1,\ldots,f_m). \\
\end{array}
\end{displaymath}
Using $(\mbox{REQ}_i)$ and the fact that $E_i$ is covariant in $R$,
we get
\begin{displaymath}
\displaylines{\qquad e_1\circ E_1(e',{\bf I},\ldots,{\bf I}) = f_1 \land
\ldots \land e_m\circ E_m(e',{\bf I},\ldots,{\bf I}) = f_m \hfill \cr
\hfill {} \implies \psi_R(e_1,\ldots,e_m)\circ e' =
\psi_R(f_1,\ldots,f_m). \qquad \cr}
\end{displaymath}
Therefore,
\begin{displaymath}
\psi_R(e_1,\ldots,e_m)\circ c' = \psi_R(e_1\circ E_1(e',{\bf
I},\ldots,{\bf I}),\ldots,e_m\circ E_m(e',{\bf I},\ldots,{\bf I})).
\rlap{\qquad \qed}
\end{displaymath}
\end{proposition}
We call this kind of right objects {\it unconditioned}.  The name
indicates that the objects are characterized without using conditional
CSL equations.  In fact, $(\mbox{REQ}_i)$, $(\mbox{REQC})$,
$(\mbox{RFEQ})$ and the following $(\mbox{REQI})$ characterize the
unconditioned right objects.\footnote{We can define unconditioned left
objects as dual.}
\begin{displaymath}
\psi_R(\alpha_1,\ldots,\alpha_m) = {\bf I}
\eqno(\mbox{REQI})
\end{displaymath}

Therefore, we assume that in $(4)$ $F$ does not appear in any of $\tilde
E'_i$.  As we have already assumed that $Y$ appears only in $\tilde
E'_1$, $(4)$ now looks like
\begin{displaymath}
\begin{tabular}{l}
right object $F(Y)$ with $\psi_F$ is \\
$\qquad \beta_1: \tilde E_1(F) \rightarrow \tilde E'_1(Y)$ \\
$\qquad \beta_2: \tilde E_2(F) \rightarrow \tilde E'_2$ \\
\multicolumn{1}{c}{$\cdots$} \\
$\qquad \beta_m: \tilde E_m(F) \rightarrow \tilde E'_m$ \\
end object \\
\end{tabular}
\end{displaymath}
($\tilde E'_2,\ldots,\tilde E'_m$ do not depend on $F$ or $Y$ in this
case, but in general $F$ might have parameters other than $Y$ and they
can appear in $\tilde E'_2,\ldots,\tilde E'_m$), and $(3)$ is
\begin{displaymath}
\displaylines{\qquad \psi_F(\hat e_1\circ \tilde E_1(\hat c),\hat
e_2\circ \tilde E_e(\hat c),\ldots,\hat e_m\circ \tilde E_m(\hat c))
\hfill \cr
\hfill {} = \psi_F(\tilde E'_1(\hat E(\psi_R(e),{\bf I}))\circ
\beta_1\circ \tilde E_1(c''),\beta_2\circ \tilde
E_2(c''),\ldots,\beta_m\circ \tilde E_m(c'')). \quad (\hat 3) \cr}
\end{displaymath}
If, furthermore, the first argument of $\psi_F$ is an element, we can
reduce the problem of solving $(3)$ into
\begin{displaymath}
\hat e_1\circ \tilde E_1(\hat c) = \tilde E'_1(\hat E(\psi_R(e),{\bf
I}))\circ \beta_1\circ \tilde E_1(c''). \eqno(5)
\end{displaymath}
Since $\hat e_1\circ \tilde E_1(\hat c)$ is an element, we can ask its
canonical element and it becomes almost like the original object $(3)$
except $E$ is replaced.  Let us see the typing rule for $\psi_F$ defined
in definition~\ref{def-CDT-CSL}.
\begin{displaymath}
\logicrule{f_1: \tilde E_1(Z) \rightarrow \tilde E'_1(Y) \qquad \ldots
\qquad}{\psi_F(f_1,\ldots): Z \rightarrow F(Y)}
\end{displaymath}
Since $\psi_F$ in $(\hat 3)$ is making an element, $Z$ above should be
the terminal object $1$.  In order that $f_1$ above is also an element,
$\tilde E_1(Z)$ (or $\tilde E_1(1)$ because $Z$ is $1$) should also the
terminal object.  Because $\tilde E_1(Z)$ cannot be independent from
$Z$, $\tilde E_1(Z)$ should be simply $Z$.  Therefore, $(5)$ is
\begin{displaymath}
\hat e_1\circ \hat c = \tilde E'_1(\hat E(\psi_R(e),{\bf I}))\circ
\beta_1\circ c''. \eqno(5')
\end{displaymath}
As $\hat e_1\circ \hat c$ is an element, we can ask its canonical
element by
\begin{displaymath}
\pair{\hat e_1,\hat c} \implies \hat c', \eqno(6)
\end{displaymath}
and we can also ask to solve
\begin{displaymath}
\hat c' = \tilde E'_1(\hat E(\psi_R(e),{\bf I}))\circ \hat c'' \eqno(7)
\end{displaymath}
by
\begin{displaymath}
\pair{\hat c',\tilde E'_1(\hat E(R,X)),R} \leadsto \pair{\psi_R(e),\hat c''}.
\end{displaymath}
If we set $c''$ to be
\begin{displaymath}
\psi_F(\hat c'',\hat e_2\circ \hat E_2(\hat c),\ldots,\hat e_m\circ \hat
E_m(\hat c)),
\end{displaymath}
$(\hat 3)$ is satisfied from $(\mbox{REQ}_i)$ in
definition~\ref{def-CDT-sem-eq}.
\begin{displaymath}
\displaylines{\qquad \tilde E'_1(\hat E(\psi_R(e),{\bf I}))\circ
\beta_1\circ \tilde E_1(c'') \hfill \cr
\qquad\llap{$=\;$} \tilde E'_1(\hat E(\psi_R(e),{\bf I}))\circ
\beta_1\circ \tilde E_1(\psi_F(\hat c'',\hat e_2\circ \hat E_2(\hat
c),\ldots,\hat e_m\circ \hat E_m(\hat c))) \hfill \cr
\qquad\llap{$=\;$} \tilde E'_1(\hat E(\psi_R(e),{\bf I}))\circ
\beta_1\circ \hat c'' \quad \dotfill \mbox{(from $(\mbox{REQ}_i)$)} \cr
\qquad\llap{$=\;$} \hat c' \quad \dotfill \mbox{(from $(7)$)} \cr
\qquad\llap{$=\;$} \hat e_1\circ \hat c \quad \dotfill \mbox{(from $(6)$)} \cr}
\end{displaymath}
\begin{displaymath}
\displaylines{\qquad \beta_i\circ \tilde E_i(c'') \quad \dotfill \mbox{($i =
2,\ldots, m$)} \cr
\qquad\llap{$=\;$} \beta_i\circ \tilde E_i(\psi_F(\hat c'',\hat e_2\circ \hat E_2(\hat c),\ldots,\hat e_m\circ \hat
E_m(\hat c))) \hfill \cr
\qquad\llap{$=\;$} \hat e_i\circ \tilde E_i(\hat c) \quad \dotfill \mbox{(from
$(\mbox{REQ}_i)$)} \cr}
\end{displaymath}
Therefore, we got the following rule.
\begin{displaymath}
\begin{array}{c}
\qquad \pair{\hat e_1,\hat c} \implies \hat c' \qquad \pair{\hat
c',\tilde E'_1(\hat E(R,X)),R} \leadsto \pair{\psi_R(e),\hat
c''} \qquad \\
\hline
\multicolumn{1}{l}{\qquad \pair{\psi_F(\hat e_1,\hat e_2,\ldots,\hat
e_m)\circ \hat c,F(\hat E(R,X)),R}} \\
\multicolumn{1}{r}{\qquad \qquad \qquad \leadsto
\pair{\psi_R(e),\psi_F(\hat c'',\hat e_2\circ \tilde E_2(\hat
c),\ldots,\hat e_m\circ \hat E_m(\hat c))} \qquad} \\
\end{array}
\eqno(\mbox{R-NAT-F})
\end{displaymath}

As an example of the rules $(\mbox{R-FACT})$, $(\mbox{R-NAT})$,
$(\mbox{R-NAT-V})$ and $(\mbox{R-NAT-F})$, let us write the rules for
the exponentials.  The CDT declaration of the exponentials is
\begin{displaymath}
\begin{tabular}{l}
right object ${\rm exp}(X,Y)$ with curry is \\
$\qquad {\rm eval}: {\rm prod}({\rm exp},X) \rightarrow Y$ \\
end object \\
\end{tabular}
\end{displaymath}
as we have seen in subsection~\ref{ssec-exponential}.  The rule
$(\mbox{R-FACT})$ is simply
\begin{displaymath}
\pair{{\rm curry}(e),c} \implies {\rm curry}(e)\circ c,
\end{displaymath}
and the rule $(\mbox{R-NAT})$ is
\begin{displaymath}
\logicrule{\pair{c,{\rm prod}({\rm exp},X),{\rm exp}} \leadsto
\pair{{\rm curry}(e),c''} \qquad \pair{e,c''} \implies c'}{\pair{{\rm
eval},c} \implies c'}.
\end{displaymath}
The rule $(\mbox{R-NAT-V})$ is simply
\begin{displaymath}
\pair{{\rm curry}(e)\circ c'',{\rm exp},{\rm exp}} \leadsto \pair{{\rm
curry}(e),c''},
\end{displaymath}
and, finally, the rule $(\mbox{R-NAT-F})$ is
\begin{displaymath}
\logicrule{\pair{\hat e_1,\hat c} \implies \hat
c' \qquad \pair{\hat c',{\rm exp},{\rm exp}} \leadsto \pair{{\rm
curry}(e),\hat c''}}{\pair{{\rm pair}(\hat e_1,\hat e_2)\circ \hat
c,{\rm prod}({\rm exp},X),{\rm exp}} \leadsto \pair{{\rm
curry}(e),{\rm pair}(\hat c'',\hat e_2\circ \hat c)}}.
\end{displaymath}
We may simplify the last three rules and have the next one instead.
\begin{displaymath}
\logicrule{\pair{e_1,c} \implies {\rm curry}(e)\circ c'' \qquad
\pair{e,{\rm pair}(c'',e_2\circ c)} \implies c'}{\pair{{\rm eval},{\rm
pair}(e_1,e_2)\circ c} \implies c'}
\end{displaymath}
We will see an example using these rules in the next section.

In order to obtain $(\mbox{R-NAT-F})$, we have put several restrictions
to the right object declaration.  To state the restrictions formally,
let us introduce the notion `{\it productive\/}'.
\begin{definition}
\label{def-prod-object}
Functorial expressions which are {\it productive} in $X$ are generated
by the following two rules.
\begin{enumerate}
\item $X$ itself is a functorial expression productive in $X$.
\item If $P(Y_1,\ldots,Y_n)$ is a functor which is productive in $Y_i$
and $E$ is a functorial expression productive in $X$,
$P(E_1,\ldots,E_{i-1},E_i,E_{i+1},\ldots,E_n)$ is productive in $X$,
where $E_1,\ldots,E_{i-1},E_{i+1},\ldots,E_n$ are functorial expressions
which do not contain $X$.
\end{enumerate}
A functor $P(Y_1,\ldots,Y_n)$ is called productive in its $i$-th
argument $Y_i$ when $P$ is declared as a right object and its
declaration
\begin{displaymath}
\begin{tabular}{l}
right object $P(Y_1,\ldots,Y_n)$ with $\psi_P$ is \\
$\qquad \alpha_{P,1}: E_{P,1}(P,Y_1,\ldots,Y_n) \rightarrow
E'_{P,1}(P,Y_1,\ldots,Y_n)$ \\
\multicolumn{1}{c}{$\cdots$} \\
$\qquad \alpha_{P,m}: E_{P,m}(P,Y_1,\ldots,Y_n) \rightarrow
E'_{P,m}(P,Y_1,\ldots,Y_n)$ \\
end object \\
\end{tabular}
\end{displaymath}
satisfies
\begin{enumerate}
\item $P$ is unconditioned (i.e.\ $P$ does not appear in
$E'_{P,1},\ldots,E'_{P,m}$),
\item $Y_i$ does not appear in $E_{P,1},\ldots,E_{P,m}$,
\item $Y_i$ appears only one of $E'_{P,1},\ldots,E'_{P,m}$, so let us
assume that it appears in $E'_{P,j}$ only,
\item $E_{P,j}$ is simply $P$, and
\item $E'_{P,j}$ is a functorial expression productive in $Y_i$.
\end{enumerate}
Therefore, the declaration above may look more like
\begin{displaymath}
\begin{tabular}{l}
right object $P(Y_1,\ldots,Y_i,\ldots,Y_n)$ with $\psi_P$ is \\
$\qquad \alpha_{P,1}: E_{P,1}(P,Y_1,\ldots,Y_{i-1},Y_{i+1},\ldots,Y_n)
\rightarrow {} \qquad$ \\
\multicolumn{1}{r}{$E'_{P,1}(Y_1,\ldots,Y_{i-1},Y_{i+1},\ldots,Y_n)$} \\
\multicolumn{1}{c}{$\cdots$} \\
$\qquad \alpha_{P,j}: P \rightarrow
E'_{P,j}(Y_1,\ldots,Y_{i-1},Y_i,Y_{i+1},\ldots,Y_n)$ \\
\multicolumn{1}{c}{$\cdots$} \\
$\qquad \alpha_{P,m}: E_{P,m}(P,Y_1,\ldots,Y_{i-1},Y_{i+1},\ldots,Y_n)
\rightarrow {} \qquad$ \\
\multicolumn{1}{r}{$E'_{P,m}(Y_1,\ldots,Y_{i-1},Y_{i+1},\ldots,Y_n)$} \\
end object \\
\end{tabular}
\eqno\mbox{(P)}
\end{displaymath}
$\alpha_{P,j}$ may be called {\it projection} to $Y_i$.
\end{definition}
The functor `prod' is a typical productive functor.  It is productive in
its both arguments.  The functor `exp' is not productive in any of its
arguments.  A functorial expression `${\rm prod}({\rm prod}(X,{\rm
exp}(Y,Z)),{\rm prod}(X,U))$' is productive in $U$ but not in $X$ or $Y$
or $Z$.

\begin{definition}
\label{def-comp-object}
A right object $R$ is called {\it computable} if its declaration
\begin{displaymath}
\begin{tabular}{l}
right object $R(X_1,\ldots,X_n)$ with $\psi_R$ is \\
$\qquad \alpha_{R,1}: E_{R,1}(R,X_1,\ldots,X_n) \rightarrow
E'_{R,1}(R,X_1,\ldots,X_n)$ \\
\multicolumn{1}{c}{$\cdots$} \\
$\qquad \alpha_{R,m}: E_{R,m}(R,X_1,\ldots,X_n) \rightarrow
E'_{R,m}(R,X_1,\ldots,X_n)$ \\
end object \\
\end{tabular}
\eqno\mbox{(RC)}
\end{displaymath}
satisfies that $E_{R,1},\ldots,E_{R,m}$ are functorial expressions
productive in $R$.  We also call a left object $L$ {\it computable} when
its declaration is
\begin{displaymath}
\begin{tabular}{l}
left object $L(X_1,\ldots,X_n)$ with $\psi_L$ is \\
$\qquad \alpha_{L,1}: E_{L,1}(L,X_1,\ldots,X_n) \rightarrow L$ \\
\multicolumn{1}{c}{$\cdots$} \\
$\qquad \alpha_{L,m}: E_{L,m}(L,X_1,\ldots,X_n) \rightarrow L$ \\
end object \\
\end{tabular}
\eqno\mbox{(LC)}
\end{displaymath}
\end{definition}
The reduction rules we have defined in this section work when all the
objects we define are computable, and all the objects declared in
section~\ref{sec-ex-CDT} are computable except the obscure ones in
subsection~\ref{ssec-obscure}.  Obviously, we did not want to have those
obscure objects in our datatype system and the computability constraint
gets rid of them.  In other words, {\it the categories which are defined
by declaring computable objects cannot be richer than cartesian closed
category with recursive objects}.  Note that we did not make the
restriction in the beginning.  We had the ability to declare a lot of
other objects, but it turned out that in order to be able to discuss the
computability in CDT, the categories should be cartesian closed with
recursive objects.  This signifies the importance of cartesian closed
categories in computer science yet again (e.g.\ the models of typed
lambda calculus correspond to cartesian closed categories).

Note that not all the productive objects are computable by the
definition~\ref{def-prod-object}, but from now on we only treat
computable objects, so productive objects mean computable productive
objects.

\begin{definition}
\label{def-reduct-rules}
Let $D_1,\ldots,D_l$ be a sequence of CDT declarations defining only
computable objects and let $\pair{\Gamma,\Delta,\Psi,\Theta}$ be the
corresponding CSL theory defined by definition~\ref{def-seq-CDT-decl}.
Then, we can have computation rules for CPL elements over the CSL
signature $\pair{\Gamma,\Delta,\Psi}$.  The computation rules are
divided into two: those in the form of
\begin{displaymath}
\pair{e,c} \implies c'
\end{displaymath}
where $c$ and $c'$ are canonical elements and $e$ and $c$ can be
composed (i.e.\ $e\circ c$ is an element), and those in the form of
\begin{displaymath}
\pair{c,E,R} \leadsto \pair{\psi_R(e_1,\ldots,e_m),c'}
\end{displaymath}
where $c$ and $c'$ are canonical elements, $R$ is a right functor name,
$E$ is a functorial expression productive in $R$ and, if $c$ is an
element of $E'$, $E$ should be more general than $E'$.
\begin{enumerate}
\item IDENT
\begin{displaymath}
\pair{{\bf I},c} \implies c
\end{displaymath}
\item COMP
\begin{displaymath}
\logicrule{\pair{e_2,c} \implies c'' \qquad \pair{e_1,c''} \implies
c'}{\pair{e_1\circ e_2,c} \implies c'}
\end{displaymath}
\item L-NAT
\begin{displaymath}
\pair{\alpha_{L,j},c} \implies \alpha_{L,j}\circ c
\end{displaymath}
\item R-FACT
\begin{displaymath}
\pair{\psi_R(e_1,\ldots,e_m),c} \implies \psi_R(e_1,\ldots,e_m)\circ c
\end{displaymath}
\item L-FACT
\begin{displaymath}
\logicrule{\pair{e_j\circ E_{L,j}[\psi_L(e_1,\ldots,e_m)/L],c}
\implies c'}{\pair{\psi_L(e_1,\ldots,e_m),\alpha_{L,j}\circ c} \implies
c'}
\end{displaymath}
\item R-NAT
\begin{displaymath}
\logicrule{
\begin{array}{c}
\pair{c,E_{R,j},R} \leadsto \pair{\psi_R(e_1,\ldots,e_m),c''} \\
\pair{E'_{R,j}[\psi_R(e_1,\ldots,e_m)/R]\circ e_j,c''} \implies c' \\
\end{array}
}{\pair{\alpha_{R,j},c} \implies c'}
\end{displaymath}
\item R-NAT-V
\begin{displaymath}
\pair{\psi_R(e_1,\ldots,e_m)\circ c'',R,R} \leadsto
\pair{\psi_R(e_1,\ldots,e_m),c''}
\end{displaymath}
\item R-NAT-F
\begin{displaymath}
\begin{array}{c}
\quad R \in E_i \qquad Y_i \in E'_{P,j} \qquad \pair{\hat e_j,c}
\implies c' \quad \\
\quad \pair{c',E'_{P,j}[E_i/Y_i],R} \leadsto
\pair{\psi_R(e_1,\ldots,e_{m'}),c''} \quad \\
\hline
\multicolumn{1}{l}{\quad \pair{\psi_P(\hat e_1,\ldots,\hat e_m)\circ
c,P(E_1,\ldots,E_n),R} \leadsto {} \quad}\\
\multicolumn{1}{l}{\quad \qquad \langle
\psi_R(e_1,\ldots,e_{m'}),\psi_P(\hat e_1\circ E_{P,1}[c/P],\ldots,
\qquad \quad} \\
\multicolumn{1}{r}{\quad \qquad \quad  \hat e_{j-1}\circ
E_{P,j-1}[c/P],c'',\hat e_{j+1}\circ E_{P,j+1}[c/P],\ldots,\hat e_m\circ
E_{P,m}[c/P]) \rangle \quad} \\
\end{array}
\end{displaymath}
\end{enumerate}
where the objects $L$, $R$ and $P$ are defined as \mbox{(LC)},
\mbox{(RC)} and \mbox{(P)}, respectively, and $R \in E_i$ means that $R$
appears in a functorial expression $E_i$.
\end{definition}
We have to show that the rules are well-formed, but we have to show
their soundness at the same time.  We will do this in
section~\ref{sec-reduct-complete} as well as showing that every
reduction terminates (in other words, every element is normalizable by
these rules).

\section{An Example of using Reduction Rules}
\label{sec-reduct-example}

In this section, we will see an example computation in CPL using the
reduction rules defined in the previous section.  Since computation by hand
is very tiresome, we do only one example, but we will see some more
examples of CPL computation done by a computer in
chapter~\ref{ch-application}.

Let us assume that we have declared the terminal object as in
subsection~\ref{ssec-terminal-initial}, products as in
subsection~\ref{ssec-product-coproduct}, exponentials as in
subsection~\ref{ssec-exponential} and natural number object as in
subsection~\ref{ssec-NNO}.  We can write down instances of the reduction
rules in definition~\ref{def-reduct-rules} as follows.
\begin{displaymath}
\pair{{\bf I},c} \implies c \eqno(\mbox{IDENT})
\end{displaymath}
\begin{displaymath}
\logicrule{\pair{e_2,c} \implies c'' \qquad \pair{e_1,c''} \implies
c'}{\pair{e_1\circ e_2,c} \implies c'} \eqno(\mbox{COMP})
\end{displaymath}
For the terminal object, we have
\begin{displaymath}
\pair{{\rm !},c} \implies {\rm !}\circ c. \eqno(\mbox{R-FACT}_{\rm !})
\end{displaymath}
For products, we have
\begin{displaymath}
\pair{{\rm pair}(e_1,e_2),c} \implies {\rm pair}(e_1,e_2)\circ c,
\eqno(\mbox{R-FACT}_{\rm pair})
\end{displaymath}
\begin{displaymath}
\logicrule{\pair{c,R,R} \leadsto \pair{{\rm pair}(e_1,e_2),c''} \qquad
\pair{e_1,c''} \implies c'}{\pair{{\rm pi1},c} \implies c'},
\eqno(\mbox{R-NAT}_{\rm pi1})
\end{displaymath}
and
\begin{displaymath}
\pair{{\rm pair}(e_1,e_2)\circ c,{\rm prod},{\rm prod}} \leadsto
\pair{{\rm pair}(e_1,e_2),c}. \eqno(\mbox{R-NAT-V}_{\rm prod})
\end{displaymath}
If we combine $(\mbox{R-NAT}_{\rm pi1})$ and $(\mbox{R-NAT-V}_{\rm
prod})$ together, we get a familiar rule
\begin{displaymath}
\logicrule{\pair{e_1,c} \implies c'}{\pair{{\rm pi1},{\rm
pair}(e_1,e_2)\circ c} \implies c'}. \eqno(\mbox{R-NAT}'_{\rm pi1})
\end{displaymath}
Similarly, for `pi2', we have
\begin{displaymath}
\logicrule{\pair{e_2,c} \implies c'}{\pair{{\rm pi2},{\rm
pair}(e_1,e_2)\circ c} \implies c'}. \eqno(\mbox{R-NAT}'_{\rm pi2})
\end{displaymath}
For exponentials, as we have seen in the previous section, we have the
following two rules.
\begin{displaymath}
\pair{{\rm curry}(e),c} \implies {\rm curry}(e)\circ c
\eqno(\mbox{R-FACT}_{\rm curry})
\end{displaymath}
\begin{displaymath}
\logicrule{\pair{e_1,c} \implies {\rm curry}(e)\circ c'' \qquad
\pair{e,{\rm pair}(c'',e_2\circ c)} \implies c'}{\pair{{\rm eval},{\rm
pair}(e_1,e_2)\circ c} \implies c'} \eqno(\mbox{R-NAT}'_{\rm eval})
\end{displaymath}
For natural number object, we have
\begin{displaymath}
\pair{{\rm zero},c} \implies {\rm zero}\circ c, \eqno(\mbox{L-NAT}_{\rm zero})
\end{displaymath}
\begin{displaymath}
\pair{{\rm succ},c} \implies {\rm succ}\circ c, \eqno(\mbox{L-NAT}_{\rm succ})
\end{displaymath}
\begin{displaymath}
\logicrule{\pair{e_1,c} \implies c'}{\pair{{\rm pr}(e_1,e_2),{\rm
zero}\circ c} \implies c'}, \eqno(\mbox{L-FACT}_{\rm zero})
\end{displaymath}
and
\begin{displaymath}
\logicrule{\pair{e_2\circ {\rm pr}(e_1,e_2),c} \implies c'}{\pair{{\rm
pr}(e_1,e_2),{\rm succ}\circ c} \implies c'}. \eqno(\mbox{L-FACT}_{\rm succ})
\end{displaymath}

Now let us try to calculate `$1+1$' in CPL.  The addition function is
defined in subsection~\ref{ssec-NNO} as
\begin{displaymath}
{\rm add} = {\rm eval}\circ {\rm prod}({\rm pr}({\rm curry}({\rm
pi2}),{\rm curry}({\rm succ}\circ {\rm eval})),{\bf I}).
\end{displaymath}
If we expand `prod' by `pair', we get
\begin{displaymath}
{\rm add} = {\rm eval}\circ {\rm pair}({\rm pr}({\rm curry}({\rm
pi2}),{\rm curry}({\rm succ}\circ {\rm eval}))\circ {\rm pi1},{\rm
pi2}).
\end{displaymath}
Therefore, the calculation `$1+1$' corresponds to the following
reduction.
\begin{displaymath}
\begin{array}{r}
\langle {\rm eval}\circ {\rm pair}({\rm
pr}({\rm curry}({\rm pi2}),{\rm curry}({\rm succ}\circ {\rm eval}))\circ
{\rm pi1},{\rm pi2}), \qquad \qquad \\
{\rm pair}({\rm succ}\circ {\rm zero},{\rm succ}\circ {\rm zero})
\rangle \implies c \\
\end{array}
\eqno(1)
\end{displaymath}
From $(\mbox{COMP})$,
\begin{displaymath}
\begin{array}{c}
\qquad
\begin{array}{r}
\langle {\rm pair}({\rm pr}({\rm curry}({\rm pi2}),{\rm curry}({\rm
succ}\circ {\rm eval}))\circ {\rm pi1},{\rm pi2}), \qquad \qquad \\
{\rm pair}({\rm succ}\circ {\rm zero},{\rm succ}\circ{\rm zero}) \rangle
\implies c_1 \\
\end{array}
\qquad \\
\qquad \pair{{\rm eval},c_1} \implies c \qquad \\
\hline
\qquad
\begin{array}{r}
\langle {\rm eval}\circ {\rm pair}({\rm pr}({\rm curry}({\rm
pi2}),{\rm curry}({\rm succ}\circ {\rm eval}))\circ {\rm pi1},{\rm
pi2}), \qquad \qquad \\
{\rm pair}({\rm succ}\circ {\rm zero},{\rm succ}\circ{\rm zero})
\rangle \implies c \\
\end{array}
\qquad \\
\end{array}
\end{displaymath}
From $(\mbox{R-FACT}_{\rm pair})$, $c_1$ is
\begin{displaymath}
{\rm pair}({\rm pr}({\rm curry}({\rm pi2},{\rm curry}({\rm
succ}\circ{\rm eval})\circ{\rm pi1},{\rm pi2})\circ{\rm pair}({\rm
succ}\circ{\rm zero},{\rm succ}\circ{\rm zero}),
\end{displaymath}
so we need to calculate
\begin{displaymath}
\begin{array}{r}
\langle {\rm eval},{\rm pair}({\rm pr}({\rm curry}({\rm pi2},{\rm curry}({\rm
succ}\circ{\rm eval})\circ{\rm pi1},{\rm pi2})\circ \qquad \qquad \\
{\rm pair}({\rm succ}\circ{\rm zero},{\rm succ}\circ{\rm zero}) \rangle
\implies c. \\
\end{array}
\end{displaymath}
From $(\mbox{R-NAT}'_{\rm eval})$,
\begin{displaymath}
\begin{array}{c}
\qquad
\begin{array}{r}
\langle {\rm pr}({\rm curry}({\rm pi2}),{\rm curry}({\rm succ}\circ{\rm
eval}))\circ{\rm pi1}, \qquad \qquad \qquad \qquad \\
{\rm pair}({\rm succ}\circ{\rm zero},{\rm succ}\circ{\rm zero}) \rangle
\implies {\rm curry}(e_1)\circ c_2 \\
\end{array}
\qquad \\
\qquad \pair{e_1,{\rm pair}(c_2,{\rm pi2}\circ{\rm pair}({\rm
succ}\circ{\rm zero},{\rm succ}\circ{\rm zero}))} \implies c \qquad \\
\hline
\qquad
\begin{array}{r}
\langle {\rm eval},{\rm pair}({\rm pr}({\rm curry}({\rm pi2},{\rm curry}({\rm
succ}\circ{\rm eval})\circ{\rm pi1},{\rm pi2})\circ \qquad \qquad \\
{\rm pair}({\rm succ}\circ{\rm zero},{\rm succ}\circ{\rm zero}) \rangle
\implies c. \\
\end{array}
\qquad \\
\end{array}
\eqno(2)
\end{displaymath}
From $(\mbox{COMP})$,
\begin{displaymath}
\begin{array}{c}
\qquad \pair{{\rm pi1},{\rm pair}({\rm succ}\circ{\rm zero},{\rm
succ}\circ{\rm zero})} \implies c_3 \qquad \\
\qquad \pair{{\rm pr}({\rm curry}({\rm pi2}),{\rm curry}({\rm
succ}\circ{\rm eval})),c_3} \implies {\rm curry}(e_1)\circ c_2 \qquad \\
\hline
\qquad
\begin{array}{r}
\langle {\rm pr}({\rm curry}({\rm pi2}),{\rm curry}({\rm succ}\circ{\rm
eval}))\circ{\rm pi1}, \qquad \qquad \qquad \qquad \\
{\rm pair}({\rm succ}\circ{\rm zero},{\rm succ}\circ{\rm zero}) \rangle
\implies {\rm curry}(e_1)\circ c_2 \\
\end{array}
\qquad \\
\end{array}
\eqno(3)
\end{displaymath}
From $(\mbox{R-NAT}'_{\rm pi1})$,
\begin{displaymath}
\logicrule{\pair{{\rm succ}\circ{\rm zero},{\rm !}} \implies
c_3}{\pair{{\rm pi1},{\rm pair}({\rm succ}\circ{\rm zero},{\rm
succ}\circ{\rm zero})} \implies c_3},
\end{displaymath}
and from $(\mbox{L-NAT}_{\rm zero})$ and $(\mbox{L-NAT}_{\rm succ})$,
$c_3$ is
\begin{displaymath}
{\rm succ}\circ{\rm zero}\circ{\rm !}.
\end{displaymath}
Going back to $(3)$, we need to calculate
\begin{displaymath}
\pair{{\rm pr}({\rm curry}({\rm pi2}),{\rm curry}({\rm succ}\circ{\rm
eval})),{\rm succ}\circ{\rm zero}\circ{\rm !}} \implies {\rm
curry}(e_1)\circ c_2.
\end{displaymath}
From $(\mbox{L-FACT}_{\rm succ})$,
\begin{displaymath}
\begin{array}{c}
\qquad \langle {\rm curry}({\rm succ}\circ{\rm eval})\circ{\rm
pr}({\rm curry}({\rm pi2}),{\rm curry}({\rm succ}\circ{\rm eval})),
\qquad \qquad \\
\multicolumn{1}{r}{{\rm zero}\circ{\rm !}\rangle \implies {\rm
curry}(e_1)\circ c_2 \qquad} \\
\hline
\pair{{\rm pr}({\rm curry}({\rm pi2}),{\rm curry}({\rm
succ}\circ{\rm eval})),{\rm succ}\circ{\rm zero}\circ{\rm !}} \implies
{\rm curry}(e_1)\circ c_2 \\
\end{array}
\end{displaymath}
and from $(\mbox{COMP})$
\begin{displaymath}
\begin{array}{c}
\qquad \pair{{\rm pr}({\rm curry}({\rm pi2}),{\rm curry}({\rm
succ}\circ{\rm eval})),{\rm zero}\circ{\rm !}} \implies c_4 \qquad \\
\qquad \pair{{\rm curry}({\rm succ}\circ{\rm eval}),c_4} \implies {\rm
curry}(e_2)\circ c_2 \qquad \\
\hline
\qquad
\begin{array}{r}
\langle {\rm curry}({\rm succ}\circ{\rm eval})\circ{\rm
pr}({\rm curry}({\rm pi2}), \qquad \qquad \qquad \qquad \\
{\rm curry}({\rm succ}\circ{\rm eval})),
{\rm zero}\circ{\rm !} \rangle \implies {\rm curry}(e_1)\circ
c_2 \\
\end{array}
\qquad \\
\end{array}
\eqno(4)
\end{displaymath}
From $(\mbox{L-FACT}_{\rm zero})$,
\begin{displaymath}
\logicrule{\pair{{\rm curry}({\rm pi2}),{\rm !}} \implies
c_4}{\pair{{\rm pr}({\rm curry}({\rm pi2}),{\rm curry}({\rm
succ}\circ{\rm eval})),{\rm zero}\circ{\rm !}} \implies c_4}
\end{displaymath}
and from $(\mbox{R-FACT}_{\rm curry})$, $c_4$ is `${\rm curry}({\rm
pi2})\circ{\rm !}$'.  Going back to $(4)$, $e_1$ is `${\rm
succ}\circ{\rm eval}$' and $c_2$ is ${\rm curry}({\rm pi2})\circ{\rm
!}$.  Therefore, going back to $(2)$, we now have to calculate
\begin{displaymath}
\pair{{\rm succ}\circ{\rm eval},{\rm pair}({\rm curry}({\rm
pi2})\circ{\rm !},{\rm pi2}\circ{\rm pair}({\rm succ}\circ{\rm
zero},{\rm succ}\circ{\rm zero}))} \implies c.
\end{displaymath}
From $(\mbox{COMP})$,
\begin{displaymath}
\begin{array}{c}
\qquad
\begin{array}{l}
\langle {\rm eval},{\rm pair}({\rm curry}({\rm pi2})\circ{\rm !}, \\
\qquad \qquad {\rm pi2}\circ{\rm pair}({\rm succ}\circ{\rm zero},{\rm
succ}\circ{\rm zero})) \rangle \implies c_5 \\
\end{array}
\qquad \\
\qquad \pair{{\rm succ},c_5} \implies c \qquad \\
\hline
\qquad
\begin{array}{l}
\langle {\rm succ}\circ{\rm eval},{\rm pair}({\rm curry}({\rm
pi2})\circ{\rm !}, \\
\qquad \qquad {\rm pi2}\circ{\rm pair}({\rm succ}\circ{\rm
zero},{\rm succ}\circ{\rm zero})) \rangle \implies c \\
\end{array}
\qquad \\
\end{array}
\eqno(5)
\end{displaymath}
From $(\mbox{R-NAT}'_{\rm eval})$,
\begin{displaymath}
\begin{array}{c}
\qquad \pair{{\rm curry}({\rm pi2})\circ{\rm !},{\rm !}} \implies {\rm
curry}(e_2)\circ c_6 \qquad \\
\qquad \pair{e_2,{\rm pair}(c_6,{\rm pi2}\circ{\rm pair}({\rm
succ}\circ{\rm zero},{\rm succ}\circ{\rm zero}))} \implies c_5 \qquad \\
\hline
\qquad \pair{{\rm eval},{\rm pair}({\rm curry}({\rm pi2})\circ{\rm
!},{\rm pi2}\circ{\rm pair}({\rm succ}\circ{\rm zero},{\rm
succ}\circ{\rm zero}))} \implies c_5 \qquad \\
\end{array}
\end{displaymath}
From $(\mbox{R-FACT}_{\rm !})$ and $(\mbox{R-FACT}_{\rm curry})$, $e_2$
is `pi2' and $c_6$ is `${\rm !}\circ{\rm !}$', so we have
\begin{displaymath}
\pair{{\rm pi2},{\rm pair}({\rm !}\circ{\rm !},{\rm pi2}\circ{\rm
pair}({\rm succ}\circ{\rm zero},{\rm succ}\circ{\rm zero}))} \implies
c_5
\end{displaymath}
From $(\mbox{R-NAT}'_{\rm pi2})$,
\begin{displaymath}
\logicrule{\pair{{\rm pi2}\circ{\rm pair}({\rm succ}\circ{\rm zero},{\rm
succ}\circ{\rm zero}),{\rm !}} \implies c_5}{\pair{{\rm pi2},{\rm
pair}({\rm !}\circ{\rm !},{\rm pi2}\circ{\rm pair}({\rm succ}\circ{\rm
zero},{\rm succ}\circ{\rm zero}))} \implies c_5}
\end{displaymath}
From $(\mbox{COMP})$,
\begin{displaymath}
\begin{array}{c}
\qquad \pair{{\rm pair}({\rm succ}\circ{\rm zero},{\rm succ}\circ{\rm
zero}),{\rm !}} \implies c_6 \qquad \\
\qquad \pair{{\rm pi2},c_6} \implies c_5 \qquad \\
\hline
\qquad \pair{{\rm pi2}\circ{\rm pair}({\rm succ}\circ{\rm zero},{\rm
succ}\circ{\rm zero}),{\rm !}} \implies c_5 \qquad \\
\end{array}
\end{displaymath}
From $(\mbox{R-FACT}_{\rm pair})$, $c_6$ is
\begin{displaymath}
{\rm pair}({\rm succ}\circ{\rm zero},{\rm succ}\circ{\rm zero})\circ{\rm
!},
\end{displaymath}
so we have
\begin{displaymath}
\pair{{\rm pi2},{\rm pair}({\rm succ}\circ{\rm zero},{\rm succ}\circ{\rm
zero})\circ{\rm !}} \implies c_5
\end{displaymath}
and $(\mbox{R-NAT}'_{\rm pi2})$
\begin{displaymath}
\logicrule{\pair{{\rm succ}\circ{\rm zero},{\rm !}} \implies
c_5}{\pair{{\rm pi2},{\rm pair}({\rm succ}\circ{\rm zero},{\rm
succ}\circ{\rm zero})\circ{\rm !}} \implies c_5}
\end{displaymath}
By using $(\mbox{L-NAT}_{\rm zero})$ and $(\mbox{L-NAT}_{\rm succ})$,
$c_5$ is
\begin{displaymath}
{\rm succ}\circ{\rm zero}\circ{\rm !}.
\end{displaymath}
Now, we go back to $(5)$ and we need to calculate
\begin{displaymath}
\pair{{\rm succ},{\rm succ}\circ{\rm zero}\circ{\rm !}} \implies c,
\end{displaymath}
but this is straightforward from $(\mbox{L-NAT}_{\rm succ})$ and $c$ is
\begin{displaymath}
{\rm succ}\circ{\rm succ}\circ{\rm zero}\circ{\rm !}.
\end{displaymath}
Therefore, the reduction $(1)$ is
\begin{displaymath}
\begin{array}{r}
\langle {\rm eval}\circ {\rm pair}({\rm
pr}({\rm curry}({\rm pi2}),{\rm curry}({\rm succ}\circ {\rm eval}))\circ
{\rm pi1},{\rm pi2}), \qquad \qquad \qquad \\
{\rm pair}({\rm succ}\circ {\rm zero},{\rm succ}\circ {\rm zero})
\rangle \implies {\rm succ}\circ{\rm succ}\circ{\rm zero}\circ{\rm !} \\
\end{array}
\end{displaymath}
that is, we have shown `$1+1$' is `2' in CPL.

\section{Well-Definedness and Normalization Theorem for Reduction Rules}
\label{sec-reduct-complete}

In section~\ref{sec-reduct-rule}, we discussed what is computation in
CPL and obtained a set of reduction rules
(definition~\ref{def-reduct-rules}).  Usual questions to be asked when
we get reduction rules are, firstly, whether they are well-defined or
not and, secondly, whether they are normalizing or not.  In this
section, we will answer both questions affirmatively.

Let us assume in this section that we are working in a CSL theory
$\pair{\Gamma,\Delta,\Psi,\Theta}$ which is obtained by a sequence of
CDT declarations, $D_1,\ldots,D_l$, each of which is a computable object
declaration.

First, we prove the well-definedness of the reduction rules.
\begin{theorem}
\label{th-reduct-sound}
{\bf Well-Definedness Theorem:} Let $e\circ c$ be an element in
$\pair{\Gamma,\Delta,\Psi}$ and $c$ be a canonical element.  If from the
rules listed in definition~\ref{def-reduct-rules} we deduce
\begin{displaymath}
\pair{e,c} \implies c',
\end{displaymath}
then $c'$ is a canonical element and $e\circ c = c'$ holds in
$\pair{\Gamma,\Delta,\Psi,\Theta}$ (or in any CSL theory structure of this).
\\
{\bf Proof:} We prove at the same time that for a canonical element $c$,
a right functor $R$, a functorial expression $E$ productive in $R$ which
is more general than the type of $c$, if the rules in
definition~\ref{def-reduct-rules} deduce
\begin{displaymath}
\pair{c,E,R} \leadsto \pair{e,c'},
\end{displaymath}
then $e$ is $\psi_R(e_1,\ldots,e_m)$ for some $e_1,\ldots,e_n$, $c'$ is
a canonical element and $c = E[\psi(e_1,\ldots,e_m)/R]\circ c'$ holds in
$\pair{\Gamma,\Delta,\Psi,\Theta}$.

The proof is done by mathematical induction on the length of reduction,
so all we have to do is to check each reduction rule.
\begin{enumerate}
\item IDENT
\begin{displaymath}
\pair{{\bf I},c} \implies c
\end{displaymath}
It is trivial from ${\bf I}\circ c = c$.
\item COMP
\begin{displaymath}
\logicrule{\pair{e_2,c} \implies c'' \qquad \pair{e_1,c''} \implies
c'}{\pair{e_1\circ e_2,c} \implies c'}
\end{displaymath}
From the induction hypothesis $c'$ is canonical and $c' = e_1\circ c'' =
e_1 \circ e_2 \circ c$.
\item L-NAT
\begin{displaymath}
\pair{\alpha_{L,j},c} \implies \alpha_{L,j}\circ c
\end{displaymath}
It is trivial since $\alpha_{L,j}\circ c$ is a canonical element.
\item R-FACT
\begin{displaymath}
\pair{\psi_R(e_1,\ldots,e_m),c} \implies \psi_R(e_1,\ldots,e_m)\circ c
\end{displaymath}
It is again trivial since $\psi_R(e_1,\ldots,e_m)\circ c$ is a canonical
element.
\item L-FACT
\begin{displaymath}
\logicrule{\pair{e_j\circ E_{L,j}[\psi_L(e_1,\ldots,e_m)/L],c}
\implies c'}{\pair{\psi_L(e_1,\ldots,e_m),\alpha_{L,j}\circ c} \implies
c'}
\end{displaymath}
If $\alpha_{L,j}\circ c$ is canonical, $c$ is canonical, and from
$(\mbox{LEQ}_j)$
\begin{displaymath}
e_j\circ E_{L,j}[\psi_L(e_1,\ldots,e_m)/L]\circ c =
\psi_L(e_1,\ldots,e_m)\circ \alpha_{L,j}\circ c.
\end{displaymath}
Therefore, the premise of the rule is well-formed, so from the
induction hypothesis $c'$ is canonical, and 
\begin{displaymath}
c' = e_j\circ E_{L,j}[\psi_L(e_1,\ldots,e_m)/L]\circ c =
\psi_L(e_1,\ldots,e_m)\circ \alpha_{L,j}\circ c.
\end{displaymath}
\item R-NAT
\begin{displaymath}
\begin{array}{c}
\qquad \pair{c,E_{R,j},R} \leadsto \pair{\psi_R(e_1,\ldots,e_m),c''}
\qquad \\
\qquad \pair{E'_{R,j}[\psi_R(e_1,\ldots,e_m)/R]\circ e_j,c''} \implies
c' \qquad \\
\hline
\qquad \pair{\alpha_{R,j},c} \implies c' \qquad \\
\end{array}
\end{displaymath}
Since $\alpha_{R,j}$ can be composed with $c$ and $\alpha_{R,j}$ has the
type $E_{R,j} \rightarrow E'_{R,j}$, $c$ is an element of a functorial
expression not more general than $E_{R,j}$.  From the induction
hypothesis, $c''$ is a canonical element and
\begin{displaymath}
E_{R,j}[\psi_R(e_1,\ldots,e_m)/R] \circ c'' = c.
\end{displaymath}
Therefore,
\begin{displaymath}
\displaylines{ \qquad \alpha_{R,j}\circ c \hfill \cr
\qquad \llap{$=\;$} \alpha_{R,j}\circ
E_{R,j}[\psi_R(e_1,\ldots,e_m)/R]\circ c'' \quad \dotfill \mbox{(from
$(\mbox{REQ}_j)$)} \cr
\qquad \llap{$=\;$} E'_{R,j}[\psi_R(e_1,\ldots,e_m)/R]\circ e_j\circ
c'' \quad \dotfill \mbox{(from hypothesis)} \cr
\qquad \llap{$=\;$} c. \quad \dotfill \mbox{(from hypothesis)} \cr}
\end{displaymath}
\item R-NAT-V
\begin{displaymath}
\pair{\psi_R(e_1,\ldots,e_m)\circ c'',R,R} \leadsto
\pair{\psi_R(e_1,\ldots,e_m),c''}
\end{displaymath}
$R$ is a variable, so it is more general than anything, so the rule is
well-formed.  Since $\psi_R(e_1,\ldots,e_m)\circ c''$ is a canonical
element, so is $c''$.
\item R-NAT-F
\begin{displaymath}
\begin{array}{c}
\quad R \in E_i \qquad Y_i \in E'_{P,j} \qquad \pair{\hat e_j,c}
\implies c' \quad \\
\quad \pair{c',E'_{P,j}[E_i/Y_i],R} \leadsto
\pair{\psi_R(e_1,\ldots,e_{m'}),c''} \quad \\
\hline
\multicolumn{1}{l}{\quad \pair{\psi_P(\hat e_1,\ldots,\hat e_m)\circ
c,P(E_1,\ldots,E_n),R} \leadsto {} \quad}\\
\multicolumn{1}{l}{\quad \qquad \langle
\psi_R(e_1,\ldots,e_{m'}),\psi_P(\hat e_1\circ E_{P,1}[c/P],\ldots,
\qquad \quad} \\
\multicolumn{1}{r}{\quad \qquad \quad  \hat e_{j-1}\circ
E_{P,j-1}[c/P],c'',\hat e_{j+1}\circ E_{P,j+1}[c/P],\ldots,\hat e_m\circ
E_{P,m}[c/P]) \rangle \quad} \\
\end{array}
\end{displaymath}
$\psi_P(\hat e_1,\ldots)\circ c = \psi_P(\hat e_1\circ
E_{P,1}[c/P],\ldots)$ and from the typing rule of $\psi_P$, $\hat
e_j\circ E_{P,j}[c/P] = \hat e_j\circ c$ is an element of a functorial
expression which is not more general than $E'_{P,j}[E_1/Y_1,\ldots]$
which is not more general than $E'_{P,j}[E_i/Y_i]$.  Therefore, the
premises of the rule are well-formed and
\begin{displaymath}
\displaylines{\qquad \hat e_j\circ E_{P,j}[c/P] \hfill \cr
\qquad \llap{$=\;$} \hat e_j\circ c \quad \dotfill \mbox{($E_{P,j}$ is simply
$P$)} \cr
\qquad \llap{$=\;$} c' \quad \dotfill \mbox{(from $\pair{\hat e_j,c}
\implies c'$)} \cr
\qquad \llap{$=\;$} E'_{P,j}[E_i/Y_i][\psi_R(e_1,\ldots)/R]\circ
c''. \quad \dotfill \mbox{(from $\pair{c',E'_{P,j}[E_i/Y_i],R} \leadsto
\ldots$)} \cr}
\end{displaymath}
Therefore,
\begin{displaymath}
\displaylines{
\qquad P(E_1,\ldots)[\psi_R(e_1,\ldots)/R]\circ \psi_P(\hat e_1\circ
E_{P,1}[c/P],\ldots,c'',\ldots) \hfill \cr
\qquad \llap{$=\;$}
\psi_P(\alpha_{P,1},\ldots,E'_{P,j}[E_i/Y_i][\psi_R(e_1,\ldots)/R],\ldots)
\hfill \cr
\qquad \qquad \qquad \qquad \qquad {} \circ \psi_P(\hat e_1\circ
E_{P,1}[c/P],\ldots,c'',\ldots) \quad \dotfill \mbox{(expand $P$)} \cr
\qquad \llap{$=\;$} \psi_P(\hat e_1\circ
E_{P,1}[c/P],\ldots,E'_{P,j}[E_i/Y_i][\psi_R(e_1,\ldots)/R]\circ
c'',\ldots) \hfill \cr
\qquad \qquad \qquad \qquad \qquad \qquad
\dotfill \mbox{(from \ref{prop-unconditioned} and $(\mbox{REQ}_k)$)} \cr
\qquad \llap{$=\;$} \psi_P(\hat e_1\circ E_{P,1}[c/P],\ldots,\hat
e_j\circ E_{P,j}[c/P],\ldots) \quad \dotfill \mbox{(from above)} \cr
\qquad \llap{$=\;$} \psi_P(\hat e_1,\ldots,\hat e_m)\circ c. \quad \dotfill
\mbox{(from \ref{prop-unconditioned})} \cr}
\end{displaymath}
Of course, $\psi_P(\hat e_1\circ E_{P,1}[c/P],\ldots,c'',\ldots)$ is a
canonical element.
\end{enumerate}
We have proved the theorem as well as shown the well-formedness of the
reduction rules.
\end{theorem}

Next, we prove the normalization, that is to prove the following
theorem.
\begin{theorem}
\label{th-reduct-complete}
{\bf Normalization Theorem:} For a canonical element $c$ and a CSL
expression $e$ whose domain is compatible with the codomain of $c$
(i.e.\ we can have $e\circ c$ as an element), there is a canonical
element $c'$ such that
\begin{displaymath}
\pair{e,c} \implies c'
\end{displaymath}
by the rules listed in definition~\ref{def-reduct-rules}.
\end{theorem}
Before proving this theorem, we need some preparation.  Note that the
theorem proves the two things together: the existence of $c'$ and the
reducibility of $\pair{e,c} \implies c'$.  Therefore, from the existence
part, we will have the following corollary.
\begin{corollary}
For any element $e$ in $\pair{\Gamma,\Delta,\Psi}$, there is a canonical
element $c$ such that $e=c$ holds in $\pair{\Gamma,\Delta,\Psi,\Theta}$.
\end{corollary}
Showing the reducibility can be regarded as showing the termination of
computation in CPL.  So, every program terminates in CPL.  This is what
we expected because we only use primitive recursions.

First, we show a property of canonical elements.
\begin{proposition}
\label{prop-cano-form}
A canonical element of $R(E_1,\ldots,E_n)$ for a right object $R$
defined by $(\mbox{RC})$ has the following form.
\begin{displaymath}
\psi_R(e_1,\ldots,e_m)\circ c
\end{displaymath}
where $c$ is another canonical element. \\
On the other hand, a canonical element of $L(E_1,\ldots,E_n)$ for a left
object $L$ defined by $(\mbox{LC})$ has the following form.
\begin{displaymath}
\alpha_{L,j}\circ c
\end{displaymath}
where $c$ is another canonical element. \\
{\bf Proof:} A canonical element only consists of natural
transformations of left objects and factorizers of right objects.
Therefore, a canonical element should look like either
\begin{displaymath}
\alpha_{L,j}\circ c \qquad \mbox{or} \qquad \psi_R(e_1,\ldots,e_m)\circ c.
\end{displaymath}
From the typing rules in section~\ref{sec-fun-calc-2}, the first one
always gives an element of $L(E_1,\ldots,E_n)$ and the second one gives
an element of $R(E_1,\ldots,E_n)$.
\end{proposition}
From this proposition, we can see that whenever we have
\begin{displaymath}
\pair{e,c} \qquad \mbox{or} \qquad \pair{c,E,R}
\end{displaymath}
we can always apply exactly one of the rules in
definition~\ref{def-reduct-rules}.  In other words, the computation in
CPL is never stuck and deterministic.  We can always proceed to {\it
the} next computation step.

We are going to prove the normalization theorem by the
{\it computability} method due to Tait \cite{tait-67}.  This method is
often used to show normalization of various systems especially that of
lambda calculi (see, for example, \cite{stenlund} and
\cite{lambek-scott-86}) where two kinds of induction are used: induction
on types and induction on structures.  The method usually goes for
lambda calculi as follows:
\begin{enumerate}
\item Define the notion of computable terms inductively on types.
\item Show that any computable term is normalizable.
\item Show that all the terms are computable by induction on terms.
\item Therefore, any term is normalizable.
\end{enumerate}
The notion of computable terms is stronger than that of normalizable
terms, but we need this stronger notion (which is a part of the essence
of the Tait computability method) to carry out the normalization proof.
The computability predicate also divide the two inductions involved in
the proof clearly.  In our case, the normalization proof goes as follows.
\begin{enumerate}
\item We define the notion of {\it computable canonical elements}
inductively on types.  Intuitively, a canonical element $c$ is
computable if
\begin{enumerate}
\item $\alpha_L\circ c$ is computable if $c$ is computable.
\item $\psi_R(e)\circ c$ is computable if there is a reduction
$\pair{\alpha_R,\psi_R(e)\circ c} \implies c'$ such that $c'$ is
computable, that is all the components of $\psi_R(e)\circ c$ are
computable.
\end{enumerate}
\item We define the notion of {\it calculability} for expressions (we
could have called it computability as well as is conventional in proofs
about lambda calculi, but we distinguish them for clarity).  An
expression $e$ is {\it calculable} if for any computable canonical
element $c$ there is a reduction $\pair{e,c} \implies c'$ such that $c'$
is computable.  Note that an expression $e$ is normalizable if for any
canonical element $e$ there is a reduction $\pair{e,c} \implies c$,
whereas $e$ is calculable if there is a reduction for any {\it
computable} canonical element.  Therefore, it should be easier to
prove that an expression is calculable than to prove it is normalizable.
\item We will prove that all the expressions are calculable by
structural induction.
\item As an easy corollary, we can show that any canonical element is
computable.
\item Finally, we prove that all the reductions are normalizing.
\end{enumerate}

First, let us assign for each $n$-ary functor $F$ a function which
given sets $C_1\ldots,C_n$ of canonical elements of type
$E_1,\ldots,E_n$ gives a set of canonical elements $C$ of type
$F(E_1,\ldots,E_n)$.  We write $\tilde F$ for the function (i.e.\ $C =
\tilde F(C_1,\ldots,C_n)$).  $\tilde F$ is monotonic in the {\it i\/}th
argument if $F$ is covariant in the {\it i\/}th argument, and $\tilde F$
is anti-monotonic in the {\it i\/}th argument if $F$ is contravariant in
the {\it i\/}th argument.
\begin{definition}
\label{def-assoc-cano-fun}
\begin{enumerate}
\item For a left object\footnote{For simplicity, we define this for a
simple left object, but the general case should be obvious.}
\begin{displaymath}
\begin{tabular}{l}
left object $L(X)$ with $\psi_L$ is \\
$\qquad \alpha_L: E_L(L,X) \rightarrow L$ \\
end object, \\
\end{tabular}
\end{displaymath}
$\tilde L(C)$ is the minimal fixed point of the following monotonic
function:
\begin{displaymath}
S \longmapsto \{\; \alpha_L\circ c \;\mid\; c \in \tilde E_L(S,C) \;\}
\end{displaymath}
The minimal fixed point can be calculated as the least upper bound of
the following ascending chain:
\begin{displaymath}
\tilde L_0(C) \subseteq \tilde L_1(C) \subseteq \cdots \subseteq \tilde
L_n(C) \subseteq \tilde L_{n+1}(C) \subseteq \cdots \subseteq \tilde
L_\omega(C) \subseteq \cdots {}
\end{displaymath}
where $\tilde L_0(C)$ is $\emptyset$ and
\begin{displaymath}
\tilde L_{\beta+1}(C) \defeq \{\; \alpha_L\circ c \;\mid\; c \in \tilde
E_L(\tilde L_\beta(C),C) \;\}.
\end{displaymath}
\item For a right object
\begin{displaymath}
\begin{tabular}{l}
right object $R(X)$ with $\psi_R$ is \\
$\qquad \alpha_R: R \rightarrow E'_R(R,X)$ \\
end object, \\
\end{tabular}
\end{displaymath}
$\tilde R(C)$ is the maximal fixed point of the following monotonic
function:
\begin{displaymath}
S \longmapsto \{\; \psi_R(e)\circ c \;\mid\;
\mbox{$\pair{\alpha_R,\psi_R(e)\circ c} \implies c'$ such that $c' \in
\tilde E'_R(S,C)$} \;\}
\end{displaymath}
The maximal fixed point can be calculated as the greatest lower bound of
the following descending chain:
\begin{displaymath}
\tilde R_0(C) \supseteq \tilde R_1(C) \supseteq \cdots \supseteq \tilde
R_n(C) \supseteq \tilde R_{n+1}(C) \supseteq \cdots \supseteq \tilde
R_\omega(C) \supseteq \cdots {}
\end{displaymath}
where $\tilde R_0(C)$ is the set of all the canonical elements of $R(E)$
(where $C$ is a set of canonical elements of type $E$) and
\begin{displaymath}
\displaylines{
\qquad \tilde R_{\beta+1}(C) \defeq \{\; \psi_R(e)\circ c \;\mid\; {}
\hfill \cr
\hfill \mbox{$\pair{\alpha_R,\psi_R(e)\circ c} \implies c'$ such that
$c' \in \tilde E'_R(\tilde R_\beta(C),C)$} \;\}. \qquad \cr}
\end{displaymath}
\item For a right object\footnote{For a right object $R$, since the
domain of $\alpha_R: E_R(R,X) \rightarrow E'_R(R,X)$ needs to be
productive in $R$, there are basically these two cases: when $E_R(R,X)$
is $R$ and when it is ${\rm prod}(R,E''(X))$.}
\begin{displaymath}
\begin{tabular}{l}
right object $R'(X)$ with $\psi_{R'}$ is \\
$\qquad \alpha_{R'}: {\rm prod}(R,E_{R'}(X)) \rightarrow E'_{R'}(R',X)$ \\
end object, \\
\end{tabular}
\end{displaymath}
$\tilde R'(C)$ is the maximal fixed point of the following monotonic
function:
\begin{displaymath}
S \longmapsto \{\; \psi_{R'}(e)\circ c \;\mid\;
\parbox[t]{3.5in}{For any $c' \in \tilde E_{R'}(C)$ $\pair{\alpha_{R'},{\rm
pair}(\psi_{R'}(e)\circ c,c')} \implies c''$ such that $c'' \in \tilde
E'_{R'}(S,C)$ $\;\}$}
\end{displaymath}
We can similarly define $\tilde R'_\beta(C)$.
\end{enumerate}
{\bf Well-definedness:} We have to show that $\tilde F$ is monotonic or
anti-monotonic according to the variance of $F$.  We prove this by
induction on the order of declaration of objects.
\begin{enumerate}
\item If $F$ is the left object $L$ above and covariant, we show that
$\tilde L_\beta$ is monotonic by induction on $\beta$.  $\tilde L_0$ is
trivially monotonic.  $\tilde L_{\beta+1}(C)$ is
\begin{displaymath}
\{\; \alpha_L\circ c \;\mid\; c \in \tilde E_L(\tilde L_\beta(C),C) \;\}.
\end{displaymath}
From the induction hypothesis $\tilde L_\beta$ is monotonic.  Since
$E_L(L,X)$ is covariant in both $L$ and $X$, from the other induction
hypothesis $\tilde E_L$ is monotonic.  Therefore, $\tilde E_L(\tilde
L_\beta(C),C)$ is monotonic in $C$, and $\tilde L_{\beta+1}$ is
monotonic.  Hence, $\tilde L$ is monotonic.  Similarly, we can show that
$\tilde L$ is anti-monotonic if $L$ is contravariant.
\item If $F$ is the right object $R$ above and covariant, we show that $\tilde
R_\beta$ is monotonic by induction on $\beta$.  $\tilde R_0$ is
trivially monotonic.  $\tilde R_{\beta+1}(C)$ is
\begin{displaymath}
\{\; \psi_R(e)\circ c \;\mid\; \pair{\alpha_R,\psi_R(e)\circ c} \implies
c' \land c' \in \tilde E'_R(\tilde R_\beta(C),C) \;\}.
\end{displaymath}
From the induction hypothesis, $\tilde R_\beta$ is monotonic, and from
the other induction hypothesis, $\tilde E'_R$ is monotonic in both
arguments.  Therefore, $\tilde R_{\beta+1}$ is monotonic, and by
induction $\tilde R$ is monotonic.  We can similarly show that $\tilde
R$ is anti-monotonic if $R$ is contravariant.
\item If $F$ is the right object $R'$ above and covariant, we show that
$\tilde R'_\beta$ is monotonic by induction on $\beta$.  $\tilde R'_0$ is
trivially monotonic.  $\tilde R'_{\beta+1}(C)$ is
\begin{displaymath}
\displaylines{
\qquad \{\; \psi_{R'}(e)\circ c \;\mid\; \forall c'\in \tilde E_{R'}(C) \;
\pair{\alpha_{R'},{\rm pair}(\psi_{R'}(e)\circ c,c')} \implies 
c'' \land {} \hfill \cr
\hfill c'' \in \tilde E'_{R'}(\tilde R'_\beta(C),C) \;\}. \qquad \cr}
\end{displaymath}
From the induction hypothesis, $\tilde R'_\beta$ is monotonic, and from
the other induction hypothesis, $\tilde E_{R'}$ is anti-monotonic and
$\tilde E'_{R'}$ is monotonic in both arguments.  Therefore, $\tilde
R'_{\beta+1}$ is monotonic, and by induction $\tilde R'$ is monotonic.
We can similarly show that $\tilde R'$ is anti-monotonic if $R'$ is
contravariant. \qed
\end{enumerate}
\end{definition}

We now define the notion of {\it computability} and {\it calculability}.
\begin{definition}
\label{def-computable}
The set $\Omega_{F(E_1,\ldots,E_n)}$ of {\it computable} canonical
elements of type $F(E_1,\ldots,E_n)$ is defined inductively by $\tilde
F(\Omega_{E_1},\ldots,\Omega_{E_n})$.
\end{definition}

\begin{definition}
\label{def-calculable}
An CSL expression $e$ of type $E \rightarrow E'$ is called {\it
calculable with respect to} $C \rightarrow C'$ for $C \subseteq
\Omega_E$ and $C' \subseteq \Omega_{E'}$ if for any $c$ in $C$ there is
a reduction $\pair{e,c} \implies c'$ such that $c'$ is in $C'$.  When
$e$ is calculable with respect to $\Omega_E \rightarrow \Omega_{E'}$, we
simply say that $e$ is {\it calculable}.
\end{definition}

\begin{example}
\label{ex-computable}
\begin{enumerate}
\item For the terminal object `1'
\begin{displaymath}
\begin{tabular}{l}
right object 1 with ! \\
end object, \\
\end{tabular}
\end{displaymath}
since there is no natural transformation, any canonical element is
computable.  Let us use $\star$ to denote an arbitrary element of $\tilde 1$.
\item For the left object `nat' of natural numbers
\begin{displaymath}
\begin{tabular}{l}
left object nat with pr is \\
$\qquad {\rm zero}: 1 \rightarrow {\rm nat}$\\
$\qquad {\rm succ}: {\rm nat} \rightarrow {\rm nat}$\\
end object, \\
\end{tabular}
\end{displaymath}
$\widetilde{\rm nat}_0$ is $\emptyset$.  $\widetilde{\rm nat}_1$ is
\begin{displaymath}
\{\; {\rm zero}\circ c,\; {\rm succ}\circ c' \;\mid\; c \in \tilde 1
\land c' \in \widetilde{\rm nat}_0 \;\} = \{\; {\rm zero}\circ \star \;\}.
\end{displaymath}
Similarly $\widetilde{\rm nat}_2$ consists of ${\rm zero}\circ\star$ and
${\rm succ}\circ{\rm zero}\circ\star$.   In general, $\widetilde{\rm
nat}_n$ is the set of $n$ elements corresponding to a set of 0, 1, 2,
\ldots, and $n-1$.  Therefore, $\widetilde{\rm nat}$ is the set of all the
canonical elements of ${\rm nat}$.
\item For the right object `prod' of products
\begin{displaymath}
\begin{tabular}{l}
right object ${\rm prod}(X,Y)$ with pair is \\
$\qquad {\rm pi1}: {\rm prod} \rightarrow X$ \\
$\qquad {\rm pi2}: {\rm prod} \rightarrow Y$ \\
end object, \\
\end{tabular}
\end{displaymath}
a canonical element ${\rm pair}(e_1,e_2)\circ c$ is computable if there
are reductions
\begin{displaymath}
\pair{{\rm pi1},{\rm pair}(e_1,e_2)\circ c} \implies c_1 \qquad
\mbox{and} \qquad \pair{{\rm pi2},{\rm pair}(e_1,e_2)\circ c} \implies
c_2
\end{displaymath}
such that $c_1$ and $c_2$ are computable, that is a canonical element of
${\rm prod}$ is computable if its components are computable.  Since the
reductions above are equivalent to $\pair{e_1,c} \implies c_1$ and
$\pair{e_2,c} \implies c_2$, if $c$ is computable and $e_1$ and $e_2$
are calculable, ${\rm pair}(e_1,e_2)\circ c$ is computable.
\item For the right object `exp' of exponentials
\begin{displaymath}
\begin{tabular}{l}
right object ${\rm exp}(X,Y)$ with curry is \\
$\qquad {\rm eval}: {\rm prod}({\rm exp},X) \rightarrow Y$ \\
end object, \\
\end{tabular}
\end{displaymath}
a canonical element ${\rm curry}(e)\circ c$ of type ${\exp}(E,E')$ is
computable, if for any computable canonical element $c'$ of type $E'$
there is a reduction
\begin{displaymath}
\pair{{\rm eval},{\rm pair}({\rm curry}(e)\circ c,c')} \implies c''
\end{displaymath}
and $c''$ is computable.  The reduction is equivalent to $\pair{e,{\rm
pair}(c,c')} \implies c''$.  Since ${\rm pair}(c,c')$ is computable,
there is such a reduction if $e$ is calculable.  Remember that ${\rm
curry}(e)$ corresponds to the closure of $e$ (or lambda closed term of
$e$) so that we can say that a closure is calculable if the application
with canonical elements always results canonical elements.  This exactly
corresponds to the definition of computability for lambda expressions
(see~\cite{stenlund}).
\item For the right object `inflist' of infinite lists
\begin{displaymath}
\begin{tabular}{l}
right object ${\rm inflist}(X)$ with fold is \\
$\qquad {\rm head}: {\rm inflist} \rightarrow X$ \\
$\qquad {\rm tail}: {\rm inflist} \rightarrow {\rm inflist}$ \\
end object, \\
\end{tabular}
\end{displaymath}
let us figure out the computable canonical elements of type ${\rm
inflist}({\rm nat})$.  $\widetilde{\rm inflist}_0(\Omega_{\rm nat})$ is
the set of all the canonical elements of type ${\rm inflist}({\rm
nat})$.  $\widetilde{\rm inflist}_1(\Omega_{\rm nat})$ is
\begin{displaymath}
\displaylines{
\qquad \{\; {\rm fold}(e_1,e_2)\circ c \;\mid\;
\begin{array}[t]{l}
\pair{{\rm head},{\rm fold}(e_1,e_2)\circ c} \implies c_1 \land c_1 \in
\Omega_{\rm nat} \land {} \\
\pair{{\rm tail},{\rm fold}(e_1,e_2)\circ c} \implies c_2 \land c_2 \in
\widetilde{\rm inflist}_0(\Omega_{\rm nat}) \;\} \\
\end{array}
\hfill \cr
\qquad \llap{${}={}$} \{\; {\rm fold}(e_1,e_2)\circ c \;\mid\;
\pair{e_1,c} \implies c_1 \land \pair{{\rm fold}(e_1,e_2)\circ e_2,c}
\implies c_2 \;\}.
\hfill \cr}
\end{displaymath}
Therefore, if $e_1$ and $e_2$ are calculable and $c$ is computable,
${\rm fold}(e_1,e_2)\circ c$ is in $\widetilde{\rm inflist}_1(\Omega_{\rm
nat})$.  We can inductively show that it is in any $\widetilde{\rm
inflist}_\beta(\Omega_{\rm nat})$, and, therefore, it is in
$\widetilde{\rm inflist}(\Omega_{\rm nat})$.
\item For the left object of ordinals
\begin{displaymath}
\begin{tabular}{l}
left object ord with pro is \\
$\qquad {\rm ozero}: 1 \rightarrow {\rm ord}$ \\
$\qquad {\rm sup}: {\rm exp}({\rm nat},{\rm ord}) \rightarrow {\rm ord}$ \\
end object, \\
\end{tabular}
\end{displaymath}
$\widetilde{\rm ord}_0$ is empty, $\widetilde{\rm ord}_1$ is $\{\; {\rm
ozero}\circ\star \;\}$, and
\begin{displaymath}
\displaylines{
\qquad \widetilde{\rm ord}_2 \hfill \cr
\qquad \llap{${}={}$} \{\; {\rm sup}\circ c, \; {\rm ozero}\circ \star
\;\mid\; c \in \widetilde{\rm exp}(\widetilde{\rm nat},\widetilde{\rm
ord}_1) \;\} \hfill \cr
\qquad \llap{${}={}$} \{\; {\rm sup}\circ{\rm curry}(e)\circ c,\; {\rm
ozero}\circ \star \;\mid\; \forall c'\in \widetilde{\rm nat} \;
\pair{e,{\rm pair}(c,c')} \implies c'' \land {} \hfill \cr
\hfill c'' \in \widetilde{\rm ord}_1 \;\} \qquad \cr
\qquad \llap{${}={}$} \{\; {\rm sup}\circ{\rm curry}(e)\circ c, \; {\rm
ozero}\circ\star \;\mid\; \forall c'\in \Omega_{\rm nat} \; \pair{e,{\rm
pair}(c,c')} \implies {\rm ozero}\circ\star \;\}. \cr}
\end{displaymath}
In general,
\begin{displaymath}
\displaylines{
\qquad \widetilde{\rm ord}_{\beta+1} = \{\; {\rm sup}\circ{\rm curry}(e)\circ
c, \; {\rm ozero}\circ\star \;\mid\; {} \hfill \cr
\hfill \forall c'\in \Omega_{\rm nat} \;
\pair{e,{\rm pair}(c,c')} \implies c'' \land c'' \in \widetilde{\rm
ord}_\beta \;\} \qquad \cr}
\end{displaymath}
Therefore, a canonical element ${\rm sup}\circ{\rm curry}(e)\circ c$ is
computable if the following reductions always exist:
\begin{displaymath}
\begin{array}{l}
\pair{e,{\rm pair}(c,c_1)} \implies {\rm sup}\circ{\rm curry}(e_1)\circ
c'_1 \\
\qquad \pair{e_1,{\rm pair}(c'_1,c_2)} \implies {\rm sup}\circ{\rm
curry}(e_2)\circ c'_2 \\
\qquad \qquad \pair{e_2,{\rm pair}(c'_2,c_3)} \implies {\rm
sup}\circ{\rm curry}(e_3)\circ c'_3 \\
\qquad \qquad \quad \cdots \\
\qquad \qquad \qquad \pair{e_\beta,{\rm pair}(c'_\beta,c_{\beta+1})}
\implies {\rm ozero}\circ\star \rlap{\qquad \qquad \qquad \qed}
\end{array}
\end{displaymath}
\end{enumerate}
\end{example}

The next proposition intuitively means that functors preserve the
structure of data.  For example, when the functor `list' (or {\tt map}
in ML and {\tt MAPCAR} in LISP) is applied to a list, it only changes
the components of the list and preserves the length.
\begin{proposition}
\label{prop-func-preserve}
Let $F$ be $n$-ary functor, and $e_1,\ldots,e_n$ be CSL
expressions calculable with respect to $C_i \rightarrow C'_i$.
Then $F(e_1,\ldots,e_n)$ is calculable with respect to $\tilde
F(C''_1,\ldots,C''_n) \rightarrow \tilde F(C'''_1,\ldots,C'''_n)$, where
$C''_i$ is $C_i$ and $C'''_i$ is $C'_i$ if $F$ is covariant in the {\it
i\/}th argument and $C''_i$ is $C'_i$ and $C'''_i$ is $C_i$ if $F$ is
contravariant in the {\it i\/}th argument. \\
{\bf Proof:} We prove this by induction on the order of declarations
of objects.
\begin{enumerate}
\item Let $F$ be a left object declared by
\begin{displaymath}
\begin{tabular}{l}
left object $L(X)$ with $\psi_L$ is \\
$\qquad \alpha_L: E_L(L,X) \rightarrow L$ \\
end object \\
\end{tabular}
\end{displaymath}
which is covariant in $X$, $e$ be a CSL expression which is calculable
with respect to $C \rightarrow C'$.  We prove that $L(e)$ is calculable
with respect to $\tilde L_\beta(C) \rightarrow \tilde L_\beta(C')$ by
induction on $\beta$.  Trivially, $L(e)$ is calculable with respect to
$\tilde L_0(C) \rightarrow \tilde L_0(C)$ because $\tilde L_0(C)$ is
empty.  Assume we have proved that $L(e)$ is calculable with respect to
$\tilde L_\beta(C) \rightarrow \tilde L_\beta(C')$.  An element of
$\tilde L_{\beta+1}(C)$ is $\alpha_L\circ c$ such that $c \in \tilde
E_L(\tilde L_\beta(C),C)$.  From L-FACT we get
\begin{displaymath}
\logicrule{
\logicrule{\pair{\alpha_L\circ E_L({\bf I},e)\circ E_L(L(e),{\bf I}),c}
\implies c'}{\pair{\alpha_L\circ E_L({\bf I},e)\circ
E_L(\psi_L(\alpha_L\circ E_L({\bf I},c)),{\bf I}),c} \implies c'}}{
\logicrule{\pair{\psi_L(\alpha_L\circ E_L({\bf I},e)),\alpha_L\circ c}
\implies c'}{\pair{L(e),\alpha_L\circ c} \implies \alpha_L\circ c'}}
\end{displaymath}
Since $E_L(L,X)$ consists of functors declared before $L$, from the
induction hypothesis, $E_L(L(e),{\bf I})$ is calculable with respect to
$\tilde E_L(\tilde L_\beta(C),C) \rightarrow \tilde E_L(\tilde
L_\beta(C'),C)$ and $E_L({\bf I},e)$ is calculable with respect to
$\tilde E_L(\tilde L_\beta(C'),C) \rightarrow \tilde E_L(\tilde
L_\beta(C'),C')$.  Therefore, there is a reduction
\begin{displaymath}
\pair{E_L({\bf I},e)\circ E_L(L(e),{\bf I}),c} \implies c'
\end{displaymath}
such that $c'$ is in $\tilde E_L(\tilde L_\beta(C'),C')$.  From
definition~\ref{def-assoc-cano-fun}, $\alpha_L\circ c'$ is in $\tilde
L_{\beta+1}(C')$.   Hence, $L(e)$ is calculable with respect to $\tilde
L_{\beta+1}(C) \rightarrow \tilde L_{\beta+1}(C')$.   By induction,
$L(e)$ is calculable with respect to $\tilde L_\beta(C) \rightarrow
\tilde L_\beta(C')$ for any $\beta$, and, therefore, $L(e)$ is
calculable with respect to $\tilde L(C) \rightarrow \tilde L(C')$.  When
$L(X)$ is contravariant, we can similarly prove that $L(e)$ is
calculable with respect to $\tilde L(C') \rightarrow \tilde L(C)$.
\item Let $F$ be a right object declared by
\begin{displaymath}
\begin{tabular}{l}
right object $R(X)$ with $\psi_R$ is \\
$\qquad \alpha_R: R \rightarrow E'_R(R,X)$ \\
end object \\
\end{tabular}
\end{displaymath}
which is covariant, and $e$ be a CSL expression which is calculable
with respect to $C \rightarrow C'$.  We prove that $R(e)$ is calculable
with respect to $\tilde R_\beta(C) \rightarrow \tilde R_\beta(C')$ by
induction on $\beta$.  From R-FACT, we have
\begin{displaymath}
\pair{R(e),c} \implies \psi_R(E'_R({\bf I},e)\circ \alpha_R)\circ c.
\end{displaymath}
Trivially, $R(e)$ is calculable with respect to $\tilde R_0(C)
\rightarrow \tilde R_0(C')$ because $\tilde R_0(C')$ is the set of all
the canonical elements of $R$.  Assume we have proved that $R(e)$ is
calculable with respect to $\tilde R_\beta(C) \rightarrow \tilde
R_\beta(C')$.  An element of $\tilde R_{\beta+1}(C)$ is $\psi_R(e')\circ
c'$ such that there is a reduction $\pair{\alpha_R,\psi_R(e')\circ c'}
\implies c''$ and $c''$ is in $\tilde E'_R(\tilde R_\beta(C),C)$.  We
will show that the following canonical element is in $\tilde
R_{\beta+1}(C')$:
\begin{displaymath}
\psi_R(E'_R({\bf I},e)\circ \alpha_R)\circ \psi_R(e')\circ c' \eqno(*)
\end{displaymath}
From R-NAT, we have
\begin{displaymath}
\logicrule{
\logicrule{\pair{E'_R(R(e),{\bf I})\circ E'_R({\bf I},e),c''} \implies
c'''}{\pair{E'_R(R(e),{\bf I})\circ E'_R({\bf I},e)\circ
\alpha_R,\psi_R(e')\circ c'} \implies c'''}
}{
\logicrule{\pair{E'_R(\psi_R(E'_R({\bf I},c)\circ \alpha_R),{\bf
I}),E'_R({\bf I},e)\circ \alpha_R,\psi_R(e')\circ c'} \implies c'''}{
\pair{\alpha_R,\psi_R(E'_R({\bf I},e)\circ \alpha_R)\circ
\psi_R(e')\circ c'} \implies c'''}
}
\end{displaymath}
Since $E'_R(R,X)$ consists of functors declared before $R$,
$E'_R(R(e),{\bf I})\circ E'_R({\bf I},e)$ is calculable with respect to
$\tilde E'_R(\tilde R_\beta(C),C) \rightarrow \tilde E'_R(\tilde
R_\beta(C'),C')$ from the induction hypothesis.  Therefore, $c'''$ is in
$\tilde E'_R(\tilde R_\beta(C'),C')$, and from
definition~\ref{def-assoc-cano-fun}, $(*)$ is in $\tilde
R_{\beta+1}(C)$.  Therefore, by induction, $R(e)$ is calculable with
respect to $\tilde R_\beta(C) \rightarrow \tilde R_\beta(C')$ for any
$\beta$, so it is calculable with respect to $\tilde R(C) \rightarrow
\tilde R(C')$.  When $R(X)$ is contravariant, we can similarly prove
that $R(e)$ is calculable with respect to $\tilde R(C') \rightarrow
\tilde R(C)$.
\item If $F$ be a right object declared by
\begin{displaymath}
\begin{tabular}{l}
right object $R'(X)$ with $\psi_{R'}$ is \\
$\qquad \alpha_{R'}: {\rm prod}(R',E_{R'}(X)) \rightarrow E'_R(R,X)$ \\
end object, \\
\end{tabular}
\end{displaymath}
we can similarly prove that $R(e)$ is calculable with respect to $\tilde
R(C) \rightarrow \tilde R(C')$ (or with respect to $\tilde R(C')
\rightarrow \tilde R(C)$ when $R(X)$ is contravariant). \qed
\end{enumerate}
\end{proposition}

In the following few lemmas, we are to prove all the expressions are
calculable.
\begin{lemma}
\label{lem-calc-I}
{\bf I} is calculable. \\
{\bf Proof:} We have to show that for any computable canonical
element $c$ there is a reduction of $\pair{{\bf I},c} \implies c'$
and that $c'$ is computable.  This is immediate from the
reduction rule {IDENT} and that $c'$ is $c$
\end{lemma}

\begin{lemma}
\label{lem-calc-comp}
If both $e_1$ and $e_2$ are calculable, so is $e_1\circ e_2$. \\
{\bf Proof:} For any computable canonical element $c$' we have the
following reduction from {COMP}:
\begin{displaymath}
\logicrule{\pair{e_2,c} \implies c'' \qquad \pair{e_1,c''} \implies
c'}{\pair{e_1\circ e_2,c} \implies c'}
\end{displaymath}
Since $e_2$ is calculable, there is a reduction for $\pair{e_2,c}
\implies c''$ so that $c''$ is computable.  Since $e_1$ is
calculable, there is a reduction for $\pair{e_1,c''} \implies c'$ so
that $c'$ is computable.  Therefore, there is a reduction for
$\pair{e_1\circ e_2,c} \implies c'$ so that $c'$ is computable.
\end{lemma}

\begin{lemma}
\label{lem-calc-l-nat}
For any natural transformation $\alpha_L$ of a left object $L$,
$\alpha_L$ is calculable. \\
{\bf Proof:} For any computable canonical element $c$, we have the
following reduction by L-NAT:
\begin{displaymath}
\pair{\alpha_L,c} \implies \alpha_L\circ c
\end{displaymath}
From definition~\ref{def-computable}, $\alpha_L\circ c$ is computable.
Therefore, $\alpha_L$ is calculable.
\end{lemma}

\begin{lemma}
\label{lem-calc-r-nat}
For any natural transformation $\alpha_R$ of a right object $R$,
$\alpha_R$ is calculable. \\
{\bf Proof:} Let $R$ be
\begin{displaymath}
\begin{tabular}{l}
right object $R(X)$ with $\psi_R$ is \\
$\qquad \alpha_R: R \rightarrow E'_R(R,X)$ \\
end object, \\
\end{tabular}
\end{displaymath}
and $c$ be a computable canonical element $R(E)$.  For any $\beta$, $c$
is in $\tilde R_{\beta+1}(\Omega_E)$.  From
definition~\ref{def-assoc-cano-fun}, there exists a reduction
$\pair{\alpha_R,c} \implies c'$ such that $c'$ is in $\tilde E'_R(\tilde
R_\beta(\Omega_E),\Omega_E)$.  Because the result of reductions does not
depend on $\beta$, $c'$ is in
\begin{displaymath}
\tilde E'_R(\bigcap_\beta \tilde R_\beta(\Omega_E),\Omega_E) = \tilde
E'_R(\tilde R(\Omega_E),\Omega_E) = \Omega_{E'_R(R(E),E)}.
\end{displaymath}
Therefore, $\alpha_R$ is calculable.  We can similarly prove that for a
right object
\begin{displaymath}
\begin{tabular}{l}
right object $R'(X)$ with $\psi_{R'}$ is \\
$\qquad \alpha_{R'}: {\rm prod}(R',E_{R'}(X)) \rightarrow E'_{R'}(R',X)$ \\
end object, \\
\end{tabular}
\end{displaymath}
$\alpha_{R'}$ is calculable.
\end{lemma}

\begin{lemma}
\label{lem-calc-l-fact}
If $e$ is calculable, so is $\psi_L(e)$ where $L$ is a left object and
$\psi_L$ is its factorizer. \\
{\bf Proof:} Let $L$ be
\begin{displaymath}
\begin{tabular}{l}
left object $L(X)$ with $\psi_L$ is \\
$\qquad \alpha_L: E_L(L,X) \rightarrow L$ \\
end object, \\
\end{tabular}
\end{displaymath}
and $e: E_L(E,E') \rightarrow E$ be a calculable CSL expression.
We will prove that $\psi_L(e)$ is calculable with respect to $\tilde
L_\beta(\Omega_{E'}) \rightarrow \Omega_E$ by induction on $\beta$.
Trivially, it is calculable with respect to $\tilde L_0(\Omega_{E'})
\rightarrow \Omega_E$ because $\tilde L_0(\Omega_{E'})$ is empty.
Assume we have proved that $\psi_L(e)$ is calculable with respect to
$\tilde L_\beta(\Omega_{E'}) \rightarrow \Omega_E$.  An element in
$\tilde L_{\beta+1}(\Omega_{E'})$ is $\alpha_L\circ c$ for $c$ which is
in $\tilde E_L(\tilde L_\beta(\Omega_{E'}),\Omega_{E'})$.  From L-FACT,
we get
\begin{displaymath}
\logicrule{
  \pair{E_L(\psi_L(e),{\bf I}),c} \implies c'' \qquad
  \pair{e,c''} \implies c'
}{
  \logicrule{
    \pair{e\circ E_L(\psi_L(e),{\bf I}),c} \implies c'
  }{
    \pair{\psi_L(e),\alpha_L\circ c} \implies c'
  }
}
\end{displaymath}
From proposition~\ref{prop-func-preserve} and the induction hypothesis,
$E_L(\psi_L(e),{\bf I})$ is calculable with respect to $\tilde
E_L(\tilde L_\beta(\Omega_{E'}),\Omega_{E'}) \rightarrow \tilde
E_L(\Omega_E,\Omega_{E'})$, and there is a reduction
$\pair{E_L(\psi_L(e),{\bf I}),c} \implies c''$.  Since $e$ is
calculable, there is a reduction $\pair{e,c''} \implies c'$ such that
$c'$ is in $\Omega_E$.  Therefore, $\psi_R(e)$ is calculable with
respect to $\tilde L_{\beta+1}(\Omega_{E'}) \rightarrow \Omega_E$, and
by induction it is calculable with respect to $\tilde
L_\beta(\Omega_{E'}) \rightarrow \Omega_E$ for any $\beta$.  Because
$\Omega_{L(E')} = \tilde L(\Omega_{E'})$ is $\bigcup_\beta \tilde
L_\beta(\Omega_{E'})$, we have proved that $\psi_L(e)$ is calculable.
\end{lemma}

\begin{lemma}
\label{lem-calc-r-fact}
If $e$ is calculable, so is $\psi_R(e)$ where $R$ is a right object and
$\psi_R$ is its factorizer. \\
{\bf Proof:} Let $R$ be
\begin{displaymath}
\begin{tabular}{l}
right object $R(X)$ with $\psi_R$ is \\
$\qquad \alpha_R: R \rightarrow E'_R(R,X)$ \\
end object, \\
\end{tabular}
\end{displaymath}
and $e$ be a CSL expression of type $E \rightarrow E'_R(E,E')$.
We are to prove that $\psi_R(e): E \rightarrow R(E')$ is calculable.
Since $\Omega_{R(E')} = \tilde R(\Omega_{E'})$ is $\bigcap_\beta \tilde
R_\beta(\Omega_{E'})$, we prove that $\psi_R(e)$ is calculable with
respect to $\Omega_E \rightarrow \tilde R_\beta(\Omega_{E'})$ by
induction on $\beta$.  Trivially, it is calculable with respect to
$\Omega_E \rightarrow \tilde R_0(\Omega_{E'})$ because for any $c\in
\Omega_E$ we have $\pair{\psi_R(e),c} \implies \psi_R(e)\circ c$ and
$\tilde R_0(\Omega_{E'})$ is the set of all the canonical elements of
type $R(E')$.  Assume we have proved that $\psi_R(e)$ is calculable with
respect to $\Omega_E \rightarrow \tilde R_\beta(\Omega_{E'})$.  For any
$c\in \Omega_E$ we have $\pair{\psi_R(e),c} \implies \psi_R(e)\circ c$.
From R-NAT, we get
\begin{displaymath}
\logicrule{\pair{e,c} \implies c'' \qquad \pair{E'_R(\psi_R(e),{\bf
I}),c''} \implies c'}{\pair{\alpha_R,\psi_R(e)\circ c} \implies c'}.
\end{displaymath}
Since $e$ is calculable, there is a reduction $\pair{e,c} \implies c''$
and $c''$ is in $\tilde E'_R(\Omega_E,\Omega_{E'})$.  As we assumed that
$\psi_R(e)$ is calculable with respect to $\Omega_E \rightarrow \tilde
R_\beta(\Omega_{E'})$, $E'_R(\psi_R(e),{\bf I})$ is calculable with
respect to $\tilde E'_R(\Omega_E,\Omega_{E'}) \rightarrow \tilde
E'_R(\tilde R_\beta(\Omega_{E'}),\Omega_{E'})$ from
proposition~\ref{prop-func-preserve}.  Therefore, there is a reduction
$\pair{E'_R(\psi_R(e),{\bf I})c''} \implies c'$ and $c'$ is in $\tilde
E'_R(\tilde R_\beta(\Omega_{E'}),\Omega_{E'})$.  From
definition~\ref{def-assoc-cano-fun}, $\psi_R(e)\circ c$ is in $\tilde
R_{\beta+1}(\Omega_{E'})$, so $\psi_R(e)$ is calculable with respect to
$\Omega_E \rightarrow \tilde R_{\beta+1}(\Omega_{E'})$, and by induction
it is calculable with respect to $\Omega_E \rightarrow \tilde
R_{\beta}(\Omega_{E'})$ for any $\beta$.  Therefore, it is
calculable with respect to $\Omega_E \rightarrow \tilde R(\Omega_{E'})$.
We can similarly prove that for a right object
\begin{displaymath}
\begin{tabular}{l}
right object $R'(X)$ with $\psi_{R'}$ is \\
$\qquad \alpha_{R'}: {\rm prod}(R',E_{R'}(X)) \rightarrow E'_{R'}(R',X)$ \\
end object, \\
\end{tabular}
\end{displaymath}
$\psi_{R'}(e)$ is calculable.
\end{lemma}

\begin{theorem}
\label{th-calculable}
Any CSL expression $e$ is calculable. \\
{\bf Proof:} This is proved by structural induction and each case
follows from the lemmas, \ref{lem-calc-I}, \ref{lem-calc-comp},
\ref{lem-calc-l-nat}, \ref{lem-calc-r-nat}, \ref{lem-calc-l-fact} and
\ref{lem-calc-r-fact}.
\end{theorem}

\begin{corollary}
\label{cor-computable}
Any canonical element is computable. \\
{\bf Proof:} As a canonical element $c$ is a CSL expression, and
therefore, from theorem~\ref{th-calculable} it is calculable.  Because
any canonical element of the terminal object 1 is computable, specially
${\bf I}$ is computable.  Therefore, there is a
reduction $\pair{c,{\bf I}} \implies c'$ such that $c'$ is computable.
Trivially, $c'$ is $c$ (using L-NAT, R-FACT and COMP), so $c$ is computable.
\end{corollary}

We now finish this section by proving the normalization theorem.
\begin{separateproof}{Normalization~\ref{th-reduct-complete}}
From theorem~\ref{th-calculable}, any expression $e$ is calculable, and
from corollary~\ref{cor-computable}, any canonical element is
computable.  Therefore, from the definition~\ref{def-calculable} of
calculable expressions, there is a reduction $\pair{e,c} \implies c'$.
\end{separateproof}

\section{Properties of Computable Objects}
\label{sec-prop-comp-object}

In section~\ref{sec-reduct-rule}, we saw that we have to restrict
ourselves to computable objects (definition~\ref{def-comp-object}) in
order to introduce our notion of computability into CDT.  Let us see in
this section some of the properties which these particular objects
enjoy.

First, we show that computable left objects are fixed points of some
domain equations.
\begin{theorem}
\label{th-left-isomorphism}
Let $L$ be a computable left object declared as follows.
\begin{displaymath}
\begin{tabular}{l}
left object $L(X_1,\ldots,X_n)$ with $\psi_L$ is \\
$\qquad \alpha_{L,1}: E_{L,1}(L,X_1,\ldots,X_n) \rightarrow L$ \\
\multicolumn{1}{c}{$\cdots$} \\
$\qquad \alpha_{L,m}: E_{L,m}(L,X_1,\ldots,X_n) \rightarrow L$ \\
end object \\
\end{tabular}
\end{displaymath}
Then, the following isomorphism holds in any CSL structure which has $L$ and
coproducts.
\begin{displaymath}
L(X_1,\ldots,X_n) \iso \sum_{j=1}^m
E_{L,j}(L(X_1,\ldots,X_n),X_1,\ldots,X_n)
\end{displaymath}
where $\sum_{j=1}^m{}$ is the $m$-ary coproduct.  Furthermore, if $A$ is
an object which satisfies
\begin{displaymath}
A \iso \sum_{j=1}^m E_{L,j}(A,X_1,\ldots,X_n),
\end{displaymath}
there is a unique morphism $h$ from $L(X_1,\ldots,X_n)$ to $A$ such that
the following diagram commutes.
\begin{displaymath}
\setlength{\unitlength}{1mm}
\begin{picture}(120,50)(0,2.5)
\put(45,10){\makebox(0,0){$\displaystyle \sum_{j=1}^m
E_{L,j}(A,X_1,\ldots,X_n)$}}
\put(85,8){\makebox(0,0)[t]{$\iso$}}
\put(110,10){\makebox(0,0){$A$}}
\put(43,25){\makebox(0,0)[r]{$\displaystyle \sum_{j=1}^m
E_{L,j}(h,X_1,\ldots,X_n)$}}
\put(80,25){\makebox(0,0){\commute}}
\put(112,25){\makebox(0,0)[l]{$h$}}
\put(45,40){\makebox(0,0){$\displaystyle \sum_{j=1}^m
E_{L,j}(L(X_1,\ldots,X_n),X_1,\ldots,X_n)$}}
\put(85,42){\makebox(0,0)[b]{$\iso$}}
\put(110,40){\makebox(0,0){$L(X_1,\ldots,X_n)$}}
\put(77.5,40){\vector(1,0){17.5}}
\put(70,10){\vector(1,0){35}}
\multiput(45,37.5)(0,-5){4}{\line(0,-1){3}}
\put(45,17.5){\vector(0,-1){5}}
\multiput(110,37.5)(0,-5){4}{\line(0,-1){3}}
\put(110,17.5){\vector(0,-1){5}}
\end{picture}
\end{displaymath}
{\bf Proof:} For simplicity, we prove the isomorphism in case $L$ does
not have any parameters (i.e.\ $n=0$).  Therefore, the isomorphism we
prove is
\begin{displaymath}
L \iso \sum_{j=1}^m E_{L,j}(L).
\end{displaymath}
Let $f$ be a morphism
\begin{displaymath}
{}[\alpha_{L,1},\ldots,\alpha_{L,m}]
\end{displaymath}
where $[~,\ldots,~]$ is the factorizer of $\sum_{j=1}^m$.  $f$ is a morphism
from $\sum_{j=1}^m E_{L,j}(L)$ to $L$.  Let $g$ be a morphism
\begin{displaymath}
\psi_L(\nu_1\circ E_{L,1}(f),\ldots,\nu_m\circ E_{L,m}(f))
\end{displaymath}
where $\nu_j$ is the $j$-th injection of $\sum_{j=1}^m{}$.  $g$ is a
morphism from $L$ to $\sum_{j=1}^m E_{L,j}(L)$.  We show that $f$ is the
inverse of $g$.  Let us first show that $f\circ g = {\bf I}$.
\begin{displaymath}
\displaylines{
\qquad f\circ g\circ \alpha_{L,j} \hfill\cr
\qquad\llap{$=\;$} f\circ \nu_j\circ E_{L,j}(f)\circ E_{L,j}(g) \dotfill
\mbox{(from $(\mbox{LEQ}_j)$)}\cr
\qquad\llap{$=\;$} \alpha_{L,j}\circ E_{L,j}(f)\circ E_{L,j}(g) \hfill\cr
\qquad\llap{$=\;$} \alpha_{L,j}\circ E_{L,j}(f\circ g) \dotfill
\mbox{($E_{L,j}(L)$ is covariant)}\cr}
\end{displaymath}
From $(\mbox{LCEQ})$,
\begin{displaymath}
\qquad f\circ g = \psi_R(\alpha_{L,1},\ldots,\alpha_{L,m}) = {\bf I}. \hfill
\end{displaymath}
The second equality holds again from $(\mbox{LCEQ})$.  Next we show that
$g \circ f = {\bf I}$.
\begin{displaymath}
\displaylines{
\qquad g\circ f \hfill \cr
\qquad\llap{$=\;$} [g\circ \alpha_{L,1},\ldots,g\circ \alpha_{L,m}] \hfill\cr
\qquad\llap{$=\;$} [\nu_1\circ E_{L,1}(f\circ g),\ldots,\nu_m\circ
E_{L,m}(f\circ g)] \hfill\cr
\qquad\llap{$=\;$} [\nu_1\circ E_{L,1}({\bf I}),\ldots,\nu_m\circ
E_{L,m}({\bf I})] \hfill\cr
\qquad\llap{$=\;$} [\nu_1,\ldots,\nu_m] \hfill\cr
\qquad\llap{$=\;$} {\bf I} \hfill\cr}
\end{displaymath}
Therefore, $L \iso \sum_{j=1}^m E_{L,j}(L)$.  For any object $A$ which
satisfies $A \iso \sum_{j=1}^m E_{L,j}(A)$, let $i$ be the isomorphism
from $\sum_{j=1}^m E_{L,j}(A)$ to $A$, then the unique morphism is given
by
\begin{displaymath}
\psi_R(i\circ\nu_1,\ldots,i\circ\nu_m).
\end{displaymath}
It is easy to see the diagram commutes from $(\mbox{LEQ}_j)$ and the
uniqueness from $(\mbox{LCEQ})$.
\end{theorem}
If we apply this theorem to the objects we defined in
chapter~\ref{ch-cdt}, we get the following isomorphisms.
\begin{displaymath}
\begin{array}{l}
{\rm nat} \iso 1 + {\rm nat} \\
{\rm list}(X) \iso 1 + {\rm prod}(X,{\rm list}(X)) \\
\end{array}
\end{displaymath}
We can see the exact correspondence to domain theory.  In domain theory,
the domain of natural numbers and that of lists are defined as the
minimal domains which satisfies the above isomorphisms.

By duality principle, we have the dual theorem of
theorem~\ref{th-left-isomorphism}.
\begin{theorem}
\label{th-right-isomorphism}
Let $R$ be a computable right object declared as follows.
\begin{displaymath}
\begin{tabular}{l}
right object $R(X_1,\ldots,X_n)$ with $\psi_R$ is \\
$\qquad \alpha_{R,1}: R \rightarrow E'_{R,1}(R,X_1,\ldots,X_n)$ \\
\multicolumn{1}{c}{$\cdots$} \\
$\qquad \alpha_{R,m}: R \rightarrow E'_{R,m}(R,X_1,\ldots,X_n)$ \\
end object \\
\end{tabular}
\end{displaymath}
Then, the following isomorphism holds in any CSL structure which has $R$ and
products.
\begin{displaymath}
R(X_1,\ldots,X_n) \iso \prod_{j=1}^m
E'_{R,j}(R(X_1,\ldots,X_n),X_1,\ldots,X_n)
\end{displaymath}
where $\prod_{j=1}^m{}$ is the $m$-ary product.  Furthermore, if $A$ is
an object which satisfies
\begin{displaymath}
A \iso \prod_{j=1}^m E'_{R,j}(A,X_1,\ldots,X_n),
\end{displaymath}
there is a unique morphism $h$ from $A$ to $R(X_1,\ldots,X_n)$ such that
the following diagram commutes.
\begin{displaymath}
\setlength{\unitlength}{1mm}
\begin{picture}(120,50)(0,2.5)
\put(10,10){\makebox(0,0){$R(X_1,\ldots,X_n)$}}
\put(10,40){\makebox(0,0){$A$}}
\put(8,25){\makebox(0,0)[r]{$h$}}
\put(35,8){\makebox(0,0)[t]{$\iso$}}
\put(35,42){\makebox(0,0)[b]{$\iso$}}
\put(40,25){\makebox(0,0){\commute}}
\put(75,10){\makebox(0,0){$\displaystyle \prod_{j=1}^m
E'_{R,j}(R(X_1,\ldots,X_n),X_1,\ldots,X_n)$}}
\put(77,25){\makebox(0,0)[l]{$\displaystyle \prod_{j=1}^m
E'_{R,j}(h,X_1,\ldots,X_n)$}}
\put(75,40){\makebox(0,0){$\displaystyle \prod_{j=1}^m
E'_{R,j}(A,X_1,\ldots,X_n)$}}
\put(25,10){\vector(1,0){17.5}}
\put(15,40){\vector(1,0){35}}
\multiput(10,37.5)(0,-5){4}{\line(0,-1){3}}
\put(10,17.5){\vector(0,-1){5}}
\multiput(75,37.5)(0,-5){4}{\line(0,-1){3}}
\put(75,17.5){\vector(0,-1){5}}
\end{picture}
\end{displaymath}
{\bf Proof:} By duality.
\end{theorem}
We can see that the infinite list defined in
subsection~\ref{ssec-final-coalg} is the maximal fixed point of the
following domain equation.
\begin{displaymath}
{\bf inflist}(X) \iso X\times {\bf inflist}(X)
\end{displaymath}

The next theorem states that productive objects define products.
\begin{theorem}
\label{th-prod-isomorphism}
Let $P(Y_1,\ldots,Y_n)$ be a functor which is productive in $Y_i$.
Then, there is a functor $F(Y_1,\ldots,Y_{i-1},Y_{i+1},\ldots,Y_n)$ such
that
\begin{displaymath}
P(Y_1,\ldots,Y_n) \iso Y_i \times F(Y_1,\ldots,Y_{i-1},Y_{i+1},\ldots,Y_n)
\end{displaymath}
{\bf Proof:} First, note that it is easy to extend the theorem to
productive functorial expressions by simply applying the theorem
repeatedly.  Let us prove the theorem by induction on the order of
declaration of productive objects.  Let the declaration of $P$ to be as
follows.
\begin{displaymath}
\begin{tabular}{l}
right object $P(Y_1,\ldots,Y_i,\ldots,Y_n)$ with $\psi_P$ is \\
$\qquad \alpha_{P,1}: E_{P,1}(P,Y_1,\ldots,Y_{i-1},Y_{i+1},\ldots,Y_n)
\rightarrow E'_{P,1}(Y_1,\ldots,Y_{i-1},Y_{i+1},\ldots,Y_n)$ \\
\multicolumn{1}{c}{$\cdots$} \\
$\qquad \alpha_{P,j}: P \rightarrow
E'_{P,j}(Y_1,\ldots,Y_{i-1},Y_i,Y_{i+1},\ldots,Y_n)$ \\
\multicolumn{1}{c}{$\cdots$} \\
$\qquad \alpha_{P,m}: E_{P,m}(P,Y_1,\ldots,Y_{i-1},Y_{i+1},\ldots,Y_n)
\rightarrow E'_{P,m}(Y_1,\ldots,Y_{i-1},Y_{i+1},\ldots,Y_n)$ \\
end object \\
\end{tabular}
\end{displaymath}
By induction hypothesis and from what we note at the beginning of the
proof, there is a functor $F'(Y_1,\ldots,Y_{i-1},Y_{i+1},\ldots,Y_n)$
such that
\begin{displaymath}
E'_{P,j}(Y_1,\ldots,Y_n) \iso Y_i \times
F'(Y_1,\ldots,Y_{i-1},Y_{i+1},\ldots,Y_n).
\end{displaymath}
Since $P$ is a computable object as well,
$E_{P,k}(P,Y_1,\ldots,Y_{i-1},Y_{i+1},\ldots,Y_n)$ is productive in $P$.
Therefore, from induction hypothesis there are functors
\begin{displaymath}
G_k(Y_1,\ldots,Y_{i-1},Y_{i+1},\ldots,Y_n)
\end{displaymath}
such that
\begin{displaymath}
E_{P,k}(P,Y_1,\ldots,Y_{i-1},Y_{i+1},\ldots,Y_n) \iso P \times
G_k(Y_1,\ldots,Y_{i-1},Y_{i+1},\ldots,Y_n).
\end{displaymath}
Using exponentials, the above definition of $P$ is essentially the same
as
\begin{displaymath}
\begin{tabular}{l}
right object $P(Y_1,\ldots,Y_i,\ldots,Y_n)$ with $\psi_P$ is \\
$\qquad \alpha_{P,1}: P \rightarrow {\rm
exp}(G_1(Y_1,\ldots,Y_{i-1},Y_{i+1},\ldots,Y_n), \qquad \qquad$ \\
\multicolumn{1}{r}{$E'_{P,1}(Y_1,\ldots,Y_{i-1},Y_{i+1},\ldots,Y_n))$} \\
\multicolumn{1}{c}{$\cdots$} \\
$\qquad \alpha_{P,j}: P \rightarrow
Y_i \times F'(Y_1,\ldots,Y_{i-1},Y_{i+1},\ldots,Y_n)$ \\
\multicolumn{1}{c}{$\cdots$} \\
$\qquad \alpha_{P,m}: P \rightarrow {\rm
exp}(G_m(Y_1,\ldots,Y_{i-1},Y_{i+1},\ldots,Y_n),$ \qquad \qquad \\
\multicolumn{1}{r}{$E'_{P,m}(Y_1,\ldots,Y_{i-1},Y_{i+1},\ldots,Y_n))$} \\
end object \\
\end{tabular}
\end{displaymath}
From theorem~\ref{th-right-isomorphism}, we have
\begin{displaymath}
P(Y_1,\ldots,Y_n) \iso Y_i \times F'(Y_1,\ldots) \times
\prod_{k=1\atop k\not= j}^m {\rm exp}(G_k(Y_1,\ldots),E'_{P,k}(Y_1,\ldots)).
\end{displaymath}
$F'(Y_1,\ldots) \times \prod_{k=1\atop k\not= j}^m {\rm
exp}(G_k(Y_1,\ldots),E'_{P,k}(Y_1,\ldots))$ does not depend on $Y_i$.
We have proved the theorem.
\end{theorem}
From this theorem, we can always make the declaration of computable objects
into an equivalent declaration to which we can apply
theorem~\ref{th-right-isomorphism}.  For example, the declaration of the
object for automata in subsection~\ref{ssec-automata} was
\begin{displaymath}
\begin{tabular}{l}
right object ${\rm dyn}'(I,O)$ with ${\rm univ}'$ is \\
$\qquad {\rm next}': {\rm prod}({\rm dyn}',I) \rightarrow {\rm dyn}'$ \\
$\qquad {\rm output}': {\rm dyn}' \rightarrow O$ \\
end object \\
\end{tabular}
\end{displaymath}
to which we cannot apply theorem~\ref{th-right-isomorphism}, but the
above declaration is equivalent to the following one.
\begin{displaymath}
\begin{tabular}{l}
right object ${\rm dyn}'(I,O)$ with ${\rm univ}'$ is \\
$\qquad {\rm next}': {\rm dyn}' \rightarrow {\rm exp}(I,{\rm dyn}')$ \\
$\qquad {\rm output}': {\rm dyn}' \rightarrow O$ \\
end object \\
\end{tabular}
\end{displaymath}
Then, from theorem~\ref{th-right-isomorphism} we can see ${\rm
dyn}'(I,O)$ as the maximal fixed point of the following domain equation.
\begin{displaymath}
{\rm dyn}'(I,O) \iso {\rm exp}(I,{\rm dyn}'(I,O)) \times O
\end{displaymath}

\section{Reduction Rules for Full Evaluation}
\label{sec-reduct-rule-full}

In section~\ref{sec-reduct-rule} we presented a set of reduction rules
which can reduce any element to a canonical element.  However, the
notion of canonical element (definition~\ref{def-cpl-canonical}) was
quite weak (or sloppy), and the canonical elements we get out of
reductions sometimes not acceptable as `canonical'.  We can define a
more refined notion of canonical elements.
\begin{definition}
\label{def-cpl-unconditioned-canonical}
A canonical element is called {\it uncondition}, if it is generated by
the following rule.
\begin{displaymath}
p \coloneq {\bf I} \mid \alpha_{L,j}\circ p \mid
\psi_R(e_1,\ldots,e_m)\circ p \mid \psi_C(\ldots,e_j,\ldots,p_k,\ldots)
\end{displaymath}
where $R$ is not a unconditioned right object, $C$ is a unconditioned object
\begin{displaymath}
\begin{tabular}{l}
right object $C(X_1,\ldots,X_n)$ with $\psi_C$ is \\
\multicolumn{1}{c}{$\cdots$} \\
$\qquad \alpha_{C,j}: E_{C,j}(C,X_1,\ldots,X_n) \rightarrow
E'_{C,j}(X_1,\ldots,X_n)$ \\
\multicolumn{1}{c}{$\cdots$} \\
$\qquad \alpha_{C,k}: C \rightarrow E'_{C,k}(X_1,\ldots,X_n)$ \\
\multicolumn{1}{c}{$\cdots$} \\
end object \\
\end{tabular}
\end{displaymath}
and if $E_{C,k}(C,X_1,\ldots,X_n)$ is simply $C$ then the {\it k\/}th
argument of $\psi_C$ needs to be a unconditioned canonical element.
\end{definition}
For example,
\begin{displaymath}
{\rm pair}({\rm succ},{\bf I})\circ {\rm zero} \qquad \mbox{and} \qquad
{\rm pair}({\rm pi1}\circ{\rm pair}({\rm succ}\circ {\rm zero},{\rm
nil}),{\rm zero})
\end{displaymath}
are canonical elements but not unconditioned.  Their equivalent
unconditioned canonical element is `${\rm pair}({\rm succ}\circ{\rm
zero},{\rm zero})$'.

We can define reduction rules which only produce unconditioned canonical
elements as result.
\begin{definition}
\label{def-reduct-rules-2}
The form of reduction rules is
\begin{displaymath}
\pair{e,p} \dimplies p'
\end{displaymath}
where $e$ is a CSL expression and $p$ is a unconditioned canonical
element whose domain is compatible with the domain of $e$.
\begin{enumerate}
\item FULL-IDENT
\begin{displaymath}
\pair{{\bf I},p} \dimplies p
\end{displaymath}
\item FULL-COMP
\begin{displaymath}
\logicrule{\pair{e_2,p} \dimplies p'' \qquad \pair{e_1,p''} \dimplies
p'}{\pair{e_1\circ e_2,p} \dimplies p'}
\end{displaymath}
\item FULL-L-NAT
\begin{displaymath}
\pair{\alpha_{L,j},p} \dimplies \alpha_{L,j}\circ p
\end{displaymath}
\item FULL-R-FACT
\begin{displaymath}
\pair{\psi_R(e_1,\ldots,e_m),p} \dimplies \psi_R(e_1,\ldots,e_m)\circ p
\end{displaymath}
where $R$ is not a unconditioned right object.
\item FULL-C-FACT
\begin{displaymath}
\logicrule{\pair{e_j,p} \dimplies e'_j \quad \mbox{or} \quad e'_j \equiv
e_j\circ E_{C,j}[p/C]}{\pair{\psi_C(e_1,\ldots,e_m),p} \dimplies
\psi_C(e'_1,\ldots,e'_m)}
\end{displaymath}
where $C$ is a unconditioned right object and $e'_j$ is either the result of
evaluating $\pair{e_j,p}$ or $e_j\circ E_{C,j}[p/C]$ depending of
whether $E_{C,j}$ is simply $C$ or not.
\item FULL-L-FACT
\begin{displaymath}
\logicrule{\pair{e_j\circ E_{L,j}[\psi_L(e_1,\ldots,e_m)/L],p}
\dimplies p'}{\pair{\psi_L(e_1,\ldots,e_m),\alpha_{L,j}\circ p} \dimplies
p'}
\end{displaymath}
\item FULL-R-NAT
\begin{displaymath}
\logicrule{\pair{E'_{R,j}[\psi_R(e_1,\ldots,e_m)/R]\circ
e_j,\psi_P(\ldots,p,\ldots)} \dimplies
p'}{\pair{\alpha_{R,j},\psi_P(\ldots,\psi_R(e_1,\ldots,e_m)\circ
p,\ldots)} \dimplies p'}
\end{displaymath}
In writing down this rule, $\psi_P(\ldots,\psi_R(e_1,\ldots,e_m)\circ
p,\ldots)$ is rather inaccurate.  It means picking up
$\psi_R(e_1,\ldots,e_m)$ according to the occurrence of $R$ in $E_{R,j}$.
$\psi_P$'s are nested as productive objects $P$'s are in $E_{R,j}$.  For
example, the rule for `pi1' of object `prod' is
\begin{displaymath}
\logicrule{\pair{p_1,{\bf I}} \dimplies p'}{\pair{{\rm pi1},{\rm
pair}(p_1,p_2)} \dimplies p'}.
\end{displaymath}
$E_{{\rm prod},1}$ is simply `prod', so there is no $\psi_C$'s.  The rule
for `eval' of object `exp' is
\begin{displaymath}
\logicrule{\pair{e,{\rm pair}({\bf I},p)} \dimplies p'}{\pair{{\rm
eval},{\rm pair}({\rm curry}(e),p)} \dimplies p'}.
\end{displaymath}
Remember that $E_{{\rm exp},1}$ is `${\rm prod}({\rm exp},X)$'.
\end{enumerate}
Let us call the new system FULL and the previous system defined in
definition~\ref{def-reduct-rules} LAZY.
\end{definition}
As we have proved theorem~\ref{th-reduct-sound}, we can easily show that
FULL system is well-defined.  In addition, we can show that
the reduction in FULL system is stronger than that in LAZY system, that
is,
\begin{proposition}
If $\pair{e,p} \dimplies p'$ in FULL system, then $\pair{e,p} \implies
c'$ in LAZY system.
\end{proposition}

On the other hand, since a FULL reduction is nothing but the repeated
application of LAZY reductions, we have the normalization theorem.
\begin{theorem}
For a unconditioned canonical element $p$ and a CSL expression $e$ whose
domain is compatible with the codomain of $p$, there is a unconditioned
canonical element $p'$ such that
\begin{displaymath}
\pair{e,p} \dimplies p'
\end{displaymath}
in FULL reduction system.
\end{theorem}

\chapter{Application of Categorical Data Types}
\label{ch-application}

In this chapter we see some applications of CDT and CPL.  We have
concentrated on category theory in the previous chapters and it is
sometimes hard to relate our results to others if they are not familiar
with category theory.  The author is not claiming that it is better to
use category theory in practice.  Category theory is used as a guiding
principle to see things without being obscured by inessentials.  Therefore,
once one establishes some results using category theory, it is very
interesting to see what it means in other terms and we might get some
deep insight.

In section~\ref{sec-imp-cpl}, we will see an implementation of CPL.  In
section~\ref{sec-lambda-calculus}, we will examine the connection
between CDT and typed lambda calculi and in section~\ref{sec-ml-cpl} we
will propose a new ML which is obtained by combining the current ML and
CPL.

\section[An implementation of Categorical Programming Language]{An
implementation of\\ Categorical Programming Language}
\label{sec-imp-cpl}

In chapter~\ref{ch-cpl}, we introduced a programming language CPL and
its computation rules.  A CPL system has been implemented using {\it
Franz Lisp}.  In the section, we will demonstrate the system and see
some examples of reductions which it can manage.

When the system is started, it prints the following message and waits
for user commands.
\begin{computer}
Categorical Programming Language (version 3)
cpl>
\end{computer}
First, we have to declare some objects because the system does not know
any objects when it is started.  The very first object we declare is the
terminal object.  We use {\tt edit} command to enter its declaration.
\begin{computer}
cpl>/edit/
| /right object 1 with !/
| /end object;/
right object 1 defined
cpl>
\end{computer}
Note that user inputs are in {\it italic} font.  We define products,
exponentials and natural number object as well.  The declarations are
exactly the same as we presented in chapter~\ref{ch-cdt} (except that to
make output shorter we use `{\tt s}' for successor and `{\tt 0}' for
zero).
\begin{computer}
cpl>/edit/
| /right object prod(a,b) with pair is/
| /  pi1: prod -> a/
| /  pi2: prod -> b/
| /end object;/
right object prod(+,+) defined
cpl>/edit/
| /right object exp(a,b) with curry is/
| /  eval: prod(exp,a) -> b/
| /end object;/
right object exp(-,+) defined
cpl>/edit/
| /left object nat with pr is/
| /  0: 1 -> nat/
| /  s: nat -> nat/
| /end object;/
left object nat defined
cpl>/edit/
| /left object coprod(a,b) with case is/
| /  in1: a -> coprod/
| /  in2: b -> coprod/
| /end object;/
left object coprod(+,+) defined
cpl>
\end{computer}
Each time we declare an object the system remembers its factorizer and
natural transformations as well as the functor associated with.  In the
above transaction, `{\tt prod(+,+)}' indicates that system recongnized
`{\tt prod}' as a covariant functor of two arguments whereas `{\tt
exp(-,+)}' indicates that `{\tt exp}' is a functor which is
contravariant in the first argument and covariant in the second.  The
variance is calculated as we formulated in section~\ref{sec-decl-CDT}.
The system can type CSL expressions using the rules in
section~\ref{sec-fun-calc-2}.  For example, we can ask the type of `{\tt
pair(pi2,eval)}'.
\begin{computer}
cpl>/show pair(pi2,eval)/
pair(pi2,ev)
    : prod(exp(*b,*a),*b) -> prod(*b,*a)
cpl>
\end{computer}
where `{\tt *a}' and `{\tt *b}' are variables for objects, or we can see
them as a kind of type variables in ML; `{\tt pair(pi2,eval)}' is a
polymorphic function in this sense.

As we have done in section~\ref{sec-reduct-example}, we can ask to the
system to calculate `1+1' using `{\tt simp}' command.
\begin{computer}
cpl>/simp eval.pair(pr(curry(pi2),curry(s.eval)).pi1,pi2).pair(s.0,s.0)/
s.s.0
    :1 -> nat
cpl>
\end{computer}
Note that the composition `$\circ$' is typed as `{\tt .}'.  The system
applied reduction rules to get the following reduction:
\begin{displaymath}
\pair{{\tt eval.pair(\ldots).pair(s.0,s.0)},{\bf I}} \implies {\tt s.s.0}.
\end{displaymath}
We can see how the system deduced the reduction by enabling the trace mode.
\begin{computer}
cpl>/set trace on/
cpl>/simp eval.pair(pr(curry(pi2),curry(s.eval)).pi1,pi2).pair(s.0,s.0)/
0:eval.pair(pr(curry(pi2),curry(s.eval)).pi1,pi2).pair(s.0,s.0)*
1:eval.pair(pr(curry(pi2),curry(s.eval)).pi1,pi2)*pair(s.0,s.0)
2:eval*pair(pr(curry(pi2),curry(s.eval)).pi1,pi2).pair(s.0,s.0)
3[1]:pr(curry(pi2),curry(s.eval)).pi1*pair(s.0,s.0)
4[1]:pr(curry(pi2),curry(s.eval)).s.0*id
5[1]:pr(curry(pi2),curry(s.eval)).s*0
6[1]:pr(curry(pi2),curry(s.eval))*s.0
7[1]:curry(s.eval).pr(curry(pi2),curry(s.eval))*0
8[1]:curry(s.eval).curry(pi2).!*
9[1]:curry(s.eval).curry(pi2)*!
10[1]:curry(s.eval)*curry(pi2).!
11[1]:*curry(s.eval).curry(pi2).!
12:s.eval*pair(curry(pi2).!,pi2.pair(s.0,s.0))
13[1]:curry(pi2).!*
14[1]:curry(pi2)*!
15[1]:*curry(pi2).!
16:s.pi2*pair(!,pi2.pair(s.0,s.0))
17:s.pi2.pair(s.0,s.0)*id
18:s.pi2*pair(s.0,s.0)
19:s.s.0*id
20:s.s*0
21:s*s.0
22:*s.s.0
s.s.0
    :1 -> nat
cpl>
\end{computer}
Each line has the following form:
\begin{sdisplaymath}
\fbox{step number}\;{\tt [}\;\fbox{depth of computation}\;{\tt
]:}\;\fbox{expression}\; {\tt *}\; \fbox{canonical element}
\end{sdisplaymath}
It indicates the following reduction:
\begin{sdisplaymath}
\pair{\;\fbox{expression}\;,\;\fbox{canonical element}\;} \implies {\ldots}
\end{sdisplaymath}
Step 0 denotes the reduction of
\begin{sdisplaymath}
\pair{{\tt
eval.pair(pr(curry(pi2),curry(s.eval)).pi1,pi2).pair(s.0,s.0)},{\bf I}}
\implies {\ldots}.
\eqno(+)
\end{sdisplaymath}
Step 1 is obtained from R-FACT rule (and R-COMP); the reduction
$(+)$ is the same as the reduction of
\begin{sdisplaymath}
\pair{{\tt
eval.pair(pr(curry(pi2),curry(s.eval)).pi1,pi2)},{\tt pair(s.0,s.0)}}
\implies {\ldots}.
\end{sdisplaymath}
Again from R-FACT, this reduction is the same as
\begin{sdisplaymath}
\pair{{\tt
eval},{\tt pair(pr(curry(pi2),curry(s.eval)).pi1,pi2).pair(s.0,s.0)}}
\implies {\ldots}.
\eqno(++)
\end{sdisplaymath}
which is step 2.  From R-NAT, we have to calculate
\begin{sdisplaymath}
\displaylines{
\qquad \langle {\tt
pair(pr(curry(pi2),curry(s.eval)).pi1,pi2).pair(s.0,s.0)}, \hfill \cr
\hfill {\tt prod(exp,a)}, {\tt exp} \rangle \leadsto \pair{{\tt
curry(\ldots)},\ldots}. \qquad \cr}
\end{sdisplaymath}
In order to do this, from R-NAT-F we have to calculate
\begin{sdisplaymath}
\pair{{\tt pr(curry(pi2),curry(s.eval)).pi1},{\tt pair(s.0,s.0)}}
\implies {\ldots}
\eqno(\dagger)
\end{sdisplaymath}
This reduction is carried out from step 3 to step 12 and we get
\begin{sdisplaymath}
\pair{{\tt pr(curry(pi2),curry(s.eval)).pi1},{\tt pair(s.0,s.0)}}
\implies {\tt curry(s.eval).curry(pi2).!}.
\end{sdisplaymath}
Note that from step 3 to 4 it did the reduction
\begin{sdisplaymath}
\pair{{\tt pi1},{\tt pair(s.0,s.0)}} \implies {\tt s.0},
\end{sdisplaymath}
and from step 6 to 7 it used L-FACT and did the reduction
\begin{sdisplaymath}
\logicrule{\pair{{\tt curry(s.eval).pr(curry(pi1),curry(s.eval))},{\tt
0}} \implies {\ldots}}{
\pair{{\tt pr(curry(pi2),curry(s.eval))},{\tt s.0}} \implies {\ldots}}.
\end{sdisplaymath}
Therefore, $(\dagger)$ is
\begin{sdisplaymath}
\pair{{\tt pr(curry(pi2),curry(s.eval)).pi1},{\tt pair(s.0,s.0)}}
\implies \pair{{\tt curry(s.eval)},{\tt curry(pi2).!}},
\end{sdisplaymath}
and from R-NAT the reduction of $(++)$ is the same as
\begin{sdisplaymath}
\pair{{\tt s.eval},{\tt pair(curry(pi2).!,pi2.pair(s.0,s.0))}} \implies
{\ldots}.
\end{sdisplaymath}
The rest of the steps are done similarly.

It is inconvenient to write down the definition of the addition
every time we want to add something.  Therefore, the system has the
facility to give names to morphisms.  For example, we can name the
addition function `{\tt add}' and use it as follows:
\begin{computer}
cpl>/let add=eval.pair(pr(curry(pi2),curry(s.eval)).pi1,pi2)/
add : prod(nat,nat) -> nat defined
cpl>/simp add.pair(s.0,s.s.0)/
s.s.s.0
    :1 -> nat
\end{computer}
We can define the multiplication and factorial functions.
\begin{computer}
cpl>/let mult=eval.prod(pr(curry(0.!),curry(add.pair(eval,pi2))),id)/
mult : prod(nat,nat) -> nat defined
cpl>/let fact=pi1.pr(pair(s.0,0),pair(mult.pair(s.pi2,pi1),s.pi2))/
fact : nat -> nat defined
cpl>/simp mult.pair(s.s.0,s.s.s.0)/
s.s.s.s.s.s.0
    :1 -> nat
cpl>/simp fact.s.s.s.s.0/
s.s.s.s.s.s.s.s.s.s.s.s.s.s.s.s.s.s.s.s.s.s.s.s.0.!
    :1 -> nat
cpl>
\end{computer}

Let us next define the object for lists.
\begin{computer}
cpl>/edit/
| /left object list(p) with prl is/
| /  nil:1->list/
| /  cons:prod(p,list)->list/
| /end object;/
left object list(+) defined
cpl>/edit/
| /let append=eval.prod(prl(curry(pi2),/
| /                         curry(cons.pair(pi1.pi1,eval.pair(pi2.pi1,pi2)))),/
| /                id);/
append : prod(list(*a),list(*a)) -> list(*a) defined
cpl>/let reverse=prl(nil,append.pair(pi2,cons.pair(pi1,nil.!)))/
reverse : list(*a) -> list(*a) defined
cpl>/let hd=prl(in2,in1.pi1)/
hd : list(*a) -> coprod(*a,1) defined
cpl>/let hdp=case(hd,in2)/
hdp : coprod(list(*a),1) -> coprod(*a,1) defined
cpl>/let tl=case(in1.pi2,in2).prl(in2,in1.pair(pi1,case(cons,nil).pi2))/
tl : list(*a) -> coprod(list(*a),1) defined
cpl>/let tlp=case(tl,in2)/
tlp : coprod(list(*a),1) -> coprod(list(*a),1) defined
cpl>/let seq=pi2.pr(pair(0,nil),pair(s.pi1,cons))/
seq : nat -> list(nat) defined
cpl>
\end{computer}
The morphism `{\tt seq}' returns a list of length $n$ for a given
natural number $n$ such that the list consists of the descending
sequence of natural numbers, $n-1, n-2, \ldots, 2, 1, 0$.  We can try it
in the system.
\begin{computer}
cpl>/simp seq.s.s.s.0/
cons.pair(s.pi1,cons).pair(s.pi1,cons).pair(0,nil).!
    :1 -> list(nat)
cpl>
\end{computer}
The result dose not look like the sequence of 2, 1 and 0, but this is
because our definition of canonical element
(definition~\ref{def-cpl-canonical}) is weak.  We can ask the system to
reduce an element to unconditioned canonical elements (see
definition~\ref{def-cpl-unconditioned-canonical}) using reduction rules
listed in definition~\ref{def-reduct-rules-2}.
\begin{computer}
cpl>/simp full seq.s.s.s.0/
cons.pair(s.s.0.!,cons.pair(s.0.!,cons.pair(0.!,nil.!)))
    :1 -> list(nat)
cpl>
\end{computer}
Now, it looks more like the sequence of 2, 1, and 0.  We may continue
to do some more reductions about lists.
\begin{computer}
cpl>/simp hd.seq.s.s.s.0/
in1.s.s.0.!
    :1 -> coprod(nat,1)
cpl>/simp hd.nil/
in2.!
    :1 -> coprod(*a,1)
cpl>/simp hdp.tl.seq.s.s.s.0/
in1.s.0.!
    :1 -> coprod(nat,1)
cpl>/simp full append.pair(seq.s.s.0,seq.s.s.s.0)/
cons.pair(s.0.!,cons.pair(0.!,cons.pair(s.s.0.!,cons.pair(s.0.!,cons.
    pair(0.!,nil.!)))))
    :1 -> list(nat)
cpl>/simp full reverse.it/
cons.pair(0.!,cons.pair(s.0.!,cons.pair(s.s.0.!,cons.pair(0.!,cons.
    pair(s.0.!,nil.!)))))
    :1 -> list(nat)
cpl>
\end{computer}
where `{\tt it}' denotes the result of the immediately-preceding
reduction.

Let us next experiment with infinite lists.
\begin{computer}
cpl>/edit/
| /right object inflist(a) with fold is/
| /  head: inflist -> a/
| /  tail: inflist -> inflist/
| /end object;/
right object inflist(+) defined
cpl>/let incseq=fold(id,s).0/
incseq : 1 -> inflist(nat) defined
cpl>/simp head.incseq/
0
    :1 -> nat
cpl>/simp head.tail.tail.tail.incseq/
s.s.s.0
    :1 -> nat
cpl>/let alt=fold(head.pi1,pair(pi2,tail.pi1))/
alt : prod(inflist(*a),inflist(*a)) -> inflist(*a)
cpl>/let infseq=fold(id,id).0/
infseq : 1 -> inflist(nat)
cpl>/simp head.tail.tail.alt.pair(incseq,infseq)/
s.0
    :1 -> nat
cpl>
\end{computer}
where `{\tt incseq}' is the infinite increasing sequence 0, 1, 2, 3, 4,
\ldots, and `{\tt infseq}' is the infinite sequence of 0s.  We can merge
two infinite lists by `{\tt alt}' which picks up elements alternatively
from the two infinite lists.

\section{Typed Lambda Calculus}
\label{sec-lambda-calculus}

In this section, we will investigate connection between CPL and typed
lambda calculi.  Lambda calculi were invented to mathematically
formalize the notion of computation.  Typed lambda calculi (first order)
are an important part of lambda calculi and are studied in various ways.
Usually a typed lambda calculus starts with a fixed number of
ground types and allows only $\rightarrow$ as type constructors.  For
example, \cite{stenlund} treats natural numbers and ordinals, and
\cite{troelstra-73} deals with one level higher ordinals.  An
interesting question is ``What kind of types can be added to lambda
calculi?''  Natural numbers, ordinals, lists, \ldots.  We will show in
this section that any data types we can define in CPL can be added into
typed lambda calculi.

We are to define a typed lambda calculus.  As CPL does not have any
ground objects to start with, our lambda calculus does not have any
ground types either.  Instead it has two ways of constructing types, one
corresponding to forming left objects and the other corresponding to
forming right objects.
\begin{definition}
The syntax of our lambda calculus is given as follows.
\begin{enumerate}
\item An enumerable set {\it TVar} of type variables.  $\rho, \nu, \ldots
\in {\rm TVar}$.
\item The set {\it Type} of types is defined by the following rules.
\begin{displaymath}
\begin{array}{c}
\logicrule{\rho \in {\rm TVar} \qquad \rho \in \Gamma}{\Gamma \vdash \rho
\in {\rm Type}}
\qquad \logicrule{\emptyset \vdash \sigma \in {\rm Type} \qquad \Gamma
\vdash \tau \in {\rm Type}}{\Gamma \vdash \sigma \rightarrow \tau \in
{\rm Type}} \\
\strut \\
\logicrule{\Gamma \cup \{\;\rho\;\} \vdash \sigma_1 \in {\rm Type}
\qquad \ldots \qquad \Gamma \cup \{\;\rho\;\} \vdash \sigma_n \in {\rm
Type}}{\Gamma \vdash \minfix{\rho}{\sigma_1,\ldots,\sigma_n} \in {\rm
Type}} \\
\strut \\
\logicrule{\Gamma \cup \{\;\rho\;\} \vdash \sigma_1 \in {\rm Type}
\qquad \ldots \qquad \Gamma \cup \{\;\rho\;\} \vdash \sigma_n \in {\rm
Type}}{\Gamma \vdash \maxfix{\rho}{\sigma_1,\ldots,\sigma_n} \in {\rm
Type}} \\
\end{array}
\end{displaymath}
We use $\sigma, \tau, \ldots$ for the meta-variables of Type.
$\minfix{\rho}{\sigma_1,\ldots,\sigma_n}$ corresponds to left objects
and $\maxfix{\rho}{\sigma_1,\ldots,\sigma_n}$ corresponds to right
objects.
\item An enumerable set {\it Var} of variables.  $x, y, z, \ldots
\in {\rm Var}$.
\item The set {\it Term} of terms and their types are defined by the
following rules.
\begin{displaymath}
\begin{array}{l}
\logicrule{x \in {\rm Var} \qquad x \;:\; \sigma \in \Gamma}{\Gamma
\vdash x \;:\; \sigma} \\
\strut \\
\logicrule{\Gamma \cup \{\; x \;:\; \sigma \} \vdash t \;:\;
\tau}{\Gamma \vdash \lambda x^\sigma.t \;:\; \sigma \rightarrow \tau} \\
\strut \\
\logicrule{\Gamma \vdash t_1 \;:\; \sigma \rightarrow \tau \qquad \Gamma
\vdash t_2 \;:\; \sigma}{\Gamma \vdash t_1 t_2 \;:\; \tau} \\
\strut \\
\Gamma \vdash C_{\minfix{\rho}{\sigma_1,\ldots,\sigma_n},i}
\;:\; \sigma_i[\minfix{\rho}{\sigma_1,\ldots,\sigma_n}/\rho] \rightarrow
\minfix{\rho}{\sigma_1,\ldots,\sigma_n} \\
\strut \\
\begin{array}{r}
\Gamma \vdash J_{\minfix{\rho}{\sigma_1,\ldots,\sigma_n},\tau} \;:\;
(\sigma_1[\tau/\rho] \rightarrow \tau) \rightarrow \ldots \rightarrow
(\sigma_n[\tau/\rho] \rightarrow \tau) \rightarrow {} \qquad \\
\minfix{\rho}{\sigma_1,\ldots,\sigma_n} \rightarrow \tau \\
\end{array}
\\
\strut \\
\Gamma \vdash D_{\maxfix{\rho}{\sigma_1,\ldots,\sigma_n},i}
\;:\; \maxfix{\rho}{\sigma_1,\ldots,\sigma_n} \rightarrow
\sigma_i[\maxfix{\rho}{\sigma_1,\ldots,\sigma_n}/\rho] \\
\strut \\
\begin{array}{r}
\Gamma \vdash P_{\maxfix{\rho}{\sigma_1,\ldots,\sigma_n},\tau} \;:\;
(\tau \rightarrow \sigma_1[\tau/\rho]) \rightarrow \ldots \rightarrow
(\tau \rightarrow \sigma_n[\tau/\rho]) \rightarrow {} \qquad  \\
\tau \rightarrow \maxfix{\rho}{\sigma_1,\ldots,\sigma_n} \\
\end{array}
\\
\end{array}
\end{displaymath}
$C_{\minfix{\rho}{\sigma_1,\ldots,\sigma_n},i}$ is the $i$-th
constructor of $\minfix{\rho}{\sigma_1,\ldots,\sigma_n}$ and
$J_{\minfix{\rho}{\sigma_1,\ldots,\sigma_n},\tau}$ is the generalized
iterator for it.  $D_{\maxfix{\rho}{\sigma_1,\ldots,\sigma_n},i}$ and
$P_{\maxfix{\rho}{\sigma_1,\ldots,\sigma_n},\tau}$ are the dual pairs. \qed
\end{enumerate}
\end{definition}

We have the usual reduction rules $\alpha$ and $\beta$ and two $delta$
rules.  We write $a \reduce b$ for the term $a$ reducing
to the term $b$ by one step reduction and write $a \reducen b$ for $a$
reducing to $b$ by some steps.  The two delta rules are:
\begin{displaymath}
J_{\minfix{\rho}{\sigma_1,\ldots,\sigma_n},\tau} a_1 \ldots a_n
(C_{\minfix{\rho}{\sigma_1,\ldots,\sigma_n},i} b)
\reduce a_i
(\sigma_i[J_{\minfix{\rho}{\sigma_1,\ldots,\sigma_n},\tau} a_1 \ldots
a_n/\rho] b)
\end{displaymath}
and
\begin{displaymath}
D_{\maxfix{\rho}{\sigma_1,\ldots,\sigma_n},i}
(P_{\maxfix{\rho}{\sigma_1,\ldots,\sigma_n},\tau} a_1 \ldots a_n b)
\reduce
\sigma_i[P_{\maxfix{\rho}{\sigma_1,\ldots,\sigma_n},\tau} a_1 \ldots 
a_n/\rho] (a_i b)
\end{displaymath}
where $\sigma[t/\rho]$ is a term of type $\sigma[\tau/\rho] \rightarrow
\sigma[\upsilon/\rho]$ when the type of $t$ is $\tau \rightarrow
\upsilon$ and is defined as follows.
\begin{enumerate}
\item If $\rho$ does not appear in $\sigma$, then $\sigma[t/\rho] \equiv
\lambda x^\sigma.x$.
\item $\rho[t/\rho] \equiv t$.
\item $(\sigma_1 \rightarrow \sigma_2)[t/\rho] \equiv \lambda
x^{\sigma_1 \rightarrow \sigma_2[\tau/\rho]}.\lambda
y^{\sigma_1}.\sigma_2[t/\rho](x y)$.
\item $\minfix{\nu}{\sigma_1,\ldots,\sigma_n}[t/\rho] \equiv
J_{\minfix{\nu}{\sigma_1[\tau/\rho],\ldots},\minfix{\nu}{\sigma_1[\upsilon/\rho],\ldots}}
a_1 \ldots a_n$ \\
where $a_i \equiv \lambda
x^{\sigma_i[\tau/\rho][\minfix{\nu}{\sigma_1[\upsilon/\rho],\ldots}/\nu]}.C_{\minfix{\nu}{\sigma_1[\tau/\rho],\ldots},i}
(\sigma_i[\minfix{\nu}{\sigma_1[\upsilon/\rho],\ldots}/\nu][t/\rho] x)$.
\item $\maxfix{\nu}{\sigma_1,\ldots,\sigma_n}[t/\rho] \equiv
P_{\maxfix{\nu}{\sigma_1[\upsilon/\rho],\ldots},\maxfix{\nu}{\sigma_1[\tau/\rho],\ldots}}
a_1 \ldots a_n$ \\
where $a_i \equiv \lambda x^{\maxfix{\nu}{\sigma_1[\tau/\rho],\ldots}}.
\sigma_i[\maxfix{\nu}{\sigma_1[\tau/\rho],\ldots}/\nu][t/\rho](
D_{\maxfix{\nu}{\sigma_1[\tau/\rho],\ldots},i} x)$.
\end{enumerate}
It looks very complicated but this is the faithful translation of
$\sigma[t/\rho]$ as $\sigma$ being a functor.

Let us see some types we can define in our lambda calculus.
\begin{example}
The empty type can be defined as $\emptyset \equiv \minfix{\rho}{}$, and
one point type can be defined as $1 \equiv \maxfix{\rho}{}$.  We denote
the element of $1$ as $\ast \equiv P_{1,1\rightarrow 1} \lambda x^1.x$.
\end{example}

\begin{example}
The product of two types, $\sigma$ and $\tau$ can be defined as
$\sigma\times \tau \equiv \maxfix{\rho}{\sigma,\tau}$.  We have two
projections.
\begin{displaymath}
\pi_1 \equiv D_{\sigma\times\tau,1} \quad :\; \sigma\times\tau \rightarrow
\sigma \qquad \qquad
\pi_2 \equiv D_{\sigma\times\tau,2} \quad :\; \sigma\times\tau \rightarrow
\tau
\end{displaymath}
If $a$ is a term of type $\sigma$ and $b$ is a term of type $\tau$, we
can define a term $\pair{a,b}$ of type $\sigma\times\tau$.
\begin{displaymath}
\pair{a,b} \equiv P_{\sigma\times\tau}(\lambda x^1.a)(\lambda x^1.b)\ast
\quad :\; \sigma\times\tau
\end{displaymath}
We have the following reduction.
\begin{displaymath}
\pi_1 \pair{a,b} \equiv
D_{\sigma\times\tau,1}(P_{\sigma\times\tau,1}(\lambda x.a)(\lambda
x.b)\ast) \reduce (\lambda x.x)((\lambda x.a)\ast) \reducen a
\end{displaymath}
Similarly, we can show that $\pi_2\pair{a,b} \reducen b$.
\end{example}

\begin{example}
Dually, the coproduct of $\sigma$ and $\tau$ is defined as $\sigma+\tau
\equiv \minfix{\rho}{\sigma,\tau}$.  Two injections are defined as
follows.
\begin{displaymath}
\iota_1 \equiv C_{\sigma+\tau,1} \quad :\; \sigma \rightarrow
\sigma+\tau \qquad \qquad
\iota_2 \equiv C_{\sigma+\tau,2} \quad :\; \tau \rightarrow
\sigma+\tau
\end{displaymath}
$J_{\sigma+\tau,\nu}$ satisfies the following reductions.
\begin{displaymath}
\begin{array}{l}
J_{\sigma+\tau,\nu} a b (\iota_1 c) \equiv J_{\sigma+\tau,\nu} a b
(C_{\sigma+\tau,1} c) \reduce a ((\lambda x.x) c) \reduce a c \\
\strut \\
J_{\sigma+\tau,\nu} a b (\iota_2 c) \reducen b c \rlap{\qquad \qquad \qed}\\
\end{array}
\end{displaymath}
\end{example}

\begin{example}
Let us define the natural numbers in our lambda calculus.  The definition of
type is
\begin{displaymath}
\omega \equiv \minfix{\rho}{1,\rho}.
\end{displaymath}
Zero and the successor function are defined by
\begin{displaymath}
0 \equiv C_{\omega,1}\ast \quad :\; \omega \qquad {\rm s} \equiv
C_{\omega,2} \quad :\; \omega \rightarrow \omega
\end{displaymath}
$J$ gives us almost the ordinary well-known iterator but its type is
\begin{displaymath}
J_{\omega,\sigma} \quad : \; (1 \rightarrow \sigma) \rightarrow (\sigma
\rightarrow \sigma) \rightarrow \omega \rightarrow \sigma.
\end{displaymath}
We can define the ordinary one by this $J_{\omega,\sigma}$ as follows.
\begin{displaymath}
\tilde J_\sigma \equiv \lambda x.\lambda y.\lambda n. J_{\omega,\sigma}
(\lambda z.x) y n \quad : \; \sigma \rightarrow (\sigma \rightarrow \sigma)
\rightarrow \omega \rightarrow \sigma
\end{displaymath}
It satisfies the usual reductions:
\begin{displaymath}
\tilde J_\sigma a b 0 \reducen J_{\omega,\sigma} (\lambda z.a) b (C_{\omega,1}
\ast) \reduce (\lambda z.a)((\lambda x.x)\ast) \reducen a
\end{displaymath}
and
\begin{displaymath}
\tilde J_\sigma a b ({\rm s} n) \reducen J_{\omega,\sigma}(\lambda z.a) b
(C_{\omega,2} n) \reduce b (J_{\omega,\sigma} (\lambda z.a) b n) \approx
b (\tilde J_\sigma a b n)
\end{displaymath}
where $\approx$ is the equivalence relation generated by $\reducen$.  Using
$\tilde J_\sigma$, we can define all the primitive recursive functions.  For
example, the addition function can be define as
\begin{displaymath}
{\rm add} \equiv \lambda n. \lambda m. \tilde J_\omega m {\rm s} n \quad
: \; \omega \rightarrow \omega \rightarrow \omega. \rlap{\qquad \qquad \qed}
\end{displaymath}
\end{example}

\begin{example}
As \cite{stenlund} and \cite{troelstra-73}, we can define the type for
ordinals by $\Omega \equiv \minfix{\rho}{1,\omega \rightarrow \rho}$.
We only check whether our definition of the iterator coincides with the
ordinary one.
\begin{displaymath}
\begin{array}{l}
\Omega \equiv \minfix{\rho}{1,\omega \rightarrow \rho} \\
\strut \\
0_\Omega \equiv C_{\Omega,1}\ast \quad : \; \Omega \\
\strut \\
\sup \equiv C_{\Omega,2} \quad : \; (\omega \rightarrow \Omega)
\rightarrow \omega \\
\strut \\
J_{\Omega,\sigma} \quad : \; (1 \rightarrow \sigma) \rightarrow ((\omega
\rightarrow \sigma) \rightarrow \sigma) \rightarrow \Omega \rightarrow
\sigma \\
\strut \\
J_{\Omega,\sigma} (\lambda x.a) b 0_\Omega \reduce (\lambda
x.a)((\lambda x.x)\ast) \reducen a \\
\strut \\
J_{\Omega,\sigma} (\lambda x.a) b (\sup t) \reduce b ((\omega
\rightarrow \rho)[J_{\Omega,\sigma} (\lambda x.a) b/\rho] t) \\
\qquad \qquad \equiv b ((\lambda y. \lambda z. J_{\Omega,\sigma}
(\lambda x.a) b (y z)) t) \reduce b (\lambda z.
J_{\Omega,\sigma}(\lambda x.a) b (t z)) \rlap{\qquad \qquad \qed}
\end{array}
\end{displaymath}
\end{example}

\begin{example}
Finally, the type for finite lists can be defined by
\begin{displaymath}
L_\sigma \equiv \minfix{\rho}{1,\sigma\times\rho}
\end{displaymath}
with
\begin{displaymath}
\begin{array}{c}
{\rm nil} \equiv C_{L_\sigma,1}\ast \quad : \; L_\sigma \qquad \qquad
{\rm cons} \equiv C_{L_\sigma,2} \quad : \; \sigma\times L_\sigma
\rightarrow L_\sigma \\
\strut \\
J_{L_\sigma,\tau} \quad : \; (1 \rightarrow \tau) \rightarrow (
\sigma\times \tau \rightarrow \tau) \rightarrow L_\sigma \rightarrow
\tau \\
\end{array}
\end{displaymath}
whereas the type for infinite lists can be defined by $I_\sigma \equiv
\maxfix{\rho}{\sigma,\rho}$ with
\begin{displaymath}
\begin{array}{c}
{\rm head} \equiv D_{I_\sigma,1} \quad : \; I_\sigma \rightarrow \sigma
\qquad \qquad 
{\rm tail} \equiv D_{I_\sigma,2} \quad : \; I_\sigma \rightarrow
I_\sigma \\
\strut \\
P_{I_\sigma,\tau} \quad : \; (\tau \rightarrow \sigma) \rightarrow (\tau
\rightarrow \tau) \rightarrow \tau \rightarrow I_\sigma \\
\strut \\
{\rm head} (P_{I_\sigma,\tau}a b c) \reducen a c \qquad \qquad
{\rm tail} (P_{I_\sigma,\tau}a b c) \reducen P_{I_\sigma,\tau}a b (b c)
\rlap{\qquad \qquad \qed} \\
\end{array}
\end{displaymath}
\end{example}

After finishing this section, the author is communicate with
\cite{mendler-86} where recursive types are introduced into
first-order and second-order typed lambda calculi.  He uses least fixed
points and greatest fixed points as we do, but their recursion
combinator $R$ has a different type from ours.
\begin{displaymath}
\logicrule{M : (\rho \arrow \tau) \arrow \sigma \arrow
\tau}{R_{\sigma,\tau}(M[\minfixop\rho.\sigma/\rho]) :
\minfixop\rho.\sigma \arrow \tau}
\end{displaymath}
The author cannot give a clear connection between our iterator and
his.  In addition, he takes fixed points over a single type
expression and, therefore, he needs some basic type constructors like
$1$ and $+$, whereas in our lambda calculus there are no basic type
constructors.

\section[ML and Categorical Programming Language]{ML and Categorical
Programming\\ Language}
\label{sec-ml-cpl}

We might say that ML is based on (first order) typed lambda calculi as we
might say that LISP is based on untyped lambda calculi.  The type
structure of ML depends on the version of ML we are talking about.  If
we are talking about the original ML developed with LCF
\cite{gordon-milner-wordsworth-79}, it had some base types, product,
disjoint sum, integer, etc.\ , and had ability to introduce new
types via recursively defined type equations.  For example, the data
type for binary trees whose leaves are integers were defined as
\begin{bcomputer}
absrectype btree = int + (btree # btree)
    with leaf n = absbtree(inl n)
    and node(t1,t2) = absbtree(inr(t1,t2))
    and isleaf t = isl(repbtree t)
    and leafvalue t = outl(repbtree t)
    and left t = fst(outr(repbtree t))
    and right t = snd(outr(repbtree t));;
\end{bcomputer}
Here, we needed the coproduct type constructor `\verb"+"' as a
primitive.  We could not do without it, whereas `int' can be defined in
terms of others primitives (ML has it as a primitive type just because
of efficiency).

At the next evolution of ML which yielded the current Standard ML
\cite{milner-84,harper-macqueen-milner-86}, we
discovered that the coproduct type constructor is no longer needed as a
primitive.  Standard ML has a `datatype' declaration mechanism by which
the coproduct type constructor can be defined.
\begin{bcomputer}
datatype 'a + 'b = inl of 'a | inr of 'b;
\end{bcomputer}
A datatype declaration lists the constructors of the defining type.  An
element of `\verb"'a + 'b"' can be obtained by either applying
`\verb"inl"' to an element of `\verb"'a"' or applying `\verb"inr"' to an
element of `\verb"'b"'.  We can define the data type for binary trees in
Standard ML as follows.
\begin{bcomputer}
datatype btree = leaf of int | node of btree * btree;
\end{bcomputer}
The symbol `\verb"|"' is just like `\verb"+"', but we shifted from the
object level of the language to the syntax level.  Note that we no
longer need the separate definition of `\verb"leaf"' or `\verb"node"'.
We can define the other functions using {\tt case} statements.
\begin{bcomputer}
exception btree;

fun isleaf t = case t of
                 leaf _ => true
               | node _ => false;

fun leafvalue t = case t of
                    leaf n => n
                  | node _ => raise btree;

fun left t = case t of
               leaf _ => raise btree
             | node(t1,t2) => t1;

fun right t = case t of
                leaf _ => raise btree
              | node(t1,t2) => t2;
\end{bcomputer}

We got rid of the coproduct type constructor from the primitives, but
Standard ML still needs the product type constructor.  From a category
theoretic point of view, we can sense asymmetry in the type structure
of Standard ML.  Let us remember that CPL (or the lambda calculus
defined in section~\ref{sec-lambda-calculus}) needs neither the
coproduct type constructor nor the product type constructor as a primitive.  We
should be able to introduce the symmetry of CPL into ML.  Let us proceed
to the next stage of the ML evolution and define Symmetric ML.
\begin{displaymath}
\begin{tabular}{||r|l|l||}
\hline
& \multicolumn{1}{c|}{Primitives} & \multicolumn{1}{c||}{Declaration
Mechanism} \\
\hline
ML & \verb"->", \verb"unit", \verb"#", \verb"+" & \verb"abstype" \\
\hline
Standard ML & \verb"->", \verb"unit", \verb"*" & \verb"datatype" \\
\hline
Symmetric ML & \verb"->" & \verb"datatype", \verb"codatatype" \\
\hline
CPL & & \verb"left object", \verb"right object" \\
\hline
\multicolumn{1}{c}{\strut} \\
\multicolumn{3}{c}{{\bf ML Evolution}} \\
\end{tabular}
\end{displaymath}
Remember that datatype declarations correspond to left object
declarations.  We list constructors for types.  In order to get rid of
the product type constructor from primitives, we should have a declaration
mechanism which corresponds to the right object declaration mechanism.
Its syntax is
\begin{bcomputer}
codatatype /TypeParam/ /TypeId/ =
      /Id/ is /TypeExp/ & ... & /Id/ is /TypeExp/;
\end{bcomputer}
A codatatype declaration introduces a type by listing its destructors.
The product type constructor can be defined as follows.
\begin{bcomputer}
codatatype 'a * 'b = fst is 'a & snd is 'b;
\end{bcomputer}
where `\verb"fst : 'a * 'b -> 'a"' gives the projection function to the
first component and `\verb"snd : 'a * 'b -> 'b"' gives the projection
function to the second component.  If the declaration is recursive, we
do not take the initial fixed point of the type equation but the final
fixed point.  This is firstly because of symmetry and secondly because
the initial fixed points are often trivial.  Because of this, we can
define infinite objects by codatatype declarations.  For example, the
following declaration gives us the data type for infinite lists.
\begin{bcomputer}
codatatype 'a inflist = head is 'a & tail is 'a inflist;
\end{bcomputer}
If we took the initial fixed point, we would get the empty data type.

Obviously we have destructors for co-data types because we declare them,
but how can we construct data for co-data types?  We had {\tt case}
statements for data types, so we have `merge' statements as dual.  Its
syntax is
\begin{bcomputer}
merge /Destructor/ <= /Exp/ & ... & /Destructor/ <= /Exp/
\end{bcomputer}
For example, the function `\verb"pair"' which makes a pair of given two
elements can be defined as follows.
\begin{bcomputer}
fun pair(x,y) = merge fst <= x & snd <= y;
\end{bcomputer}
As a more complicated example, we might define a function which combines
two infinite lists together.
\begin{bcomputer}
fun comb(l1,l2) = merge head <= head l1
                      & tail <= comb(l2,tail l1);
\end{bcomputer}
It is now clear that, if elements of co-data types are just records and
`{\tt merge}' creates records after evaluating expressions, this `{\tt
comb}' function never terminates because it tries to sweep the entire
infinite lists which cannot be done in finite time.  We need lazyness in
the evaluation mechanism.  An element of `{\tt inflist}' is a record of
two components but each component is a closure whose computation leads to
a value.  A `{\tt merge}' statement creates a record consisting of these
records.   Therefore, the declaration of `{\tt inflist}' is not like
\begin{bcomputer}
datatype 'a inflist = something of 'a * 'a inflist;
\end{bcomputer}
but is closer to
\begin{bcomputer}
datatype 'a inflist = something of (unit -> 'a) *
                                   (unit -> 'a inflist);
\end{bcomputer}
and `{\tt head}', `{\tt tail} and `{\tt comb}' are like
\begin{bcomputer}
fun head(something(x,l)) = x();

fun tail(something(x,l)) = l();

fun comb(l1,l2) = something(fn () => head l1,
                            fn () => comb(l2,tail l1));
\end{bcomputer}
Note that, as we use pattern matching to declare functions over data
types, we can also use it to declare functions over co-data types.  For
example, an alternative definition of `{\tt comb}' may be
\begin{bcomputer}
fun head comb(l1,_) = head l1
  & tail comb(l1,l2) = comb(l2,tail l1);
\end{bcomputer}

\chapter*{Conclusions}
\label{ch-conclusions}
\addcontentsline{toc}{chapter}{Conclusions}

We have looked at a categorical approach to the theory of data types.
The goal of this thesis was to develop CPL (Categorical Programming
Language) which is a programming language in a categorical style and
which has a categorical way of defining data types.

CSL (Categorical Specification Language) was actually developed later
than CDT (Categorical Data Types) and {CPL}.  At first, CDT was given
its semantics without depending on {CSL}.  We could have carried out the
thesis without CSL, but CSL provides the syntactic materials for CDT and
CPL so we would have still needed those parts.  CSL is very much like an
ordinary algebraic specification language, but it is not trivial in two
senses: the treatment of functors and the treatment of natural
transformations.  Functors are very similar to functions but variances
make them special and interesting.  Natural transformations are
essentially polymorphic functions, so if there had been a specification
language for polymorphic functions, we might not have needed to struggle
for developing {CSL}.  It might be interesting to investigate what {\it
polymorphic algebraic specification languages} can be.

CSL is equational.  Much of category theory can be presented
equationally so that CSL is good enough in this sense, but presenting
categorical concepts equationally loses half of the essential meaning.
For example, although the adjoint situation can be explained
equationally, its essence is something more.  This is why, the author
believes, there are so many equivalent forms of defining the adjoint
situation.  Therefore, it is nice to have a specification language which
can naturally express categorical concepts.
Sketches~\cite{barr-wells-85} are more categorical than equations, so it
might be an idea to use sketches in {CSL}.

CDT is the heart of the thesis.  It was developed after the author first
studied category theory and tried to express categorical definitions in
algebraic specification languages.  As we have seen in chapter~\ref{ch-cdt},
algebraic specification languages can express categorical definitions but
not naturally.  CDT succeeded to define some basic categorical
definitions like products, coproducts, exponentials, natural numbers and
so on more naturally, but it cannot define, for example, pullbacks or more
complicated categorical concepts.  One of the suggestions to extend CDT
is to allow equations inside the CDT declarations.  In this way, we may
define pullbacks as follows:
\begin{displaymath}
\begin{tabular}{l}
right object ${\rm pullback}(f:A \rightarrow C,g:B \rightarrow C)$ with
pbpair is \\
$\qquad \pi_1: {\rm pullback} \rightarrow A$ \\
$\qquad \pi_2: {\rm pullback} \rightarrow B$ \\
{} \quad where \\
$\qquad f\circ \pi_1 = g\circ \pi_2$ \\
end object
\end{tabular}
\end{displaymath}
The declaration should be read as follows:
\begin{enumerate}
\item For any morphisms $f: A \rightarrow C$ and $g: B \rightarrow C$,
${\rm pullback}(f,g)$ is an object and it is associated with two
morphisms
\begin{displaymath}
\pi_1: {\rm pullback}(f,g) \rightarrow A \qquad \mbox{and} \qquad \pi_2:
{\rm pullback}(f,g) \rightarrow B
\end{displaymath}
such that $f\circ \pi_1 = g\circ \pi_2$.
\item For any morphisms $h: D \rightarrow A$ and $k: D \rightarrow B$
such that $f \circ h = g \circ k$, there exists a unique morphism ${\rm
pbpair}(h,k): D \rightarrow {\rm pullback}(f,g)$ such that
\begin{displaymath}
\pi_1\circ {\rm pbpair}(h,k) = h \qquad \mbox{and} \qquad
\pi_2\circ {\rm pbpair}(h,k) = k
\end{displaymath}
\end{enumerate}
Note that `pullback' is no longer a simple functor but takes two
morphisms.  We can similarly define pushouts, equalizers, co-equalizers
and so on.  In fact, we can define any finite limit or colimit.  Since
limits and colimits are something to do with diagrams, it seems natural
to introduce the declaration mechanism of diagrams.  For example, we may
have a diagram consisting of three objects and two morphisms as follows:
\begin{displaymath}
\begin{tabular}{l}
diagram el is \\
{} \qquad objects $A,B,C$ \\
{} \qquad morphisms $f: A \rightarrow C, g: B \rightarrow C$ \\
end diagram \\
\end{tabular}
\end{displaymath}
Then, `pullback' can be regarded as taking an `el' diagram as its
parameter, and it is a functor from the category of `el' diagrams.  This
extension is becoming very much similar to the parametrization mechanism
in algebraic specification languages.  Diagrams correspond to so-called
loose specifications, and object declarations correspond to parametrized
specifications (or procedures in CLEAR's terminology) which take a
specification which matches as a parameter and return a new
specification.  It is very interesting to investigate the possibility of
CDT with equations along this line as a first class specification
language.  CDT we presented in this thesis was bounded by the
restriction of computability.  If we introduce equations, it becomes
increasingly difficult to connect them to computing.  If we had
`pullback' in CDT, we would have to prove $f\circ k = g\circ h$ before
using ${\rm pbpair}(k,h)$.  Therefore, the programming would involve some
proving.

CDT and CSL are essentially one sorted systems (here sort = category),
and some would like to extend them to many sorted systems.  We could
have extended them here, but since our main goal in this thesis was to
understand data types, we were interested in only one category, the
category of data types, and so CDT and CSL were single sorted.  Our
approach is very close to that of domain theory which mainly deals only
one category, the category of domains.  On the other hand, algebraic
specification methods deal with many categories.  Each specification is
associated with a category.  However, they are still related in some
sense because they all are algebras over the category of sets (or some
other underlying category).  As we mentioned above, if we extend CDT
with the diagram declarations, we will have to deal with a lot of
different categories of diagrams, and it will be interesting to find out
what $F,G$-dialgebras can give us in this context.

CPL is a functional programming language without variables.  It may look
like FP proposed by John Backus because FP also has no variables.
However, CPL is based on category theory and it has an ability to
declare data types by means of CDT.  CPL does not need any primitives to
start with.  One of the reasons for not having variables is that
category theory is abstract in the sense that objects are simply points
and only their outer behaviour is concerned.  However, we could have
variables for morphisms.  For example, we might want to have
\begin{displaymath}
{\rm twice}(f) \defeq f\circ f
\end{displaymath}
which takes a morphism $f:A \rightarrow A$ and returns a morphism of $A
\rightarrow A$.  The current CPL system cannot handle it and we have to
write it like
\begin{displaymath}
{\rm twice} \defeq {\rm eval}\circ {\rm pair}(\pi_1,{\rm eval})
\end{displaymath}
which is a morphism from ${\rm exp}(A,A)$ to ${\rm exp}(A,A)$.  This
definition is not self-explanatory.  It is evident that we need morphism
variables in CPL for easier use.  Note that ${\rm twice}(f)$ can simply
be a macro because definitions can never be recursive.

We proposed in chapter~\ref{ch-application} to make CPL more like an
ordinary functional programming language.  It has datatype declarations
as well as co-datatype declarations.  It is left for the future to
actually implement the language.  It is interesting to see how to handle
(or represent) lazy data types.

Since CPL is an applicative language and has the possibility of
executing programs in parallel as well as the possibility of partial
evaluation, some kind of special hardware can be invented to execute CPL
programs fast.

The future plan of CDT and CPL would be to extend CDT to cope with
equations and to develop a total programming environment in which users
can define things categorically, reason (or prove) their properties
categorically,  execute some programs categorically.

\chapter*{Declaration}
\thispagestyle{empty}

This thesis has been written by myself, and the work is my own.

\leftline{}
\leftline{}
\leftline{Edinburgh, 1 June 1987}
\vskip 5em
\leftline{\hphantom{Edinburgh, 00 June 0000}\hfil Tatsuya Hagino}

\end{document}